\documentclass[a4paper,12pt]{article}
\usepackage[pdftex]{graphicx}

\usepackage{amsmath}
\usepackage{amsfonts}
\usepackage{color}
\usepackage{array}
\usepackage{stmaryrd}
\newcolumntype{I}{|{\vrule width 3pt}}
\newlength \savewidth
\newcommand \whline{\noalign{\global \savewidth \arrayrulewidth \global \arrayrulewidth 1.4pt}%
\hline
\noalign{\global \arrayrulewidth \savewidth}}
\numberwithin{equation}{section}
\begin{document}

\def\bea{\begin{eqnarray}}
\def\eea{\end{eqnarray}}
\def\nn{\nonumber\\}
\def\rt{\rightarrow}

\def\cby{\colorbox{yellow}}
\def\cbr{\colorbox{red}}
\def\cbc{\colorbox{blue}}
\def\cbg{\colorbox{green}}
\def\tcy{\textcolor{yellow}}
\def\tcr{\textcolor{yred}}
\def\tcb{\textcolor{blue}}

\setlength{\baselineskip}%
{1.2\baselineskip}

\definecolor{db}{rgb}{0,0.08,0.45}
\definecolor{brick}{rgb}{0.6,0.1,0.3}

\definecolor{zz}{cmyk}{0,0,0,1}
\definecolor{yy}{cmyk}{0,0,0,1}
\definecolor{ww}{rgb}{0.6,0.1,0.3}

\definecolor{rr}{cmyk}{0,0,0,1}
\definecolor{vv}{rgb}{0.5,0,0.5}
\definecolor{ss}{cmyk}{0,0,0,1}

\def\zz{\textcolor{zz}}
\def\yy{\textcolor{yy}}
\def\rr{\textcolor{rr}}

\begin{titlepage}

\textbf{{\LARGE TeV Scale Strings and Scattering Amplitudes at the LHC}}

	\bigskip

	\begin{center}
		{\large \textbf{Dean Carmi}}
	\end{center}
	
	{\small Raymond and Beverly Sackler Faculty of Exact Sciences School of Physics and Astronomy Tel-Aviv University, Ramat-Aviv 69978, Israel}

\abstract{We study aspects of TeV string scale models of intersecting D-branes. The gauge bosons arise from strings ending on stacks of D-branes, whereas chiral matter arises from strings stretched between intersecting D-branes. Our focus is on scattering amplitudes (at tree-level), \emph{Regge states} (string excitations), and collider phenomenology.
	
	Achieving a low string scale is possible in models of Large extra dimensions. At the LHC, a low enough string scale implies that cross sections will deviate from their standard model predictions. Moreover, \emph{Regge states} as well as \emph{Kaluza-Klein states} and \emph{winding states} may be produced. 
	
	In a large class of intersecting D-brane models, the quark-gluon amplitudes with at most 2 quarks turn out to be independent of the geometry of the extra dimensions. Therefore these type of amplitudes, which we call \emph{universal amplitudes}, are model independent.  The \emph{universal amplitudes} involve exchanges of Regge states only, whereas amplitudes with more then 2 quarks also involve exchanges of KK and winding states. 
	
	The main computational part of this work is concerned with suggesting methods to calculate the decay widths of the Regge states, and with the formalism for treating amplitudes containing exchanges of higher spin particles.}
	
\end{titlepage}

\clearpage

\tableofcontents


\section{\textcolor{ww}{Content Summary}}

This work is based on the author's MSc thesis.

\textbf{Section \ref{sec:intro}} is an introduction to the framework and to some of the concepts which arise in this work.

\textbf{Section \ref{sec:revamp8}} is a review of methods to calculate tree level amplitudes in field theory and string theory.
Section \ref{sec:a3} reviews tree level field theory quark-gluon amplitude calculations via the color decomposition and helicity techniques. 
Section \ref{subsec:fds4} reviews lowest order amplitude calculations in string theory. Quark-gluon amplitudes in string theory are introduced. The important concept of \emph{universal amplitudes} is discussed. Also discussed is the property of the \emph{equal form factors} of the two classes of universal amplitudes.

\textbf{Section \ref{sec:4point8}} then focuses on the lowest order scattering amplitudes at the LHC: the 4-point amplitudes.
Section \ref{subsec:ftsa} presents the squared amplitudes in field theory (QCD+EW).
Section \ref{subsec:string57} starts with a presentation of the Veneziano amplitude, then the string theory squared amplitudes are presented, and collider phenomenology and LHC constraints discussed.

\textbf{Section \ref{sec:dec}} presents a procedure for calculating the decay widths of the exchanged Regge excitations of the gluon and quarks. 

\textbf{Section \ref{sec:5point57} }is basically a review of the 5-point squared amplitudes. Section \ref{subsec:5pointfield} presents the squared amplitudes in field theory (QCD+EW). Section \ref{subsec:5pointstring} presents the squared amplitudes in string theory.

\textbf{Section \ref{sec:multi-gluon amplitudes}} discusses the generalization to higher point string theory amplitudes.

\textbf{Section \ref{sec:a2}} is basically a review of the 4-point squared amplitudes for direct production of the first excited ($n=1$) Regge states.

\textbf{Appendices:}

Appendix \ref{sec:apphelamp} is important in that it gives the 4-point amplitudes which are used in the calculation of the decay widths of section \ref{sec:dec}.

Appendix \ref{sec:b8} gives the $\mathcal{M}(ggqq)$ amplitude as an example of a full calculation of a 4-point amplitude of the type that appear in Appendix \ref{sec:apphelamp}.

Appendices \ref{sec:appa} and \ref{sec:appcorr} list vertex operators and correlation functions which are used in the calculation of the string amplitudes.

Appendix \ref{sec:appb} introduces the helicity formalism used in section \ref{sec:revamp8} and throughout this work.

Appendix \ref{sec:appc} lists formulas related to the color part of the amplitudes.

Appendix \ref{sec:phenapp} is a short introduction to hadron collider phenomenology.

Appendix \ref{sec:appe} reviews the mathematical functions used in this work.

Appendix \ref{sec:appd} deals with the Wigner $d$-functions which are used often in this work.

Appendix \ref{sec:appf} contains tables of $C$ coefficients which are calculated via approach 1 of section \ref{subsubsec:approach1}.

\bigskip

\underline{\textbf{The parts which are either original or at least partly original.}}

\begin{itemize}
\item In section \ref{subsubsec:a7}, the squared amplitudes from \rr{\cite{lust}} are expanded as a sum of s-channel poles. This exhibits some of the properties of these squared amplitudes, e.g  the vanishing of $n=even$ poles for some of the amplitudes.

\item In section~\ref{subsubsec:softened14} the Breit-Wigner form of the squared amplitudes for exchanges of Regge states with arbitrary $(n,J)$ is given.

\item In section \ref{subsubsec:low89} the low energy limit of the string squared amplitudes is taken: the first stringy correction to the standard model squared amplitudes is obtained. 

\item Section \ref{sec:dec} contains many of the calculations of this work. The procedure for calculating the decay widths of the exchanged Regge states is given in sections \ref{subsec:sett}-\ref{subsec:d'scal}. This is done for arbitrary quantum numbers $n$ and $J$, and generalizes the treatment done in \rr{\cite{anc5}, \cite{dong}} for $n=1, 2$. 
The procedure gives formulas for the decay widths in terms of the unknown $C_{m,m'}^{n,J}$ coefficients. Then section \ref{subsec:calc14} introduces a number of approaches to calculate these $C$'s.

\item In section \ref{subsubsec:6gluesquare} a partial treatment of the squaring of the color factor of the 6-point gluon amplitude is presented.

\item Appendix \ref{sec:apphelamp} gives the 4-point string amplitudes expanded near a pole. This form is used for extracting the properties of the exchanged Regge states, in particular the decay widths.

\item Appendix \ref{sec:appd} (together with section \ref{sec:dec}) deals with the Wigner $d$-functions and their relation to higher spin scattering amplitudes. Also some $d$'s are calculated and used for the calculation of $C_{m,m'}^{n,J}$ coefficients in the tables of Appendix \ref{sec:appf}.

\item Appendix \ref{sec:appf} contains tables of the calculated $C_{m,m'}^{n,J}$ coefficients.
\end{itemize}

\clearpage

\section{\textcolor{ww}{Introduction}}
\label{sec:intro}
The LHC has collected about 5 $fb^{-1}$ of data, and there are first hints of a Higgs boson. So far there aren't any significant deviations from the standard model. Expectations of discovering new physics in the next year(s) are still very high.

The scenario with which this work deals with, while probably not the most likely scenario, is certainly very interesting. If it is correct, some of the the following astonishing phenomena can be reality:

\begin{itemize}
\item String theory at the LHC.

\item Discovery of higher spin particles: Regge states.

\item Quantum gravity at the LHC.

\item Extra dimensions, KK gravitons, Black holes, Hawking radiation..

\end{itemize}

So these scenarios are extremely interesting.\\

We will concentrate on issues related to string theory in the open sector, and less on gravity and extra dimensions.
In particular, we will focus on scattering amplitudes in string theory with regards to their collider phenomenology.

String theory is a high energy completion of the standard model. It is both a quantum gravity theory and a unification theory. At low energies, the spectrum and interactions of string theory must reduce to the standard model. Indeed we will see that the standard model matter and gauge fields can arise as ground states of the open string.
String scattering amplitudes in the low energy limit must equal those of the standard model. 
The string scale, quite generally, is the scale at which stringy phenomena start to appear.
Near this scale, scattering amplitudes begin to deviate from the standard model ones. 

We will see that there are classes of string amplitudes which are model independent (for a large class of intersecting D-brane models.): they are completely independent of the geometry of the extra dimensions. These amplitudes are the $n$-gluon amplitude, and the $n$-gluon plus two quark amplitudes. From now on we shall call these two types: \emph{universal amplitudes}.
Thus by measuring these amplitudes one can discern string theory regardless of compactification and landscape issues.
These universal amplitudes are purely stringy since they contain only exchanges of \emph{Regge states} (string excitations) and not KK states (caused by the presence of extra dimensions) or winding states (strings or branes wound around the extra dimensions.). Quark-gluon amplitudes with four or more quarks are \emph{non-universal} since they are dependent on the compactification and can contain exchanges of KK and winding states.

In open string theory the analogues of tree amplitudes are called \emph{disk amplitudes}. Roughly speaking, when calculating for example the 4 gluon disk amplitude, one obtains the field theory tree result multiplied by a \emph{Veneziano amplitude}.
The Veneziano amplitude $V_t$ is basically a beta function of the Mandelstam variables (see section \textcolor{ss}{\ref{sec:1}}). $V_t$ goes to 1 in the low energy limit, causing the string amplitudes to match the field theory amplitudes at that limit. When the scattering energy approaches the string scale, $V_t$ deviates from 1 and stringy effects become noticeable. The Veneziano amplitude has an infinite number of poles at a constant interval of $E_{CM}^2$. This gives rise to an infinite tower of resonances called \emph{Regge states}. The Regge states are excitations of the string. 
At colliders, these resonances can be discovered directly as peaks in the cross section at equal intervals of the energy squared. The standard model matter and gauge fields occupy the ground state of the string, and each one of them has an infinite tower of Regge excitations.

Regardless of if string theory describes nature, there is no doubt that string amplitude techniques have been extremely fruitful to the understanding and calculation of field theory amplitudes. To name some of the techniques studied over the years: The KLT \cite{kawai} and BCFW \cite{britto, britto2} relations, the works of Bern-Kosower-Dixon, and AdS/CFT techniques \cite{maldacena}.

Usually, the string scale and the quantum gravity scale are assumed to be at around the Planck scale ($\sim 10^{19}\ GeV$). In this case it is very difficult to discern stringy effects at present collider energies.  
In the mid nineteen-nineties, studies on D-branes, Large extra dimensions, and related issues, made it possible to consider string and gravity scales even as low as a TeV in type I or II string theory. This makes it possible to observe the wonderful phenomena discussed before, in the near future.

The types of models to be considered are \emph{intersecting D-brane models} of type II orientifolds with \emph{Large extra dimensions}. These models can realize the standard model gauge group and matter fields.

\subsection{Higher spins, compositeness,  and duality}
\label{subsec:higher}

\emph{References:} ~\textcolor{rr}{\cite{gsw}}.\\

The standard model (plus gravity) contains particles of spin 1/2, spin 1, spin 0, and spin 2. These are the quarks and leptons, gauge bosons, Higgs boson, and graviton, respectively. The last two are yet to be discovered. Hypothesized extensions of the standard model have additional particles. A spin 0 axion is added to solve the strong CP problem.
Supersymmetry has SUSY partners of spin 1/2 and spin 0. Supergravity has a spin 3/2 partner to the graviton (the gravitino). Extra dimensional theories have (in the simplest case) KK gravitons of spin 2 and possibly a spin 0 radion.

The point is, that all these particles have spin $J \leq 2$. No consistent theory is known for a \underline{finite} number of \underline{interacting} particles with spin larger then 2. From Eq. (\textcolor{zz}{\ref{eq:high1}}) we see that tree level amplitudes for spins $J>1$ will diverge at high energies, and this creates problems with unitarity. 

That being said, composite higher spin particles are abundant. This is because composite particles have orbital angular momentum, which is unbounded and also discrete because of quantum mechanics. For example the electron in a hydrogen atom is electromagnetically bound to the proton. At a given energy state $E=-\frac{E_I}{n^2}$, the orbital angular momentum has the possible values $l=0,1,\ldots ,n-1$ \footnote{Likewise, we will see later that the Veneziano amplitude has a similar relation between $l$ and $n$. The difference will be that $E \propto \sqrt{n}$.}. At the center of atoms there is the nucleus, which is a composite system of nucleons bound by the strong force. The nucleon in turn, and hadrons in general, are composite quark systems bound by the strong force. Not only do quarks tend to form composite systems (as electrons and protons do), but they \underline{must}. Quarks (and gluons) are \emph{confined} inside the hadrons, and were never observed as free particles. Quarks (and leptons) are point particles as far as experiments can tell, though there has been theoretical work done on \emph{quark compositeness}.  Another hypothesized theory based on confinement is \emph{technicolor}, in which confinement generates the electroweak scale.

Returning to hadrons, in the 1950's and 1960's a multitude of them, with increasingly higher masses and spins, were discovered at accelerators. It seemed as if more and more particles will be discovered as the energy will increase.  Theorists struggled making sense of the results. Quantum field theory, which was so succesfull at explaining electrodynamics, appeared not very useful for explaining the dynamics of these particles. First, a way of dealing with high spin particles was unknown (largely true till this day. String theory is an exception). Second, Putting by hand a large number of different fields in a lagrangian seems awkward. Third, the particles are strongly interacting and QFT calculations were very difficult\footnote{We now know that the correct explanation is a non-abelian gauge theory of spin 1/2 quarks and spin 1 gluons, called QCD. There are 6 types of quarks, and they are the fields which enter the lagrangian. The hadrons are composed of quarks. In experiments we never see quarks since a strong force bounds them together inside hadrons. The dynamics of hadrons is complicated, just like the dynamics of atoms is complicated.
Hadrons are numerous because of the different possible combinations of quarks, and because of the existence of excited states.}. 
These difficulties stimulated different approaches based on an S-matrix or a scattering ampltude, instead of a lagrangian as the starting point. Many of these models, being of phenomenological nature, tried directly to take into account various properties of the measured scattering amplitudes: correct high/low energy behavior, crossing symmetry, duality, Regge trajectories etc.

In 1968 Veneziano attempted a model of meson scattering. He introduced an ansatz for the amplitude, now called the \emph{Veneziano amplitude}. This model predicted an infinite number of higher spin states: \emph{Regge states}. The string scale was assumed at the $GeV$ scale for the Regge states to be identified with the discovered mesons. A few years later it was realized that the Veneziano amplitude can be derived from a more fundamental theory: a strange type of field theory in which there are 1-dimensional objects (strings) instead of 0-dimensional particles. Meanwhile two things were realized: the Veneziano model was not very successful in explaining hadrons and the strong force, and a gauge theory named QCD emerged as the correct theory. Soon enough though, string theory was revived as a quantum theory of gravity, and the estimate of the string scale naturally jumped 20 orders of magnitude to Planck scale territory. With the string scale being so high, research on phenomenology of Regge states became sparse. 

\begin{figure}
\centering
\includegraphics[width= 90mm]{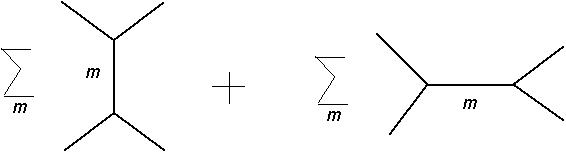}
\caption{Field theory. Sum of s and t-channels.\label{stchan}}
\end{figure}

We now discuss the \emph{DHS duality} \cite{dolen} also known as \emph{worldsheet duality}. Consider a field theory with two types of particles $\phi$ and $\sigma$. If $\sigma$ is a scalar (spin 0), then an interaction may be of the form $\phi^* \phi \sigma$. If $\sigma$ is of spin $J$ type, then it will have $J$ indices and the interaction must be $\phi^* \partial_{\mu_1}\partial_{\mu_2}\ldots \partial_{\mu_J} \phi \cdot \sigma^{\mu_1\ldots\mu_J}$. The lowest order scattering of $\phi$ particles has (inequivalent) $s$ and $t$-channel exchanges, Fig.~\textcolor{yy}{\ref{stchan}}. A scalar $\sigma$ gives the following $t$-channel amplitude
\begin{eqnarray}
A(s,t)\ =\ -\frac{g^2}{t-m^2}
\end{eqnarray}

\begin{figure}
\centering
\includegraphics[width= 160mm]{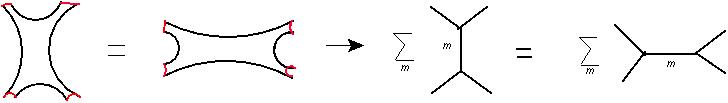}
\caption{String theory. Left: string diagrams and duality. Right: DHS duality between s and t-channel particle exchanges. \label{stringst8}}
\end{figure}

The exchange of a spin $J$ particle gives $2J$ momenta coming from the derivatives in the interaction vertex.
The $t$-channel amplitude is 
\begin{eqnarray}
\label{eq:high1}
A(s,t)\ =\  \frac{g^2(-s)^J}{t-m^2}
\end{eqnarray}

If we have exchange of particles with different masses and spins, then
\begin{eqnarray}
\label{eq:ych1}
A(s,t)\ =\  \sum_J \frac{g_J^2(-s)^J}{t-m_J^2}
\end{eqnarray}

Likewise, the $s$-channel amplitude is 
\begin{eqnarray}
\label{eq:sch1}
A^t(s,t)\ =\  \sum_J \frac{g_J^2(-t)^J}{s-m_J^2}
\end{eqnarray}

\emph{Duality} is the hypothesis that the $s$ and $t$-channel amplitudes are equal (Fig.~\textcolor{yy}{\ref{stringst8}}). This may be done by cleverly choosing the $g_J$'s and $m_j$'s, and was achieved by Veneziano in 1968\footnote{Looking at the Veneziano amplitude Eq.~(\textcolor{zz}{\ref{eq:venred}}) we see that it can be written as $V_u= \sum_{n,J} \frac{g_{J,n}^2(-t)^J}{s-m_n^2}$}. Notice that this is only possible if the sum is infinite since if it was finite, Eq.~(\textcolor{zz}{\ref{eq:sch1}}) would be an analytic function in the complex $t$-plane meaning that there are no $t$-channel poles, and it couldn't possibly equal Eq.~(\textcolor{zz}{\ref{eq:ych1}}).

An infinite sum also has consequences on the high energy behavior of the amplitude. With a finite number of particles, the high energy behavior is determined by the highest spin particle. On the other hand, an infinite sum can make the high energy behavior much better, just like the function $e^{-x}=\sum_{n=0}^\infty (-x)^n/n!$ has a much better high energy behavior then any of it's individual terms. Indeed, we will see that Veneziano amplitude can be written as in Eq.~(\textcolor{zz}{\ref{eq:sch1}}), and that at high energies it is exponentially suppressed. This is at the heart of string theory's ultra soft UV behavior.

\subsection{Large extra dimensions}

\emph{References:} \textcolor{rr}{\cite{lust, lust2, hewett, giudice}}, see also \textcolor{rr}{\cite{arkani, antoniadis}}.\\

The main motivation for the LED scenario is that it offers a solution to the \emph{hierarchy problem}.
The hierarchy problem is the unnaturalness of the large energy gap between the Planck scale ($M_{Planck}=G^{-1/2}\sim 10^{19}\ GeV$) and the electroweak scale ($M_{EW} \sim 10^3\ GeV$).

Extra dimensions are postulated and the D-dimensional quantum gravity scale is assumed to be low near the EW scale ($M_D\sim 1\ TeV$). 
So between these two scales there is no hierarchy.
The effective 4 dimensional gravity scale is determined after compactifying the extra dimensions. The large hierarchy between the 4-d QG scale and the EW scale is explained by assuming that the compactification volume $V_{\delta}$, and hence some of the extra dimensions, are large. In this scenario gravity is allowed to move in the extra dimensions, and this "leakage" of gravity causes it's weakness from the point of view of a 4d observer. 

In the basic scenario, one assumes the existence of $\delta$ extra spatial dimensions so that space-time now has $D=\delta+4$ dimensions. Only gravity may propagate in the extra dimensions while the standard model fields are restricted to regular 3d space. The extra dimensions must be of finite extent (compactified) in order to have been avoided in experiments which test Newton's gravitational force law.

The effective 4d gravity scale $M_{Planck}$ is related to $M_D$ by Gauss's law:
\begin{eqnarray}
\label{eq:a}
M_{Planck}^{2}\ =\ V_\delta \ M_D^{2+\delta}
\end{eqnarray}
We see that $V_\delta$ needs to be large, meaning that some of the extra dimensions ought to be large. 

We now discuss \emph{KK particles}. If there exists extra dimensions of finite extent and gravity is allowed to propagate in them, there will be new particles. This is completely analogous to the text book problem of a particle in a box which will have it's momentum quantized. Similarly, the graviton will have its momentum quantized in the direction of the extra dimensions.
The energy-momentum relation for a massless graviton in a space with one extra dimension is :
\begin{eqnarray} 
E^2\ =\ p_x^2+p_y^2+p_z^2+p_{ED}^2\ =\ p_x^2+p_y^2+p_z^2+\frac{n^2}{R^2}
\end{eqnarray}
Where $R$ is the size of the extra dimension and $n$ is an integer.

This shows that to an observer in 4-d, the quantized graviton momenta effectively appear as an infinite tower of new particles (\emph{KK gravitons}) with the same quantum numbers as the graviton.
These spin 2 resonances\footnote{More generally, every particle which is allowed to move in the extra dimensions will have a corresponding tower of states.} couple gravitationally with equal strength to all the standard model particles, and are evenly spaced in their mass:
\bea
m_{KK}\ =\ \frac{n}{R}
\eea

In the large extra dimension scenario the spacing in mass of the KK states is small (Possibly $10^{-3} \ eV$). 
Regarding collider signals, it turns out that this spacing is smaller then the resolution of the detectors, so that a continuum signal is expected as opposed to resonance peaks. The two most important processes are \emph{virtual exchange} and \emph{direct production} of KK gravitons (These are also the two most important processes for Regge states, as will be discussed).
Table~\textcolor{yy}{\ref{ledkk}} lists a few of these processes at hadron colliders. The signal for virtual exchange will be a smooth deviation from the standard model cross sections. In direct production, (at least) one of the final state particles is a KK graviton which is not detected since it is gravitationally interacting. Thus the signal is missing energy. 

\begin{table}[h]
\centering
\begin{tabular}{|c|c|}
\hline
Direct production   & Virtual exchange\\ 
\whline
 $gg\to gG$ &  $gg \to q\bar{q}$  \\  
\hline
        $q\bar{q}\to gG$ &  $gg \to \gamma \gamma$  \\  
\hline
       $q\bar{q}\to \gamma G$ &  $q\bar{q}\to q'\bar{q}'$ \\ 
\hline
 $qg\to qG$ & $q\bar{q}\to \gamma \gamma$  \\ 
\hline
\end{tabular}
\caption{Examples of direct production and virtual exchange of KK gravitons $G$.\label{ledkk}} 
\end{table}

One thing is very clear: The possibility of having quantum gravity effects at LHC energies is exciting.
 
\subsection{Brane world and the string scale} 

\emph{References:} ~\textcolor{rr}{\cite{gsw, lust, lust3, ibanez3, cic}}.\\

Extra dimensions necessarily arise in  string theory in order to get a consistent and realistic theory. Cancellation of the conformal anomaly requires the spactime dimension to be equal to the critical dimension. In bosonic string theory the critical dimension is $D=26$ ($\delta=22$), and in superstring theory it is $D=10$ ($\delta=6$). 
 
In string theory there is an additional scale called the \emph{string scale} $M_s$, which quite generally is the energy at which stringy phenomena start to appear. The string scale is related to the \emph{Regge slope}, \emph{string length}, and \emph{string tension} through $M_s^2= \frac{1}{\alpha'}=\frac{1}{l_s^2}=2\pi T$. This parameter enters the string action as: 
\begin{eqnarray}
S\ =\ -\frac{T}{2}\int d^2\sigma \sqrt{-h}h^{\alpha \beta}\partial_\alpha X^\mu \partial_\beta X_\mu
\end{eqnarray} 
In the conformal gauge $h_{\alpha \beta}= \eta_{\alpha \beta}e^{\phi}$, and the action becomes that of D free scalar fields: 
\begin{eqnarray}
S\ =\ -\frac{T}{2}\int d^2\sigma \partial_\alpha X^\mu \partial^\alpha X_\mu
\end{eqnarray} 
Where $\eta_{\alpha \beta}$ is the flat metric.

The equation of motion is:
\begin{eqnarray}
\label{eq:eom8}
\frac{\partial^2X^\mu}{\partial \sigma^2}- \frac{\partial^2 X^\mu}{\partial \tau^2}\ =\ 0
\end{eqnarray}

and the constraint equations $\frac{\delta S}{\delta h_{\alpha \beta}}=0$ are: 
\begin{eqnarray}
\label{eq:eom9}
T_{10}\ =\ T_{01}\ =\ \dot{X}\cdot X'\ =\ 0
\nonumber\\[5pt]
T_{00}\ = \ T_{11}\ =\ \frac{1}{2}(\dot{X}^2+X'^2)\ =\ 0
\end{eqnarray}

The energy and angular momentum are:
\begin{eqnarray}
\label{eq:angener}
P^{\mu}\ =\ T\int_0^\infty d\sigma \frac{dX^\mu}{d\tau}
\end{eqnarray}
\begin{eqnarray}
\label{eq:angener2}
J^{\mu \nu}\ =\ T\int_0^\infty d\sigma \Big(X^\mu \frac{dX^\nu}{d\tau}- X^\nu \frac{X^\mu}{d\tau}\Big)
\end{eqnarray}

As an example, we can now use this information to show (classically) that $\alpha'$ is the \emph{Regge slope} for the open string. The Regge slope is defined as the maximum angular momentum per energy squared. This is satisfied by a spinning string with a stationary center of mass. The spinning string is described by:
\begin{eqnarray}
x\ =\ A\cos \tau \cos \sigma\ ,\ \ y\ =\ A\sin \tau \cos \sigma\ ,\ \ t\ =\ A\tau
\end{eqnarray}
Which can easily be shown to satisfy Eqs.~(\textcolor{zz}{\ref{eq:eom8}}), (\textcolor{zz}{\ref{eq:eom9}}). Plugging this solution in Eqs.~(\textcolor{zz}{\ref{eq:angener}}), (\textcolor{zz}{\ref{eq:angener2}}) gives:
\begin{eqnarray}
E\ =\  P^0\ =\  \pi A T
\end{eqnarray}
\begin{eqnarray}
J_z\ =\  \pi A^2T/2
\end{eqnarray}

So that,
\begin{eqnarray}
J_{z}/E^2\ =\ 1/(2\pi T)\ =\  \alpha'
\end{eqnarray}
Note that we will obtain this same result $J_{max}< \alpha'E^2+\text{constant}$, from the poles of the Veneziano amplitude.

We now discuss the brane-world scenario.
Following \rr{\cite{lust}}, we consider type II superstring theory in $D=10$ space-time with the Dp-branes wrapped around p-3-cycles. The remaining 3 dimensions being the regular three dimensional space. The 6 internal (compactified) dimensions are decomposed into $d_{||}=p-3$ directions parallel to the Dp-brane, and $d_\bot=9-p$  directions transverse to the Dp-brane. We denote the transverse and parallel radii as $R_j^{\bot}$ and $R_i^{||}$ respectively.
The total internal volume is:
\begin{eqnarray}
\label{eq:c3}
V_6\ =\  (2\pi)^6 \ \prod_{i=1}^{d_{||}} R_i^{||}\ \prod_{j=1}^{d_{\bot}} R_j^{\bot} 
\end{eqnarray}

The relation between the Planck scale and the string scale is given by\footnote{Compare this with the heterotic string in which $M_s= g_s M_{planck}$. Since there is no dependence on the compactification volume, the string scale cannot be lowered by enlarging the volume. This difference between the type I or II strings and the heterotic string can be traced back to the fact that the type I or II gauge fields arise from open strings whereas in the heterotic theory both the gauge and gravity fields arise from closed strings. 
These relations for $M_{Planck}$ are obtained by considering the $d=4$ low energy effective string action and comparing it to the Einstein-Hilbert action. Similarly, the relation for the gauge coupling constant Eq.~(\textcolor{zz}{\ref{eq:c2}}) is obtained by comparing the gauge part of the effective action to the Yang-Mills action.}:
\begin{eqnarray}
\label{eq:c1}
M_{Planck}^{2}\ =\  \frac{8}{(2\pi)^6}\ g_s^{-2}\ M_{s}^8\ V_6
\end{eqnarray}
Where $g_s= e^{\phi_{10}}$ is the string coupling and $\phi_{10}$ the dilaton field.
We can have $M_s \sim TeV$ and $V_6\sim 10^{32}$, or $M_s \sim M_{Planck}$ and $V_6\sim 1$, or any intermediate case.
We consider the first case in which the string scale is accessible at present colliders.

From Eqs.~(\textcolor{zz}{\ref{eq:a}}) and (\textcolor{zz}{\ref{eq:c1}}):
\begin{eqnarray}
M_D\ =\  \bigg(\frac{8}{(2\pi)^6}\bigg)^{1/8} g_s^{-1/4} M_{s}
\end{eqnarray}
These two scales are tied together and $M_{s}\sim M_D \sim TeV$.

The gauge fields are confined to the Dp-brane and are free to propagate in the $d_{||}$ directions. Since colliders measure the gauge interactions, they constrain $R_i^{||}$ to be smaller than about a TeV. The gauge couplings will depend on the radii of parallel directions according to:

\begin{eqnarray}
\label{eq:c2}
g_{D_p}^{-2}\ =\  \frac{1}{2\pi}\ g_s^{-1}\ M_{s}^{p-3}\ \prod_{i=1}^{d_{||}} R_i^{||}
\end{eqnarray}

Combining Eqs. (\textcolor{zz}{\ref{eq:c1}}), (\textcolor{zz}{\ref{eq:c3}}) and (\textcolor{zz}{\ref{eq:c2}}) gives:
\begin{eqnarray}
g_{D_p}^{2}\ M_{Planck}\ =\  2^{\frac{5}{2}}\pi \ M_{s}^{7-p}\ \bigg(\prod_{j=1}^{d_{\bot}} R_j^{\bot}\bigg) ^{\frac{1}{2}}\ \bigg(\prod_{i=1}^{d_{||}} R_i^{||}\bigg) ^{-\frac{1}{2}}
\end{eqnarray}

For $p<7$ we see that enlarging $R_j^{\bot}$, decreases $M_{s}$ ($M_{Planck}$ is constant).\\
We can give a rough estimate of $R^{\bot}$ as follows:
We assume that all of the $R_j^{\bot}$ are equal, and that all of the $R_j^{||}$ are separately equal. Eqs. (\textcolor{zz}{\ref{eq:c1}}) and (\textcolor{zz}{\ref{eq:c3}}) give:

\begin{eqnarray}
M_{Planck}^{2}\ =\ 8g_s^{-2}M_{s}^8\ \big(R^{||}\big)^{6-d_\bot}\ \big(R^{\bot}\big)^{d_\bot}
\end{eqnarray}
 \\
Further assuming $R^{||}\approx M_{s}^{-1}=1\ TeV^{-1}$ and $g_s= 1/25$, we get the estimates in Table~\textcolor{yy}{\ref{led1}}.

\begin{table}[h]
\centering
\begin{tabular}{|c||c|c|c|c|c|c|}
\hline
$ $ & $d_\bot=1$ & $d_\bot=2$ & $d_\bot=3$ & $d_\bot=4$ & $d_\bot=5$ & $d_\bot=6$  \\  
\whline
$R^{\bot}[GeV^{-1}]$ & $1.6\cdot10^{26}$ & $4\cdot10^{11}$ & $5.4\cdot10^{6}$ & $2\cdot10^{4}$ & $693$ & $74$ \\  
\hline
$R^{\bot}[m]$ & $1.6\cdot10^{11}$ & $4\cdot10^{-4}$ & $5.4\cdot10^{-9}$ & $2\cdot10^{-11}$ & $7\cdot10^{-13}$ & $7\cdot10^{-14}$  \\   
\hline
$E_R[MeV]$ & $7.7\cdot10^{-24}$ & $3\cdot10^{-9}$ & $2\cdot10^{-4}$ & $0.06$ & $1$ & $16$ \\   
\hline 
\end{tabular} 
\caption{$R^{\bot}$ estimations. From~\rr{\cite{lust}}. \label{led1}} 
\end{table}
Experimental constraints on $R^{\bot}$ come from Cavendish experiments testing Newton's inverse square law. These give roughly $R^{\bot}< 1\ [\text{mm}]$, therefore $d_{\bot}=1$ is ruled out. If $d_{\bot}=2$, then this estimate gives an $R_{\bot}$ of the millimeter size.

In the previous section we mentioned KK gravitons, which are excitations of the graviton (closed string state). Their masses will be determined by all the extra dimensions (including the large ones $R_{\bot}$) since they move in the bulk. At low energies though, only KK momenta from the \underline{large} extra dimensions will be excited:
\bea
m_{KK}^{\bot}\ =\ \frac{n^{\bot}}{R^{\bot}}
\eea

The open string ground states (the standard model gauge bosons)\footnote{In intersecting brane models, the fermions are placed at the intersection of the D-branes, thus they move in a smaller number of dimensions compared to the gauge bosons which move on the branes.} will have corresponding KK excitations since they propagate in the small extra dimensions $R_{||}$:
\bea
m_{KK}^{||}\ =\ \frac{n^{||}}{R^{||}}
\eea

In addition to KK states there will also be towers of particles called \emph{winding states}, caused by the wrapping of strings around the extra dimensions:
\bea
m_{wind.}\ =\ WR^{||}M_s^2
\eea
$W$ is an integer called the winding number.

As opposed to Regge states, KK states and winding states are clearly dependent on the geometry of the extra dimensions.
Note that closed string states, such as the transverse KK states, interact at 1-loop level and hence their effects are suppressed.

\subsection{Intersecting D-brane models}

\label{subsec:inter}
\emph{References:} ~\textcolor{rr}{\cite{lust, blumenhagen, blumenhagen2, blumenhagen3, blumenhagen4, ibanez, ibanez2, becker, cve23, anc122}}.\\

String phenomenology is the study of how to embed the standard model into superstring theory.
Intersecting D-brane models make such an attempt in the framework of type I or II superstring theory.
In these models it was shown possible to achieve the following:

\begin{enumerate}
\item Contain the standard model gauge group $SU(3)\times SU(2)\times U(1)$.

\item Have chiral fermions: the quarks and leptons.

\item Family replication: the 3 families.

\item $\mathcal{N}=1$ SUSY  or no SUSY.
\end{enumerate}

\subsubsection{Generalities}

\begin{figure}
\centering
\includegraphics[width= 110mm]{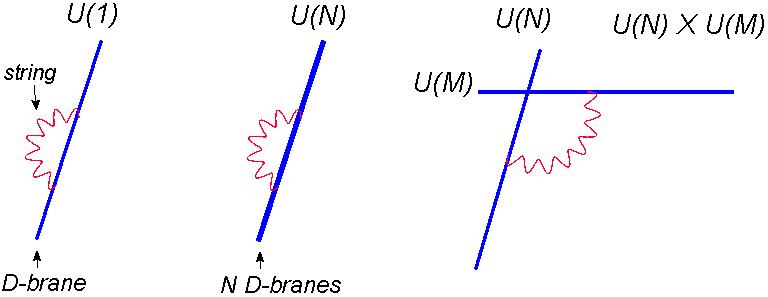}
\caption{Left: $U(1)$ gauge field. Middle: $U(N)$ gauge field. Right: chiral fermion in bi-fundamental of $U(N)\times U(M)$.\label{intersecting3}}
\end{figure}

A Dp-brane is a non-perturbative extended object with $p$ space dimensions. The fluctuations of a D-brane are described by a string theory. Open strings attach to D-branes and their ends satisfy Dirichlet boundary conditions transverse to the D-brane, and Neumann boundary conditions along the D-brane. D-branes couple to the gauge fields of the R-R sector in type II string theory. Type IIA contains stable $p=$even D$p$-branes that couple to the $n=$odd n-forms of the theory. Likewise type IIB has $p=$odd and $n=$even. These so called BPS branes are stable since they carry a conserved charge.
Type I (II) string theory contains $\mathcal{N}=1 (2)$ supersymmetry in 10 dimensions. Introducing D-branes into type I or II string theory breaks some of the supersymmetry. In this way, model building can result in different amounts of supersymmetry.

Closed strings are not attached to D-branes and can propagate in the bulk. Gravity arises from the massless sector of the closed string, so it too can propagate in the bulk.
An open string with both ends attached to a single D-brane gives rise to a $U(1)$ gauge field that is confined to the brane, Fig.~\textcolor{yy}{\ref{intersecting3}}. $N$ copies of this configuration gives of course $U(1)^N$. If the N D-branes are brought close together and stacked on top of each other, the gauge fields will be in the adjoint representation of $U(N)$. It is thus possible to realize the gauge group $SU(N)$ via $U(N)\sim SU(N)\times U(1)$, and necessarily there will be extra $U(1)$'s.

On top of the gauge and gravity interactions, D-branes also make it possible to realize chiral matter. One way to do this is by placing the D-branes on orbifold or conifold singularities. We will consider a different method: D-branes intersecting at angles.
Consider a stack of $N$ D-branes intersecting a stack of $M$ D-branes. An open string  stretched between the two stacks can give rise to a chiral Weyl fermion in the bi-fundamental representation of $U(N)\times U(M)$, see Fig.~\textcolor{yy}{\ref{intersecting3}}. 

A D-brane is described by the DBI action, and it has tension (a positive contribution to the vacum energy\footnote{This is one way to see that D-branes break SUSY. In a supersymmetric theory the vacum energy is zero.}). Negative tension objects called \emph{orientifold planes} are introduced to cancel the D-brane tension.
These models are then called \emph{orientifold models}. A new feature is that now the $SP(2N)$ and $SO(2N)$ gauge groups are possible for the gauge fields, in addition to $U(N)$. $SP(2N)$ and $SO(2N)$ appear if the 3-cycle is invariant under the anti-holomorphic involution $\bar{\sigma}$, whereas $U(N)$ appears if it is not invariant. We consider the second case in this work.
In addition to the bi-fundamental, the symmetric and anti-symmetric representations for the chiral fermions are now also possible. These new representations arise from strings stretched between a D-brane and its image. Since these exotic chiral fermions do not appear in the standard model they are usually unwanted, and indeed they do not appear in the model we consider in the next section.

Family replication is achieved as follows. The number of chiral fermions at an intersection of two branes is determined by the \emph{intersection number}. In a flat non-compact space, the intersection numbers obviously can be only $\pm 1$. But we must consider compact extra dimensions, and this enables multiple intersections between the branes. 
Consider the compact space to be a 6-torus $T^6= T^2\times T^2\times T^2$ and that D6-branes cover a 1-dimensional cycle on each $T^2$. Each $T^2$ is then described by a pair of wrapping numbers $(n^i,m^i)$ along the cycles $[x^i]$ and $[y^i]$. A 3-cycle can then be written as product of three 1-cycles:
\bea
\pi_a\ =\ \prod_{i=1}^{3}(n_a^i[x^i]+m_a^i[y^i])
\eea
Since $[x^i]\circ [y^i]= -1$, the intersection number between branes $a$ and $b$ is: 
\begin{eqnarray}
\label{eq:intnum1}
I_{ab}= \pi_a \circ \pi_b\ =\  \prod_{i=1}^{3}(n_a^im_b^i-m_a^in_b^i)
\end{eqnarray}

The mirror cycles $\pi_a'$ have wrapping numbers $(n^i,-m^i)$, therefore:
\begin{eqnarray}
\label{eq:intnum2}
I_{a'b}\ =\ \pi_a' \circ \pi_b\ =\  \prod_{i=1}^{3}(n_a^im_b^i+m_a^in_b^i)
\end{eqnarray}

The chiral spectrum of many orientifold models can be read from Table~\textcolor{yy}{\ref{dbrane1}}.  
In the next section the intersection numbers and chiral spectrum of a 4-stack model with D6-branes will be shown.

\begin{table}[h]
\centering
\begin{tabular}{|c||c|c|}
\hline
sector & representation & intersection number $I$ \\  
\whline
$a'\ a$ & $A_a$ & $\frac{1}{2}(\pi_a' \circ \pi_a+\pi_{O_6} \circ \pi_a)$  \\  
\hline
$a'\ a$ & $S_a$ & $\frac{1}{2}(\pi_a' \circ \pi_a-\pi_{O_6} \circ \pi_a)$ \\   
\hline
$a\ b$ & $(\bar{N}_a,N_b)$ & $\pi_a \circ \pi_b$ \\   
\hline
$a'\ b$ & $(N_a,N_b)$ & $\pi_a' \circ \pi_b$ \\   
\hline       
\end{tabular}  
\caption{Intersection of 3-cycles. $\pi_{O_6}$ is the 3-cycle of the orientifold plane. The existence of the symmetric and anti-symmetric representations can be seen. From~\rr{\cite{lust}}. \label{dbrane1}}.
\end{table}

\subsubsection{4 stack D-brane models}
We now describe the important class of 4 stack D-brane models. These will be our prototype models.
Consider type II orientifolds with D6-branes wrapping compact homology 3-cycles $\pi_a$ of the internal space. The massless gauge fields live in the subspaces $\mathbf{R}^{1,3}\times \pi_a$. There are also $O6$-planes $\pi_{O6}$, and for each stack of D6-branes there is an orientifold mirror stack wrapped around the reflected cycles $\pi_a'$. The Intersection numbers $I$ fix the chiral spectrum.  

\begin{figure}
\centering
\includegraphics[width= 60mm]{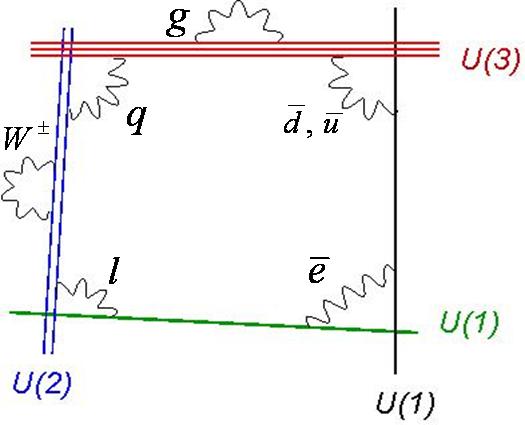}
\caption{4-stack D6-brane model, giving rise to the standard model spectrum.\label{47stacks}}
\end{figure}

Fig.~\textcolor{yy}{\ref{47stacks}} shows the intersection pattern of the four D6-branes.

The gauge group is: 
\begin{eqnarray}
\boxed{
U(3)_a\times U(2)_b\times U(1)_c\times U(1)_d}  
\end{eqnarray}
Which is equivalent to:
\bea
\label{eq:gg9}
SU(3)_a\times SU(2)_b\times U(1)_a\times U(1)_b \times U(1)_c\times U(1)_d
\eea

The $SU(3)$ and $SU(2)$ groups correspond to the strong and weak gauge groups.
The four $U(1)$'s of Eq.~(\zz{\ref{eq:gg9}}) generally mix to form the physical particles. Three of these so called $Z$' bosons will receive masses of the order of the string scale, via the generalized Green-Schwarz mechanism, see e.g \cite{blumenhagen, blumenhagen2}.
The remaining $U(1)$ field stays massless and is identified with the hypercharge. 

In general, the hypercharge can be written as a linear combination of all the $U(1)'s$:
\begin{eqnarray}
\label{eq:hypcomb}
U(1)_Y\ =\  \sum_ic_iU(1)_i
\end{eqnarray}
so that
\bea
Q_Y\ =\ \sum_i c_iQ_i
\eea
\bea
\frac{1}{\alpha_Y}\ =\ \sum_i \frac{N_ic_i^2}{2}\frac{1}{\alpha_i}
\eea

In our model we assume the so called \emph{Madrid hypercharge embedding}:
\begin{eqnarray}
Q_Y\ =\ \frac{1}{6}Q_a+\frac{1}{2}Q_c+\frac{1}{2}Q_d 
\end{eqnarray}

We see that the abelian gauge boson from the color stack $U(1)_a$ mixes with the hypercharge, and hence with the photon and $Z$ boson. This can be viewed as mixing of the gluon with photon and $Z$.
As we will see this gives rise to tree level amplitudes which are forbidden in the standard model (e.g. scattering of gluons into photons).

We now turn to the chiral spectrum of this model. In \rr{\cite{ibanez2}} a general solution for the wrapping numbers $(n_a^i,m_a^i)$ which give the standard model spectrum was found. One example for such a solution is given in  Table~\textcolor{yy}{\ref{wrappingnum1}}. From Eqs.~(\zz{\ref{eq:intnum1}}) and (\zz{\ref{eq:intnum2}}) These wrapping numbers give rise to the following intersection numbers:
\bea
I_{ab}\ =\ 1\ \ \ \ ,\ \ \ \ I_{ab'}\ =\ 2
\nn[5pt]
I_{ac}\ =\ -3\ \  ,\ \ I_{ac'}\ =\  -3
\nn[5pt]
I_{bd}\ =\  0\ \ \ ,\ \ \ I_{bd'}\ =\  -3
\nn[5pt]
I_{cd}\ =\  -3\ \ \ ,\ \ \ I_{dc'}\ =\  3
\eea
and these intersection numbers give rise to the standard model spectrum (plus extra $U(1)$'s) as shown in Table~\textcolor{yy}{\ref{intersecnum1}}. The 3 quark families comes from $I_{ab}+I_{ab'}=3$, and so on..

\begin{table}[h]
\centering
\begin{tabular}{|c||c|c|c|}
\hline
$N$ & $(n^1,\ m^1)$ & $(n^2,\ m^2)$ & $(n^3,\ m^3)$ \\[6pt]  
\whline
$N_a= 3$ & $(1\ ,\ 0)$  & $(2\ ,\ 1)$ & $(1\ ,\ \frac{1}{2})$ \\[6pt]  
\hline
$N_b= 2$ & $(0\ ,\ -1)$ & $(1\ ,\ 0)$ & $(1\ ,\ \frac{3}{2})$ \\[6pt]   
\hline
$N_c= 1$ & $(1\ ,\ 3)$ & $(1\ ,\ 0)$ & $(0\ ,\ 1)$ \\[6pt]   
\hline
$N_d= 1$ & $(1\ ,\ 0)$ & $(0\ ,\ -1)$ & $(1\ ,\ \frac{3}{2})$ \\[6pt]   
\hline       
\end{tabular}  
\caption{Wrapping numbers. From \rr{\cite{ibanez2}} \label{wrappingnum1} }
\end{table}

\begin{table}[h]
\centering
\begin{tabular}{|c||c|c|c|c|c|c|c|c|c|c|}
\hline
Intersection & Matter fields & {} & $Q_a$ & $Q_b$ & $Q_c$& $Q_d$ & $Y$\\[3pt]
\whline
$(a,b)$  & $Q_L$ & $(3,2)$    &    $1$ &    $-1$ &  $0$ &     $0$  & $\ \frac{1}{6}$ \\[3pt]
\hline
$(a,b')$  & $q_L$ & $2(3,2)$  &    $1$ &     $1$ &   $0$ &    $0$  & $\ \frac{1}{6}$ \\[3pt]
\hline
$(a,c)$  & $U_R$ & $3(\bar{3},1)$ &$-1$ &    $0$ &  $1$ &   $0$ & $-\frac{2}{3}$ \\[3pt]
\hline
$(a,c')$  & $D_R$ & $3(\bar{3},1)$&$-1$ &    $0$ &  $-1$ &   $0$ & $\ \frac{1}{3}$ \\[3pt] 
\hline 
$(b,d')$  & $L$ & $3(1,2)$       & $0$ &    $-1$ &  $0$ &   $-1$ & $-\frac{1}{2}$ \\[3pt] 
\hline 
$(c,d)$  & $E_R$ & $3(1,1)$   &   $0$ &     $0$ &    $-1$   &    $1$   &   $\ 1$ \\[3pt]  
\hline  
$(c,d')$  & $N_R$ & $3(1,1)$   &   $0$ &   $0$ &   $1$    &   $1$  &   $\ 0$ \\[3pt]  
\hline        
\end{tabular}  
\caption{Standard model spectrum and $U(1)$ charges. From \rr{\cite{ibanez2}}  \label{intersecnum1}.}
\end{table}

We note the following issues:
\begin{itemize}

\item Different types of models are possible. For example, more then 4 stacks of D-branes, D5 instead of D6-branes, gauge groups of grand unified theories, etc. 

\item $U(1)$ anomalies are canceled by the Green-Schwarz mechanism, see e.g \cite{blumenhagen, blumenhagen2}. The corresponding $Z'$ bosons receive masses from Chern-Simons terms even if they are not anomalous. The $U(1)$'s survive as perturbative global symmetries and can be identified with baryon and lepton number. This leads to proton stability and prevents Majorana neutrino masses. That being said, the condition for the hypercharge to remain massless is (see Eq.~(\textcolor{zz}{\ref{eq:hypcomb}})):
\begin{eqnarray}
\sum_ic_iN_i(\pi_i-\pi_i')\ =\  0
\end{eqnarray}

\item These models contain 3 right handed neutrinos.

\item intersecting D-brane models are classified as either supersymmetric or non-supersymmetric. Supersymmetric models usually assume a high string scale near the planck scale, whereas non-supersmmetric models usually assume a low string scale $\sim 1-100$ TeV. Our model is non-supersymmetric.

\item It has been argued that the effects of four-fermi operators on FCNC's, EDM's (electric dipole moments), and supernova cooling, constrain the string scale to be above $\sim 10^4$ TeV. This implies that non-supersymmetric intersecting brane models suffer a severe fine tuning problem.
\end{itemize}

\clearpage

\section{\textcolor{ww}{Review of amplitude calculations}}
\label{sec:revamp8}

\subsection{Field theory}
\label{sec:a3}

\emph{References:} Mainly follows~\textcolor{rr}{\cite{mangano}}. See also~\textcolor{rr}{\cite{dixon, peskin, mangano2, berends3}} and Appendix \ref{sec:appb} where the formalism and notation is presented.\\

This section is a review of amplitude calculation in field theory via the helicity amplitude technique and the trace color decomposition.

\begin{table}[h]
\centering
\begin{tabular}{|c||c|c|c|c|c|c|c|}
\hline
$ n $ & $4$ & $5$ & $6$ & $7$ & $8$ & $9$ & $10$  \\  
\whline
no. of diagrams & $4$ & $25$ & $220$ & $2485$ & $34300$ & $559405$ & $10525900$ \\  
\hline      
\end{tabular}  
\caption{The number of Feynman diagrams for the $n$-gluon amplitude at tree level. From \rr{\cite{mangano}}. \label{nofeyn}}
\end{table}

\begin{figure}
\centering
\includegraphics[width= 90mm]{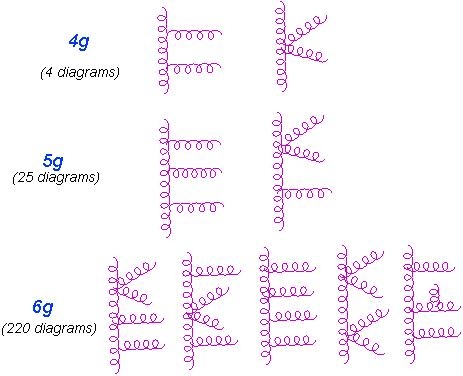}
\caption{Classes of diagrams for 4, 5 and 6 gluons. Each drawing represents all of the Feynman diagrams which can be obtained from it by permutation of the external particles.\label{56gluons}}
\end{figure}

\begin{figure}
\centering
\includegraphics[width= 90mm]{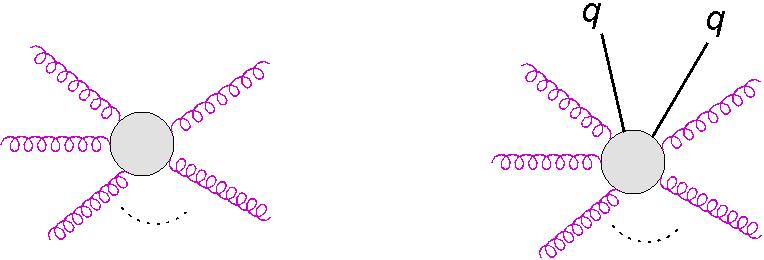}
\caption{The universal amplitudes.\label{ngluon}}
\end{figure}

We use the words "quarks" and "gluons" but the gauge symmetry is $SU(N)$ or $U(N)$ and not necessarily $SU(3)$.
Evaluation of amplitudes via text book methods for calculating Feynman diagrams, becomes very complex as one goes to higher loops or as one adds external particles. In this paper we do not deal with loop diagrams. The complexity arises from the large number of Feynman diagrams (see Table~\textcolor{yy}{\ref{nofeyn}} and Fig.~\textcolor{yy}{\ref{56gluons}}) and from the fact that the non-abelian vertices give rise to a large number of terms.  Usually at the end of the calculation there is a large amount of cancellation between the terms, giving rise to a relatively simple answer. This suggests the existence of a formalism which may simplify the procedure by taking better account of the symmetries of the amplitude. 
Since the perturbative expansion in Feynman diagrams is not gauge invariant, a major step forward was identifying what combination of feynman diagrams can give rise to a gauge invariant basis in which to expand. A particularly useful color decomposition was discovered via analogy with the Chan-Paton structure of string amplitudes: The \emph{trace color decomposition}. 
Other usefull calculation techniques include finding simple representations for the polarization vectors in terms of massless spinors, spinor products, recursion relations among the amplitudes, and Supersymmetric Ward identities.

The helicity amplitude technique consists of calculating the amplitudes with definite helicities for the external particles. There are two types of amplitudes which are particulary simple: the $n$-gluon amplitude and the $n$-gluon plus a quark anti-quark pair amplitude, Fig.~\textcolor{yy}{\ref{ngluon}}. We will refer to these amplitudes as \emph{universal amplitudes} for reasons which will become clear in the string theory section \ref{subsec:fds4}. 
We will see shortly that the universal amplitudes have a simple color decomposition on a color basis which is orthogonal at leading order in $1/N$. Written in this basis, a closed formula for the MHV sub-amplitudes exists.

All amplitudes of the form $(1^+, 2^+\ldots n^+)$ and $(1^-, 2^+\ldots n^+)$ vanish. In the $ng+q\bar{q}$ case, also the amplitudes with $q$ and $\bar{q}$ having the same helicity vanish. 
Explicitly:
\begin{eqnarray}
\label{eq:T9}
m(g_1^+,g_2^+,\ldots,g_n^+)\ =\  0
\nonumber\\[6pt]
m(g_1^-,g_2^+,g_3^+,\dots,g_n^+)\ =\  0
\end{eqnarray}
\begin{eqnarray}
\label{eq:T8}
m(q^+,\bar{q}^-,g_1^+, g_2^+, \dots,g_n^+)\ =\  m(q^-,\bar{q}^+,g_1^+, g_2^+, \dots,g_n^+)\ =\ 0
\nonumber\\[6pt]
m(q^+,\bar{q}^+,g_1,\dots,g_n)\ =\  0\ \ \ \ \ \ \ \ \ \ \ \ \ \ \ \ \ \ \ \ \ \ \ \ \ \ \ \ \ \ \ \ \ \ \ \ \ \ \ \ \ \ \ \ \ 
\end{eqnarray}
And of course, amplitudes obtained by reversing all helicities or permuting identical particles, vanish as well.

The \emph{Maximally Helicity Violating amplitude} (\emph{MHV})\footnote{We saw that the "would be" two most violating helicity amplitudes vanish, hence they are not called MHV.} is an amplitude with 2 particles having a certain helicity, and the rest having the opposite helicity: $(1^+, 2^+, 3^-, 4^-\ldots n^-)$. For the universal amplitudes, the MHV amplitudes do not vanish but they have a simple closed formula for arbirary $n$. For $n=4$ or $n=5$, Eqs. (\textcolor{zz}{\ref{eq:T9}}) and (\textcolor{zz}{\ref{eq:T8}}) imply that the MHV are the only non-vanishing helicity configuration. This trend ends at $n=6$ which has also the non-vanishing $m(1^+, 2^+,3^+,4^-,5^-,6^-)$.

\subsubsection{$n$ gluons}
\label{subsubsec:ngluyo}

An $n$- gluon amplitude can be decomposed as follows\footnote{\underline{\textbf{Proof}:}
 An $n$-gluon feynman diagram contains only gluon lines. A 3-gluon vertex contains $f^{abd}$ which can be written as  $f^{abc}= -iTr\big[T^aT^bT^c-T^cT^bT^a\big]$, by using $\big[ T^a,T^b \big]= if^{abc}$. Now each leg attached to this vertex has a $T$ attached to it. Each of these legs goes either to an external gluon or to another vertex. In the latter case, the $f^{cde}$ from the second vertex can be combined with $T^c$ from the first vertex to give: $T^cf^{cde}= -i[T^d, T^e]$. So we got for the two vertices: $f^{abc}f^{cde}= -Tr\big[T^aT^b[T^d, T^e]-[T^d, T^e]T^bT^a\big]$, which is of the required form. The 4-gluon vertex has $f^{abc}f^{cde}$ which is already in the required form. This process can easily be seen to continue by iteration. }:

\begin{equation} 
\label{eq:gluen}
\boxed{\mathcal{M}(g_1,\dots ,g_n)\ =\ \sum_{\{1,\ldots , n\}'} Tr\big(T^{a_1}T^{a_2}\cdot \cdot \cdot T^{a_n}\big)\ m(g_1,\dots ,g_n)}
\end{equation}
\\
Where the color matrices are in the fundamental representation. The \emph{subamplitudes} (Sometimes called \emph{colored ordered amplitudes} $m(1,\dots,n)\equiv m(g_1,\ldots,g_n)\equiv m(p_1,\epsilon_1,\ldots p_n,\epsilon_n)$ contain the kinematics: polarization vectors from the external legs and momentum vectors from the vertices. It is seen that they multiply a Chan-Paton color factor. The sum $\{1,\ldots , n\}'$ is over the $(n-1)!$ cylic inequivalent permutations .

Eq.~(\textcolor{zz}{\ref{eq:gluen}}), when squared and summed over colors and permutations gives:
\begin{equation} 
\label{eq:slam}
|\mathcal{M}|^2(g_1,\dots ,g_n)\ =\  \sum_{\lambda, \lambda '}m_{\lambda}\mathcal{S}_{ \lambda \lambda'}m^*_{\lambda '}
\end{equation}
Where $m_{\lambda}\equiv m(g_{1_\lambda},\dots ,g_{n_\lambda})$ is a given permutation, and

\begin{equation} 
\label{eq:slam1}
\mathcal{S}_{ \lambda \lambda'}\ \equiv \ \sum_{a_1,...a_n}Tr \big(T^{a_{1_\lambda}}\cdots T^{a_{n_\lambda}}\big)\  \big[Tr \big(T^{a_{1_{\lambda'}}} \cdots T^{a_{n_{\lambda'}}}\big)\big]^*
\end{equation}

The subamplitudes satisfy the following properties\footnote{The first two properties follow from the linear independence of the Chan-Paton factor (to leading order in $1/N$, see Eq.~(\zz{\ref{eq:sq12}})). Since  $Tr[T^{a_1}\cdots T^{a_n}]$ is cyclic invariant, so will the subamplitude be. Since the full amplitude is gauge invariant, so will the subampltudes be.}:

\begin{enumerate}
\item $m(1,\ldots,n)$ is gauge invariant.
\item $m(1,\ldots,n)$ is invariant under cyclic permutations of $1,\ldots,n$ 
\item Reflection:\ \ $m(n,\ldots,1)\ =\ (-1)^n m(1,\ldots,n)$
\item The \emph{dual ward identity} (Also called sub-cyclic identity or photon decoupling identity): \\ 
\begin{equation}
\label{eq:dwi7}
m(1,2,\ldots,n)+m(2,1,3,\ldots,n)+m(2,3,1,\ldots,n)+\ldots + m(2,3,\ldots,1,n)=0
\end{equation} 

\item Factorization of $m(1,\ldots,n)$ on multi-gluon poles 
\end{enumerate}

The \emph{MHV} amplitude has a simple closed formula: 
\begin{equation} 
\label{eq:b5yu1}
m(g_1^+,g_2^+,g_3^-,\ldots,g_n^-)\ =\  ig^{n-2}\frac{\langle 12\rangle^4}{\langle12\rangle \langle23\rangle \cdots \langle n1\rangle}
\end{equation}

So that,
\begin{equation} 
|m|^2(g_1^+,g_2^+,g_3^-,\ldots,g_n^-)\ =\ g^{2n-4}\frac{s_{12}^4}{s_{12}\cdots s_{n1}}
\end{equation}
Where $s_{ij}=(k_i+k_j)^2$,  $\langle ij\rangle=\sqrt{|s_{ij}|}e^{i\phi_{ij}}$,  $[ij]=-\sqrt{|s_{ij}|}e^{-i\phi_{ij}}$ , see Appendix \ref{sec:appb}.

In most colliders (in particular hadron colliders), the helicities of the particles are not measured. Hence, after the  amplitude is squared it should be summed over possible helicities and also over colors. Squaring and summing over colors (see Eq.~(\textcolor{zz}{\ref{eq:sq12}})) gives:
\begin{equation}
\label{eq:b7} 
\sum_{colors} |\mathcal{M}_n|^2\ =\ N^{n-2}(N^2-1)\sum_{\{1,\ldots , n\}'} \Big[\ |m|^2(1,\ldots,n)\ +\ \mathcal{O}(1/N^2)\ \Big] 
\end{equation} 
 
When summed over colors (see Eq.~(\textcolor{zz}{\ref{eq:b7}})) and MHV configurations one gets the \emph{Parke-Taylor amplitudes}: 
\begin{equation} 
\label{eq:parke13}
\sum_{helicities}\sum_{colors}|\mathcal{M}|^2(g_1,\ldots,g_n)= 2g^{2n-4}N^{n-2}(N^2-1)\sum_{i>j} s_{ij}^4\sum_{\{1,\ldots , n\}'} \Big[ \frac{1}{s_{12}\cdots s_{n1}}+\mathcal{O}(1/N^2)\Big]
\end{equation}
The factor of 2 comes from the sum over $(++-\ldots -)$ and $(--+\ldots +)$, and it is absent for $n=4$.

It turns out that for $4$ and $5$ gluons \footnote{For $6$ gluons Eq.~(\textcolor{zz}{\ref{eq:b7}}) will be:
\begin{eqnarray}
\label{eq:6gluons8}
\sum_{colors} |\mathcal{M}_6|^2= N^4(N^2-1)\sum_{\{2,3,4,5,6\}} \Big[\ |m|^2(1,2...,6)+\frac{2}{N^2}m^*(123456)\Big(m(135264)+m(153624)+m(136425)\Big)\ \Big] \nonumber\\
\end{eqnarray} }
the $\mathcal{O}(1/N^2)$ correction vanishes in Eq.~(\textcolor{zz}{\ref{eq:b7}}), so that  Eq.~(\textcolor{zz}{\ref{eq:parke13}}) becomes:
\begin{equation} 
\label{eq:4gs}
\sum_{helicities} \sum_{colors} |\mathcal{M}|^2(g_1,g_2,g_3,g_4)\ =\  N^2(N^2-1)g^4\sum_{i>j} s_{ij}^4\sum \frac{1}{s_{12}s_{23}s_{34}s_{41}}
\end{equation}

\begin{equation} 
\label{eq:5gs}
\sum_{helicities} \sum_{colors} |\mathcal{M}|^2(g_1,g_2,g_3,g_4,g_5)\ =\  2g^6N^3(N^2-1)\sum_{i>j}\ s_{ij}^4\sum \frac{1}{s_{12}s_{23}s_{34}s_{45}s_{51}}
\end{equation}

\subsubsection{$n$ gluons + $2$ quarks}

The $2$ quark plus $n$-gluons amplitude can be decomposed in the following way

\begin{equation} 
\label{eq:qqbarn}
\boxed{\mathcal{M}(q, \bar{q}, g_1,\ldots, g_n)\ =\  \sum_{\{1,\ldots n\}}\big(T^{a_1}T^{a_2}\cdot \cdot \cdot T^{a_n}\big)_{ij}\ m(q, \bar{q}, g_1,\ldots, g_n)}
\end{equation}

Eq.~(\textcolor{zz}{\ref{eq:qqbarn}}), when squared and summed over colors and permutations gives:
\begin{equation} 
\label{eq:plam}
|\mathcal{M}|^2(q, \bar{q}, g_1,\ldots, g_n)\ =\ \sum_{\lambda, \lambda '}m_{\lambda}\mathcal{P}_{ \lambda \lambda'}m^*_{\lambda '}
\end{equation}

Where,
\begin{equation} 
\label{eq:plam1}
\mathcal{P}_{ \lambda \lambda'}\ \equiv \ \sum_{a_1,...a_n} \big( T^{a_{1_\lambda}}\cdots T^{a_{n_\lambda}}\big)_{ij}\  \big(T^{a_{1_{\lambda'}}} \cdots T^{a_{n_{\lambda'}}}\big)^*_{ij}
\end{equation}

and $m_{\lambda}\equiv m(q,\bar{q},g_{1_\lambda},\dots ,g_{n_\lambda})$

The \emph{MHV} amplitude has a simple closed formula: 
\begin{equation} 
\label{eq:b5yu2}
m(\bar{q}^+, q^-,g_1^-,g_2^+,\ldots,g_n^+)\ =\ ig^n\ \frac{\langle q1\rangle ^3\langle \bar{q}1\rangle}{\langle \bar{q}q\rangle}\ \frac{1}{\langle q1\rangle \langle12\rangle \cdot \cdot \cdot \langle n\bar{q}\rangle}
\end{equation}

So that,
\begin{equation} 
|m|^2(\bar{q}^+,q^-,g_1^-,g_2^+,\ldots,g_n^+)\ =\ g^{2n}\ \frac{s_{q1}^3s_{\bar{q}1}}{s_{q \bar{q}}}\ \frac{1}{s_{q1}s_{12}\cdot \cdot \cdot s_{n\bar{q}}}
\end{equation}

Squaring and summing over colors and MHV configurations:
\begin{eqnarray} 
\sum_{helicities}\sum_{colors}|\mathcal{M}|^2(q,\bar{q},g_1,\ldots,g_n)\ =\  2g^{2n}N^{n-1}(N^2-1)\sum_{i=1} ^n (s_{qi}^3s_{\bar{q}i}+s_{qi}s_{\bar{q}i}^3)
\nonumber\\
\times \frac{1}{s_{q \bar{q}}}  \sum_{\{1,\ldots ,n\}} \frac{1}{s_{q1}s_{12}\cdot \cdot \cdot s_{n\bar{q}}}\ +\  \mathcal{O}(1/N^2)
\end{eqnarray}

\subsection{String theory}
\label{subsec:fds4}

\emph{References:} We mainly follow ~\textcolor{rr}{\cite{tong, lust, lust2}}, see also ~\textcolor{rr}{\cite{ gsw, polchinski, becker, schwarz, sta1, kir1, bed22}}.\\

\subsubsection{Generalities}
\label{subsubsec:b1}

We review in this section the techniques used to calculate amplitudes in string theory. We obtain the equations which enable us to calculate the leading order amplitudes (disk and sphere amplitudes).\\

\begin{figure}[h]
\centering 
\includegraphics[width= 65mm]{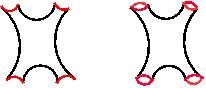}
\caption{Open and closed string diagrams with 4 external particles. The legs should be imagined to continue to infinity.\label{stringint}}	
\end{figure}

\begin{figure}[h]
\centering
\includegraphics[width= 90mm]{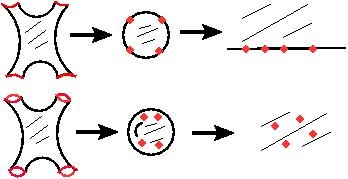}
\caption{Open strings: transforming to the disk and to the complex half-plane. closed strings: transforming to the sphere and to the complex plane.\label{stringinter}}
\end{figure}

\begin{figure}[h]
\centering
\includegraphics[width= 80mm]{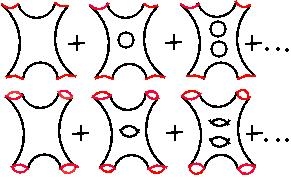}
\caption{Perturbation expansion of the 4-point string amplitudes. Top figure: open strings, bottom figure: closed strings.\label{closedpert}}
\end{figure}

\begin{figure}[h]
\centering
\includegraphics[width= 80mm]{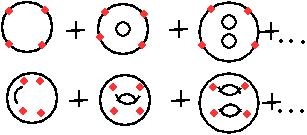}
\caption{The previous figure after a conformal transformation. For the open string the vertex operators are on the boundary.\label{closedpert2}}
\end{figure}

In quantum field theory one calculates correlation functions $\langle \phi(x_1)\cdot \cdot \cdot \phi(x_n)\rangle$. To get scattering amplitudes, the correlation functions are put on-shell. In a quantum theory of gravity it is not so clear how to deal with off-shell correlation functions. Instead we can calculate the S-matrix by taking the limit $x_i \to \infty$ in the correlation functions.
In string theory, a drawing of the lowest order interaction of strings looks as in Fig.~\textcolor{yy}{\ref{stringint}}. Unlike QFT there are no interaction vertices, and locally it is a free theory. Only when observed globally the interactions are seen. Taking $x_i \to \infty$ amounts to taking the legs of the diagram to infinity. The state-operator map says that a state at infinity is equivalent to an insertion of a vertex operator $V_i$ on the world sheet. A conformal transformation can transform our 2 diagrams into a disk and a sphere. A mobius transformation can transform the disk to the upper half plane, and a stereographic projection takes the sphere to the plane (Fig.~\textcolor{yy}{\ref{stringinter}}). The vertex operators will be placed on the boundary of the disk and on the sphere (The sphere obviously has no boundary). Weyl invariance enforces the vertex operators to be on-shell. Higher order diagrams are possible by considering holes and handles in the diagrams for open and closed strings respectively. A scattering amplitude will therefore consist of an expansion in the topology of the world sheet (Fig.~\textcolor{yy}{\ref{closedpert}}). Fig.~\textcolor{yy}{\ref{closedpert2}} shows the same expansion after performing the conformal transformation. 

We start with the following expression for the scattering amplitude for $n$ external particles:

\begin{equation}
\boxed{\mathcal{M}_n\ =\  \sum_{\text{topologies}}g_s^{-\chi}\ \frac{1}{\text{Vol.}}\int \mathcal{D}X\mathcal{D}g\ e^{-S_{Poly}}\ \ V_1\cdot\cdot\cdot V_n}
\end{equation}

Since we are only dealing with tree-level amplitudes, we safely ignored the Faddeev-Popov ghost fields.
This equation has the form of an expansion in the string coupling $g_s$. There is a functional integration over the coordinate $X$, an integration over all possible worldsheet metrics $g$, and a sum over the different topologies weighted by $g_s^{-\chi}$. It is therefore crucial to note that \textbf{we assume weak coupling}.

The \emph{Polyakov action} in the conformal gauge is:
\begin{equation}
S_{Poly}\ =\  \frac{1}{2\pi\alpha'}\int d^2z\ \partial X\cdot \bar{\partial} X
\end{equation}

$\chi$ is a topological invariant know as the \emph{Euler number}:
\begin{equation}
\chi(M)\ =\ \frac{1}{4\pi}\int d^2 \sigma \sqrt {g}R\ =\  2-2n_h-n_b-n_c
\end{equation}
Where $n_h, n_b, n_c$ are the number of handles, boundaries, and cross-caps of the worldsheet. For sphere topology $\chi=2$, while for the disk $\chi=1$.

We now focus on the lowest order, sphere and disk amplitudes. We need now to integrate over all metrics $g$. We transform to the flat metric and recall the remnant global transormations: the conformal killing group $PSL(2;C)\equiv SL(2;C)/Z_2$ for the sphere, and $PSL(2;R)\equiv SL(2;R)/Z_2$ for the disk. Technically this means making the following replacements in the path integral:

\begin{eqnarray}
\label{eq:voli}
\frac{1}{\text{Vol.}} \int \mathcal{D}g\ \ \ \  \to \ \ \ \  \frac{1}{\text{Vol.}(SL(2;C)}\ \int \prod_{i=1}^n d^2z_i\ \ \ \ \ \ \ 
\nonumber\\[7pt]
\frac{1}{\text{Vol.}} \int \mathcal{D}g\ \ \ \  \to \ \ \ \  \frac{1}{\text{Vol.}(SL(2;R)}\  \sum_{{\{1,\ldots ,n\}'}}\int \prod_{i=1}^n dz_i
\end{eqnarray}
For open strings the vertex operators are on the boundary of the disk, and hence have a given ordering to them. A cylic permutation of the vertex operators (which is just a rotation) gives an equivalent configuration because of the reparametrisation invariance of the string action. Hence there should be a sum over the $(n-1)!$ cylic inequivalent permutations. Also note that for open strings there is a single and not a double integration. 

We define $V_i\equiv V(z_i,k_i)$ and:
\begin{eqnarray}
\label{eq:b3}
\langle V_1\cdot \cdot \cdot V_n\rangle \ \equiv \ \int \mathcal{D}X \exp \bigg(-\frac{1}{2\pi \alpha'}\int d^2z\partial X\cdot \bar{\partial} X\bigg)\ \prod_{i=1}^n  V_1\cdot \cdot \cdot V_n
\end{eqnarray}
and we note that The $PSL(2,R)$ and $PSL(2,C)$ Symmetries on the disk and sphere respectively, allow to fix 3 of the insertion points $z_i$. This leaves $n-3$ integrations in Eq.~(\textcolor{zz}{\ref{eq:voli}}). The usual choice is $z_{k}=0$ , $z_{l}=1$, $z_{m}=\infty$.

We can finally write our master formulas for the closed and open leading order string amplitudes:
\begin{equation}
\label{eq:a99}
\boxed{\mathcal{M}^{(closed)}_n\ =\  \frac{g_s^{n-2}}{\text{Vol.}(SL(2;C))}\ \int \prod_{i=1}^n d^2z_i  
\ \langle V_1\cdot \cdot \cdot V_n\rangle}
\end{equation}

\begin{equation}
\label{eq:a9}
\boxed{\mathcal{M}^{(open)}_n\ =\ \sum_{{\{1,\ldots,n\}'}} \frac{g_s^{n-1}}{\text{Vol.}(SL(2;R))}\ \int \prod_{i=1}^n dz_i \ \langle V_1\cdot \cdot \cdot V_n\rangle}
\end{equation}

These formulas are valid for any type and any number of external particles. For each particle there corresponds a vertex operator and an integration.

To summarize, the procedure for calculating a scattering amplitude is:
\begin{enumerate}
\item Write the vertex functions of the external particles.  
\item Calculate the correlator of the vertex functions.
\item Fix 3 of the $z_i$'s, and perform the remaining $n-3$ integrations.
\item For open strings, sum over permutations of the $z_i$'s.
\end{enumerate}

\subsubsection{Tachyon amplitudes}
\label{subsubsec:tach7}
A simple example is the \emph{tachyon} (lowest state scalar) scattering amplitude. 
The vertex function of a tachyon is $V(z_j,k_j)= e^{ik_j\cdot X(z_j)}$. Defining $z_{ij}\equiv z_i-z_j$, Eq. (\textcolor{zz}{\ref{eq:appbc2}}) gives the correlation function for open string tachyons:

\begin{eqnarray}
\langle e^{ik_1\cdot X(z_1)}\cdots e^{ik_n\cdot X(z_n)}\rangle \ =\  \prod_{i<j}^n|z_{ij}|^{2\alpha'k_i\cdot k_j}  
\end{eqnarray}

We now use our freedom to fix 3 $z_i$'s. We start by choosing $z_1=\text{constant} >>z_i$. This gives: 
\begin{eqnarray}
|z_{1}|^{2\alpha'k_1\cdot(k_2+\ldots +k_n)}\prod_{1< i<j}^n|z_{ij}|^{2\alpha'k_i\cdot k_j}\ =\  \text{const.} \prod_{1< i<j}^n|z_{ij}|^{2\alpha'k_i\cdot k_j}  
\end{eqnarray}
Where we used $k_2+\dots +k_n=-k_1$ and $k_1^2=m_1^2=\text{constant}$.

Further choosing $z_2=0,\ z_3=1$, we get:
\begin{eqnarray}
\label{eq:opr1}
\mathcal{M}^{(open)}_n \ =\  \text{const.} \int \prod_{l=4}^n dz_l\ |z_l|^{2\alpha'k_2\cdot k_l}|1-z_l|^{2\alpha'k_3\cdot k_l}\  \prod_{4\leq i<j}^n|z_{ij}|^{2\alpha'k_i\cdot k_j} 
\end{eqnarray}

For closed strings the correlation function is Eq.~(\textcolor{zz}{\ref{eq:appbc1}}), so it is easy to see that:
\begin{eqnarray}
\mathcal{M}^{(closed)}_n \ =\  \text{const.} \int \prod_{l=4}^n d^2z_l\ |z_l|^{\alpha'k_2\cdot k_l}|1-z_l|^{\alpha'k_3\cdot k_l}\  \prod_{4\leq i<j}^n|z_{ij}|^{\alpha'k_i\cdot k_j} 
\end{eqnarray}

For $n=3$ tachyons we get $\mathcal{M}_3=\text{const}$. Defining $\hat{s}_{ij}\equiv 2\alpha' k_i\cdot k_j$, we have for $n=4$:
\begin{eqnarray}
\label{eq:ven7}
\boxed{\mathcal{M}^{(open)}_4\ \propto \ \int dz|z|^{\hat{u}}|1-z|^{\hat{s}}\ =\  B(\hat{s}+1, \hat{u}+1)}
\end{eqnarray}
where $B(a,b)= \frac{\Gamma(a)\Gamma(b)}{\Gamma(a+b)}$ is the Beta function.

For the closed string:
\begin{eqnarray}
\mathcal{M}^{(closed)}_4\ \propto \ \int d^2z|z|^{\hat{u}/2}|1-z|^{\hat{s}/2}\ =\  \frac{\Gamma(-1-\frac{\hat{s}}{4})\ \Gamma(-1-\frac{\hat{u}}{4})\ \Gamma(3+\frac{\hat{s}}{4}+\frac{\hat{u}}{4})}
{\Gamma(2+\frac{\hat{s}}{4})\ \Gamma(2+\frac{\hat{u}}{4})\ \Gamma(-2-\frac{\hat{s}}{4}-\frac{\hat{u}}{4})}
\end{eqnarray}

We can write the last equation in a symmetric form. Closed string tachyons have a negative mass of $m=-4M_s$, hence $\hat{s}+\hat{t}+\hat{u}= \frac{1}{M_s^2}\sum_{1}^4m_j^2=-16$. This gives
\begin{eqnarray}
\label{eq:vir7}
\boxed{\mathcal{M}^{(closed)}_4\ \propto \  \frac{\Gamma(-1-\frac{\hat{s}}{4})\ \Gamma(-1-\frac{\hat{t}}{4})\ \Gamma(-1-\frac{\hat{u}}{4})}
{\Gamma(2+\frac{\hat{s}}{4})\ \Gamma(2+\frac{\hat{t}}{4})\ \Gamma(2+\frac{\hat{u}}{4})}}
\end{eqnarray}

Eqs.~(\textcolor{zz}{\ref{eq:ven7}}) and (\textcolor{zz}{\ref{eq:vir7}}) are the \emph{Veneziano amplitude} and the \emph{Virasoro-Shapiro amplitude} respectively.


\subsubsection{Quark-gluon amplitudes}
\label{subsubsec:b2}

In the previous section we gave tachyons as an example of a scattering amplitude in string theory. Although tachyons are present in bosonic string theory, they are eliminated from the spectrum of string theories which contain world-sheet fermions: superstring theories. In the models we consider, massless fermions and gauge bosons occupy the ground state of the spectrum of the open string. In particular we are interested in quark and gluon scattering amplitudes, and these are calculated quite similarly to tachyons. We recall from section \ref{sec:a3} that quark-gluon amplitudes may be classified according to weather they are \emph{universal} or \emph{non-universal}. Universal amplitudes are defined as those containing 0 or 2 quark (or squark) fields, and non-universal amplitudes are those with more quark fields.

As an example, in order to calculate the $n$-point \emph{universal} amplitudes we need the following correlation functions.
\begin{eqnarray}
\label{eq:un81}
\langle \ V_A(z_1)\ \ \cdots \ V_A(z_n)\ \rangle\ \ \ \ \ \ \ \ \ \ \ \ \ \ \ \ \ \ \ \ \ \ \ \ 
\\[5pt]
\langle \ V_A(z_1)\ \cdots \ V_A(z_{n-2})\ \ V_\psi(z_{n-1})\ V_{\bar{\psi}}(z_{n})\ \rangle
\end{eqnarray}

Appendix~\ref{sec:appa} lists the vertex functions of gluons and quarks in terms of the fields from the underlying SCFT. Appendix~\ref{sec:appcorr} lists correlation functions of SCFT fields needed in order to calculate the above correlation functions. In Appendix~\ref{sec:b8} we present a full calculation of the $\mathcal{M}(ggqq)$ amplitude.
The vertex functions contain the color matrices in such a way that a chain of vertex operators (e.g Eq.~(\textcolor{zz}{\ref{eq:un81}})) gives the \emph{Chan-Paton} color structure. The Chan-Paton structure is identical to the color decomposition that was done for the field theory amplitudes (Eqs.~(\textcolor{zz}{\ref{eq:gluen}}), (\textcolor{zz}{\ref{eq:qqbarn}})). Hence the universal string amplitudes will now be written as\footnote{The Chan-Paton structure relates to open string diagrams as shown in Fig.~\textcolor{yy}{\ref{chanpaton3}}. Each string is assumed to carry a "`quark"' at one end and an "`anti-quark"' at the other end. The quarks being charged and transform as $N\times \bar{N}$ under  a $U(N)$ symmetry. To each string there corresponds an $N\times N$ matrix $T^i_{\ j}$. An $n$-point amplitude is obtained by contracting in cylic order the "`anti-quark"' index of a string with the "`quark"' index of the next string, and so will contain the factor:
$\big(T^{a_1}\big)^p_{\ \ l}\big(T^{a_2}\big)^l_{\ k}\cdots \big(T^{a_n}\big)^i_{\ p}=Tr\big[T^{a_1}T^{a_2}\cdots T^{a_n} \big]$. Historically, this picture of a string with a quark and anti-quark at its ends was introduced as a model for mesons, so that $U(N)$ was the flavor group. This picture remains approximately correct, with the QCD flux tube acting as the string.}:

\begin{figure}
\centering
\includegraphics[width= 35mm]{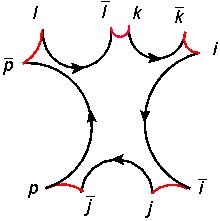}
\caption{Chan-Paton structure.\label{chanpaton3}}
\end{figure}

\begin{equation} 
\boxed{\mathcal{M}_{string}(g_1,\ldots,g_n)\ =\  \sum_{\{1,\ldots,n\}'} Tr \big(T^{a_1}\cdot \cdot \cdot T^{a_n}\big)\ m_{string}(g_1,\dots ,g_n)}
\end{equation}

\begin{equation} 
\boxed{\mathcal{M}_{string}(q,\bar{q},g_1,\ldots,g_n)\ =\  \sum_{\{1,\ldots ,n\}}\big(T^{a_1}\cdot \cdot \cdot T^{a_n}\big)_{ij}\ m_{string}(q,\bar{q},g_1,\ldots,g_n)}
\end{equation}
We put the label "string" because shortly these will be compared to the field theory amplitudes.
From Eq.~(\textcolor{zz}{\ref{eq:a9}}) the subamplitude is:
\begin{equation}
m_{string}\ =\  \frac{g_s^{n-1}}{\text{Vol.}(SL(2;R))}\ \int \prod_{i=1}^n dz_i\ \langle \widetilde{V}_1\cdot \cdot \cdot \widetilde{V}_n\rangle
\end{equation}
and $\widetilde{V}_i$ is is just $V_i$ after stripping it from it's color matrices.

Following \rr{\cite{lust, lust2}},  we define $m_{QCD}$ as the sub-amplitude in field theory ($m$ of section~\ref{sec:a3}). We can write $m_{string}$ as $m_{QCD}$ times a function (form factor) that needs to be calculated. Focusing on the $4$ and $5$-point  helicity amplitudes, the non-MHV subamplitudes vanish also in string theory (Eqs.~(\textcolor{zz}{\ref{eq:T9}}) and (\textcolor{zz}{\ref{eq:T8}})). So we write (and explain afterwords..) for the MHV sub-amplitudes:

\begin{itemize}
\item {\large 4 partons}
\end{itemize}

\begin{equation}
\label{eq:b6} 
\boxed{m_{string}(g_1, g_2, g_3, g_4)\ \ =\ \ \mathfrak{B}_4(k_j)\ m_{QCD}(g_1, g_2, g_3, g_4)}
\end{equation}

\begin{equation} 
\label{eq:b759}
\boxed{m_{string}(g_1,g_2,q_3,\bar{q}_4)\ \ =\ \ \mathfrak{B}_4(k_j)\ m_{QCD}(g_1,g_2,q_3,\bar{q}_4)}
\end{equation}

\begin{equation} 
\label{eq:b789}
\boxed{m_{string}(q_1,\bar{q}_2,q_3,\bar{q}_4)\ \ =\ \ \widehat{\mathfrak{B}}_4(k_j, \theta _i)\  m_{QCD}(q_1,\bar{q}_2,q_3,\bar{q}_4)}
\end{equation}

Where\footnote{The kinematical factor in front of the integral ensures that $\mathfrak{B}_4\to 1$ as $\hat{s},\hat{u}\to 1$ so that at low energies $m_{string}\to m_{QCD}.$}
\begin{eqnarray}
\label{eq:bfrak4}
\mathfrak{B}_4(\hat{s},\hat{u})\ \equiv \ V_t\ =\  \frac{\hat{s}\hat{u}}{\hat{t}} \int_0^1dz\ z^{s-1}(1-z)^{u-1}
\end{eqnarray}

\begin{eqnarray}
\widehat{\mathfrak{B}}_4\ =\  \frac{\hat{s}\hat{u}}{\hat{t}} \int_0^1dz\ z^{s-1}(1-z)^{u-1}\ I_\rho (z_i,\theta_j)
\end{eqnarray}
Where $I_\rho$ is a function which depends on the D-brane setup, and therefore is model dependent.

\begin{itemize}
\item {\large 5 partons}
\end{itemize}

\begin{equation} 
\label{eq:b758}
\boxed{m_{string}(g_1,g_2,g_3,g_4,g_5)\ \ =\ \  \mathfrak{B}_5(k_j)\ m_{QCD}(g_1,g_2,g_3,g_4,g_5)}
\end{equation}

\begin{equation} 
\label{eq:b9}
\boxed{m_{string}(g_1,g_2,g_3,q_4,\bar{q}_5)\ \ =\ \  \mathfrak{B}_5(k_j)\ m_{QCD}(g_1,g_2,g_3,q_4,\bar{q}_5)}
\end{equation}

\begin{equation} 
\label{eq:b919}
\boxed{m_{string}(g_1,g_2,g_3,q_4,\bar{q}_5)\ \ =\ \  \widehat{\mathfrak{B}}_5(k_j, \theta _i)\   m_{QCD}(g_1,q_2,\bar{q}_3,q_4,\bar{q}_5)}
\end{equation}
$\mathfrak{B}_5$ will be given in section~\ref{subsubsec:genven5}.
\\

There are two things to be learned from these equations, showing the special properties of the universal amplitudes.
\begin{enumerate}
\item 
\textbf{Equality of form factors}:
 a string helicity amplitude is equal to the corresponding field theory amplitude times a stringy form factor. Eqs.~(\textcolor{zz}{\ref{eq:b6}}) and (\textcolor{zz}{\ref{eq:b759}}) have the \underline{same} form factor $\mathfrak{B}_4$, and likewise Eqs.~(\textcolor{zz}{\ref{eq:b758}}) and (\textcolor{zz}{\ref{eq:b9}})) have the \underline{same} $\mathfrak{B}_5$.
\item 
\textbf{Universality}:
The form factors $\mathfrak{B}_4$ and $\mathfrak{B}_5$ depend \underline{only} on the kinematics and hence are universal or model independent. On the other hand, the form factors $\widehat{\mathfrak{B}}_4$ and $\widehat{\mathfrak{B}}_5$ depend on the setup of D-branes and geometry of the extra dimensions, hence they are model dependent or non-universal.
\end{enumerate}

These two properties generalize to $n$-point universal amplitudes.
The claim is that an $n$-point universal helicity amplitude can
be written as:
\begin{eqnarray}
\label{eq:formfactor16}
m_{string}(g_1\ldots g_n)\ =\  \mathfrak{B}_n(k_j)\  m_{QCD}(g_1\ldots g_n)\ \ \ \ \ \ \ \ \ \ \ \ \ \ \ \ \ \ \ \ \ \ \ \ \ \ \ 
\\[6pt]
m_{string}(g_1\ldots g_{n-2}, q_{n-1}, \bar{q}_{n})\ =\  \mathfrak{B}_n(k_j)\  m_{QCD}(g_1\ldots g_{n-2}, q_{n-1}, \bar{q}_{n})
\end{eqnarray}
with the same form factors which depend only on the external momenta, and not on the compactification. Furthermore,
$\mathfrak{B}_n$ can be expressed in terms of generalized hypergeometric functions, and there are $(n-3)!$ independent sub-amplitudes, see Section~\ref{sec:multi-gluon amplitudes}. 

Including the supersymmetric gluino and squark ($\chi$ and $\phi$), the $n$-point universal amplitudes are:
\begin{eqnarray}
1.\ \ m(g^{a_1}\ldots g^{a_n})\ \ \ \ \ \ \ \ \ \ \ \ \ \   ,\ \ \ \ 3.\ m(q^{a_1}, \bar{q}^{a_2}, g^a_3\ldots  g^{a_n})
\nonumber\\[6pt]
2.\ m(\chi^{a_1}, \bar{\chi}^{a_2}, g^a_3\ldots  \ g^{a_n})\ \ \ ,\ \ \ \ 4.\ m(\phi^{a_1}, \bar{\phi}^{a_2}, g^a_3\ldots g^{a_n})
\end{eqnarray}
The universal amplitudes 1 and 2 are related through supersymmetric ward identities, as are amplitudes 3 and 4. In addition, amplitudes 1 and 3 have equal stringy form factors.

\subsubsection{Discussion}
\label{subsubsec:disc15}

We further explain the two properties of the universal amplitudes, see \rr{\cite{lust2}}.  

\begin{figure}[h]
\centering
\includegraphics[width= 110mm]{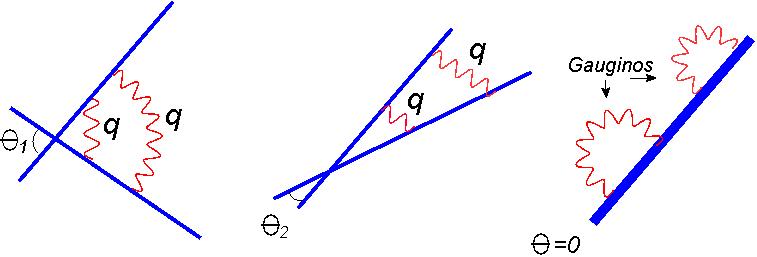}
\caption{Explanation of the equality of form factors. The 2 quarks of the $n$-gluon plus 2-quark amplitude are shown. From left to right: changing the intersection angle until it is zero.\label{equalffproof2}}
\end{figure}

\begin{figure}[h]
\centering
\includegraphics[width= 150mm]{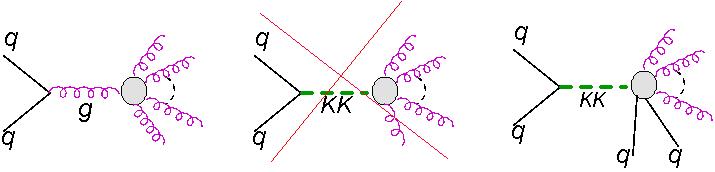}
\caption{Explanation of universality. The conservation of internal charge forbids the middle diagram. The same argument holds for the $n$-gluon amplitude.\label{lala7}}
\end{figure}

\begin{enumerate}
\item 
\textbf{Equality of form factors}:
The explanation is as follows, see Fig.~\textcolor{yy}{\ref{equalffproof2}}. Consider an helicity amplitude  with $k$ quarks and $n$ gluons: $m'(q_1\ldots q_k, g_1\ldots g_n)$ The quarks arise from strings stretched between between 2 stacks of D-branes intersecting at an angle $\theta$. If the angle $\theta$ is gradually changed and taken to zero, the quarks will appear as gluinos of the enhanced gauge group $T^a\oplus T^b$, since the stacks are on top of each other. This new configuration describes an amplitude with $k$ gluinos and $n$ gluons: $m''(\chi_1\ldots \chi_k, g_1\ldots g_n)$. 
Supersymmetry relates gluinos and gluons, so that $m''$ has the same form factor as the all-gluon amplitude  $m'''(g_1\ldots g_{k+n})$: $\mathfrak{B}''= \mathfrak{B}'''$.
Finally, if $k=2$ then $m'$ is a universal amplitude and in particular independent of $\theta$. Hence in this case $m' = m''$, and thus $\mathfrak{B}'=\mathfrak{B}''=\mathfrak{B}'''$. We have thus proved that the two types of universal amplitudes have equal form factors.    

\item
\textbf{Universality}:
The most interesting property of these amplitudes is that they are \textit{universal} or \textit{model independent}. They are the same in many different models of string theory, because they do not depend on the compactification of the extra dimensions. \textit{Universal} amplitudes contain only Regge states and not KK or Winding states, which appear in amplitudes with more quarks. KK states arise from the compactification of extra dimensions. Winding states arise when extended objects such as strings or D-branes wrap around the extra dimensions. Regge states are pure string states independent of the extra dimensions.

Mathematicaly the reason KK and winding states do not appear in the universal amplitudes is the following.
KK and winding states only appear in amplitudes  constructed from correlators of the boundary changing operators $\Xi^{a \cap b}$, and only when there are 4 or more $\Xi^{a \cap b}$'s : 
\begin{eqnarray} 
\langle \  \Xi^{a \cap b}(z_1)\ \bar{\Xi}^{a \cap b}(z_2)\ \rangle \ =\  z_{12}^{-3/4}\ \ \ \ \ \ \ \ \ \ \ \ \ \ \ \ \ \ \ \ \ \ \ \ \ \ \ \ \ \ \ \ \ \ \ \ \ \ \ \ \ \ \ \ \ \ \ \ \ \ \ \ \ \ \  
\nonumber\\[7pt]
\langle \  \Xi^{a \cap b}(z_1)\ \bar{\Xi}^{b \cap d}(z_2)\ \Xi^{d \cap c}(z_3)\ \bar{\Xi}^{c \cap a}(z_4)\ \rangle \ =\  \Big(\frac{z_{13}z_{24}}{z_{12}z_{14}z_{23}z_{34}}\Big)^{3/4}\ I_\rho\ (\{ z_i\};\theta^j)
\end{eqnarray} 
$I_\rho$ depends on the compactification and intersections of the D-branes. It includes exchanges of KK and winding states.
Since $\Xi^{a \cap b}$ appears only in the quark vertex function and not in the gluon vertex function (see Eqs.~(\textcolor{zz}{\ref{eq:appb1}})-(\textcolor{zz}{\ref{eq:appb4}})), KK and winding states will appear only in amplitudes with 4 or more quarks.

This property can also be seen diagramatically. Fig.~\textcolor{yy}{\ref{lala7}} shows the difference, in this respect, between an amplitude containing 2 quarks and an amplitude with 4 quarks. KK and winding states carry internal charge, and charge conservation requires quark pairs on both sides of a KK/winding state line in the diagram. So an amplitude with one quark pair can not have KK/winding state exchange.
\end{enumerate}

\clearpage


\section{\textcolor{ww}{4-point amplitudes}}
\label{sec:4point8}

\begin{figure}
\centering
\includegraphics[width= 60mm]{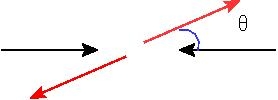}
\caption{4 particle kinematics. The scattering angle $\theta$.\label{scattangle}}
\end{figure}
 
4 particle amplitudes are $2\to2$ processes at colliders. At the LHC, important signals of $2\to2$ processes include:  $pp\to$ 2 jets, $pp\to$ jet + EW gauge boson and $pp\to$ 2 EW gauge bosons. 

The Mandelstam variables are
\begin{eqnarray}
s\ \equiv \ (k_1+k_2)^2\ =\ 2k_1 k_2
\nonumber\\
t\ \equiv \ (k_1+k_3)^2\ =\ 2k_1 k_3
\nonumber\\
u\ \equiv \ (k_1+k_4)^2\ =\ 2k_1 k_4
\end{eqnarray}
The kinematics are shown in Fig.~\textcolor{yy}{\ref{scattangle}}.

Energy-momentum conservation for massles quarks and gluons gives
\begin{equation}
s+t+u\ =\ \sum m_i^2\ =\ 0
\end{equation}

The hatted Mandelstam variables $\hat{s}$, $\hat{t}$, $\hat{u}$ are defined as the Mandelstam variables in units of $M_s$:
\begin{eqnarray}
\hat{s}\ \equiv \ s/M_s^2\ =\  \alpha's
\nonumber\\
\hat{t}\ \equiv \ t/M_s^2\ =\  \alpha't
\nonumber\\
\hat{u}\ \equiv \ u/M_s^2\ =\  \alpha'u
\end{eqnarray}
Since this is a scale transformation of the external momenta and since massles QCD/QED are scale invariant, the QCD/QED amplitudes will be invariant under $s,t,u \to \hat{s},\hat{t},\hat{u}$. Amplitudes with $W$ and $Z$ bosons are not scale invariant so that their form will change when passing to the hatted variables. For $s,t,u >> M_Z$, the EW amplitudes become scale invariant as well.
The string amplitudes to be discussed later, will obviously change under this scale transformation. These properties are easily seen by looking at the squared amplitudes in sections \ref{subsec:ftsa} and \ref{subsubsec:a7}

\subsection{Field theory}
\label{subsec:ftsa}
\emph{References:} ~\textcolor{rr}{\cite{eichten, ellis, campbell}}.\\

Using the techniques and results of section \ref{sec:a3}, the 4-point squared amplitudes may be computed. 
As an example, consider the $gg\to gg$ squared amplitude. Looking at Eq.~(\textcolor{zz}{\ref{eq:4gs}}) and using $\sum s^4_{ij}\propto (s^4+t^4+u^4)$ and $\sum \frac{1}{s_{12}s_{23}s_{34}s_{41}}\propto \frac{1}{s^2t^2}+\frac{1}{s^2u^2}+\frac{1}{u^2t^2}$, we get:
\begin{eqnarray}
|\mathcal{M}|^2(gg\to gg)\ \propto \ g^4(s^4+t^4+u^4)\frac{s^2+t^2+u^2}{s^2t^2u^2}\ =\  4g^4\Big[3-\frac{tu}{s^2}-\frac{su}{t^2}-\frac{st}{u^2}\Big]
\end{eqnarray}
Which is given also in Eq.~(\textcolor{zz}{\ref{eq:ggg9}}). In the last equation we used $s+t+u=0$

\begin{figure}
\centering
\includegraphics[width= 100mm]{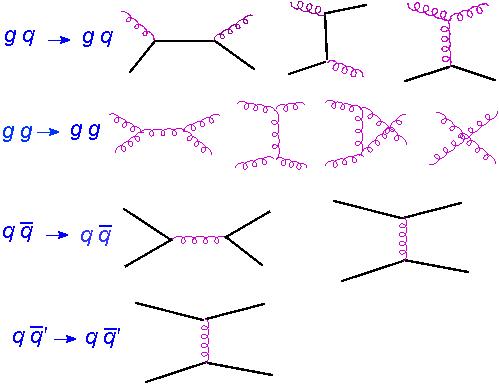}
\caption{QCD Feynman diagrams for 4-particles. The other processes can be obtained by crossing.\label{2g2Q}}
\end{figure}

\begin{figure}
\centering
\includegraphics[width= 90mm]{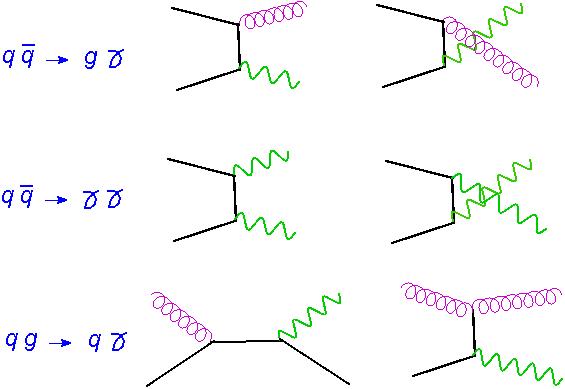}
\caption{Feynman diagrams for partons + photons. The same diagrams describe Z bosons.\label{feynphot}}
\end{figure}

In the next section we list the squared amplitudes for $2\to2$ amplitudes in terms of the Mandelstam variables. Some of the Feynman diagrams are shown in Figs.~\textcolor{yy}{\ref{2g2Q}}, \textcolor{yy}{\ref{feynphot}}. We consider the processes which are the most important in a hadron collider, namely initial states with a quark or a gluon.

\subsubsection{The squared amplitudes}

\begin{itemize}
\item {\large $gg$ initial state}
\end{itemize}

\begin{equation}
\label{eq:ggg9}
|\mathcal{M}|^2(\textcolor{blue}{gg\to gg})\ =\ \frac{3}{2}g^4\Big[3-\frac{tu}{s^2}-\frac{su}{t^2}-\frac{st}{u^2}\Big]
\end{equation}

\begin{equation}
|\mathcal{M}|^2(\textcolor{blue}{gg \to g\gamma})\ =\ 0
\end{equation}

\begin{equation}
|\mathcal{M}|^2(\textcolor{blue}{gg \to \gamma \gamma})\ =\ 0
\end{equation}

\begin{equation}
|\mathcal{M}|^2(\textcolor{blue}{gg \to q\bar{q}})\ =\ \frac{3}{8}g^4(t^2+u^2)\Big[\frac{4}{9tu}-\frac{1}{s^2}\Big]
\end{equation}

\begin{itemize}
\item {\large $gq$ initial state}
\end{itemize}

\begin{equation}
|\mathcal{M}|^2(\textcolor{blue}{gq\to gq})\ =\ -g^4(s^2+u^2)\Big[\frac{4}{9su}-\frac{1}{t^2}\Big]
\end{equation}

\begin{equation}
|\mathcal{M}|^2(\textcolor{blue}{gq\to \gamma q})\ =\  -\frac{1}{3}g^2 e^2\frac{s^2+u^2}{us}
\end{equation}

\begin{equation}
|\mathcal{M}|^2(\textcolor{blue}{gq\to Wq'})\ =\  
\frac{\pi \alpha \alpha_s}{12x_W}\ \frac{s^2+u^2+2M_W^2t}{-s^2u}
\end{equation}

\begin{itemize}
\item {\large $q\bar{q}$ initial state}
\end{itemize}

\begin{equation}
|\mathcal{M}|^2(\textcolor{blue}{q\bar{q}\to gg})\ =\ \Big(\frac{8}{3}\Big)^2\ |\mathcal{M}|^2(gg\to q\bar{q})
\end{equation}

\begin{equation}
|\mathcal{M}|^2(\textcolor{blue}{q\bar{q}\to \gamma g})\ =\ \frac{8}{9}g^2 e^2\ \frac{t^2+u^2}{ut}
\end{equation}

\begin{equation}
|\mathcal{M}|^2(\textcolor{blue}{q\bar{q}\to \gamma\gamma})\ =\ \frac{3}{4}\frac{e^4}{g^4}\ |\mathcal{M}|^2(q\bar{q}\to \gamma g)
\end{equation}

\begin{equation}
|\mathcal{M}|^2(\textcolor{blue}{q\bar{q}'\to Wg})\ =\ \frac{2\pi \alpha \alpha_s}{9x_W}\ \frac{(t-M_W^2)^2+(u-M_W^2)^2}{tu}
\end{equation}

\begin{equation}
|\mathcal{M}|^2(\textcolor{blue}{q\bar{q'}\to q\bar{q'}})\ =\ \frac{4}{9}g^4\frac{s^2+u^2}{t^2}
\end{equation}

\begin{equation}
|\mathcal{M}|^2(\textcolor{blue}{q\bar{q}\to q'\bar{q'}})\ =\ \frac{4}{9}g^4\frac{t^2+u^2}{s^2}
\end{equation}

\begin{equation}
|\mathcal{M}|^2(\textcolor{blue}{q\bar{q}\to q\bar{q}})\ =\  \frac{4}{9}g^4\Big[\frac{t^2+u^2}{s^2}+\frac{s^2+u^2}{t^2}-\frac{2u^2}{3st}\Big]
\end{equation}

\begin{itemize}
\item {\large $qq$ initial state}
\end{itemize}

\begin{equation}
|\mathcal{M}|^2(\textcolor{blue}{qq\to qq})\ =\  \frac{4}{9}g^4\Big[\frac{s^2+t^2}{u^2}+\frac{s^2+u^2}{t^2}-\frac{2s^2}{3tu}\Big]
\end{equation}

\begin{equation}
|\mathcal{M}|^2(\textcolor{blue}{qq'\to qq'})\ =\ \frac{4}{9}g^4\frac{s^2+u^2}{t^2}
\end{equation}

\clearpage

\subsection{String theory}
\label{subsec:string57}
\subsubsection{The Veneziano amplitude}
\label{sec:1}

The \emph{Veneziano amplitude} $V_t$, which enters the 4-point open string scattering amplitudes, is a fantastically rich object.
$V_t$ is the form factor connecting the string and field theory 4-point universal sub-amplitudes, Eqs.~(\zz{\ref{eq:b6}}), (\zz{\ref{eq:b759}}), (\zz{\ref{eq:bfrak4}}):
\bea
\label{eq:ff780}
m_{string}= V_t\ m_{QCD}
\eea
 
Explicitly, the Veneziano amplitude is:
\begin{eqnarray}
\label{eq:vent4}
V_t\ \equiv \ \frac{\hat{s}\hat{u}}{\hat{t}}B(-\hat{s},-\hat{u})\ =\  \frac{\hat{s}\hat{u}}{\hat{t}} \frac{\Gamma(-\hat{s})\Gamma(-\hat{u})}{\Gamma(\hat{t})}
\ =\  \frac{\Gamma(1-\hat{s})\Gamma(1-\hat{u})}{\Gamma(1+\hat{t})}\ \ \ \ \ \ \ \ \ \ \ \ \ \ \ \ \ \  
\end{eqnarray}

and by crossing:
\begin{eqnarray}
V_u \  \equiv \ V_t(\hat{t}\leftrightarrow \hat{u})
\nonumber\\[6pt] 
V_s \  \equiv \ V_t(\hat{t}\leftrightarrow \hat{s})
\end{eqnarray}

The beta function has an integral representation:
\begin{eqnarray}
B(a,b) \ \equiv \ \int_0^1dx\ x^{a-1}(1-x)^{b-1}\ =\ \frac{\Gamma{(a)}\Gamma{(b)}}{\Gamma{(a+b)}}
\end{eqnarray}

One thing to notice is that whereas the field theory tree amplitudes are independent of the collision energy (For massless fermions and gauge bosons.),  the string amplitudes do depend on it through the Veneziano amplitude. This means e.g. that the angular distribution of the scattered particles changes as the energy changes.\\

\underline{Properties of the Veneziano factor:}

\begin{enumerate}

\item Low energy expansion ($\hat{s}, \hat{t}, \hat{u} << 1$):
\bea
V_t= 1+ \frac{\pi^2}{6}\hat{s}\hat{u}+ \ldots
\eea
In this limit $V_t$ is 1 plus corrections in inverse powers of the string scale.
See also section~\ref{subsubsec:low89}.

\item 
High energy limit ($\hat{s}>>1$).
 
There are two types of high energy limits that are usually considered: the \emph{fixed scattering angle limit} and the \emph{Regge limit}. In the first case $\theta= \text{constant}$ and $|\hat{s}|, |\hat{t}| \to \infty$. The Regge limit is $|\hat{s}| \to \infty$, $\hat{t}= \text{constant}$. From Eqs.~(\textcolor{zz}{\ref{eq:betaf1}})-(\textcolor{zz}{\ref{eq:betaf3}}) we have:

\textbf{Fixed angle limit}:
\begin{eqnarray}
V_t, V_u, V_s\ \ \  \longrightarrow\ \ \  e^{-f(\theta)s}
\end{eqnarray}

\textbf{Regge limit}:
\begin{eqnarray}
V_u, V_s\ \ \  \longrightarrow\ \ \  s^{-t}
\nonumber\\
V_t\ \ \  \longrightarrow\ \ \  e^{-f(\theta)s}
\end{eqnarray}

Where,
\begin{eqnarray}
f(\theta)\ =\ \Big(\frac{1-\cos\theta}{2}\Big)\ln \Big(\frac{1-\cos\theta}{2}\Big)\ +\ \Big(\frac{1+\cos\theta}{2}\Big)\ln \Big(\frac{1+\cos\theta}{2}\Big)
\end{eqnarray}
In the fixed angle limit, $V_t$ is exponentially decreasing. This is extremely soft UV behavior.

\item s-channel pole expansion:

$V_t$ can be expanded on s-channel poles, giving rise to the most useful equation of this work:
\begin{equation}
\label{eq:venred}
\boxed{V_t\ =\ -\frac{\hat{s}\hat{u}}{\hat{t}}\ \sum_{n=0}^\infty \frac{1}{n!}\ \frac{1}{\hat{s}-n}\ \prod_{K=1}^n(\hat{u}+ K)}
\end{equation}

There are simple s-channel poles at each integer $n$:
\begin{eqnarray}
\label{eq:esh78}
s\ =\ nM_s^2
\end{eqnarray}
Notice that the residue is a function of $\hat{u}$ only (mind the simple factor in front of the sum..).

\item
D.H.S duality: 
\bea
V_t(-\hat{s},-\hat{u})= V_t(-\hat{u},-\hat{s})
\eea

Thus Eq.~(\zz{\ref{eq:venred}}) can then be written as a sum of u-channel poles.
\begin{equation}
V_t\ =\ -\frac{\hat{s}\hat{u}}{\hat{t}}\ \sum_{n=0}^\infty \frac{1}{n!}\ \frac{1}{\hat{u}-n}\ \prod_{K=1}^n(\hat{s}+ K)
\end{equation}
We will almost always use the s-channel pole expansion though.

\item
Positivity of the residues: "The no-ghost theorem":

For $V_t$ to describe a scattering amplitude, the residues of the poles must be positive.
This is difficult to prove, and it is correct only if the dimensions of space-time are  
$D=26$ or $D=10$, for the bosonic and super-string respectively.

\item
Polynomial residues and spins:

The residue of the pole contains the angular part of the amplitude, which determines the spins of the exchanged resonances. If the residue is a polynomial in $\cos\theta$ of degree $k$, then there can be exchanges of spins from $0$ to $k$. It is seen from Eq.~(\zz{\ref{eq:venred}}) that the residue is a polynomial of degree $n$ in $\hat{u}$ or equivalently in $\cos\theta$.
\bea
Res(V_t) \propto \prod_{K=0}^{n-1}(\hat{u}+ K)= \hat{u}(\hat{u}+1)\cdots (\hat{u}+n-1)= \sum_{p=1}^n a_p\hat{u}^p=  \sum_{p=1}^n b_p\cos ^p \theta
\eea
Where $a_p$ and $b_p$ are constants.

$V_t$ is the form factor which multiplies $m_{QCD}$ (Eq.~(\zz{\ref{eq:ff780}})), which itself depends on $\cos\theta$. The angular dependency in $m_{QCD}$ can shift the minimum and maximum spins, so that in general: 
\bea
0+J_0\  \leq  \ J \   \leq \ n+ J'_0
\eea

\end{enumerate}

\begin{figure}
\centering
\includegraphics[width= 80mm]{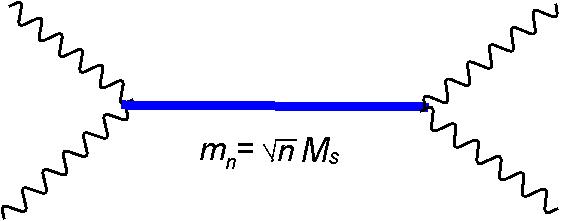}
\caption{s-channel Regge state exchange.\label{r11}}
\end{figure}

From Eq.~(\zz{\ref{eq:esh78}}), there are exchanges of of an infinite number of resonances with masses (Fig.~\textcolor{yy}{\ref{r11}}):
\bea
m_n\ \ =\ \ \sqrt{n}\ M_s
\eea
These are string excitations called \emph{Regge states}.

\begin{figure}
\centering
\includegraphics[width= 140mm]{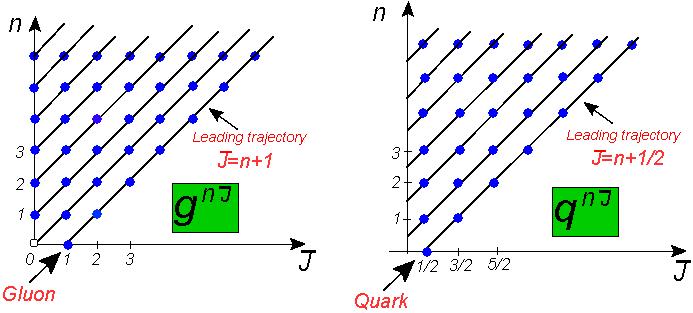}
\caption{Left: The spectrum of Regge excitations of the gluon: $g^{nJ}$. Right: Regge excitations of the quark: $q^{nJ}$. Linear Regge trajectories are also shown.\label{chewblue1}}
\end{figure}

We will see that the Regge excitations of the gluon (which we denote by $g^{nJ}$) have, at a mass level $n$, spins in the range:
\bea
0 \leq \ J \ \leq  n+1
\eea
These $g^{nJ}$ particles are exchanged e.g. in $gg \to gg$.
The spectrum of $g^{nJ}$ states can be plotted on the $n-J$ plane as in Fig.~\textcolor{yy}{\ref{chewblue1}}.

In sections \ref{subsubsec:b2}, \ref{subsubsec:disc15} we mentioned that the processes $gg \to gg$ and $gq \to gq$ have \underline{the same form factor} $V_t$. So the amplitude for $gq \to gq$ will also have an infinity of poles, corresponding to the Regge excitations of the quark $q^{nJ}$ (Fig.~\textcolor{yy}{\ref{chewblue1}}).
The figure also shows Regge trajectories which have the form $J=J_0+\alpha'm_n^2= J_0+n$, where $J_0= 0 \text{ or } 1/2$ for $g^{nJ}$ and $q^{nJ}$ respectively.

Near a resonance, one term in the sum of Eq.~(\zz{\ref{eq:venred}}) is dominant: 
\begin{eqnarray}
\label{eq:b4}
V_t\ \  \xrightarrow{\hat{s} \to n}\ \  \frac{1}{(n-1)!}\ \frac{1}{\hat{s}-n}\ \hat{u} \prod_{K=1}^{n-1}(\hat{u}+ K)
\nonumber\\[9pt]
V_u\ \  \xrightarrow{\hat{s} \to n}\ \  \frac{(-1)^{n-1}}{(n-1)!}\ \frac{1}{\hat{s}-n}\ \hat{t} \prod_{K=1}^{n-1}(\hat{u}+ K)
\end{eqnarray}

In contrast to $V_t$ and $V_u$, $V_s$ is finite (has no s-channel poles):
\begin{eqnarray}
\label{eq:b46}
V_s\ \  \xrightarrow{\hat{s} \to n}\ \  \frac{\Gamma(1-\hat{t})\Gamma(1-\hat{u})}{\Gamma(1+n)}\ =\ \text{finite}
\end{eqnarray}
so near a resonance of the amplitude we can neglect $V_s$ terms.

An interesting property can be seen: 
\bea
\frac{V_u}{V_t}= \frac{\Gamma(1-\hat{t})\ \Gamma(1+\hat{t})}{\Gamma(1-\hat{u})\ \Gamma(1+\hat{u})}=\frac{\hat{t}\ \sin (\pi \hat{u})}{\hat{u}\ \sin (\pi \hat{t})}
\eea
near a pole this becomes:
\bea
\label{eq:int76}
\frac{V_u}{V_t} \ = \ (-1)^{n-1}\frac{\hat{u}}{\hat{t}} \ \ \ ,\ \ \ \text{when } \hat{s}\to n
\eea
From Eq.~(\zz{\ref{eq:b4}}) the previous equation is seen to be equivalent to:
\bea
\label{eq:denden2}
\prod_{K=1}^{n-1}(\hat{u}+ K)\ =\ (-1)^{n-1} \prod_{K=1}^{n-1}(\hat{t}+ K) \ \ \ ,\ \ \ \text{when } \hat{s}\to n
\eea

\subsubsection{The squared amplitudes}
\label{subsubsec:a7}

\emph{References:} squared amplitudes from~\textcolor{rr}{\cite{lust}}.\\

In Appendix~\ref{sec:b8} we give an example of a full calculation of a squared amplitude.
In this section we write down the string squared amplitudes just as we did in the field theory case.
The Veneziano factors will be written explicitly in terms of the Mandelstam variables. It is then immediately seen that some the processes have only $n=$odd resonances (the residues vanish for $n=$even). This is explained in section \ref{subsubsec:exglue1} and Appendix \ref{sec:apphelamp}.

We denote by $A$ the $U(1)$ gauge boson from the stack $a$ (the color stack), and by $B$ the non-abelian gauge boson from stack $b$.

\begin{itemize}
\item {\large $gg$ initial state}
\end{itemize}

\begin{eqnarray}
|\mathcal{M}|^2(\textcolor{blue}{gg\to gg})\ \ =\ \ \ \ \ \ \ \ \ \ \ \ \ \ \ \ \ \ \ \ \ \ \ \ \ \ \ \ \ \ \ \ \ \ \ \ \ \ \ \ \  \ \ \ \ \ \ \ \ \ \ \ \ \ \ \ \ \ \ \ \ \ \ 
\nn[7pt] g^4\Big(\frac{1}{s^2}+\frac{1}{t^2}+\frac{1}{u^2}\Big)\Big[\frac{9}{4}\big(s^2V_s^2+t^2V_t^2+u^2V_u^2\big)\ -\ \frac{1}{3}\big(sV_s+tV_t+uV_u\big)^2\Big]
\nonumber\\[9pt]
= \ \frac{n^4+\hat{u}^4+\hat{t}^4}{n^2(\hat{s}-n)^2}\ \frac{g^4}{(n-1)!^2}\ \Bigg[ \prod_{K=1}^{n-1}(\hat{u}+ K)^2\Bigg]   
\cdot \begin{Bmatrix} 19/12 , & \textrm{odd}\ n\\ 9/4, & \textrm{even}\ n \end{Bmatrix}. 
\end{eqnarray}

\begin{eqnarray}
\label{eq:gon10}
|\mathcal{M}|^2(\textcolor{blue}{gg\to gA})= \frac{5}{6}g^4 Q_A^2\ \Big(\frac{1}{s^2}+\frac{1}{t^2}+\frac{1}{u^2}\Big)\big(sV_s+tV_t+uV_u\big)^2
\nonumber\\[9pt]
=\frac{5g^4Q_A^2}{3}\ \frac{n^4+\hat{u}^4+\hat{t}^4}{n^2(\hat{s}-n)^2}\ \frac{1}{(n-1)!^2}\ \Bigg[ \prod_{K=1}^{n-1}(\hat{u}+ K)^2\Bigg]\cdot   \begin{Bmatrix} 1, & \textrm{odd $n$}\\ 0, & \textrm{even $n$}  \end{Bmatrix}. 
\end{eqnarray}

\begin{eqnarray}
\label{eq:gon11}
|\mathcal{M}|^2(\textcolor{blue}{gg\to AA})\ \ =\ \ \frac{12}{5}Q_A^2\ |\mathcal{M}|^2(gg\to gA)
\end{eqnarray}

\begin{eqnarray}
\label{eq:treeate2}
|\mathcal{M}|^2(\textcolor{blue}{gg\to q\bar{q}})\ =\  g^4 \frac{t^2+u^2}{s^2}\ \Big[\frac{1}{6ut}\big(tV_t+uV_u\big)^2-\frac{3}{8}V_t V_u\Big]
\nonumber\\[10pt]
=\frac{\hat{u}\hat{t}(\hat{u}^2+\hat{t}^2)}{n^2(\hat{s}-n)^2}\ \frac{g^4}{(n-1)!^2}\ \Bigg[\prod_{K=1}^{n-1}(\hat{u}+ K)^2\Bigg]  \cdot \begin{Bmatrix} 7/24, & \textrm{odd $n$}\\  25/24, & \textrm{even $n$} \end{Bmatrix}. 
\end{eqnarray}

\begin{itemize}
\item {\large $gq$ initial state}
\end{itemize}

\begin{eqnarray}
|\mathcal{M}|^2(\textcolor{blue}{gq\to gq})\ =\  g^4 \frac{s^2+u^2}{t^2}\ \Big[V_s V_u-\frac{4}{9}\frac{1}{su}\ \big(sV_s+uV_u\big)^2\Big]
\nonumber\\[8pt]
=-\frac{4g^4}{9}\ \frac{\hat{u}(\hat{u}^2+n^2)}{n(\hat{s}-n)^2}\ \frac{1}{(n-1)!^2}\ \prod_{K=1}^{n-1}(\hat{u}+ K)^2
\end{eqnarray}

\begin{eqnarray}
|\mathcal{M}|^2(\textcolor{blue}{gq\to Aq})\ =\  \frac{-g^4}{3}Q_A^2\ \frac{s^2+u^2}{sut^2}\ \big(sV_s^2+uV_u^2\big)
\nonumber\\[8pt]
=\frac{-1}{3}g^4Q_A^2\ \frac{\hat{u}(\hat{u}^2+n^2)}{n(\hat{s}-n)^2}\ \frac{1}{(n-1)!^2}\ \prod_{K=1}^{n-1}(\hat{u}+ K)^2
\end{eqnarray}

\begin{eqnarray}
|\mathcal{M}|^2(\textcolor{blue}{gq\to Bq'})\ =\  \frac{-g^4}{6}|T^B_{q\bar{q}'}|^2\ \frac{s^2+u^2}{su}\ V_t^2
\nonumber\\[8pt]
=\frac{-g^4}{6}|T^B_{q\bar{q}'}|^2\ \frac{\hat{u}(\hat{u}^2+n^2)}{n(\hat{s}-n)^2}\ \frac{1}{(n-1)!^2}\ \prod_{K=1}^{n-1}(\hat{u}+ K)^2
\end{eqnarray}

\begin{itemize}
\item {\large $q\bar{q}$ initial state}
\end{itemize}

\begin{eqnarray}
|\mathcal{M}|^2(\textcolor{blue}{q\bar{q}\to gg})\ \ =\ \  \Big(\frac{8}{3}\Big)^2\ |\mathcal{M}|^2(gg\to q\bar{q})
\end{eqnarray}

\begin{eqnarray}
\label{eq:gon12}
|\mathcal{M}|^2(\textcolor{blue}{q\bar{q}\to gA})\ \ =\ \  g^4\frac{8}{9}Q_A^2\ \frac{t^2+u^2}{tus^2}\ \big(tV_t+uV_u\big)^2=\ \ \ \ \ \ \ \ \ \ \ \ \ \ 
\nonumber\\[8pt]
\frac{32}{9}g^4Q_A^2\ \frac{\hat{u}\hat{t}(\hat{u}^2+\hat{t}^2)}{n^2(\hat{s}-n)^2}\ \frac{1}{(n-1)!^2} \ \Bigg[\prod_{K=1}^{n-1}(\hat{u}+K)^2\Bigg]\cdot  \begin{Bmatrix} 1, & \textrm{odd $n$}\\  0, & \textrm{even $n$} \end{Bmatrix}. 
\end{eqnarray}

\begin{eqnarray}
\label{eq:gon13}
|\mathcal{M}|^2(\textcolor{blue}{q\bar{q}\to AA})\ \ =\ \  \frac{3}{4}Q_A^2\ |\mathcal{M}|^2(q\bar{q}\to gA)
\end{eqnarray}

\begin{equation}
\label{eq:bnlmp}
|\mathcal{M}|^2(\textcolor{blue}{q\bar{q}\to gB})\ \ =\ \  g^4\frac{4}{9}|T^B_{q\bar{q}}|^2Q_A^2\ \frac{t^2+u^2}{tu}\ V_s^2
\end{equation}

\begin{equation}
|\mathcal{M}|^2(\textcolor{blue}{q\bar{q}\to BA})\ \ =\ \  \frac{3}{4}Q_A^2\ |\mathcal{M}|^2(q\bar{q}\to gB)
\end{equation}

\begin{itemize}
\item {\large $qq$ initial state}
\end{itemize}

\begin{eqnarray}
|\mathcal{M}|^2(\textcolor{blue}{qq\to qq})
\ =\  g^4 \bigg\{ \frac{2}{9}\ \frac{1}{t^2}\Big[(sF_{tu}^{bb})^2+(sF_{tu}^{cc})^2+(uG_{tu}^{'bc})^2+(uG_{tu}^{'cb})^2\Big]+ 
\nonumber\\[6pt]
\frac{2}{9}\ \frac{1}{u^2}\Big[(sF_{ut}^{bb})^2+(sF_{ut}^{cc})^2+(tG_{ut}^{'bc})^2+(tG_{ut}^{'cb})^2\Big]- 
\frac{4}{27}\ \frac{s^2}{tu} \big(F_{tu}^{bb}F_{ut}^{bb}+F_{tu}^{cc}F_{ut}^{cc}\big) \bigg\}
\end{eqnarray}

\begin{equation}
|\mathcal{M}|^2(\textcolor{blue}{qq'\to qq'})\ = \ \frac{2g^4}{9}\ \frac{1}{t^2}\Big[(sF_{tu}^{bb})^2+(sG_{tu}^{cc'})^2+(uG_{tu}^{'bc})^2+(uG_{tu}^{'bc'})^2\Big]
\end{equation}
See \rr{\cite{lust}} for further details on quark-quark scattering.\\

Let us note a few things that can be seen from these squared amplitudes:
\begin{itemize}
\item As seen in Eqs.~(\zz{\ref{eq:gon10}}), (\zz{\ref{eq:gon11}}), (\zz{\ref{eq:gon12}}), (\zz{\ref{eq:gon13}}), amplitudes with a $gA$ or $AA$ in the final state do not have $n=even$ poles. This happens because of the vanishing of the following color factors: $f^{0ab}= f^{00a}= f^{000}=0$. See also section~\ref{subsubsec:exglue1} and Appendix~\ref{sec:apphelamp}.

\item Taking the leading term near a pole, we see that there are 3 classes of amplitudes:
\begin{enumerate}
\item $gq \to gq$ , $gq \to Aq$ , $gq \to Bq'$ are proportional near a pole.
\item $gg \to gg$ , $gg \to gA$ , and $gg \to AA$ are proportional near a pole (the latter two vanish at $n=even$ poles.). 
\item $gg \to q\bar{q}$ , $q\bar{q} \to gg$ , $q\bar{q} \to gA$ , and $q\bar{q} \to AA$ are proportional near a pole (the latter two vanish at $n=even$ poles.). 
\end{enumerate}

Moreover, these 3 classes differ (near a pole) only by a simple kinematic factor: $\hat{u}(\hat{u}^2+n^2)$ , $(n^4+\hat{u}^4+\hat{t}^4)$ , $\hat{u}\hat{t}(\hat{u}^2+\hat{t}^2)$ for the first, second, and third class respectively.
\end{itemize}

\subsubsection{The softened squared amplitudes}
\label{subsubsec:softened14}

The simple poles of the amplitudes are given finite widths via the Breit-Wigner form as in section~\ref{subsubsec:bw}.
Following for example \rr{\cite{anc9}}, we write the softened squared amplitudes for exchange of Regge states from the first excited state $n=1$. The Regge states with quantum numbers $(n,J)$ are written as: $g^{n,J}, A^{n,J}, q^{n,J}$, recall also Fig.~\yy{\ref{chewblue1}}. 

\begin{itemize}
\item {\large $gg$ initial state}
\end{itemize}

\begin{eqnarray}
|\mathcal{M}|^2(\textcolor{blue}{gg\to gg})
= \frac{19g^4}{12}\Bigg\{ \ W_{g^{n,J}}^{gg\to gg}\bigg[\frac{1}{(\hat{s}-1)^2+(\Gamma_{g^{1,0}}/M_s)^2}+\frac{\hat{t}^4+\hat{u}^4}{(\hat{s}-1)^2+(\Gamma_{g^{1,2}}/M_s)^2}\bigg]
\nonumber\\[8pt]
+W_{A^{n,J}}^{gg\to gg}\bigg[\frac{1}{(\hat{s}-1)^2+(\Gamma_{A^{1,0}}/M_s)^2}+\frac{\hat{t}^4+\hat{u}^4}{(\hat{s}-1)^2+(\Gamma_{A^{1,2}}/M_s)^2}\bigg]\ \Bigg\}
\end{eqnarray}

\begin{eqnarray}
|\mathcal{M}|^2(\textcolor{blue}{gg\to gA})\ =\  \frac{5}{3}g^4Q_A^2 \Bigg[\frac{1}{(\hat{s}-1)^2+(\Gamma_{g^{1,0}}/M_s)^2}+\frac{\hat{u}^4+\hat{t}^4}{(\hat{s}-1)^2+(\Gamma_{g^{1,2}}/M_s)^2}\Bigg]
\end{eqnarray}

\begin{eqnarray}
|\mathcal{M}|^2(\textcolor{blue}{gg\to AA})\ =\  4g^4Q_A^4 \Bigg[\frac{1}{(\hat{s}-1)^2+(\Gamma_{A^{1,0}}/M_s)^2}+\frac{\hat{u}^4+\hat{t}^4}{(\hat{s}-1)^2+(\Gamma_{A^{1,2}}/M_s)^2}\Bigg]
\end{eqnarray}

\begin{eqnarray}
|\mathcal{M}|^2(\textcolor{blue}{gg\to q\bar{q}})= \frac{7}{24}g^4 N_f \bigg\{W_{g^{n,J}}^{gg\to q\bar{q}}  \frac{\hat{u}\hat{t}(\hat{u}^2+\hat{t}^2)}{(\hat{s}-1)^2+(\Gamma_{g^{1,2}}/M_s)^2}+W_{A^{n,J}}^{gg\to q\bar{q}}\frac{\hat{u}\hat{t}(\hat{u}^2+\hat{t}^2)}{(\hat{s}-1)^2+(\Gamma_{A^{1,2}}/M_s)^2}\bigg\}
\nonumber\\
\end{eqnarray}

\begin{itemize}
\item {\large $gq$ initial state}
\end{itemize}

\begin{eqnarray}
|\mathcal{M}|^2(\textcolor{blue}{gq\to gq})= -\frac{4g^4}{9} \Bigg[\frac{\hat{u}}{(\hat{s}-1)^2+(\Gamma_{q^{1,\frac{1}{2}}}/M_s)^2}+\frac{\hat{u}^3}{(\hat{s}-1)^2+(\Gamma_{q^{1,\frac{3}{2}}}/M_s)^2}\Bigg]
\end{eqnarray}

\begin{eqnarray}
|\mathcal{M}|^2(\textcolor{blue}{gq\to Aq})= -\frac{1}{3}g^4Q_A^2 \Bigg[\frac{\hat{u}}{(\hat{s}-1)^2+(\Gamma_{q^{1,\frac{1}{2}}}/M_s)^2}+\frac{\hat{u}^3}{(\hat{s}-1)^2+(\Gamma_{q^{1,\frac{3}{2}}}/M_s)^2}\Bigg]
\end{eqnarray}

\begin{eqnarray}
|\mathcal{M}|^2(\textcolor{blue}{gq\to B q'})= -\frac{1}{6}g^4|T^B_{q\bar{q}'}|^2 \Bigg[\frac{\hat{u}}{(\hat{s}-1)^2+(\Gamma_{q^{1,\frac{1}{2}}}/M_s)^2}+\frac{\hat{u}^3}{(\hat{s}-1)^2+(\Gamma_{q^{1,\frac{3}{2}}}/M_s)^2}\Bigg]
\end{eqnarray}

\begin{itemize}
\item {\large $q\bar{q}$ initial state}
\end{itemize}

\begin{eqnarray}
|\mathcal{M}|^2(\textcolor{blue}{q\bar{q}\to gg})= \frac{56}{27}g^4 \bigg\{W_{g^{n,J}}^{ q\bar{q}\to gg}\frac{\hat{u}\hat{t}(\hat{u}^2+\hat{t}^2)}{(\hat{s}-1)^2+(\Gamma_{g^{1,2}}/M_s)^2}+W_{A^{n,J}}^{q\bar{q}\to gg}\frac{\hat{u}\hat{t}(\hat{u}^2+\hat{t}^2)}{(\hat{s}-1)^2+(\Gamma_{A^{1,2}}/M_s)^2}\bigg\}
\nonumber\\
\end{eqnarray}

\begin{eqnarray}
|\mathcal{M}|^2(\textcolor{blue}{q\bar{q}\to gA})= \frac{32}{9}g^4Q_A^2\frac{\hat{u}\hat{t}(\hat{u}^2+\hat{t}^2)}{(\hat{s}-1)^2+(\Gamma_{g^{1,2}}/M_s)^2}
\end{eqnarray}

\begin{eqnarray}
|\mathcal{M}|^2(\textcolor{blue}{q\bar{q}\to AA})= \frac{8}{3}g^4Q_A^4\frac{\hat{u}\hat{t}(\hat{u}^2+\hat{t}^2)}{(\hat{s}-1)^2+(\Gamma_{A^{1,2}}/M_s)^2}
\end{eqnarray}

Where the widths are: 
\bea
\Gamma_{g^{1,0}} \ =\ \frac{g^2}{4\pi}\ \frac{N}{4}\ M_s \ \approx\ 0.075\ M_s
\eea
\bea
\Gamma_{A^{1,0}} \ =\ \frac{g^2}{4\pi}\ \frac{N}{2}\ M_s \ \approx\ 0.15\ M_s
\eea
\bea
\Gamma_{g^{1,2}} \ =\ \frac{g^2}{4\pi}\ \Big(\frac{N}{10}+\frac{N_f}{40}\Big)\ M_s \ \approx\ 0.045\ M_s
\eea
\bea
\Gamma_{A^{1,2}} \ =\ \frac{g^2}{4\pi}\ \Big(\frac{N}{5}+\frac{N_f}{40}\Big)\ M_s \ \approx\ 0.075\ M_s
\eea
\bea
\Gamma_{q^{1,\frac{1}{2}}} \ =\ \Gamma_{q^{1,\frac{3}{2}}} \ =\  \frac{g^2}{4\pi}\ \frac{N}{8}\ M_s \ \approx\ 0.038\ M_s
\eea
and the right hand side was obtained by setting $\alpha_s= g^2/4\pi \approx 0.1$, $N=3$, and $N_f=6$.

We now write the relative weights between exchange of an $SU(N)$ and $U(1)$ gauge bosons.
\begin{eqnarray}
W_{g^{nJ}}^{ gg\to gg}=\frac{(N^2-1)(\Gamma_{g^{nJ}\to gg})^2}{(N^2-1)(\Gamma_{g^{nJ}\to gg})^2+(\Gamma_{A^{nJ}\to gg})^2}\ =\  \begin{Bmatrix} \frac{1}{1+\frac{4(N^2-1)}{(N^2-4)^2}}=\frac{25}{57}\ ,\  n=odd\\  \ \ \ \ \ \ \ \ \ \ \ \ \ \ \ \ \ 1\ ,\ n=even
\end{Bmatrix}
\end{eqnarray}

\begin{eqnarray}
W_{A^{nJ}}^{ gg\to gg}= 1-W_{g^{nJ}}^{ gg\to gg} = \frac{(\Gamma_{A^{nJ}\to gg})^2}{(N^2-1)(\Gamma_{g^{nJ}\to gg})^2+(\Gamma_{A^{nJ}\to gg})^2} \ =\  \begin{Bmatrix} \frac{32}{57}\ ,\  n=odd \\  \ 0\ ,\  n=even \end{Bmatrix}
\end{eqnarray}

\begin{eqnarray}
W_{g^{nJ}}^{ gg\to q\bar{q}}= W_{g^{nJ}}^{q\bar{q}\to gg} = \frac{(N^2-1)\Gamma_{g^{nJ}\to gg}\Gamma_{g^{nJ}\to q\bar{q}}}{(N^2-1)\Gamma_{g^{nJ}\to gg}\Gamma_{g^{nJ}\to q\bar{q}}+\Gamma_{A^{nJ}\to gg}\Gamma_{A^{nJ}\to q\bar{q}}} \ =\ 
\nonumber\\[10pt]
\begin{Bmatrix} \frac{1}{1+\frac{2}{N^2-4}}= \frac{5}{7}\ ,\  n=odd \\  \ \ \ \ \ \ \ \ \ \ \ \ \ 1\ ,\  n=even \end{Bmatrix}
\end{eqnarray}

\begin{eqnarray}
W_{A^{nJ}}^{ gg\to q\bar{q}}= W_{A^{nJ}}^{  q\bar{q}\to gg} = 1-W_{g^{nJ}}^{gg\to q\bar{q}} = \frac{\Gamma_{A^{nJ}\to gg}\Gamma_{A^{nJ}\to q\bar{q}}}{(N^2-1)\Gamma_{g^{nJ}\to gg}\Gamma_{g^{nJ}\to q\bar{q}}+\Gamma_{A^{nJ}\to gg}\Gamma_{A^{nJ}\to q\bar{q}}}\ =\ 
\nonumber\\[10pt]
\begin{Bmatrix} \frac{2}{7}\ ,\  n=odd \\  0\ ,\  n=even \end{Bmatrix}
\end{eqnarray}
Where we put $N=3$ on the right hand side, and the decay widths are taken from sections \ref{subsubsec:exglue1} and \ref{subsubsec:excquarks7}.
The weights are independent of $n$ and $J$, they only differ for $n=odd$ and $n=even$.

Now we that we have seen the $n=1$ case, we jump a little bit ahead of time and give the prescription for arbitrary $n$ and $J$.
A general helicity amplitude will have the following form near a resonance of mass squared $m_n^2=\hat{s}=n$, see Eq.~(\zz{\ref{eq:a4}}):

\begin{eqnarray} 
\mathcal{M}_{m,m'}\ =\    \frac{1}{(\hat{s}-n)+i(\Gamma_{g^{nJ}}/M_s)}\ A \ \sum_{J}C^{n,J}_{m,m'}\ d^J_{m,m'}(\theta)
\end{eqnarray}
We wrote this with $g^{nJ}$, but the same form will hold also for $q^{nJ}$.

The amplitude can then be squared:
\begin{eqnarray} 
|\mathcal{M}_{m,m'}|^2\ =\  
A^2 \sum_{J_1 , J_2}C^{n,J_1}_{m,m'}\ C^{n,J_2}_{m,m'}\ 
\frac{d^{J_1}_{m,m'}\ d^{J_2}_{m,m'}}{(\hat{s}-n)^2+\Gamma_{g^{n J_1}}\ \Gamma_{g^{n J_2}}/M_s^2}
\end{eqnarray}

Eq.~(\zz{\ref{eq:orthds}}) shows that $d$'s are orthogonal, therefore the interference terms vanish in the total cross section:
\begin{eqnarray}
\sigma_{m,m'}\ =\  \int_{-1}^1\frac{d\cos\theta}{32\pi s}|\mathcal{M}_{m,m'}|^2\ =\  \frac{A^2}{32\pi s} \sum_J \big(C^{n,J}_{m,m'}\big)^2\ \frac{\frac{4}{2J+1}}{(\hat{s}-n)^2+(\Gamma_{g^{nJ}}/M_s)^2}
\end{eqnarray}

The amplitudes having exchanges of $g^{nJ}$ (as opposed to $q^{nJ}$) have also the opposite helicity configuration $\mathcal{M}_{m,-m'}$. In this case, when the two squared amplitudes are added we get:
\begin{eqnarray} 
|\mathcal{M}_{m,m'}|^2+|\mathcal{M}_{m,-m'}|^2= A^2 \sum_{J_1 , J_2} C^{n,J_1}_{m,m'}\ C^{n,J_2}_{m,m'}\ 
\bigg[ \frac{d^{J_1}_{m,m'} d^{J_2}_{m,m'} + (-1)^{J_1+J_2}\ d^{J_1}_{m,-m'} d^{J_2}_{m,-m'}}{(\hat{s}-n)^2+\Gamma_{g^{nJ_1}}\ \Gamma_{g^{nJ_2}}/M_s^2}\bigg]
\end{eqnarray}
Where Eqs.~(\zz{\ref{eq:dean331}}), (\zz{\ref{eq:dean332}}) were used.

\subsubsection{Low energy limit}
\label{subsubsec:low89}

\emph{References:} ~\textcolor{rr}{\cite{lust, chandia, collinucci}}.\\

When the center of mass energy is significantly lower than the string scale, $s<<M_s^2$, the string amplitudes coincide with the standard model ones. In this section we calculate the first stringy correction to the standard model amplitudes.\\

The Veneziano amplitudes can be expanded in powers of $\alpha'= 1/M_s^2$. To order $\alpha '^7$:

\begin{eqnarray}
\frac{1}{st}V_u\ =\ \frac{1}{st}-\alpha'^2 \frac{\pi^2}{6}-\alpha'^3(s+t)\zeta (3)-\alpha'^4 \frac{\pi^4}{360}(4s^2+st+4t^2)+\ \ \ \ \ \ \ \ \ \ \ \ \ \ \ \ \ \ \ \ \ \ \ \ \ \ \ \ \ 
\nonumber\\[5pt]
\alpha'^5\Big[\frac{\pi^2}{6}st(s+t)\zeta(3)-(s+t)(s^2+st+t^2)\zeta(5)\Big]-\ \ \ \ \ \ \ \ \ \ \ \ \ \ \ \ \ \ \ \ \ \ \ \ \ \ \ \ \ \ \ \ \ \ \ \ \ \ \ \ \ \ \ \ \ \ \ \  
\nonumber\\[5pt]
\alpha'^6\Big[\frac{\pi^6}{15120}16s^4+12s^3t+23s^2t^2+12st^3+16t^4)-\frac{1}{2}st(s+t)^2\zeta(3)^2\Big]+\ \ \ \ \ \ \ \ \ \ \ \ \ \ \ \ \ \ \ \ \ \ 
\nonumber\\[5pt]
\alpha'^7 \Big[ \frac{\pi^4}{360}st(s+t)(4s^2+st+4t^2)\zeta(3)+\frac{\pi^2}{6}st(s^2+st+t^2)\zeta(5)-(s^2+st+t^2)\zeta(7)\Big]+...\nonumber\\
\end{eqnarray}
We note that the $\alpha'$ correction vanishes.
We will need only up to $\alpha'^3$ order\footnote{In this equation, the $\alpha'^2$ correction can be shown to arise from the following effective lagrangian: 
\begin{eqnarray}
\mathcal{I}_{F^4}= -\frac{\alpha'^2\pi^2}{6}Tr\Big[F_{\mu_1\mu_2}F_{\mu_2\mu_3}F_{\mu_3\mu_4}F_{\mu_4\mu_1}+
2F_{\mu_1\mu_2}F_{\mu_3\mu_4}F_{\mu_2\mu_3}F_{\mu_4\mu_1}-
\nonumber\\
\frac{1}{4}F_{\mu_1\mu_2}F_{\nu_1\nu_2}F_{\mu_2\mu_1}F_{\nu_2\nu_1}-\frac{1}{2}F_{\mu_1\mu_2}F_{\mu_2\mu_1}F_{\nu_1\nu_2}F_{\nu_2\nu_1}\Big]
\end{eqnarray} }:

\begin{eqnarray}
\label{eq:ghgp2}
V_t \ =\  1-\Big[\frac{\pi^2}{6}su\Big]\alpha'^2-[\zeta(3)stu]\alpha'^3\ldots
\nonumber\\[5pt]
V_u \ =\  1-\Big[\frac{\pi^2}{6}st\Big]\alpha'^2-[\zeta(3)stu]\alpha'^3\ldots
\nonumber\\[5pt]
V_s \ =\  1-\Big[\frac{\pi^2}{6}tu\Big]\alpha'^2-[\zeta(3)stu]\alpha'^3\ldots
\end{eqnarray}

We write $|\mathcal{M}|^2=|\mathcal{M}|^2_{SM}+\Delta |\mathcal{M}|^2$, where $|\mathcal{M}|^2_{SM}$ is the standard model squared amplitude and $\Delta |\mathcal{M}|^2$ is the first correction.
The first corrections to the squared amplitudes of section \ref{subsubsec:a7} are:

\begin{itemize}
\item {\large $gg$ initial state}
\end{itemize}

\begin{equation}
\Delta |\mathcal{M}|^2(\textcolor{blue}{gg\to gg})\ =\  -2g^4\zeta(3)stu \frac{9}{2}\Big[3-\frac{tu}{s^2}-\frac{us}{t^2}-\frac{ts}{u^2}\Big]\ \alpha'^3
\end{equation}

\begin{equation}
\Delta |\mathcal{M}|^2(\textcolor{blue}{gg\to gA})\ =\ \frac{5\pi^4g^4Q_A^2}{24}\ (s^2t^2+s^2u^2+u^2t^2)\ \alpha'^4
\end{equation}

\begin{equation}
\Delta |\mathcal{M}|^2(\textcolor{blue}{gg\to AA})\ =\ \frac{\pi^4g^4Q_A^4}{2}\ (s^2t^2+s^2u^2+u^2t^2)\ \alpha'^4
\end{equation}

\begin{equation}
\Delta |\mathcal{M}|^2(\textcolor{blue}{gg\to q\bar{q}})\ =\ \frac{g^4\pi^2}{3}\ (u^2+t^2)\ \alpha'^2
\end{equation}

\begin{itemize}
\item {\large $gq$ initial state}
\end{itemize}

\begin{equation}
\Delta |\mathcal{M}|^2(\textcolor{blue}{gq\to gq})\ =\ \frac{\pi^2g^4Q_A^4}{3}\ (u^2+s^2)\ \alpha'^2
\end{equation}

\begin{equation}
\Delta |\mathcal{M}|^2(\textcolor{blue}{gq\to Aq})\ =\ \frac{-2g^4\pi^2}{9}\ (u^2+s^2)\ \alpha'^2
\end{equation}

\begin{equation}
\Delta |\mathcal{M}|^2(\textcolor{blue}{gq\to Bq'})\ =\ \frac{g^4\pi^2}{18}|T^B_{q\bar{q}}|^2\ (u^2+s^2)\ \alpha'^2
\end{equation}

\begin{itemize}
\item {\large $q\bar{q}$ initial state}
\end{itemize}

\begin{equation}
\Delta |\mathcal{M}|^2(\textcolor{blue}{q\bar{q}\to gg})\ =\ \frac{64g^4\pi^2}{27}\ (u^2+t^2)\ \alpha'^2
\end{equation}

\begin{equation}
\Delta |\mathcal{M}|^2(\textcolor{blue}{q\bar{q}\to gA})\ =\ \frac{8\pi^4g^4Q_A^2}{81}\ ut(u^2+t^2)\ \alpha'^4
\end{equation}

\begin{equation}
\Delta |\mathcal{M}|^2(\textcolor{blue}{q\bar{q}\to AA})\ =\ \frac{2\pi^4g^4Q_A^4}{27}\ ut(u^2+t^2)\ \alpha'^4
\end{equation}

\begin{equation}
\Delta |\mathcal{M}|^2(\textcolor{blue}{q\bar{q}'\to gB})\ =\ \frac{-4g^4\pi^4|T^B_{q\bar{q}}|^2}{27}\ ut(u^2+t^2)\ \alpha'^2
\end{equation}

\begin{equation}
\Delta |\mathcal{M}|^2(\textcolor{blue}{q\bar{q}'\to BA})\ =\ \frac{-g^4\pi^4|T^B_{q\bar{q}}|^2Q_A^2}{9}\ ut(u^2+t^2)\ \alpha'^2
\end{equation}

We note that for all the processes with a final state $A$, the first correction is $\alpha'^4$.
The correction for gluon scattering is $\alpha'^3$, and all the rest are $\alpha'^2$ .

\subsection{Collider phenomenology}
\label{subsec:phen4}
\emph{References:} ~\textcolor{rr}{\cite{perelstein, eichten, cms, cms2, cms3, atlas, atlas2, atlas3, lust, lust2, anc, anc2, anc3, anc4, anc5, anc6, anc7, anc8, anc9, dong, feng, kit, kit2, bar, per23, harris22}}.\\

The calculation of a cross section $d\widetilde{\sigma}$ is done by convoluting the partonic cross section $d\sigma (ij\to kl)$ with the \emph{parton distribution functions} of the two colliding protons:
\bea
d\widetilde{\sigma} = \int_0^1 \int_0^1 dx_a\ dx_b \sum_{ijkl}\ f_i(x_a, M)\ f_j(x_b, M)\ d\sigma (ij\to kl)
\eea

The partonic cross section and the squared amplitude are related:
\begin{equation} 
|\mathcal{M}(ij \to kl)|^2\ =\  64\pi^2 s \ \frac{d\sigma}{d \Omega} \ =\ 16\pi s^2\ \frac{d\sigma}{d t}
\end{equation}

A useful form for the dijet cross section is given in Eq.~(\zz{\ref{eq:dijetcross}}).
The dijet cross section can thus be calculated in field theory and in string theory, using the squared amplitudes written earlier. As is well known, the field theory cross section is a smooth power-law decreasing function. The string theory cross section exhibits bumps at $s=nM_s^2$, and these are clear signals of new physics. If the string scale is higher then the collision energy then these bumps cannot be seen, but smooth deviations from the field theory cross section can still be searched for (e.g. contact interaction searches).
 
Another useful type of analysis are dijet angular distributions. Angular distributions are a sensitive probe of new physics since QCD dijets are more central (because of the t-channel poles) whereas new physics tend to be more isotropic. Most importantly, angular distributions are a way to probe exchanges of different spins. Therefore they can be used to differentiate e.g a bump coming from a spin 2 KK graviton, from Regge state exchange of different spins.

The ratio $R$ is a useful measure of angular distributions:
\begin{equation} 
R\ =\  \frac{\frac{d\sigma}{dM}(|y_1|,|y_2|<0.5)}{\frac{d\sigma}{dM}(0.5<|y_1|,|y_2|<1)}
\end{equation}
It also has the benefit that systematic uncertainties, such as the jet energy scale (JES), tend to cancel in the ratio.

It is very important that the following 4-fermion processes, which are non-universal amplitudes, are suppressed at the LHC:
\bea
q\ \bar{q} \to q\ \bar{q}
\nn
q\ q \to q\ q
\eea
The first process is suppressed because $\bar{q}$ has low luminosity in proton collisions, and the second process does not have s-channel Regge state exchange.
Therefore, the universal (model independent) amplitudes will dominate the dijet signal.

Looking at the squared amplitudes of section~\ref{subsubsec:a7}, we note the following things:
\begin{enumerate}
\item The process $p+p \rightarrow \gamma /Z+ Jet$ has even resonances only from the partonic process $gq \rightarrow Aq$.
This means that the even resonances of a $\gamma/Z + Jet$ signal are a probe of the Regge excitations of the quarks.

\item 
The process $p+p \rightarrow W+Jet$ has resonances only from the partonic process $gq \rightarrow Bq'$.

\item 
The process  $p+p \rightarrow \gamma /Z+ \gamma/Z$ has only odd resonances.
\end{enumerate}

Possibilities for collider phenomenology other then virtual exchange of Regge states include:
\begin{itemize}
\item Direct production of a Regge state in the final state, section~\ref{sec:a2} and \rr{\cite{feng}}.

\item Multi-jets in the final state, beginning with the 3-jet signal. The squared amplitudes of \rr{\cite{lust2}} and section~\ref{subsubsec:5squared} can be used for this purpose.  

\item Phenomenology at a lepton collider or photon collider, \rr{\cite{anc6}}. For example the process $\gamma + \gamma \to \gamma + \gamma$ exhibits tree level Regge state exchange.

\item Signals other then Regge states include: production of $Z$' bosons coming from the extra $U(1)$'s of D-brane models (section~\ref{subsec:inter} and \rr{\cite{anc8}}). Also signals arising from the presence of extra dimensions:  KK and winding state exchange and production, miniature black hole production and Hawking radiation etc..

\end{itemize}

\subsubsection{Constraints from the LHC}

The most directly related limit is from the CMS experiment \rr{\cite{cms4}}: exclusion of string resonances from dijet mass distribution with $1\ fb^{-1}$:

\bea
\boxed{M_s\ >\ 4\ TeV}
\eea

We list some additional constraints which have some relevance for us.
\begin{itemize}

\item \rr{\cite{cms4}} CMS limits from dijets searches with $1\ fb^{-1}$ :

- Bound on the mass of excited quarks:

$M_{q^*}>2.49\ TeV$ \ .

- Bound on the mass of axigluons:

$M>2.47\ TeV$ \ .

\item \rr{\cite{cms2}}, \rr{\cite{cms3}} CMS lower limit on quark contact interaction scale for left handed quarks via dijet angular distributions with $36\ pb^{-1}$ :

$\Lambda > 5.6-6.7\ TeV$ \ .

\item  \rr{\cite{ab321}} ATLAS limits from dijet searches with $0.81\ fb^{-1}$:

- Bound on excited quarks: 

$M_{q^*}>2.91\ TeV$ \ .

-Bound on axigluons: 

$M>3.21\ TeV$ \ .

\item \rr{\cite{atlas}}, \rr{\cite{atlas2}}, \rr{\cite{atlas3}} ATLAS limits from dijet mass and angular distributions with $36\ pb^{-1}$ (distributions measured up to $\sim 3.5\ TeV$):

- Exclusion of quantum gravity scales from Randall-Meade quantum black holes: 

$0.75< m_D < 3.67\ TeV$ \ .

- Limit on quark contact interactions:

 $\Lambda > 9.5\ TeV$ \ .

\end{itemize}

Another way to discover new physics is by measuring the couplings of the Higgs boson, which (Maybe) was recently discovered at the LHC \rr{\cite{atlas71,cms71}} and Tevatron\rr{\cite{tevatron71}}. See for example \rr{\cite{car}}.

\clearpage


\section{\textcolor{ww}{Decay widths }}
\label{sec:dec}
\emph{References:} ~\textcolor{rr}{\cite{lust, anc5, dong, perelstein}}. After the submission of this article, \cite{feng22} appeared which deals with related issues.

In this section we suggest several methods to compute decay widths of Regge states. The basic idea is that of \cite{anc5}, in which a tree level amplitude is factorized into two trilinear couplings connected  by an s-channel resonance (see e.g the s-channel diagram of Fig.~\textcolor{yy}{\ref{stchan}}). This is e.g similar to the tree level production of a standard model $Z$ boson. In field theory, 1-loop corrections give an imaginary part to the amplitude which causes the resonance to decay. As we will shortly see, the optical theorem enables to compute the decay widths from tree level amplitudes.

This technique is basically field theoretical. 1-loop amplitudes can also be computed in string theory. This is beyond the scope this work.

\subsection{Setting the stage}
\label{subsec:sett}

\subsubsection{The Breit-Wigner form}
\label{subsubsec:bw}

In order to compare a squared amplitude to scattering experiments, the decay width of the exchanged particles must be taken in to account. Recall that the string amplitudes exhibit an infinite sum of poles of the type $\sim \frac{1}{s-nM_s^2}$ corresponding to exchange of Regge states with zero decay width. Higher order corrections will produce a finite decay width $\Gamma$, causing a \emph{Breit-Wigner} softening of the poles:

\begin{eqnarray}
\frac{1}{s-nM_s^2}\to \frac{1}{s-nM_s^2+i\Gamma M_s}
\end{eqnarray}

The squared amplitude will then be
\begin{eqnarray}
\sim \frac{1}{(s-nM_s^2)^2+(\Gamma M_s)^2}
\end{eqnarray}

\begin{figure}
\centering
\includegraphics[width= 130mm]{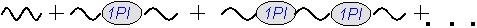}
\caption{The full propagator.\label{prop2}}
\end{figure}

We now derive this, following \rr{\cite{perelstein}}. In field theory (for example $\phi^3$ theory) a tree level s-channel exchange of a resonance is of the form ~$\frac{1}{s-M_0^2}$. Radiative corrections will remove the pole at $s=M_0^2$. If $\Pi(s)$ is the 1PI radiative correction to the tree propagator then the full propagator will be Fig.~\textcolor{yy}{\ref{prop2}}:

\begin{eqnarray}
\frac{1}{s-M_0^2}\ \ \longrightarrow \ \  \frac{1}{s-M_0^2}+\frac{1}{s-M_0^2}\Pi(s) \frac{1}{s-M_0^2}+\ldots \  =
\nonumber\\[5pt] 
\frac{1}{s-M_0^2}\bigg[1-\frac{\Pi(s)}{s-M_0^2}+\Big(\frac{\Pi(s)}{s-M_0^2}\Big)^2+\ldots \bigg]=
\frac{1}{s-M_0^2-\Pi(s)}
\end{eqnarray}

The physical mass is determined by:
\begin{eqnarray}
M^2-M_0^2 - Re\ \Pi(M^2)=0
\end{eqnarray}

Using the fact that near the pole we have $s\approx M^2$, we get
\begin{eqnarray}
M^2-M_0^2 -\Pi(s)= s-M_0^2 -\big[Re\ \Pi(M^2)+Re\ \Pi(s)'|_{s=M^2}(s-M^2)\big]+iIm\ \Pi(M^2)=  
\nonumber\\[5pt]
\big[1+Re\ \Pi(s)'|_{s=M^2}\big](s-M^2)+iIm\ \Pi(M^2)=
\nonumber\\[5pt]
\mathcal{Z}^{-1}(s-M^2)+iIm\ \Pi(M^2)= \mathcal{Z}^{-1}\big[(s-M^2)+iM \Gamma \big] 
\end{eqnarray}

Where $\mathcal{Z}$ is the field strength renormalization and in the last equation we used:
\begin{eqnarray}
Im\ \Pi(M^2)= -\frac{\mathcal{Z}^{-1}}{2}\sum_{f_1,f_2}\int \frac{d^3k_1}{(2\pi)^32E_1}\ \frac{d^3k_2}{(2\pi)^32E_2} |\overline{\mathcal{M}}|^2(Z\to f_1f_2)= -\mathcal{Z}^{-1}M\Gamma
\end{eqnarray}

We finally achieved:
\begin{eqnarray}
\frac{1}{s-M_0^2}\ \ \longrightarrow \ \   \frac{\mathcal{Z}}{s-M^2+iM\Gamma }
\end{eqnarray}

\subsubsection{Amplitudes in terms of the $d$-functions}
\label{subsubsec:d's}

In order to exhibit the exchange of resonances, a given amplitude should be expanded in terms of the physical states, i.e states with a definite spin. Put differently, the amplitude needs to be expanded on the basis of \emph{Wigner $d$-functions}.

Therefore we write an helicity amplitude as:
\begin{equation}
\mathcal{M}_{m,m'}=\mathcal{M}(12\to 34)= \sum_{J^*}\mathcal{M}(12\to J^*\to 34)
\end{equation}

or,
\begin{equation}
\mathcal{M}(12\to 34)\equiv \langle34;\theta|\mathcal{M}|12;0\rangle= \sum_{a,J}\langle34;\theta|\mathcal{M}^{aJ}|12;0\rangle
\end{equation}

Where, 

\begin{eqnarray}
\label{eq:n4}
\langle34;\theta|\mathcal{M}^{aJ}|12;0\rangle \ =\  \frac{1/M_s^2}{\hat{s}-n}\ F^{aJ}_{\lambda_3 \lambda_4;a_3a_4}\ F^{aJ}_{\lambda_1 \lambda_2;a_1a_2}\ d^{J}_{\lambda_1-\lambda_2;\lambda_3-\lambda_4} (\theta)
\end{eqnarray}
This equation can be viewed as the definition of the $F$'s which are called \emph{collinear amplitudes}. 
 
A general helicity amplitude will have the following form near a resonance of mass squared $m_n^2=\hat{s}=n$:
\begin{eqnarray} 
\label{eq:a4}
\mathcal{M}_{m,m'}\ =\    \frac{1}{\hat{s}-n}\ \kappa \ \Big( \sum_a \eta^a_{a_3a_4a_1a_2}\Big)  \ \sum_{J}C^{n,J}_{m,m'}\ d^J_{m,m'}(\theta)
\end{eqnarray}
Where $\kappa$ contains constants, and $\eta$ is the color factor of the amplitude. The tree level string helicity amplitudes will be written in this form in Eqs.~(\zz{\ref{eq:gface1}})-(\zz{\ref{eq:gfacelast}}).

\subsubsection{Decay widths}
\label{subsubsec:dec1}

We now show how the decay width can be calculated from the $F$'s, following \rr{\cite{anc5}}. 

\begin{figure}
\centering
\includegraphics[width= 60mm]{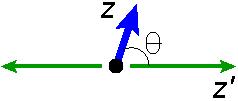}
\caption{A particle (with spin in the direction of the blue arrow) decaying to 2 particles.\label{dec1}}	
\end{figure}

Consider a particle at rest with mass $M$, spin $J$, and $J_z= \Lambda$, decaying in to two particles with helicities $\lambda_3$ and $\lambda_4$ moving in opposite directions along the $z'$ axis (see Fig.~\textcolor{yy}{\ref{dec1}}).

The S-matrix element for the decay is:
\begin{equation}
S=i(2\pi)^4\delta^4(P-p_3-p_4)\ \ \langle \vec{p}_3\lambda_3a_3\ ;\ \vec{p}_4\lambda_4a_4|\ \mathcal{L}\ |0,\Lambda,a\rangle
\end{equation}

The partial decay width into two particles with definite helicities and colors is:
\begin{eqnarray}
\Gamma^{aJ}_{\lambda_3\lambda_4;a_3a_4}\ =\  \frac{1}{2M}\frac{(2\pi)^4}{(2\pi)^6}\int d^4p_3d^4p_4\ \delta ^4 (P-p_3-p_4) \ \delta ^+ (p_3^2 -m_3^2)\ \delta ^+ (p_4^2 -m_4^2)
\nonumber\\[7pt]
 \times \ \ |\langle \vec{p}_3\lambda_3a_3\ ;\ \vec{p}_4\lambda_4a_4|\ \mathcal{L}\ |0,\Lambda,a\rangle|^2
\nonumber\\[7pt]
=\frac{p^*}{32\pi^2M^2}\int d\Omega_3 \ |\langle -\vec{p}_3\lambda_3a_3\ ;\ \vec{p}_4\lambda_4a_4|\ \mathcal{L}\ |0,\Lambda,a\rangle|^2
\end{eqnarray}

Where in the c.m frame:
\begin{equation}
p^*\ =\ |\vec{p}_3|\ =\ |\vec{p}_4|\ =\ \frac{M}{2}
\end{equation}

Now expand $|0,\Lambda \rangle$ on spin states in the $z'$ direction:
\begin{equation}
|0,\Lambda\rangle \ =\  \sum_{\Lambda'}|0,\Lambda '\rangle \ \langle 0,\Lambda '|0,\Lambda \rangle \ =\   \sum_{\Lambda'}\ d^J_{\Lambda \Lambda'}(\theta) \ |0,\Lambda '\rangle
\end{equation}

From angular momentum conservation:
\begin{equation}
\Lambda' \ =\ \lambda_3-\lambda_4
\end{equation}

So we get
\begin{eqnarray}
\Gamma^{aJ}_{\lambda_3\lambda_4;a_3a_4}\ =\ 
\frac{1}{64\pi^2M}|\langle\vec{p}_3\lambda_3a_3\ ;\ -\vec{p}_3\lambda_4a_4|\ \mathcal{L}\ |0,\Lambda',a\rangle|^2
\int d\Omega_3\ |d^{J}_{\Lambda, \lambda_3-\lambda_4}(\theta)|^2 
\nonumber\\[7pt]
\equiv \frac{1}{64\pi^2M} \ |F^{aJ}_{\lambda_3\lambda_4;a_3a_4}|^2\ \int d\Omega \ |d^{J}_{\Lambda, \lambda_3-\lambda_4}(\theta)|^2 
\end{eqnarray}

Since
\begin{equation}
\int d\Omega \ |d^J_{\Lambda,\lambda_3-\lambda_4}(\theta)|^2\ =\  \frac{4\pi}{2J+1}
\end{equation}

We finally get,
\begin{eqnarray}
\label{eq:parwidth}
\boxed{
\Gamma^{aJ}_{\lambda_3\lambda_4;a_3a_4}\ =\  \frac{1}{16(2J+1)\pi M}\ |F^{aJ}_{\lambda_3\lambda_4;a_3a_4}|^2  }
\end{eqnarray} 

Summing over colors and helicites,
\begin{equation}
\label{eq:totwidth}
\Gamma^{aJ}\ =\  \sum_{\lambda_3,\lambda_4}\sum_{a_3,a_4}\Gamma^{aJ}_{\lambda_3\lambda_4;a_3a_4}\ =\  \frac{1}{16(2J+1)\pi M}\sum_{\lambda_3,\lambda_4}\sum_{a_3,a_4}|F^{aJ}_{\lambda_3\lambda_4;a_3a_4}|^2
\end{equation}

The total decay width of a particle is the sum, over all allowed final states, of the partial decay widths.

\subsection{Calculations of decay widths}
\label{subsec:d'scal}

In this section, expressions for the decay widths of the quark and gluon Regge excitations will be derived in terms of the coefficients $C_{m,m'}^{n,J}$. In section~\ref{subsec:calc14}, four methods to calculate the $C$'s will be suggested.

We note that since we consider only initial and final states which are $n=0$ ground states (the standard model particles), the calculations do not include decays of Regge states into lower lying Regge states.

\subsubsection{Amplitudes in terms of the $d$-functions}

Consider the following 7 expansions, which will enable us to write our string helicity amplitudes on a basis of angular functions which exhibit exchanges of particles with a definite spin. 
\begin{equation}
\label{eq:a1}
\prod_{K=1}^{n-1}(\hat{u}+K)\ \ =\ \  \sum_{J=0}^{n-1}C_{0,0}^{n,J}\ d^J_{0,0}(\theta)
\end{equation}
\begin{equation}
\frac{\hat{t}^2}{\hat{s}^2}\prod_{K=1}^{n-1}(\hat{u}+K)\ \ =\ \  \sum_{J=2}^{n+1}C_{2,-2}^{n,J}\ d^J_{2,-2}(\theta)
\end{equation}
\begin{equation}
\frac{\hat{u}^2}{\hat{s}^2}\prod_{K=1}^{n-1}(\hat{u}+K)\ \ =\ \  \sum_{J=2}^{n+1}C_{2,2}^{n,J}\ d^J_{2,2}(\theta)
\end{equation}
\begin{equation}
\frac{\hat{t}^{\frac{3}{2}}\hat{u}^{\frac{1}{2}}}{\hat{s}^2}\ \prod_{K=1}^{n-1}(\hat{u}+K)\ \ =\ \  \sum_{J=2}^{n+1}C_{2,-1}^{n,J} \ d^J_{2,-1}(\theta)
\end{equation}
\begin{equation}
\frac{\hat{u}^{\frac{3}{2}}\hat{t}^{\frac{1}{2}}}{\hat{s}^2}\ \prod_{K=1}^{n-1}(\hat{u}+K)\ \ =\ \  \sum_{J=2}^{n+1}C_{2,1}^{n,J} \ d^J_{2,1}(\theta)
\end{equation}
\begin{equation}
\frac{\hat{u}^{\frac{1}{2}}}{\hat{s}^{\frac{1}{2}}}\ \prod_{K=1}^{n-1}(\hat{u}+K)\ \ =\ \  \sum_{J=\frac{1}{2}}^{n-\frac{1}{2}}C_{\frac{1}{2},\frac{1}{2}}^{n,J}\ d^J_{\frac{1}{2},\frac{1}{2}}(\theta)
\end{equation}
\begin{equation}
\label{eq:denden3}
\frac{\hat{t}^{\frac{3}{2}}}{\hat{s}^{\frac{3}{2}}}\ \prod_{K=1}^{n-1}(\hat{u}+K)\ \  =\ \  \sum_{J=\frac{3}{2}}^{n+\frac{1}{2}}C_{\frac{3}{2},-\frac{3}{2}}^{n,J}\  d^J_{\frac{3}{2},-\frac{3}{2}}(\theta)
\end{equation}

We can immediately obtain the following relations from Eqs.~(\zz{\ref{eq:denden}}), (\zz{\ref{eq:denden2}}):
\bea
\label{eq:dean331}
C_{2,2}^{n,J}\ =\ (-1)^{J+n-1}\ C_{2,-2}^{n,J}
\eea
\bea
\label{eq:dean332}
C_{2,1}^{n,J}\ =\ (-1)^{J+n-1}\ C_{2,-1}^{n,J}
\eea
For this reason, in the following we will not explicitly consider $C_{2,2}^{n,J}$ and $C_{2,1}^{n,J}$.

We use the expansions Eqs.~(\zz{\ref{eq:a1}})-(\zz{\ref{eq:denden3}}) in order to rewrite the string helicity amplitudes (near a pole $\hat{s}\to n$) on the basis of $d$ functions.
Eqs.~(\zz{\ref{eq:apphela5}}), (\zz{\ref{eq:appheld1}}), (\zz{\ref{eq:appheld2}}), (\zz{\ref{eq:appheld3}}), (\zz{\ref{eq:appheld4}}), (\zz{\ref{eq:appheld6}}) , (\zz{\ref{eq:appheld68}}) then become:
\begin{eqnarray}
\label{eq:gface1}
\mathcal{M}_{0,0}\ =\ 4g^2 \frac{n}{(n-1)!}\ \frac{1}{\hat{s}-n}\ \Bigg[\sum_{J=0}^{n-1}C_{0,0}^{nJ}\ d^J_{0,0}(\theta)\Bigg]\cdot  \begin{Bmatrix} 8\sum_a \texttt{d}^{a_1a_2a}\texttt{d}^{a_3a_4a}\\ \sum_a \texttt{f}^{a_1a_2a}\texttt{f}^{a_3a_4a} \end{Bmatrix}
\end{eqnarray}
\begin{eqnarray}
\mathcal{M}_{2,-2}\ =\ 4g^2 \frac{n}{(n-1)!}\ \frac{1}{\hat{s}-n}\ \Bigg[ \sum_{J=2}^{n+1}C_{2,-2}^{nJ}\ d^J_{2,-2}\Bigg]\cdot  \begin{Bmatrix} 8\sum_a \texttt{d}^{a_1a_2a}\texttt{d}^{a_3a_4a} \\  \sum_a \texttt{f}^{a_1a_2a}\texttt{f}^{a_3a_4a} \end{Bmatrix}
\end{eqnarray}
\begin{eqnarray}
\label{eq:gon8}
\mathcal{M}_{2,-1}= 2g^2\delta^{\beta_4}_{\beta_3}\frac{n}{(n-1)!}\ \frac{1}{\hat{s}-n}\ \Bigg[\sum_{J=2}^{n+1}C_{2,-1}^{nJ}\  d^J_{2,-1}(\theta)\Bigg]
\cdot \begin{Bmatrix} 4\sum_a \texttt{d}^{a_1a_2a}T^a_{\alpha_3\alpha_4} \\  \sum_a \texttt{f}^{a_1a_2a}T^a_{\alpha_3\alpha_4}  \end{Bmatrix}
\end{eqnarray}
\begin{eqnarray}
\mathcal{M}_{\frac{1}{2},\frac{1}{2}}\ =\  (-1)^{n-1}2g^2\delta^{\beta_4}_{\beta_3} \frac{n}{(n-1)!}\ \frac{1}{\hat{s}-n}\ \Bigg[\sum_{J=\frac{1}{2}}^{n-\frac{1}{2}}C_{\frac{1}{2},\frac{1}{2}}^{nJ}\ d^J_{\frac{1}{2},\frac{1}{2}}(\theta)\Bigg]
(T^{a_2}T^{a_1})^{\alpha_3}_{\alpha_4} 
\end{eqnarray}
\begin{eqnarray}
\mathcal{M}_{\frac{3}{2},-\frac{3}{2}}\ =\  2g^2\delta^{\beta_4}_{\beta_3} \frac{n}{(n-1)!}\ \frac{1}{\hat{s}-n}\ \Bigg[\sum_{J=\frac{3}{2}}^{n+\frac{1}{2}}C_{\frac{3}{2},-\frac{3}{2}}^{nJ}\ d^J_{\frac{3}{2},-\frac{3}{2}}(\theta)\Bigg]
(T^{a_1}T^{a_2})^{\alpha_3}_{\alpha_4}
\end{eqnarray}
\begin{eqnarray}
\mathcal{M}^{(B)}_{\frac{1}{2},\frac{1}{2}}\ =\  2g_{D_{P_b}}g \frac{n}{(n-1)!}\ \frac{1}{\hat{s}-n}\ \Bigg[\sum_{J=\frac{1}{2}}^{n-\frac{1}{2}}C_{\frac{1}{2},\frac{1}{2}}^{nJ}\ d^J_{\frac{1}{2},\frac{1}{2}}(\theta)\Bigg]
(T^{a_1})_{\alpha_4}^{\alpha_3}(T^{a_2})_{\beta_3}^{\beta_4} 
\end{eqnarray}
\begin{eqnarray}
\label{eq:gfacelast}
\mathcal{M}^{(B)}_{\frac{3}{2},-\frac{3}{2}}\ =\  2g_{D_{P_b}}g \frac{n}{(n-1)!}\ \frac{1}{\hat{s}-n}\ \Bigg[\sum_{J=\frac{3}{2}}^{n+\frac{1}{2}}C_{\frac{3}{2},-\frac{3}{2}}^{nJ}\ d^J_{\frac{3}{2},-\frac{3}{2}}(\theta)\Bigg]
(T^{a_1})_{\alpha_4}^{\alpha_3}(T^{a_2})_{\beta_3}^{\beta_4} 
\end{eqnarray}
Where in the first three amplitudes, the curly brackets use shorthand notation in which we do not write that the upper row is $n=\text{odd}$ and the lower row is $n=\text{even}$.

All of these amplitudes contain: the simple pole $\frac{1}{\hat{s}-n}$, angular dependence in the square brackets, and color factors on the right.

Regarding Eq.~(\zz{\ref{eq:a1}}), we would like to note that $\prod_{K=1}^{n-1}(\hat{u}+K)$ when expanded as a polynomial in $\cos\theta$ has a definite parity (it is an even (odd) polynomial when $n$ is odd (even), see e.g Eq.~(\zz{\ref{eq:parityui}})). Likewise, $d^J_{00}(\theta)$ has a definite parity since it is the Legendre polynomial $P_J(\cos\theta)$, see Eq.~(\zz{\ref{eq:ylm7}}). It follows that:
\bea
\label{eq:pardire}
C_{0,0}^{n=\text{odd}, J=\text{odd}}\ =\ C_{0,0}^{n=\text{even}, J=\text{even}}\ =\ 0
\eea

\subsubsection{Plan for extracting the $F$'s}

As we saw, the amplitudes have the following the general form: 
\begin{eqnarray} 
\mathcal{M}_{m,m'}\ =\    \frac{1}{\hat{s}-n}\ \kappa \ \Big( \sum_a \eta^a_{a_3a_4a_1a_2}\Big)  \ \sum_{J}C^{n,J}_{m,m'}\ d^J_{m,m'}(\theta)
\end{eqnarray}
Where $\kappa$ contains constants, and $\eta$ is the color factor of the amplitude.
In order to calculate the decay widths, we must first calculate the $F$'s defined by:
\begin{eqnarray}
\mathcal{M}_{m,m'}\ =\  \frac{1}{M_s^2}\ \frac{1}{\hat{s}-n}\ \sum_{a, J}F^{aJ}_{m;a_3a_4}\ F^{aJ}_{m';a_1a_2}\ d^{J}_{m,m'} (\theta)
\end{eqnarray}
This was written in Eq.~(\zz{\ref{eq:n4}}), but now we use a slightly different notation in terms of $m,m'$ instead of $\lambda_1, \lambda_2, \lambda_3,\lambda_4$.

Comparing the last two equations we get
\begin{eqnarray}
\label{eq:ffeta}
\boxed{
F^{aJ}_{m';a_3a_4}F^{aJ}_{m;a_1a_2}\ =\ M_s^2 \  \kappa \ \eta^a_{a_3a_4a_1a_2}\  C^{n,J}_{m,m'} }
\end{eqnarray} 
We need to extract the $F$'s from this equation.
If we have initial and final states which are identical, then $m=\pm m'$ and the color part factorizes into two equal parts: $\eta^a_{a_3a_4a_1a_2}= \widetilde{\eta}^a_{a_3a_4}\ \widetilde{\eta}^a_{a_1a_2}$. Then the two $F$'s are equal and can be extracted:
\begin{eqnarray}
\label{eq:tyui}
\boxed{
F^{aJ}_{m;a_3a_4}\ =\  M_s\ \sqrt{\kappa_1}\ \widetilde{\eta}^a_{a_3a_4}\ \sqrt{C^{n,J}_{m,\pm m}}}
\end{eqnarray} 
This formula applies for $\mathcal{M}_{0,0}$ , $\mathcal{M}_{2,-2}$ , $\mathcal{M}_{\frac{1}{2},\frac{1}{2}}$ , $\mathcal{M}_{\frac{3}{2},-\frac{3}{2}}$, $\mathcal{M}^{(B)}_{\frac{1}{2},\frac{1}{2}}$ , $\mathcal{M}^{(B)}_{\frac{3}{2},-\frac{3}{2}}$.

Now we deal with $\mathcal{M}_{2,-1}$ which describes $g^-g^+\to q^-\bar{q}^+$, and obviously has different initial and final states. From Eq.~(\zz{\ref{eq:ffeta}}) we write:
\begin{eqnarray}
\label{eq:ariy78}
F^{aJ}_{-1;a_3a_4}\ F^{aJ}_{2;a_1a_2}\ =\ M_s^2\ \kappa_2\ \eta '^a_{a_3a_4a_1a_2}\ C^{n,J}_{2,-1} 
\end{eqnarray} 

In this equation, we know $F^{aJ}_{2;a_3a_4}$ from Eq.~(\zz{\ref{eq:tyui}}) when applied to $\mathcal{M}_{2,-2}$.

So we divide Eq.~(\zz{\ref{eq:ariy78}}) by $F^{aJ}_{2;a_3a_4}$:
\begin{eqnarray}
\label{eq:tyui2}
\boxed{
F^{aJ}_{-1;a_3a_4}\ =\ \frac{M_s^2}{F^{aJ}_{2;a_1a_2}} \ \kappa_2\  \eta '^a_{a_3a_4a_1a_2}\ C^{n,J}_{2,-1}\ =\ M_s \  \frac{\kappa_2}{\sqrt{\kappa_1}} \ \frac{\eta '^a_{a_3a_4a_1a_2}}{\widetilde{\eta}^a_{a_1a_2}} \ \frac{C^{n,J}_{2,-1}}{\sqrt{C^{n,J}_{2, -2}}}          }
\end{eqnarray} 

After we calculate the $F$'s from Eqs.~(\zz{\ref{eq:tyui2}}) and (\zz{\ref{eq:tyui}}), the decay widths can then be found using Eq.~(\zz{\ref{eq:parwidth}}):
\begin{eqnarray}
\Gamma^{aJ}_{(m,m');a_3a_4}= \frac{|F^{aJ}_{m,m';a_3a_4}|^2}{16(2J+1)\pi \sqrt{n}M_s}
\end{eqnarray} 

Where we put: $M=\sqrt{n}M_s$.

\subsubsection{Extracting the $F$'s}

Using the techniques above we now extract the $F$'s and $\Gamma$'s.

All of the decay widths will depend on the following combination of constants:
\begin{eqnarray}
\mathcal{Y}\ \equiv \ \frac{g^2M_s}{(2J+1)\pi }\ \frac{\sqrt{n}}{(n-1)!}
\end{eqnarray}
We also use the short hand notation: $F_m$ and $\Gamma_{m}$ thus omitting the obvious dependence on the other indices.
Inserting the relevant $\eta$'s and $\kappa$'s from Eqs.~(\zz{\ref{eq:gface1}})-(\zz{\ref{eq:gfacelast}}) we get the following $F's$ and $\Gamma$'s:
\begin{eqnarray}
F_{m=0}\ =\  2gM_s\sqrt{\frac{n}{(n-1)!}}\ \sqrt{C_{0,0}^{nJ}}\cdot  \begin{Bmatrix} \sqrt{8}\texttt{d}^{a_1a_2a} \\  \texttt{f}^{a_1a_2a} \end{Bmatrix}
\end{eqnarray}
\begin{eqnarray}
\label{eq:gammaw1}
\Gamma_{m=0}\ =\  2\mathcal{Y}\ C_{0,0}^{nJ} \cdot \begin{Bmatrix} \texttt{d}^{a_3a_4a}\texttt{d}^{a_3a_4a} \\ \frac{1}{8} \texttt{f}^{a_3a_4a}\texttt{f}^{a_3a_4a}  \end{Bmatrix}. 
\end{eqnarray}

\begin{eqnarray}
F_{m=2}\ =\ 2gM_s\sqrt{\frac{n}{(n-1)!}}\ \sqrt{C_{2,-2}^{nJ}}\cdot \begin{Bmatrix} \sqrt{8}\texttt{d}^{a_1a_2a} \\  \texttt{f}^{a_1a_2a}  \end{Bmatrix}
\end{eqnarray}
\begin{eqnarray}
\label{eq:gammaw2}
\Gamma_{m=2}\ =\  2\mathcal{Y}\ C_{2,-2}^{nJ}\cdot \begin{Bmatrix} \texttt{d}^{a_3a_4a}\texttt{d}^{a_3a_4a} \\ \frac{1}{8} \texttt{f}^{a_3a_4a}\texttt{f}^{a_3a_4a}  \end{Bmatrix}. 
\end{eqnarray}

\begin{eqnarray} 
F_{m=-1}\ =\  gM_s\sqrt{\frac{n}{(n-1)!}}\ \frac{C_{2,-1}^{nJ}}{\sqrt{C_{2,-2}^{nJ}}}\ \ \begin{Bmatrix} \sqrt{2}T^{a}_{\alpha_1\alpha_2} \\  T^{a}_{\alpha_1\alpha_2}  \end{Bmatrix}
\end{eqnarray}
\begin{eqnarray}
\label{eq:gammaw3}
\Gamma_{m=-1}\ =\  \frac{\mathcal{Y}}{16}\ \frac{\big(C_{2,-1}^{nJ}\big)^2}{C_{2,-2}^{nJ}}\ \begin{Bmatrix} 2T^{a}_{\alpha_1\alpha_2}T^{a}_{\alpha_1\alpha_2} \\  T^{a}_{\alpha_1\alpha_2}T^{a}_{\alpha_1\alpha_2}  \end{Bmatrix}
\end{eqnarray}

\begin{eqnarray} 
F_{m=\frac{1}{2}}\ =\ \sqrt{2}gM_s\sqrt{\frac{n}{(n-1)!}}\ \sqrt{C_{\frac{1}{2},\frac{1}{2}}^{nJ}}\  T^{a_4}_{\alpha_3\alpha}
\end{eqnarray} 
\begin{eqnarray}
\label{eq:gammaw4}
\Gamma_{m=\frac{1}{2}}\ =\ \frac{\mathcal{Y}}{8}\ C_{\frac{1}{2},\frac{1}{2}}^{nJ}\ T^{a_4}_{\alpha\alpha_3}T^{a_4}_{\alpha_3 \alpha}
\end{eqnarray}

\begin{eqnarray} 
F_{m=\frac{3}{2}}\ =\ \sqrt{2}gM_s\sqrt{\frac{n}{(n-1)!}}\ \sqrt{C_{\frac{3}{2},-\frac{3}{2}}^{nJ}} \ T^{a_4}_{\alpha_3\alpha}
\end{eqnarray}
\begin{eqnarray}
\label{eq:gammaw5}
\Gamma_{m=\frac{3}{2}}\ =\  \frac{\mathcal{Y}}{8}\ C_{\frac{3}{2},-\frac{3}{2}}^{nJ}\ T^{a_4}_{\alpha \alpha_3}T^{a_4}_{\alpha_3 \alpha}
\end{eqnarray}

\begin{eqnarray} 
F^{(B)}_{m=\frac{1}{2}}\ =\ \sqrt{2}g_{D_{P_b}}M_s\sqrt{\frac{n}{(n-1)!}}\ \sqrt{C_{\frac{1}{2},\frac{1}{2}}^{nJ}}\  T^{a_4}_{\alpha_3\alpha}
\end{eqnarray} 
\begin{eqnarray}
\label{eq:gammaw15}
\Gamma^{(B)}_{m=\frac{1}{2}}\ =\ \frac{g_{D_{P_b}}}{g}\frac{\mathcal{Y}}{8}\ C_{\frac{1}{2},\frac{1}{2}}^{nJ}\ T^{a_4}_{\alpha\alpha_3}T^{a_4}_{\alpha_3 \alpha}
\end{eqnarray}

\begin{eqnarray} 
F^{(B)}_{m=\frac{3}{2}}\ =\ \sqrt{2}g_{D_{P_b}}M_s\sqrt{\frac{n}{(n-1)!}}\ \sqrt{C_{\frac{3}{2},-\frac{3}{2}}^{nJ}} \ T^{a_4}_{\alpha_3\alpha}
\end{eqnarray}
\begin{eqnarray}
\label{eq:gammaw16}
\Gamma^{(B)}_{m=\frac{3}{2}}\ =\  \frac{g_{D_{P_b}}}{g} \frac{\mathcal{Y}}{8}\ C_{\frac{3}{2},-\frac{3}{2}}^{nJ}\ T^{a_4}_{\alpha \alpha_3}T^{a_4}_{\alpha_3 \alpha}
\end{eqnarray}

\subsubsection{Decay widths of the excited gluons}
\label{subsubsec:exglue1}

We denote the Regge excitations of the gluon by $\tilde{g}^{anJ}$, see also Fig.~\yy{\ref{chewblue1}}.

The color index "$a$" will be omitted sometimes. 
The gauge symmetry is $U(N)$ which is decomposed as $U(N)\sim SU(N)\times U(1)$, hence we put a tilde in $\tilde{g}^{anJ}$ to remind that it is $U(N)$.

We denote the Regge excitations of the $SU(N)$ gluon $g$ as:
\bea
g^{anJ}\equiv \tilde{g}^{anJ} \text{(without the tilde), for } a=1,\ldots ,N^2-1.
\eea
Likewise, we denote the Regge excitations of the $U(1)$ partner of the gluon $A\equiv \tilde{g}^{001}$ as:
\bea
A^{nJ}\equiv \tilde{g}^{0nJ}, \text{ for } a=0.
\eea

The decay width for the process $\tilde{g}^{anJ}\to \tilde{g}\tilde{g}$ is the sum from all the helicity states,   Eq~(\textcolor{zz}{\ref{eq:totwidth}}):
\begin{eqnarray}
\label{eq:dean43}
\Gamma_{\tilde{g}^{anJ}\to \tilde{g}\tilde{g}}\ =\  \frac{1}{2}\sum_{\lambda_3,\lambda_4}\sum_{a_3,a_4} \Gamma^{aJ}_{\lambda_3\lambda_4;a_3a_4} 
\end{eqnarray}
The $\frac{1}{2}$ is because of double counting or identical final state particles, see \rr{\cite{anc5}}.

We get from Eqs.~(\zz{\ref{eq:gammaw1}}), (\zz{\ref{eq:gammaw2}}), (\zz{\ref{eq:dean43}}):
\begin{eqnarray}
\label{eq:gluedec1}
\boxed{\Gamma_{\tilde{g}^{anJ}\to \tilde{g}\tilde{g}}\ =\ \mathcal{Y}\ \Big[ C_{0,0}^{nJ}+2C_{2,-2}^{nJ}\Big]\cdot \begin{Bmatrix} \sum_{a_3,a_4} \texttt{d}^{a_3a_4a}\texttt{d}^{a_3a_4a}\ \ , & \textrm{odd $n$}\\ \frac{1}{8} \sum_{a_3,a_4}\texttt{f}^{a_3a_4a}\texttt{f}^{a_3a_4a}, & \textrm{even $n$} \end{Bmatrix} }
\end{eqnarray}
The factor of $2$ in front of $C_{2,-2}^{nJ}$ is because it also counts $C_{2,2}^{nJ}$.

\begin{figure}
\centering
\includegraphics[width= 80mm]{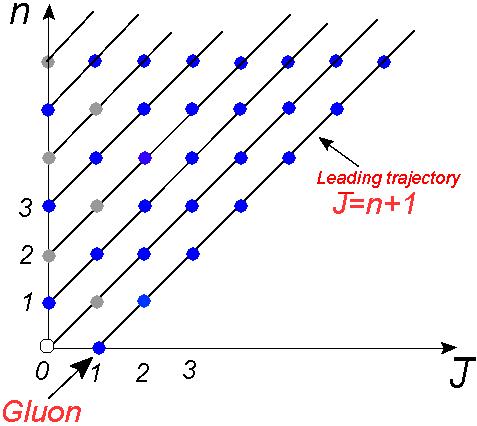}
\caption{A plot in the $n-J$ plane of the gluon Regge excitations $g^{nJ}$ exchanged in $gg \to gg$. The gray dots are not exchanged in this process.\label{chew}}
\end{figure}

\begin{figure}
\centering
\includegraphics[width= 80mm]{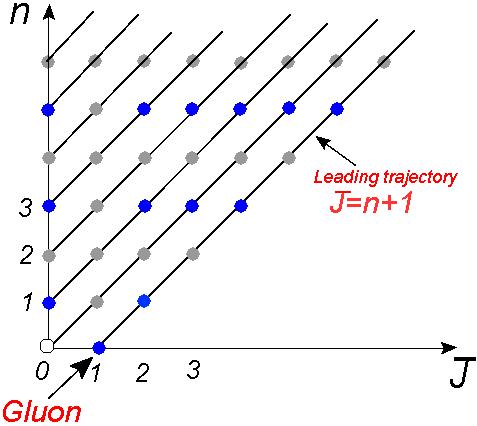}
\caption{A plot in the $n-J$ plane of the gluon Regge excitations $g^{nJ}$ exchanged in $gg \to gA$. The gray dots are not exchanged.\label{chewga}}
\end{figure}

In Eq.~(\zz{\ref{eq:gluedec1}}), for a given $n$ it should be understood that: $C_{2,-2}^{n,J=0}= C_{2,-2}^{n,J=1}=0$ and $C_{0,0}^{n,J=n+1}=C_{0,0}^{n,J=n}= 0$. Therefore, from Eq.~(\zz{\ref{eq:pardire}}) the exchange of any $J=0$ particle occurs only at $n=\text{odd}$, and that of $J=1$ only  at $n=\text{even}$, see Figs.~\textcolor{yy}{\ref{chew}, \ref{chewga}}. \\

Using the identities from Appendix~\ref{sec:appc}, we plug in the color factors for the six different combinations of $SU(N)$ and $U(1)$ fields:
\begin{eqnarray}
\label{eq:ggg71}
\Gamma_{g^{nJ}\to gg}\ =\ \mathcal{Y}\ \Big[ C_{0,0}^{nJ}+2C_{2,-2}^{nJ}\Big]\cdot \begin{Bmatrix}
 \frac{N^2-4}{16N} \\ \frac{N}{8}  \end{Bmatrix}
\end{eqnarray}
\begin{eqnarray}
\label{eq:ggc71}
\Gamma_{g^{nJ}\to gA}\ =\ \mathcal{Y}\ \Big[ C_{0,0}^{nJ}+2C_{2,-2}^{nJ}\Big]\cdot \begin{Bmatrix}
 \frac{1}{4N}  \\ \ 0   \end{Bmatrix} 
\end{eqnarray}
\begin{eqnarray}
\label{eq:gcc71}
\Gamma_{g^{nJ}\to AA}\ =\ 0
\end{eqnarray}
\begin{eqnarray}
\Gamma_{A^{nJ}\to gg}\ =\ \mathcal{Y}\ \Big[ C_{0,0}^{nJ}+2C_{2,-2}^{nJ}\Big]\cdot \begin{Bmatrix} \frac{N^2-1}{8N} \\ \ \ 0   \end{Bmatrix}
\end{eqnarray}
\begin{eqnarray}
\Gamma_{A^{nJ}\to gA}\ =\ 0
\end{eqnarray}
\begin{eqnarray}
\Gamma_{A^{nJ}\to AA}\ =\ \mathcal{Y}\ \Big[ C_{0,0}^{nJ}+2C_{2,-2}^{nJ}\Big]\cdot \begin{Bmatrix} \frac{1}{8N} \\ \ 0   \end{Bmatrix}
\end{eqnarray}

Figs.~\textcolor{yy}{\ref{chew}, \ref{chewga}} show the spectrum of the gluon Regge excitations $g^{nJ}$ which are exchanged in $gg\to gg$ and $gg\to gA$ respectively. In the process $gg\to gA$, $A^{nJ}$ are not exchanged at all. In the process $gg\to AA$ only $A^{nJ}$ are exchanged and $g^{nJ}$ are not. This happens because $d^{a00}=0$, Eq.~(\zz{\ref{eq:du12}}). Processes with a $gA$ or $AA$ in the final state do not have $n=even$ poles since $f^{0ab}= f^{00a}= f^{000}=0$, see Appendix~\ref{sec:apphelamp}.

Similar reasoning applied to $q\bar{q} \to gg$ and $q\bar{q} \to gA$ will yield a spectrum which is different only by the fact that now there no exchanges of spin $J=0,1$ (except for exchange of a massless gluon). This happens of course because of the absence of $C_{0,0}^{nJ}$, Eqs.~(\zz{\ref{eq:gon8}}) and (\zz{\ref{eq:gon7}}).

Proceeding now to the decay into $q\bar{q}$, we have from Eq.~(\zz{\ref{eq:gammaw3}}):
\begin{eqnarray}
\label{eq:gon7}
\boxed{\Gamma_{\tilde{g}^{anJ}\to q\bar{q}}\ =\ \frac{\mathcal{Y}}{8}\ \frac{\big(C_{2,-1}^{nJ}\big)^2}{C_{2,-2}^{nJ}}\ \begin{Bmatrix} 2Tr(T^{a}T^{a}) \\  Tr(T^{a}T^{a})  \end{Bmatrix}  }
\end{eqnarray}

For the $SU(N)$ and $U(1)$ fields this gives an equal result since $Tr(T^{a}T^{a})=\frac{1}{2}$ in both cases:
\begin{eqnarray}
\label{eq:gqq71}
\Gamma_{g^{nJ}\to q\bar{q}} \ =\  \Gamma_{A^{nJ}\to q\bar{q}} \ =\  \frac{\mathcal{Y}}{8}\ \frac{\big(C_{2,-1}^{nJ}\big)^2}{C_{2,-2}^{nJ}}\ \begin{Bmatrix} 1 \\ \frac{1}{2}  \end{Bmatrix} 
\end{eqnarray}

The total $g^{nJ}$ width is the sum from the four channels. After taking into account that there may be $N_f$ quark flavors, we get:
\begin{eqnarray}
\Gamma_{g^{nJ}}=
\bigg[\ \mathcal{Y}\Big(C_{0,0}^{nJ}+2C_{2,-2}^{nJ}\Big) \begin{Bmatrix} \frac{N}{16} \\ \frac{N}{8}  \end{Bmatrix} +  
\frac{\mathcal{Y}}{8}N_f\ \frac{\big(C_{2,-1}^{nJ}\big)^2}{C_{2,-2}^{nJ}}\ \begin{Bmatrix} 1 \\ \frac{1}{2} \end{Bmatrix} \ \bigg]
\end{eqnarray}

Similarly for $A^{nJ}$:
\begin{eqnarray}
\Gamma_{A^{nJ}}=
\bigg[\ \mathcal{Y}\Big(C_{0,0}^{nJ}+2C_{2,-2}^{nJ}\Big) \begin{Bmatrix} \frac{N}{8} \\ 0 \end{Bmatrix} +  
\frac{\mathcal{Y}}{8}N_f\ \frac{\big(C_{2,-1}^{nJ}\big)^2}{C_{2,-2}^{nJ}}\ \begin{Bmatrix} 1 \\ \frac{1}{2} \end{Bmatrix} \ \bigg]
\end{eqnarray}

\subsubsection{Decay widths of the excited quarks}
\label{subsubsec:excquarks7}

We denote the Regge excitations of the quark by $q^{nJ}$, see also Fig.~\yy{\ref{chewblue1}}.

From Eqs.~(\zz{\ref{eq:gammaw4}}), (\zz{\ref{eq:gammaw5}}) we have:
\begin{eqnarray}
\boxed{\Gamma_{q^{nJ}\to q\tilde{g}}\ =\ \frac{\mathcal{Y}}{8}\ \Big[ C_{\frac{1}{2},\frac{1}{2}}^{nJ}+2C_{\frac{3}{2},-\frac{3}{2}}^{nJ}\Big]\ \sum_{\alpha_3,a_4} T^{a_4}_{\alpha \alpha_3}T^{a_4}_{\alpha_3 \alpha}}
\end{eqnarray}
It is understood that:
$C_{\frac{3}{2},-\frac{3}{2}}^{n,J=\frac{1}{2}}=0$ and $C_{\frac{1}{2},\frac{1}{2}}^{n,J=n+\frac{1}{2}}=0$. The factor of $2$ in front of $C_{\frac{3}{2},-\frac{3}{2}}^{nJ}$ is because it also counts $C_{\frac{3}{2},\frac{3}{2}}^{nJ}$.

The gluon in the decay $q^{nJ}\to \tilde{g}q$ can be either a $SU(N)$ or a $U(1)$.
Eqs.~(\zz{\ref{eq:dire7}}) ,(\zz{\ref{eq:dire77}}) give:

\begin{eqnarray}
\Gamma_{q^{nJ}\to qg}\ =\ \frac{\mathcal{Y}}{8}\ \Big[ C_{\frac{1}{2},\frac{1}{2}}^{nJ}+2C_{\frac{3}{2},-\frac{3}{2}}^{nJ}\Big]\ \frac{N^2-1}{2N}
\end{eqnarray}

\begin{eqnarray}
\Gamma_{q^{nJ}\to qA}\ =\ \frac{\mathcal{Y}}{8}\ \Big[ C_{\frac{1}{2},\frac{1}{2}}^{nJ}+2C_{\frac{3}{2},-\frac{3}{2}}^{nJ}\Big]\ \frac{1}{2N}
\end{eqnarray}

Proceeding now to the decay into $q'\tilde{B}$ (where $\tilde{B}$ is the $U(N_b)$ gauge boson from stack $b$), we have from Eqs.~(\zz{\ref{eq:gammaw15}}), (\zz{\ref{eq:gammaw16}}):
\begin{eqnarray}
\boxed{ \Gamma_{q^{nJ}\to q'\tilde{B}}\ =\ \frac{g_{D_{P_b}}}{g} \frac{\mathcal{Y}}{8}\ \Big[ C_{\frac{1}{2},\frac{1}{2}}^{nJ}+2C_{\frac{3}{2},-\frac{3}{2}}^{nJ}\Big]\ \sum_{\alpha_3,a_4} T^{a_4}_{\alpha \alpha_3}T^{a_4}_{\alpha_3 \alpha}  }
\end{eqnarray}

For the $SU(N_b)$ and $U(1)$ particles
\begin{eqnarray}
\Gamma_{q^{nJ}\to q'B}\ =\ \frac{g_{D_{P_b}}}{g} \frac{\mathcal{Y}}{8}\ \Big[ C_{\frac{1}{2},\frac{1}{2}}^{nJ}+2C_{\frac{3}{2},-\frac{3}{2}}^{nJ}\Big]\ \frac{N_b^2-1}{2N_b}
\end{eqnarray}

\begin{eqnarray}
\Gamma_{q^{nJ}\to q'B^0}\ =\ \frac{g_{D_{P_b}}}{g} \frac{\mathcal{Y}}{8}\ \Big[ C_{\frac{1}{2},\frac{1}{2}}^{nJ}+2C_{\frac{3}{2},-\frac{3}{2}}^{nJ}\Big]\ \frac{1}{2N_b}
\end{eqnarray}

The total $q^{nJ}$ width is the sum from all channels:
\begin{eqnarray}
\Gamma_{q^{nJ}}\ =\  \Gamma_{q^{nJ}\to qg}+ \Gamma_{q^{nJ}\to qA}+ \Gamma_{q^{nJ}\to q'B}+ \Gamma_{q^{nJ}\to q'B^0}= 
\nn[8pt]
\frac{\mathcal{Y}}{8}\ \bigg[ C_{\frac{1}{2},\frac{1}{2}}^{nJ}+2C_{\frac{3}{2},-\frac{3}{2}}^{nJ}\bigg]\ \bigg(\frac{N}{2}+ \frac{g_{D_{P_b}}}{g} \frac{N_b}{2} \bigg) 
\end{eqnarray}

\subsection{Calculation of the coefficients $C_{m,m'}^{n,J}$}
\label{subsec:calc14}

In this section we suggest four methods to calculate the coefficients $C_{m,m}^{nJ}$ of Eqs.~(\zz{\ref{eq:a1}})-(\zz{\ref{eq:denden3}}). The $C$'s can then be plugged into the expressions for the decay widths in sections~\ref{subsubsec:exglue1} and \ref{subsubsec:excquarks7}. 

\subsubsection{Approach 1}
\label{subsubsec:approach1}
In this approach both sides of Eqs.~(\zz{\ref{eq:a1}})-(\zz{\ref{eq:denden3}}) are expanded in powers of $\cos\theta$ for a given value of $n$.
For the left hand side this is done by plugging $\hat{u}= -\frac{n}{2}[1+\cos\theta]$ and $\hat{t}= -\frac{n}{2}[1-\cos\theta]$, and expanding $\prod_{K=1}^{n-1}(\hat{u}+K)$. In the right hand side, the $d$'s are calculated using Eq.~(\zz{\ref{eq:zz1}}) and using trigonometric identities are written as a power series in $\cos\theta$. 
Then one compares both sides, and the $C$'s can be extracted recursively starting from the highest power of $\cos\theta$ and proceeding to lower powers. 

The calculated $d$ functions are given in the tables of Appendix~\ref{sec:appd}.

The process of calculating the $C_{m,m}^{nJ}$'s for $n=1\ldots 5$ is shown in the tables of Appendix~\ref{sec:appf}.

Tables~\yy{\ref{c1n}}, \yy{\ref{c2n}}, \yy{\ref{c3n}}, \yy{\ref{c4n}}, \yy{\ref{c5n}}  collect all ofthe C's.

\begin{itemize}
\item \textbf{Notes on expanding the $\prod_{K=1}^{n-1}(\hat{u}+K)$ .}
\end{itemize}

The factor $\prod_{K=1}^{n-1}(\hat{u}+K)$ appears in each of the $5$ equations, and is a polynomial in $\cos\theta$ of degree $n-1$. For high $n$ it may be helpful to simplify this factor as follows.
\begin{eqnarray}
\label{eq:poch7}
\prod_{K=1}^{n-1}(\hat{u}+K)= \Big(-\frac{n}{2}[1+\cos\theta]+1\Big)\Big(-\frac{n}{2}[1+\cos\theta]+2\Big) \cdots \Big(-\frac{n}{2}[1+\cos\theta]+n-1\Big)=
\nonumber\\
\Big(\frac{-1}{2}\Big)^{n-1}\Big(n\cos\theta+[n-2]\Big)\Big(n\cos\theta+[n-4]\Big) \cdots \Big(n\cos\theta-[n-4]\Big) \Big(n\cos\theta-[n-2]\Big)\nonumber\\
\end{eqnarray}

The first two terms with highest powers are easy to sum, giving:
\begin{eqnarray}
\label{eq:parityui}
\Big(\frac{-1}{2}\Big)^{n-1}\ \bigg[\ n^{n-1}\ (\cos\theta)^{n-1}\ \  -\ \ \frac{(n-2)(n-1)n}{6}\ (\cos\theta)^{n-3} \ +\  \ldots \ \ \   \bigg] \nonumber\\
\end{eqnarray}

In Eq.~(\textcolor{zz}{\ref{eq:poch7}}) we can cut the number of terms in half
by multiplying the first term with the last, the second term with the next to last, etc...

We get for $n=$odd:
\begin{eqnarray}
\boxed{\prod_{K=1}^{n-1}(\hat{u}+K)= \Big(\frac{-1}{2}\Big)^{n-1}\Big(n^2\cos^2\theta-[n-2]^2\Big) \Big(n^2\cos^2\theta-[n-4]^2\Big)\ \cdots \ \Big(n^2\cos^2\theta-1^2\Big)}
\nonumber\\
\end{eqnarray}

for $n=$even:
\begin{eqnarray}
\boxed{\prod_{K=1}^{n-1}(\hat{u}+K)= \Big(\frac{-1}{2}\Big)^{n-1}\Big(n^2\cos^2\theta-[n-2]^2\Big) \Big(n^2\cos^2\theta-[n-4]^2\Big)\ \cdots \ \Big(n^2\cos^2\theta-2^2\Big)\cdot n\cos\theta}
\nonumber\\
\end{eqnarray}

\subsubsection{Approach 2}

In essence, this approach similar to the previous approach but takes into account some things that simplify the computations. The procedure for expanding the right hand side is simplified by using known properties of the $d$ functions, Namely that the $d$ functions are simply proportional to Jacobi polynomials, and the power expansions of the Jacobi polynomials are known. As for the left hand side, the coefficients of the power expansion of $\prod_{K=1}^{n-1}(\hat{u}+K)$ are known to be the Stirling numbers.

We start by noting that the power expansions of Eqs.~(\zz{\ref{eq:a1}})-(\zz{\ref{eq:denden3}}) are now done in terms of $\hat{u}$ instead of $\cos\theta$. Plugging Eqs.~(\zz{\ref{eq:d0000}})-(\zz{\ref{eq:d3232}}) in Eqs.~(\zz{\ref{eq:a1}})-(\zz{\ref{eq:denden3}}) we get:         
\begin{eqnarray}
\prod_{K=1}^{n-1}(\hat{u}+K)\ \  =\ \ \sum_{J=0}^{n-1}C_{0,0}^{n,J}\ P^{(0,0)}_{J}(\hat{u}) \ \ = \ \ 
\sum_{J=2}^{n+1}C_{2,-2}^{n,J}\ P^{(4,0)}_{J-2}(\hat{u})\ \ \ =\ \ \ \ \ \ \ \ \ \ \ \ \ \ \ \ \ \  
\nn[8pt]
 \sum_{J=2}^{n+1}\sqrt{\frac{J+2}{J-1}}\ C_{2,-1}^{n,J}\ P^{(3,1)}_{J-2}(\hat{u})\ \ =\ \  
 \sum_{J=\frac{1}{2}}^{n-\frac{1}{2}}C_{\frac{1}{2},\frac{1}{2}}^{n,J}\ P^{(0,1)}_{J-1/2}(\hat{u})\ \ =\ \  \sum_{J=\frac{3}{2}}^{n+\frac{1}{2}}C_{\frac{3}{2},-\frac{3}{2}}^{n,J}\  P^{(3,0)}_{J-3/2}(\hat{u})
\end{eqnarray}
All the factors on the left hand side that multiplied $\prod_{K=1}^{n-1}(\hat{u}+K)$ got canceled. Therefore, the problem of finding the $C$'s reduces to comparing $\prod_{K=1}^{n-1}(\hat{u}+K)$ with the five different Jacobi polynomials above.

This procedure should be possible for for amplitudes with general helicity states $(m,m')$. We have using Eq.~(\zz{\ref{eq:dean11}}):
\begin{eqnarray}
\label{eq:knopf6}
\boxed{
\prod_{K=1}^{n-1}(\hat{u}+K)\ =\ \sum_{J=m}^{n-1+m} \Omega^{(J-m)}\ C_{m,m'}^{n,J}\ P^{(m-m',m+m')}_{J-m}(\hat{u})   }
\end{eqnarray}
This can be used to calculate decay widths for a decay of a particle into excited states, as in the direct production of section~\ref{sec:a2}.

Lets continue to show how one can extract the $C$'s.
The power expansion of the l.h.s of Eq.~(\zz{\ref{eq:knopf6}}) is from Eq.~(\zz{\ref{eq:pochstir1}}):
\bea
\label{eq:stir98}
\prod_{K=1}^{n-1}(\hat{u}+K)\ =\ \frac{(\hat{u})_n}{\hat{u}}\ =\  \sum_{i=0}^{n-1}(-1)^{n-i-1}\ S_n^{(i+1)}\ \hat{u}^i
\eea

As for the r.h.s, we recall Eq.~(\zz{\ref{eq:dean12}}):
\bea
\label{eq:mont99}
P_k^{(m-m',m+m')}(\hat{u}) \ =\ \sum_{p=0}^{k}\Delta^{(p)}_{m,m'}\ \hat{u}^p 
\eea
Defining, $k=J-m$, the r.h.s of Eq.~(\zz{\ref{eq:knopf6}}) is now:
\begin{eqnarray}
\sum_{k=0}^{n-1}\ \Omega^{(k)}\ C_{m,m'}^{n,k+m}\ P^{(m-m',m+m')}_{k}(\hat{u})\ =\ \sum_{k=0}^{n-1}\ \sum_{p=0}^k \Omega^{(k)} \ C_{m,m'}^{n,k+m}\ \Delta^{(p)}_{m,m'}\ \hat{u}^p   
\end{eqnarray}

Using $\sum_{k=0}^{n-1}\ \sum_{p=0}^k = \sum_{p=0}^{n-1}\ \sum_{k=p}^{n-1}$ we get:
\bea
\sum_{p=0}^{n-1}\ \sum_{k=p}^{n-1}\ \Omega^{(k)} \ C_{m,m'}^{n,k+m}\ \Delta^{(p)}_{m,m'}\ \hat{u}^p
\eea
Comparing this with Eq.~(\zz{\ref{eq:stir98}}), we get:
\bea
\boxed{
\sum_{k=p}^{n-1} \Omega^{(k)} \ C_{m,m'}^{n,k+m} \ =\  (-1)^{n-p-1}\ \frac{S_n^{(p+1)}}{\Delta^{(p)}_{m,m'}}    }
\eea

These are equations for the $C$'s. We can begin with $p=n-1$ and proceed to extract the $C$'s recursively:
\bea
\label{eq:dean14}
C_{m,m'}^{n,n-1+m} \ =\ \frac{S_n^{(n)}}{\Omega^{(n-1)}\ \Delta^{(n-1)}_{m,m'}}\ = \ \frac{1}{\Omega^{(n-1)}\ \Delta^{(n-1)}_{m,m'}}
\eea

\bea
C_{m,m'}^{n,n-2+m} \ =\ \frac{1}{\Omega^{(n-2)}} \ \Big[-\frac{S_n^{(n-1)}}{\Delta^{(n-2)}_{m,m'}}\ - \ \Omega^{(n-1)}\ C_{m,m'}^{n,n-1+m} \Big]
\eea
And so on... We see that the leading Regge trajectories are obtained first.

So let us write explicitly the $C$'s for the leading trajectory.

From Eq.~(\zz{\ref{eq:dean13}}) we have:
\bea
\Delta^{(n-1)}_{m,m'}\ =\  (-1)^{n-1}\ \frac{(2n-2+2m)!}{(n-1)!\ (n-1+2m)!\ n^{n-1}}
\eea
Then Eq.~(\zz{\ref{eq:dean14}}) gives for the 5 combinations of $(m,m')$:
\bea
C_{0,0}^{n,n-1} \ =\ \frac{1}{\Delta^{(n-1)}_{0,0}}\ =\  (-1)^{n-1}\ \frac{(n-1)!^2\ n^{n-1}}{(2n-2)!}
\eea
\bea
C_{2,-2}^{n,n+1} \ =\  (-1)^{n-1}\ \frac{(n-1)!\ (n+3)!\ n^{n-1}}{(2n+2)!}
\eea
\bea
C_{2,-1}^{n,n+1} \ =\  (-1)^{n-1}\ \frac{(n-1)!\ (n+3)!\ n^{n-1}}{(2n+2)!}\ \sqrt{\frac{n}{n+3}}
\eea
\bea
C_{\frac{1}{2},\frac{1}{2}}^{n,n-\frac{1}{2}} \ =\  (-1)^{n-1}\ \frac{(n-1)!\ n!\ n^{n-1}}{(2n-1)!}
\eea
\bea
C_{\frac{3}{2},-\frac{3}{2}}^{n,n+\frac{1}{2}} \ =\  (-1)^{n-1}\ \frac{(n-1)!\ (n+2)!\ n^{n-1}}{(2n+1)!}
\eea

For consistency, we checked these formulas against the corresponding leading trajectory results (up to $n=5$) from approach 1 (given in Tables~\yy{\ref{c1n}}, \yy{\ref{c2n}}, \yy{\ref{c3n}}, \yy{\ref{c4n}}, \yy{\ref{c5n}} ). Agreement was found in all cases.

Using these $C$'s we can write the $n$ dependence of the decay widths for the leading trajectory resonances. These are given in terms of the combination $\Gamma \sim \mathcal{Y}C_{m,m'}^{n,J}$, as seen in e.g Eq.~(\zz{\ref{eq:gluedec1}}). So all that is needed is to multiply the previous 5 equations by $\mathcal{Y}\sim \frac{\sqrt{n}}{(2J+1)(n-1)!}$, with $J=n+const.$ and the constant is different for each  one of the five helicity states. The large $n$ dependence is seen to be dominated by the $n^n$ dependence of the $C$'s.\\

We also note the possibility to obtain a formal expression for the $C$'s by reversing Eq.~(\zz{\ref{eq:knopf6}}) using the orthogonality of the Jacobi polynomials:
\bea 
C_{m,m'}^{n,J}= \sqrt{\frac{(2J+1)(J-m)!(J+m)!}{2^{2J+1}(J-m')!(J+m')!}}  \ \ \ \times \ \ \ \ \ \ \ \ \ \ \ \ \ \ \ \ \ \ \ \ \ \ \ \ \ \ \ \ \ \ \ \ \ \ \ \ 
\nn[8pt]
\int_{-1}^{1}dx\  (1-x)^{m-m'}(1+x)^{m+m'}P^{(m-m',m+m')}_{J-m}(x)\prod_{K=1}^{n-1}\Big(-\frac{\hat{s}}{2}\big[1-x\big]+K\Big) 
\eea
It is not clear though if this equation is useful for calculations.

\subsubsection{Approach 3}
This approach was inspired by \cite{dobson}.

Imagine that we have in our disposal the following expansion in which we know the $\lambda_b$'s:
\bea
\label{eq:fknopfler7}
\hat{u}^i= \sum_{b=0}^i \lambda_b^{(i)}\ P_b^{(m-m',m+m')}(\hat{u})
\eea
This is the reverse of Eq.~(\zz{\ref{eq:mont99}}).
We plug this expansion into Eq.~(\zz{\ref{eq:stir98}}) and get:
\bea
\prod_{K=1}^{n-1}(\hat{u}+K)\ =\  \sum_{i=0}^{n-1}(-1)^{n-i-1}\ S_n^{(i+1)}\ \hat{u}^i=
\nn[8pt]
\sum_{i=0}^{n-1}\sum_{b=0}^i (-1)^{n-i-1}\ S_n^{(i+1)}\lambda_b^{(i)}\ P_b^{(m-m',m+m')}(\hat{u})
\eea
Using $\sum_{i=0}^{n-1}\sum_{b=0}^i = \sum_{b=0}^{n-1}\sum_{i=b}^{n-1}$, we get:
\bea
\sum_{b=0}^{n-1}\sum_{i=b}^{n-1} (-1)^{n-i-1}\ S_n^{(i+1)}\ \lambda_b^{(i)}\ P_b^{(m-m',m+m')}(\hat{u})
\eea
Changing dummy variable to $J=b+m$ we get:
\bea
\sum_{J=m}^{n-1+m}\sum_{i=J-m}^{n-1} (-1)^{n-i-1}\ S_n^{(i+1)}\ \lambda_{J-m}^{(i)}\ P_{J-m}^{(m-m',m+m')}(\hat{u})
\eea
Now this form can finally be compared with the right hand side of Eq.~(\zz{\ref{eq:knopf6}}), yielding:
\bea
\boxed{
C_{m,m'}^{n,J}= \sqrt{\frac{(J+m')!(J-m')!}{(J+m)!(J-m)!}}\ \sum_{i=J-m}^{n-1}(-1)^{n-i-1}\ S_n^{(i+1)} \lambda_{J-m}^{(i)}  }
\eea
This is a closed expression for $C_{m,m'}^{n,J}$ for any $n,J,m,m'$.
Obviously, the question is if we can find an expansion of the form of Eq.~(\zz{\ref{eq:fknopfler7}}).

\subsubsection{Approach 4}
We take this from \rr{\cite{oh}}.
In this approach only the $(m,m')=(0,0)$ helicity state was taken into account, therefore only the Legendre polynomial appears. It might be possible to generalize such an approach to arbitrary helicity states, but we only have time to write what was done in \cite{oh}.

Putting $\alpha(s)=as+b$, The partial waves are: 
\begin{eqnarray}
\label{eq:parw5}
a(J,s)=\frac{1}{2}\int_{-1}^1 dzA(s,t)P_J(z)= \pi^{\frac{1}{2}}2^{-J-1}\alpha(s)(2aq^2)^{s}\int_0^\infty dxx^{J-\alpha(s)-1}f(x)e^{-x}
\end{eqnarray}

Where the definition
\begin{eqnarray}
f(2aq^2y)= \Big[\frac{1-e^{-y}}{y}\Big]^{-\alpha(s)-1}(aq^2y)^{-J-\frac{1}{2}}I_{J+\frac{1}{2}}(2aq^2y)e^{b-1}y
\end{eqnarray}
and $I_l(x)$ is the modified Bessel function.

The partial width is found by taking the residue:
\begin{eqnarray}
\Gamma(n,J)=-\frac{q_n}{2M_{n}^2}\text{res}\ \big[ a(J,s)\big]=\ \ \ \ \ \ \ \ \ \ \ \ \ \ \ \ \ \ \ \ \ 
\nonumber\\[5pt]
\pi^{\frac{1}{2}}2^{-J-1}n(2aq_n^2)^n\frac{1}{(n-J)!}\Big(\frac{d}{dx}\Big)^{n-J}\big[f(x)e^{-x}\big]_{x=0}
\end{eqnarray}
So that $\Gamma(n,J)$ can be calculated by taking derivatives.

Defining
\begin{eqnarray}
r_n(x)\equiv \frac{n!}{(2n+1)!}(2x)^n\ \ \ \ ,\ \ \ \ 
\end{eqnarray}

The first 4 trajectories were calculated:
\begin{eqnarray}
\Gamma(n,n)\ =\ \frac{M_n^2}{q_n}\ r_n(b_n)
\end{eqnarray}
\begin{eqnarray}
\Gamma(n,n-1)\ =\ \frac{M_n^2}{q_n}\ r_{n-1}(b_n)\Big[a_n+\frac{n-1}{2}\Big]
\end{eqnarray}
\begin{eqnarray}
\Gamma(n,n-2)=\  \frac{M_n^2}{q_n}\ r_{n-2}(b_n)\frac{1}{2}\bigg[\frac{b_n^2}{2n-1}+a_n^2+a_n(n-1)+\frac{(3n-1)(n-2)}{12}\bigg]
\end{eqnarray}
\begin{eqnarray}
\Gamma(n,n-3)\ =\ \frac{M_n^2}{q_n}\ r_{n-3}(b_n)\frac{1}{2}\bigg[\frac{b_n^2}{2n-3}(a_n+\frac{n-1}{2})+
\nonumber\\[6pt]
\frac{1}{3}\Big\{a_n^3+\frac{3}{2}a_n^2(n-1)+\frac{(3n-1)(n-2)}{4}a_n+\frac{n}{8}(n-1)(n-3)\Big\}\bigg]
\end{eqnarray}

\clearpage


\section{\textcolor{ww}{5-point amplitudes}}
\label{sec:5point57}
\emph{References:} ~\textcolor{rr}{\cite{berends, berends2, gottschalk, lust2}}.\\

5 particle amplitudes are $2\to3$ processes at colliders. Important signals at the LHC are $pp\to$ 3 jets, $pp\to$ 2 jets + EW gauge boson, $pp\to$ jet + 2 EW gauge bosons and $pp\to$ 3 EW gauge boson . These amplitudes are one order higher than the $2\to2$ processes. As in the 4 amplitude case, it is interesting to study the effect of a low string scale on the energy and angular dependence of such amplitudes.
The ultimate goal of a collider is to help uncover the underlying theory. After the initial discovery of a resonance, for instance in the dijet signal, it will be useful to check other types of signals for confirmation, and for measuring the properties of the resonance. The $2\to3$ amplitudes can be helpful in this way.
In this section we present the tree squared amplitudes in field theory and string theory, making it comfortable to compare the two. Then we write down the low energy corrections to the string amplitudes.

\subsection{Kinematics and definitions}

We define
\begin{eqnarray}
s_{ij}\ \equiv \ (k_i+k_j)^2\ =\  2k_ik_j 
\end{eqnarray}

then
\begin{eqnarray}
s_1\ \equiv \ s_{12}\ \ ,\ \ s_2\ \equiv \ s_{23}\ \ ,\ \  s_3\ \equiv \ s_{34} \ \ ,\ \  s_4\ \equiv \ s_{45}\ \ ,\ \ s_5\ \equiv \ s_{51}
\end{eqnarray}

Introducing the dimensionless units:
\begin{eqnarray}
\hat{s}_{ij}\ \equiv \ \alpha' s_{ij}
\nonumber\\
\hat{s}_{i}\ \equiv \ \alpha' s_{i}
\end{eqnarray}

We define the following kinematic functions that will appear in the squared amplitudes:
\begin{eqnarray}
\mathcal{S}_4 \ \equiv \ \sum_{i<j} s_{ij}^4\ \ \ \ \ \ \ \ \ \ \ \ \ \ \ \ \ \ 
\nonumber\\[5pt]
\mathcal{S}_3 \ \equiv \ \sum_{i=1,2,3} (s_{i4}^3 s_{i5}+s_{i4} s_{i5}^3)
\end{eqnarray}

Energy-momentum conservation:

\begin{eqnarray}
\sum^5_{i=1}k_i\ =\ 0 
\end{eqnarray}

\begin{figure}
\centering
\includegraphics[width= 70mm]{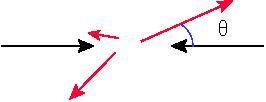}
\caption{5 particle kinematics. Energy-momentum conservation enforces the 3 outgoing particles (in red) to lie in a plane.\label{3jetkin}}
\end{figure}

The momentum 4-vectors are usually parametrized as (See Fig.~\textcolor{yy}{\ref{3jetkin}}):
\begin{eqnarray}
k_1\ =\ \frac{\sqrt{s}}{2}(1, \sin\theta \cos\varphi, \sin\theta \sin\varphi, \cos\theta)
\nonumber\\
k_5\ =\ \frac{\sqrt{s}}{2}(1, -\sin\theta \cos\varphi, -\sin\theta \sin\varphi, -\cos\theta)
\nonumber\\
k_4\ =\ \frac{x_1\sqrt{s}}{2}(1, 1, 0, 0)
\nonumber\\
k_2\ =\ \frac{x_2\sqrt{s}}{2}(1, \cos\theta_{12}, \sin\theta_{12}, 0)
\nonumber\\
k_3\ =\ \frac{x_3\sqrt{s}}{2}(1, \cos\theta_{13}, -\sin\theta_{13}, 0)
\end{eqnarray}

Where
\begin{eqnarray}
\cos\theta_{12}\ =\ 1-2\frac{x_1+x_2-1}{x_1 x_2}
\nonumber\\
\cos\theta_{13}\ =\ 1-2\frac{x_1+x_3-1}{x_1 x_3}
\end{eqnarray}

We define the following permutations (see \rr{\cite{lust2}}):
\begin{eqnarray}
\label{eq:peryu3}
\Pi_5\equiv \bigg\{\ \  (1,2,3,4,5)_1, (1,2,4,3,5)_2, (1,3,4,2,5)_3, (1,3,2,4,5)_4, (1,4,2,3,5)_5, (1,4,3,2,5)_6,
\nonumber\\ \nonumber\\ 
              (2,1,3,4,5)_7, (2,1,4,3,5)_8, (2,3,1,4,5)_9, (2,4,1,3,5)_{10}, (3,1,2,4,5,)_{11}, (3,2,1,4,5)_{12}\ \bigg\}
\nonumber\\ \nonumber\\ 
\Pi_q\equiv \bigg\{\ (1,3,2,4,5)_1, (1,2,4,3,5)_4, (2,1,3,4,5)_7, (2,3,1,4,5)_9, (3,1,2,4,5)_{11}, (3,2,1,4,5)_{12}\ \bigg\}
\nonumber\\ \nonumber\\ 
\Pi_q\equiv \bigg\{\ (1,2,4,3,5)_2, (2,1,4,3,5)_8\ \bigg\} \ \ \ \ \ \ \ \ \ \ \ \ \ \ \ \ \ \ \ \ \ \ \ \ \ \ \ \ \ \ \ \ \ \ \ \ \ \ \ \ \ \ \ \ \ \ \ \ \ \ \ \ \ \ \ \ \ \ \ \ \ \ \ \ \ \ \ \ \ \ \ \ 
\end{eqnarray}
The index running from 1 to 12 labels the permutations, and will help refering to them in a concise manner.
The 12 permutations of $\Pi_5$ appear in $\mathcal{M}(ggggg)$. They come from $(n-1)!/2=12$.
The 6 permutations of $\Pi_5$ appear in $\mathcal{M}(gggq\bar{q})$ and they come from $3!$.

\subsection{Field theory}
\label{subsec:5pointfield}

The 5-point amplitudes have historical significance as the gluon was discovered via $e^+e^-\to 3\ jets$ at \emph{PETRA}.
But also for a different reason.

At the end of the 1970's the 5-point tree level QCD and QED amplitudes have been computed using brute force Feynman diagram techniques. The calculation is very difficult especially due to the large number of terms that have to be controlled. Later it was found that these amplitudes can be algebraically manipulated into a simple form which also exhibits factorization. This fact stimulated research during the 1980's which led to insights into the cause behind the simplicity, and led to the discovery of powerful methods for calculating amplitudes with loops and higher point amplitudes (See section~\ref{sec:a3}). 

\begin{figure}
\centering
\includegraphics[width= 105mm]{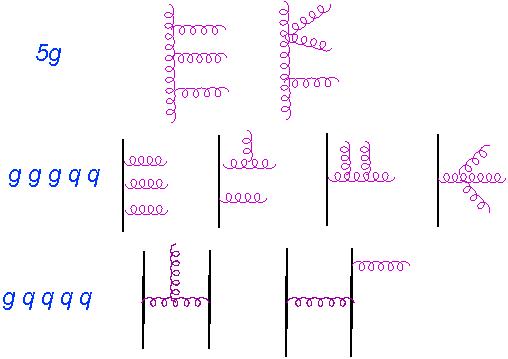}
\caption{Classes of Feynman diagrams for the 5 particle QCD amplitudes. Each drawing represents all of the diagrams that can be obtained from it by permuting the external legs.\label{5gluonsb}}
\end{figure}

Considering QCD and QED processes not including leptons, we have the 9 processes in Table~\textcolor{yy}{\ref{9processes}}. The diagrams for the QCD processes are given in Fig.~\textcolor{yy}{\ref{5gluonsb}}. All of the other processes can be obtained by crossing.

\begin{table}[h]
\centering
\begin{tabular}{|c|c|c|c|c|c|c|c|c|}
\hline
QCD& QED & QCD$+$QED  \\  
\whline
$gg \to ggg$ & $q\bar{q} \to \gamma\gamma\gamma$ &$q\bar{q} \to gg\gamma $ \\  
\hline
$q\bar{q} \to ggg$ & $q\bar{q} \to q\bar{q}\gamma $ &$q\bar{q} \to g\gamma\gamma$ \\  
\hline
$q\bar{q} \to q\bar{q}g$ & $q \bar{q}' \to q\bar{q}'\gamma$ & $-$\\  
\hline
$q\bar{q}' \to q\bar{q}'g$ & $-$ & $-$\\  
\hline
\end{tabular}  
\caption{9 QCD and QED processes.\label{9processes} }
\end{table}

Using the techniques and results of section \ref{sec:a3}, the 5-point squared amplitudes
may be computed. For example, looking at Eq.~(\textcolor{zz}{\ref{eq:5gs}}), the $gg\to ggg$ squared amplitude can be written at once:

\begin{eqnarray}
\sum_{\text{helicities}} \sum_{\text{colors}} |\mathcal{M}|^2(g_1,g_2,g_3,g_4,g_5)\ \sim \ g^6\  \mathcal{S}_4\ \sum \frac{1}{s_{12}s_{23}s_{34}s_{45}s_{51}}
\end{eqnarray}

In the next section we list the squared amplitudes in the compact form that was achieved in \rr{\cite{berends2}}. 

\subsubsection{The squared amplitudes}

\begin{itemize}
\item {\large QCD amplitudes}
\end{itemize}
We follow \rr{\cite{eichten}}, and write the QCD amplitudes.

\begin{eqnarray}
|\mathcal{M}|^2 (\textcolor{blue}{g(k_i)\ g(k_j) \to g(k_1)\ g(k_2)\ g(k_3)})\ \equiv \ |\mathcal{M}_1|^2(k_i, k_j, k_1, k_2, k_3)=
\nonumber\\[10pt]
g^6 \frac{27\cdot 2}{160} \frac{\mathcal{S}_4}{\prod_{m<n}s_{mn}}\sum_{perm.}s_{ij} s_{j1}s_{12}s_{23}s_{3i}
\end{eqnarray}

\begin{eqnarray} 
|\mathcal{M}|^2(\textcolor{blue}{q(k_i)\ \bar{q}(k_j) \to g(k_1)\ g(k_2)\ g(k_3)})\ \equiv \ |\mathcal{M}_2|^2(k_i, k_j, k_1, k_2, k_3)=\ \ \ \ \ \ \ \ \ \ \ \ \ \ \ \ 
\nonumber\\[9pt]  
\frac{2g^6}{81}\ \frac{\mathcal{S}_3}{s_{i1}s_{i2}s_{i3}s_{j1}s_{j2}s_{j3}}\Bigg[5s_{ij}-9\bigg( \frac{s_{i1}s_{j2}+s_{i2}s_{j1}}{s_{12}}+\frac{s_{i1}s_{j3}+s_{i3}s_{j1}}{s_{13}}+\frac{s_{i2}s_{j3}+s_{i3}s_{j2}}{s_{23}}\bigg)   
\nonumber\\ \nonumber\\ 
+\frac{81}{s_{ij}}\bigg(\frac{2s_{i3}s_{j3}(s_{i1}s_{j2}+s_{i2}s_{j1})}{s_{13}s_{23}}+\frac{2s_{i1}s_{j1}(s_{i2}s_{j3}+s_{i3}s_{j2})}{s_{12}s_{13}}+\frac{2s_{i2}s_{j2}(s_{i1}s_{j3}+s_{i3}s_{j1})}{s_{12}s_{23}}\bigg)\Bigg]
\nonumber\\ \nonumber\\ 
\end{eqnarray}

\begin{eqnarray} 
|\mathcal{M}|^2(\textcolor{blue}{q(k_i)q'(k_j) \to q(k_1)q'(k_2)g(k_3)})\equiv |\mathcal{M}_3|^2((k_i, k_j, k_1, k_2, k_3)= g^6\frac{2(s_{ij}^2+s_{12}^2+s_{i2}^2+s_{j1}^2)}{s_{i1}s_{j2}s_{i3}s_{j3}s_{13}s_{23}}
\nonumber\\[8pt]   \times  \bigg\{C_1\Big[(-s_{i2}-s_{j1})(s_{ij}s_{12}+s_{i1}s_{j2}-s_{i2}s_{j1})-s_{i2}(-s_{ij}s_{i1}-s_{12}s_{j2})-s_{j1}(-s_{ij}s_{j2}-s_{i1}s_{12})\Big]
\nonumber\\
-C_2\Big[(s_{ij}+s_{12})(s_{ij}s_{12}-s_{i1}s_{j2}-s_{i2}s_{j1})+2s_{i1}s_{j2}(-s_{i2}-s_{j1})+2s_{i2}s_{j1}(-s_{i1}-s_{j2})\Big] \bigg\} 
\nonumber\\ \nonumber\\
\end{eqnarray}
Where $C_1= \frac{16}{27}$ and $C_2= \frac{2}{27}$.

\begin{eqnarray} 
|\mathcal{M}|^2(\textcolor{blue}{q(k_i)\ q(k_j) \to q(k_1)\ q(k_2)\ g(k_3)})\ \equiv \ |\mathcal{M}_4|^2(k_i, k_j, k_1, k_2, k_3)= \ \ \ \ \ \ \ \ \ \ \ \ \ \ \ \ \ \ \ \ \ \  
\nonumber\\[10pt]
|\mathcal{M}_3|^2(k_1, k_2, k_3, k_4, k_5)+|\mathcal{M}_3|^2(k_1, k_2, k_4, k_3, k_5)+
g^6\frac{2(s_{ij}^2+s_{12}^2)(s_{ij}s_{12}-s_{i1}s_{j2}-s_{i2}s_{j1})}{s_{i1}s_{j2}s_{i2}s_{j1}s_{i3}s_{j3}s_{13}s_{23}}\nonumber\\[7pt] \times \bigg\{C_3\Big[(s_{ij}+s_{12})(s_{ij}s_{12}-s_{i1}s_{j2}-s_{i2}s_{j1})+2s_{i1}s_{j2}(-s_{i2}-s_{j1})+2s_{i2}s_{j1}(-s_{i1}-s_{j2})\Big]
\nonumber\\[6pt]
+C_4\Big[(s_{ij}+s_{12})(s_{ij}s_{12}-s_{i1}s_{j2}-s_{i2}s_{j1})-2s_{i1}s_{j2}(-s_{i2}-s_{j1})-2s_{i2}s_{j1}(-s_{i1}-s_{j2})
\nonumber\\
-2s_{ij}(s_{i1}s_{i2}+s_{j2}s_{j1})-2s_{12}(s_{i1}s_{j1}+s_{j2}s_{i2})
\Big] \bigg\} 
\nonumber\\ \nonumber\\
\end{eqnarray}
Where $C_3=\frac{10}{81}$ and $C_4=\frac{8}{81}$.

All of the QCD 5-point squared amplitudes can be calculated from $|\mathcal{M}_1|^2$, $|\mathcal{M}_2|^2$, $|\mathcal{M}_3|^2$, and $|\mathcal{M}_4|^2$ by crossing, as shown in Table~\textcolor{yy}{\ref{crossy7}}.

\begin{table}[h]
\centering
\begin{tabular}{|c||c|c|c|c|c|c|c|c|}
\hline
 &Process &   \\  
 & $ij\to 123$ & $|\mathcal{M}|^2$ \\
\whline
$qqq'q'g$ &$ qq'\to qq'g$ & $|\mathcal{M}_3|^2\ (k_i,k_j,k_1,k_2,k_3)$  \\    
$ $ & $q\bar{q}'\to q\bar{q}'g$ & $|\mathcal{M}_3|^2\ (k_i,k_2,-k_1,-k_j,k_3)$  \\  
$ $ & $q\bar{q}\to \bar{q}'q'g$ & $|\mathcal{M}_3|^2\ (k_i,-k_1,-k_j,k_2,k_3)$  \\  
$ $ & $qg\to qq'\bar{q}'$ & $(-\frac{3}{8})|\mathcal{M}_3|^2\ (k_i,-k_3,k_1,k_2,-k_j)$  \\[5pt]  
\hline
$qqqqg$ & $qq\to qqg$ & $|\mathcal{M}_4|^2\ (k_i,k_j,k_1,k_2,k_3)$  \\ 
$ $ & $q\bar{q}\to q\bar{q}g$ & $|\mathcal{M}_4|^2\ (k_i,-k_2,k_1,-k_j,k_3)$  \\  
$ $ & $qg\to qq\bar{q}$ & $(-\frac{3}{8})|\mathcal{M}_4|^2\ (k_i,-k_3,k_1,k_2,-k_j)$  \\[5pt] 
\hline
$ggggg$ & $gg\to ggg$ & $|\mathcal{M}_1|^2\ (k_i,k_j,k_1,k_2,k_3)$  \\[5pt]
\hline  
$qqggg$ & $q\bar{q}\to ggg$ & $|\mathcal{M}_2|^2\ (k_i,k_j,k_1,k_2,k_3)$  \\  
$ $ & $qg\to qgg$ & $(-\frac{3}{8})|\mathcal{M}_2|^2\ (k_i,-k_1,-k_j,k_2,k_3)$  \\  
$ $ & $gg\to q\bar{q}g$ & $(\frac{9}{64})|\mathcal{M}_2|^2\ (-k_2,-k_1,-k_j,-k_i,k_3)$  \\[5pt] 
\hline       
\end{tabular}  
\caption{The QCD processes which are related by crossing.\label{crossy7} }
\end{table}

\begin{itemize}
\item {\large QED amplitudes}
\end{itemize}
We follow \rr{\cite{berends}}, and write the QED amplitudes.
 
\begin{eqnarray} 
|\mathcal{M}|^2(\textcolor{blue}{q(k_i)\ \bar{q}(k_j) \to \gamma(k_1)\ \gamma(k_2)\ \gamma(k_3)})\ =\   2e^6 S_3\ \frac{s_{ij}}{s_{i1}s_{i2}s_{i3}s_{j1}s_{j2}s_{j3}}
\end{eqnarray}

\begin{eqnarray} 
|\mathcal{M}|^2(\textcolor{blue}{q(k_i)\ \bar{q}(k_j) \to q'(k_1)\ \bar{q}'(k_2)\ \gamma(k_3)})= \ \ \ \ \ \ \ \ \ \ \ \ \ \ \ \ \ \ \ \ \ \ \ \ \ \ \ \ \ \ \ \ \ \ \ \ \ \ \ \ \ \ \ \ \ \ \ \ 
\nonumber\\[10pt] -\frac{e^6}{2}\bigg(-\frac{s_{12}}{s_{23}s_{13}}-\frac{s_{j2}}{s_{23}s_{j3}}+\frac{s_{21}}{s_{23}s_{i3}}+\frac{s_{j1}}{s_{13}s_{j3}}+\frac{s_{i1}}{s_{13}s_{i3}}-\frac{s_{ij}}{s_{j3}s_{i3}} \bigg) \frac{s_{i2}^2+s_{j1}^2+s_{j1}^2+s_{i2}^2}{s_{ij}s_{12}}
\end{eqnarray}

\begin{eqnarray} 
|\mathcal{M}|^2(\textcolor{blue}{q(k_i)\ \bar{q}(k_j) \to q(k_1)\ \bar{q}(k_2)\ \gamma(k_3)})=\ \ \ \ \ \ \ \ \ \ \ \ \ \ \ \ \ \ \ \ \ \ \ \ \ \ \ 
\nonumber\\[10pt]
-\frac{e^6}{2}\bigg(-\frac{s_{12}}{s_{23}s_{13}}-\frac{s_{j2}}{s_{23}s_{j3}}+\frac{s_{21}}{s_{23}s_{i3}}+\frac{s_{j1}}{s_{13}s_{j3}}+\frac{s_{i1}}{s_{13}s_{i3}}-\frac{s_{ij}}{s_{j3}s_{i3}} \bigg)
\nonumber\\[10pt]
\times \  \frac{s_{i2}s_{j1}(s_{i2}^2+s_{j1}^2)+s_{j1}s_{i2}(s_{j1}^2+s_{i2}^2)+s_{ij}s_{12}(s_{ij}^2+s_{12}^2)}{s_{ij}s_{12}s_{i2}s_{j1}}
\end{eqnarray}

\begin{itemize}
\item {\large QCD+QED amplitudes}
\end{itemize}

\begin{eqnarray} 
|\mathcal{M}|^2(\textcolor{blue}{q\ q\ \to \ g \ \gamma \ \gamma})\ =\ \frac{4}{3}\frac{g^2}{e^2} |\mathcal{M}|^2(\textcolor{blue}{q\ q\  \to \gamma \ \gamma \ \gamma})
\end{eqnarray}

\begin{eqnarray} 
|\mathcal{M}|^2(\textcolor{blue}{q(k_i)\ \bar{q}(k_j) \to g(k_1)\ \gamma(k_2)\ \gamma(k_3)})\ = \ 2g^4e^2 S_3\ \frac{1}{s_{i1}s_{12}s_{2j}s_{i3}s_{3j}}
\end{eqnarray}

\begin{eqnarray} 
|\mathcal{M}|^2(\textcolor{blue}{g\ g \to g \ \gamma \ \gamma})\ =\  |\mathcal{M}|^2(\textcolor{blue}{g\ g \to g\ g \ \gamma})\ =\  0
\end{eqnarray}

\subsection{String theory}
\label{subsec:5pointstring}
\emph{References:} We follow \textcolor{rr}{\cite{lust2}}.
\subsubsection{Generalization of the Veneziano amplitude}
\label{subsubsec:genven5}

The 5-point string amplitudes have been calculated (see \rr{\cite{lust2}}). The two universal amplitudes were explicitly shown to have the same stringy form factor. These amplitudes are expressed in terms of 2 generalized hypergeometric functions. 

The two hypergeometric functions are:
\begin{eqnarray}
f_1\ =\ \int_0^1 dx \int_0^1 dy\ x^{\hat{s}_2-1} y^{\hat{s}_5-1} (1-x)^{\hat{s}_3} (1-y)^{\hat{s}_4} (1-xy)^{\hat{s}_1-\hat{s}_3-\hat{s}_4} 
\end{eqnarray}
\begin{eqnarray}
f_2\ =\ \int_0^1 dx \int_0^1 dy\ x^{\hat{s}_2} y^{\hat{s}_5} (1-x)^{\hat{s}_3} (1-y)^{\hat{s}_4} (1-xy)^{\hat{s}_1-\hat{s}_3-\hat{s}_4-1} 
\end{eqnarray}

These two functions can be written as a sum:
\begin{eqnarray}
f_1\ =\ \sum_{n=0}^\infty \frac{1}{n!}\frac{\Gamma(-\hat{s}_{35}+n)}{\Gamma(-\hat{s}_{35})}\ B(\hat{s}_{23}+n, \hat{s}_{34}+1)\ B(\hat{s}_{45}+1,\hat{s}_{51}+n)
\end{eqnarray}
\begin{eqnarray}
f_2\ =\ \sum_{n=1}^\infty \frac{1}{(n-1)!}\frac{\Gamma(-\hat{s}_{35}+n)}{\Gamma(-\hat{s}_{35}+1)}\ B(\hat{s}_{23}+n, \hat{s}_{34}+1)\ B(\hat{s}_{45}+1, \hat{s}_{51}+n)
\end{eqnarray}

\begin{figure}
\centering
\includegraphics[width= 60mm]{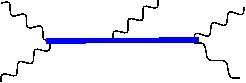}
\caption{Regge states exchange in the 5-point function.\label{r5}}
\end{figure}

The beta functions can be expanded as a sum of their poles (Eq.~(\zz{\ref{eq:venred}})):
\begin{eqnarray}
f_1\ =\ \sum_{n=0}^\infty \sum_{n'=0}^\infty \sum_{j=0}^{min\{n,n'\}}
\gamma(s_{35},j)\ \frac{\gamma(\hat{s}_{34},n-j)}{\hat{s}_{23}+n}\ \frac{\gamma(\hat{s}_{45},n'-j)}{\hat{s}_{51}+n'}
\end{eqnarray}
\begin{eqnarray}
f_2\ =\ \sum_{n=1}^\infty \sum_{n'=1}^\infty \sum_{j=1}^{min\{n,n'\}}
\gamma(\hat{s}_{35}-1,j-1)\ \frac{\gamma(\hat{s}_{34},n-j)}{\hat{s}_{23}+n}\ \frac{\gamma(\hat{s}_{45},n'-j)}{\hat{s}_{51}+n'}
\end{eqnarray}
Where $\gamma(\hat{s},n)\equiv \frac{1}{n!}\frac{\Gamma(-\hat{s}+n)}{\Gamma(-\hat{s})}$. So there is a sum over double poles implying exchanges of two resonances, Fig.~\textcolor{yy}{\ref{r5}}.

We make the following definitions:
\begin{eqnarray}
V^{(5)}(s_j)\ \equiv \ s_2 s_5 f_1 \ +\  \frac{1}{2}\Big(s_2 s_3 + s_4 s_5 - s_1 s_2 - s_3 s_4- s_1 s_5 \Big)f_2
\end{eqnarray}
\begin{eqnarray}
P^{(5)}(s_j)\ \ \equiv \ \ f_2 
\end{eqnarray}
\begin{eqnarray}
\epsilon(i,j,m,n)\ \equiv \ \alpha'^2 \epsilon_{\alpha\beta\mu\nu}k_i^{\alpha}k_j^{\beta}k_m^{\mu}k_n^{\nu} 
\end{eqnarray}

Then a universal sub-amplitude can be written as (see also section~\ref{subsubsec:b2}):

\begin{eqnarray} 
\label{eq:b5v56}
\boxed{
m_{string}\ =\ \mathfrak{B}_5\ m_{QCD}\ =\  \Big[V^{(5)}(s_i)-2i \epsilon(1,2,3,4) P^{(5)}(s_i)\Big]m_{QCD}   }
\end{eqnarray}
Compare this with the 6-gluon case of Eq.~(\textcolor{zz}{\ref{eq:b5v767}}).

One difference from the 4-point amplitude, is the inclusion of the antisymmetric tensor.
It will be convenient to write the amplitudes in terms of:
\begin{eqnarray}
\mathcal{C}(k_1, k_2,k_3, k_4, k_5)\ \equiv \ \frac{V^{(5)}(s_j)- 2iP^{(5)}(s_j) \epsilon(1, 2, 3, 4)}{[12][23][34][45][51]}
\end{eqnarray}
\begin{eqnarray}
\mathcal{C}_\lambda \ \equiv \ \mathcal{C}(k_{1_{\lambda}}, k_{2_{\lambda}},k_{3_{\lambda}}, k_{4_{\lambda}}, k_{5_{\lambda}})
\end{eqnarray}

From Eqs.~(\textcolor{zz}{\ref{eq:b5v56}}), (\textcolor{zz}{\ref{eq:b5yu1}}),(\textcolor{zz}{\ref{eq:b5yu2}}) the two universal amplitudes can then be written as:
\begin{eqnarray} 
m_{string}\ =\  ig^3 \langle IJ\rangle^4\  \mathcal{C}_{\lambda}
\end{eqnarray}
\begin{eqnarray} 
m_{string}\ =\  ig^3 \langle IJ\rangle^4 \langle qI\rangle ^3\langle \bar{q}I\rangle \ \mathcal{C}_{\lambda} 
\end{eqnarray}
Where $I,J$ stand for the gluons with negative helicity.

\subsubsection{Squaring the amplitudes}

Written below are the two matrices that arise from squaring and summing over colors. The $12\times 12$ matrix $\mathcal{S}_{ \lambda \lambda'}$ corresponds to the permutations $\Pi_5$, and the $6\times 6$ matrix $\mathcal{P}_{ \lambda \lambda'}$ to $\Pi_q$.
The entries for the matrices are given in Tables~\textcolor{yy}{\ref{groupfact}} and \textcolor{yy}{\ref{su3group}}. 

From Eqs.~(\textcolor{zz}{\ref{eq:slam1}}) and (\textcolor{zz}{\ref{eq:plam1}}),
\begin{eqnarray}
\label{eq:spi5}
\setcounter{MaxMatrixCols}{14}
\mathcal{S}_{ \lambda \lambda'}\ \equiv \ \sum_{a_1,\ldots ,a_5} t^{a_{1_\lambda}a_{2_\lambda}a_{3_\lambda}a_{4_\lambda}a_{5_\lambda}}\  \big[t^{a_{1_{\lambda'}}a_{2_{\lambda'}}a_{3_{\lambda'}}a_{4_{\lambda'}}a_{5_{\lambda'}}}\big]^*\ \ \ \ \ \ \ \ \ \ \ \ \ \ \ \ \ \ \ \ \ \ \ \ \ \ \ \ \ \ \ \ 
\nonumber\\ \nonumber\\
= \begin{bmatrix}
D&X&Y&X&Y&-X&X&-Y&Y&0&Y&-X\\
X&D&-X&Y&X&Y&-Y&X&0&Y&X&-Y\\
Y&-X&D&X&Y&X&X&-Y&-Y&-X&-Y&0\\
X&Y&X&D&-X&Y&Y&0&-X&Y&X&Y\\
Y&X&Y&-X&D&X&0&Y&Y&-X&-Y&-X\\ 
-X&Y&X&Y&X&D&-Y&X&-X&-Y&0&Y\\
X&-Y&X&Y&0&-Y&D&X&X&Y&-X&Y\\ 
-Y&X&-Y&0&Y&X&X&D&Y&X&-Y&X\\
Y&0&-Y&-X&Y&-X&X&Y&D&-X&Y&X\\ 
0&Y&-X&Y&-X&-Y&Y&X&-X&D&-X&-Y\\ 
Y&X&-Y&X&-Y&0&-X&-Y&Y&-X&D&X\\ 
-X&-Y&0&Y&-X&Y&Y&X&X&-Y&X&D\\ 
\end{bmatrix}
\end{eqnarray}

\begin{eqnarray}
\label{eq:ppi5}
\mathcal{P}_{ \lambda \lambda'}\ \ \equiv \ \ 
\begin{bmatrix}
D_q&X_q&X_q&Y_q&Y_q&Z_q\\
X_q&D_q&Y_q&Z_q&X_q&Y_q\\
X_q&Y_q&D_q&X_q&Z_q&Y_q\\
Y_q&Z_q&X_q&D_q&Y_q&X_q\\
Y_q&X_q&Z_q&Y_q&D_q&X_q\\
Z_q&Y_q&Y_q&X_q&X_q&D_q\\
\end{bmatrix}
\end{eqnarray}

\begin{table}[h]
\centering
\begin{tabular}{|c||c|c|c|c|c|c|c|}
\hline
$Group$ & $C_A$ & $C_F$ & $N_A$ & $D/N_A$ & $X/N_A$ & $Y/N_A$ \\  
\whline 
   $SU(N)$ &  $N$ &      $\frac{N^2-1}{2N}$ & $N^2-1$ & $\frac{N^4-4N^2+10}{16N}$ & $\frac{2-N^2}{8N}$ & $\frac{1}{8N}$  \\[4pt]  
\hline
$SU(3)$ &  $3$ &      $\frac{4}{3}$ & $8$ & $\frac{55}{48}$ & $-\frac{7}{24}$ & $\frac{1}{24}$  \\[4pt]  
\hline  
$SU(2)$ &  $2$ &      $\frac{1}{2}$ & $3$ & $\frac{5}{16}$ & $-\frac{1}{8}$ & $\frac{1}{24}$  \\[4pt]
\hline 
$SO(N)$ &  $\frac{N-2}{2}$ &  $\frac{N-1}{4}$ & $\frac{N(N-1)}{2}$ & $\frac{(N-2)(N^2-2N+2)}{128}$ & $\frac{(N-2)^2}{128}$ & $\frac{N-2}{64}$ \\[4pt] 
\hline
$SP(N)$ & $\frac{N+2}{2}$ & $\frac{N+1}{4}$ & $\frac{N(N+1)}{2}$ & $\frac{(N+2)(N^2+2N+2)}{128}$ & $-\frac{(N+2)^2}{128}$ & $\frac{N+2}{64}$ \\[4pt] 
\hline         
\end{tabular} 
\caption{Group factors for Eq.~(\zz{\ref{eq:spi5}}). The $SO(N)$ and $SP(N)$ gauge groups are also included. From \rr{\cite{lust2}}. \label{groupfact}} 
\end{table}

\begin{table}[h]
\centering
\begin{tabular}{|c||c|c|c|c|c|c|c|c|c|c|c|}
\hline
$ $ & $D_q$ & $X_q$ & $Y_q$ & $Z_q$ \\[3pt]  
\whline 
$\scriptstyle [N]_a\ ,\ [N]_b$ & $\scriptstyle \big[NC_F^3\big]_a[N]_b$ & $\scriptstyle \big[NC_F^2(C_F-\frac{C_A}{2})\big]_a[N]_b$ & $\scriptstyle \big[NC_F(C_F-\frac{C_A}{2})^2\big]_a[N]_b$ & $\scriptstyle \big[NC_F(C_F-\frac{C_A}{2})(C_F-C_A)\big]_a[N]_b$  \\[3pt]  
\hline    
$\scriptstyle [N]_a=3\ ,\ [N]_b=2$  & $\frac{128}{9}$ & $-\frac{16}{9}$ & $\frac{2}{9}$ & $\frac{20}{9}$  \\[3pt]  
\hline        
\end{tabular}
\caption{Group factors for Eq.~(\zz{\ref{eq:ppi5}}). From \rr{\cite{lust2}}. \label{su3group}}  
\end{table}

\subsubsection{The squared amplitudes}
\label{subsubsec:5squared}

We want to use a notation similar to the 4-point case, so we write the squared amplitudes with the $U(1)$ gauge boson from the color stack $A$, whereas \rr{\cite{lust2}} used $\gamma$. We also write $B$ for the non-abelian gauge boson from stack $b$.

\begin{itemize}
\item {\large 5 gauge bosons}
\end{itemize}

\begin{eqnarray}
|\mathcal{M}|^2(\textcolor{blue}{g_1, g_2, g_3, g_4, g_5})\ =\  64 g^6\ \mathcal{S}_4\sum_{\lambda,\lambda ' \in \Pi_5} \mathcal{C}_{\lambda} \mathcal{S}_{\lambda\lambda '} \mathcal{C}^*_{\lambda '}
\end{eqnarray}
See also Eq.~(\textcolor{zz}{\ref{eq:slam}}).

\begin{eqnarray}
\label{eq:77a1}
|\mathcal{M}|^2(\textcolor{blue}{A_1, g_2, g_3, g_4, g_5})\ =\  8g^6 Q_A^2(N^2-1)(N^2-4)\ \mathcal{S}_4 \ \ \ \ \ \ \ \ \ \ \ \ \ \ \ \ \ \ \ \ \ \ \  
\nonumber\\[7pt] \times
\Big[\ \big|\mathcal{C}_1-\mathcal{C}_6+\mathcal{C}_7+\mathcal{C}_9\big|^2+\big|\mathcal{C}_2-\mathcal{C}_3+\mathcal{C}_8+ \mathcal{C}_{10}\big|^2+\big|\mathcal{C}_4-\mathcal{C}_5+\mathcal{C}_{11}+\mathcal{C}_{12}\big|^2\Big] 
\end{eqnarray}

\begin{eqnarray}
\label{eq:77a2}
|\mathcal{M}|^2(\textcolor{blue}{A_1, A_2, g_3, g_4, g_5})\ =\ 16g^6 Q_A^4(N^2-1)\ \mathcal{S}_4 \ \ \ \ \ \ \ \ \ \ \ \ \ \ \ \ \ \ \ \ \ \ \ \ \ \ \ \ \ \ 
\nonumber\\[7pt] \times 
\Big[\ \big|\mathcal{C}_1-\mathcal{C}_6+\mathcal{C}_7+\mathcal{C}_9-\mathcal{C}_2+\mathcal{C}_3-\mathcal{C}_8-\mathcal{C}_{10}+\mathcal{C}_4-\mathcal{C}_5+\mathcal{C}_{11}+\mathcal{C}_{12}\big|^2\ \Big] \nonumber\\
\end{eqnarray}

\begin{eqnarray}
|\mathcal{M}|^2(\textcolor{blue}{A_1, A_2, A_3, g_4, g_5})= |\mathcal{M}|^2(\textcolor{blue}{A_1, A_2, A_3, A_4, g_5})= |\mathcal{M}|^2(\textcolor{blue}{A_1, A_2, A_3, A_4, A_5})= 0
\end{eqnarray}

\begin{itemize}
\item {\large 3 gauge bosons + 2 quarks}
\end{itemize}

\begin{eqnarray}
|\mathcal{M}|^2(\textcolor{blue}{g_1, g_2, g_3, q_4, \bar{q}_5})\ =\  16g^6 \ \mathcal{S}_3 \sum_{\lambda,\lambda \in \Pi_q} \mathcal{C}_{\lambda} \mathcal{P}_{\lambda\lambda} \mathcal{C}^*_\lambda
\end{eqnarray}
See also Eq.~(\textcolor{zz}{\ref{eq:plam}}).

\begin{eqnarray}
|\mathcal{M}|^2(\textcolor{blue}{A_1, g_2, g_3,  q_4, \bar{q}_5})\ =\  4g^6 Q_A^2[N]_b\Big[\frac{N^2-1}{N}\Big]_a\mathcal{S}_3\ \ \ \ \ \ \ \ \ \ \ \ \ \ \ \ \ \ \ \ \ \ \ \ \ \ \ \ \ \ \ \ \ \ \ \ \ \ \ \ \ \ \ \ \ \ \ \ \ \ 
\nonumber\\[7pt] \times \bigg[[N^2-1]_a\Big(\big|\mathcal{C}_1+\mathcal{C}_7+\mathcal{C}_9\big|^2+\big|\mathcal{C}_4+\mathcal{C}_{11}+\mathcal{C}_{12}^2\big|^2\Big) -2Re\Big\{\big(\mathcal{C}_1+\mathcal{C}_{7}+\mathcal{C}_9\big)\big(\mathcal{C}_4^*+\mathcal{C}_{11}^*+\mathcal{C}_{12}^*\big)\Big\}\bigg] \nonumber\\
\end{eqnarray}

\begin{eqnarray}
|\mathcal{M}|^2(\textcolor{blue}{A_1, A_2, g_3, q_4, \bar{q}_5})\ =\  8g^6 Q_A^4[N]_b[N^2-1]_a\ \mathcal{S}_3\ \Big|\sum_{\lambda \in \Pi_q} \mathcal{C}_{\lambda}\Big|^2
\end{eqnarray}

\begin{eqnarray}
|\mathcal{M}|^2(\textcolor{blue}{A_1, A_2, A_3, q_4, \bar{q}_5})\ =\  16g^6 Q_A^6[N]_b[N]_a\ \mathcal{S}_3\ \Big|\sum_{\lambda \in \Pi_q} \mathcal{C}_{\lambda}\Big|^2
\end{eqnarray}

\begin{eqnarray}
|\mathcal{M}|^2(\textcolor{blue}{g_1, g_2, B_3, q_4, \bar{q}_5})\ =\  16g^4 g_b^4[NC_F]_b[NC_F]_a\ \mathcal{S}_3\ \ \ \ \ \ 
\nonumber\\[7pt]
\times \bigg\{[C_F]_a\Big(|\mathcal{C}_2|^2+|\mathcal{C}_8|^2\Big)+\Big[C_F-\frac{C_A}{2}\Big]_a\Big(\mathcal{C}_2 \mathcal{C}_8^*+\mathcal{C}_2^* \mathcal{C}_8\Big)\bigg\}
\end{eqnarray}

\begin{eqnarray}
|\mathcal{M}|^2(\textcolor{blue}{g_1, g_2, B^0_3, q_4, \bar{q}_5})\ =\ |\mathcal{M}|^2(\textcolor{blue}{g_1, g_2, B_3, q_4, \bar{q}_5})\ \Big|_{\ [NC_F]_b \to [NQ^2]_b}
\end{eqnarray}

\begin{eqnarray}
|\mathcal{M}|^2(\textcolor{blue}{A_1, g_2, B_3, q_4,  \bar{q}_5})\ =\  16g^4g_b^4 Q_A^2 [NC_F]_b[N C_F]_a\ \mathcal{S}_3\ |\mathcal{C}_2+\mathcal{C}_8|^2
\end{eqnarray}

\begin{eqnarray}
|\mathcal{M}|^2(\textcolor{blue}{A_1, g_2, B^0_3, q_4,  \bar{q}_5})\ =\ |\mathcal{M}|^2(\textcolor{blue}{A_1, g_2, B_3, q_4,  \bar{q}_5})\ \Big|_{\ [NC_F]_b \to [NQ^2]_b}
\end{eqnarray}

\subsubsection{Low energy limit}

From the expansions 

\begin{eqnarray}
f_1=\frac{1}{s_2 s_5}-\zeta(2)\Big(\frac{s_3}{s_5}+\frac{s_4}{s_2}\Big)+\zeta(3)\Big(-s_1+s_3+s_4+\frac{s_4^2+s_4s_5}{s_2}+\frac{s_3^2+s_2s_3}{s_5}\Big)+\mathcal{O}(\alpha'^2) \nn
\end{eqnarray}

\begin{eqnarray}
f_2\ =\  \zeta(2)-\zeta(3)\big(s_1+s_2+s_3+s_4+s_5\big)\ +\ \mathcal{O}(\alpha'^2)
\end{eqnarray}

we have:
\begin{eqnarray}
\label{eq:c1low}
\mathcal{C}_1\equiv\mathcal{C}(k_1, k_2,k_3, k_4, k_5)\approx \frac{1-\frac{\zeta(2)}{2}\big(s_1s_2+s_2s_3+s_3s_4+s_4s_5+s_5s_1+4i\epsilon (1234)\big)}{[12][23][34][45][51]}
\end{eqnarray}

\begin{eqnarray}
\mathcal{C}_1-\mathcal{C}_6+\mathcal{C}_7+\mathcal{C}_9\ \ \ \ \ \  \stackrel{s_i\rightarrow 0}{\longrightarrow}\ \ \ \ \ \ \ 0
\end{eqnarray}
\begin{eqnarray}
\mathcal{C}_2-\mathcal{C}_3+\mathcal{C}_8+\mathcal{C}_{10}\ \ \ \ \ \ \stackrel{s_i\rightarrow 0}{\longrightarrow}\ \ \ \ \ \ 0
\end{eqnarray}
\begin{eqnarray}
\mathcal{C}_4-\mathcal{C}_5+\mathcal{C}_{11}+\mathcal{C}_{12}\ \ \ \ \ \ \stackrel{s_i\rightarrow 0}{\longrightarrow}\ \ \ \ \ \  0
\end{eqnarray}

So that Eqs.~(\textcolor{zz}{\ref{eq:77a1}}) and (\textcolor{zz}{\ref{eq:77a2}}) can be seen to vanish in the field theory limit as required.
\clearpage

\section{\textcolor{ww}{Higher point amplitudes}}
\label{sec:multi-gluon amplitudes}
\emph{References:} In large parts we follow~\textcolor{rr}{\cite{stieberger2}}.\\

From section~\ref{subsec:fds4} it is seen that higher point universal amplitudes can be computed by generalizing the integral representation of the Veneziano amplitude to multiple integrals. As we have seen for the 5-point case,  
these integrals have the form of multiple hypergeometric functions. The $n$-gluon amplitude can be expressed in terms of generalized hypergeometric functions of the form:

\begin{eqnarray}
\label{eq:hyph1}
F
\bigg[\begin{array}{c}
\scriptstyle n_a \\ \scriptstyle n_{ab}
\end{array} \bigg]\ =\ \int_0^1 dx_1\ldots \int_0^1 dx_{n-3}\ \prod_{a=1}^{N-3}x_a^{1+a-n+n_a} \ \prod_{b=a}^{n-3}x_a^{2\alpha'k_{b+3}\left( k_1+\Sigma^{b+2}_{j=a+3} k_j \right)} 
\nonumber\\[6pt]
\times \left(1- \prod_{j=a}^{b}x_j\right)^{2\alpha'k_{2+a}k_{3+b}+n_{ab}}
\end{eqnarray}
There are $(n-3)!$ independent hypergeometric functions\footnote{(n-3)! is also the number of independent string subamplitudes. The relations between the string subamplitudes reduce to the Kleiss-Kuijf and BCJ relations in the field theory limit. This gives a neat explanation to why there are precisely $(n-3)!$ independent \underline{field} theory subamplitudes. See \rr{\cite{stieberger4, stieberger5, lust2}}.}. The number $n-3$ comes, due to the $PSL(2,R)$ invariance, from fixing three coordinates.

For example for $n=4,5$ we have:
\begin{eqnarray}
F
\bigg[\begin{array}{c}
\scriptstyle n_1 \\ \scriptstyle n_{11}
\end{array} \bigg]\ =\  
\int_0^1 dx_1\ x_1^{-2+s_{23}+n_1}(1-x_1)^{s_{12}+n_{11}}\ = \ 
\frac{{_2F_1}\bigg[\begin{array}{c}
\scriptstyle s_{23}+n_1-1, -s_{12}-n_{11}
\\
\scriptstyle s_{23}+n_{1}
\end{array} ; 1 \bigg]}{s_{23}+n_1-1}
\nonumber\\
\end{eqnarray}

and
\begin{eqnarray}
F
\bigg[\begin{array}{c}
\scriptstyle n_1, n_2 \\ \scriptstyle n_{11}, n_{12}, n_{22}
\end{array} \bigg]=\ \ \ \ \ \ \ \ \ \ \ \ \ \ \ \ \ \ \ \ \ \ \ \ \ \ \ \ \ \ \ \ \ \ \ \ \ \ \ \ \ \ \ \ \ \ \ \  \ \ \ \ \ \ \ \ \ \ \ \ \ \ \ \ \ \ \ \ \ \ \ \ \ \ \ \ \ \ \ \ \ \ \ \ 
\nonumber\\[8pt] 
\int_0^1 \int_0^1 dx_1dx_2\  x_1^{-3+s_{23}+n_1}x_2^{-2+s_{15}+n_2}(1-x_1)^{s_{34}+n_{11}}(1-x_2)^{s_{45}+n_{22}}(1-x_1x_2)^{s_{35}+n_{12}}= 
\nonumber\\[8pt]
\frac{\Gamma(s_{23}+n_1-2)\ \Gamma(s_{15}+n_2-1)\ \Gamma(s_{34}+n_{11}+1)\ \Gamma(s_{45}+n_{22}+1)}{\Gamma(s_{23}+s_{34}+n_1+n_{11}-1)\ \Gamma(s_{15}+s_{45}+n_2+n_{22})} 
\nonumber\\[8pt]
\times \ {_3F_2}\bigg[\begin{array}{c}
\scriptstyle s_{23}+n_1-2, s_{15}+n_2-1, -s_{35}-n_{12} \\ \scriptstyle s_{23}+s_{34}+n_1+n_{11}-1, s_{15}+s_{45}+n_2+n_{22}-1
\end{array} ; 1 \bigg]
\end{eqnarray}

In section~\ref{subsubsec:genven5} we used the definitions:
\begin{eqnarray}
f_1\ =\  F
\bigg[\begin{array}{c}
\scriptstyle 2, 1 \\ \scriptstyle 0, 0, 0
\end{array} \bigg]\ \ \ \ , \ \ \ \ 
f_2\ =\  F
\bigg[\begin{array}{c}
\scriptstyle 3, 2 \\ \scriptstyle 0, -1, 0
\end{array} \bigg]
\end{eqnarray}

\subsection{$6$ gluons}

Eq.~(\textcolor{zz}{\ref{eq:hyph1}}) gives:
\begin{eqnarray}
F
\bigg[\begin{array}{c}
\scriptstyle n_1, n_2, n_3 \\ \scriptstyle n_{11}, n_{12}, n_{22}, n_{13}, n_{23}, n_{33}
\end{array} \bigg]=
\int_0^1\int_0^1\int_0^1dx_1 dx_2 dx_3 x_1^{-4+s_{23}+n_1}x_2^{-3+\alpha'(k_2+k_3+k_4)^2+n_2}x_3^{-2+s_{16}+n_3}
\nonumber\\[8pt] \times\  
(1-x_1)^{s_{34}+n_{11}}(1-x_2)^{s_{45}+n_{22}}(1-x_3)^{s_{56}+n_{33}}
\nonumber\\[8pt] \times\ 
(1-x_1x_2)^{s_{346}+n_{23}}(1-x_2x_3)^{s_{46}+n_{23}}(1-x_1x_2x_3)^{s_{36}+n_{13}}\nonumber\\
\end{eqnarray}

As in QCD, the case of 6 gluons is more complicated then 4 and 5 gluons due to the non-vanishing of the non-MHV amplitude $(1^+,2^+,3^-,4^-,5^-,6^-)$

The MHV amplitude can be written as:
\begin{eqnarray} 
\label{eq:b5v767}
\boxed{m_{string}(1^-,2^-,3^+,4^+,5^+,6^+)\ =\  \Big[V^{(6)}(s_i,t_i)-2i\sum _{k=1}^{k=5} \epsilon_k P_k^{(6)}(s_i,t_i)\Big]\ m_{QCD}^{(6)}}
\end{eqnarray}
In other words the stringy form factor is (see Eq.~(\zz{\ref{eq:formfactor16}})): 
\begin{eqnarray}
\mathfrak{B}_6^{(MHV)}\ =\  V^{(6)}(s_i,t_i)-2i\sum _{k=1}^{k=5} \epsilon_k P_k^{(6)}(s_i,t_i)
\end{eqnarray}
Compare this with the 5-gluon case of Eq.~(\textcolor{zz}{\ref{eq:b5v56}}).

Where the six functions $P_k^{(6)},\ V^{(6)}$ can be written in terms of six generalized hypergeometric functions:
\begin{eqnarray} 
P_1^{(6)}\ =\ s_6F_2+ (s_6+s_2+s_5-t_1-t_2)F_4+(s_2+s_1-s_6-t_1)F_3
\end{eqnarray}
\begin{eqnarray}
P_2^{(6)}\ =\ s_6F_2 -(s_1-s_3+s_5-t_1)F_5+ (s_2+s_5+s_3-t_2)F_4+
\nonumber\\[5pt]
(s_2-s_3+s_5-s_6-t_1+t_3)F_3+(s_1+s_3-s_5-t_3)F_6
\end{eqnarray}
\begin{eqnarray}
P_3^{(6)}\ =\ s_6F_2+ (s_2-s_6-s_3-t_1+t_3)F_3-(s_1-s_3+s_5-t_1)F_5+
\nonumber\\[5pt]
(s_4+s_5-t_1)F_4+(s_3+s_1-t_3)F_6
\end{eqnarray}
\begin{eqnarray}
P_4^{(6)}\ =\ s_6F_2+(-s_6+s_1+s_2-t_1)F_3+(s_4+s_5-t_1)F_4-(s_1-s_3+s_5-t_1)F_5
\end{eqnarray}
\begin{eqnarray}
P_5^{(6)}\ =\ s_6F_2+(s_1+s_4-t_1-t_3)(F_3-F_5)+(s_4+s_5-t_1)F_4+ (s_3+s_4-s_5-t_3)F_5
\nonumber\\
\end{eqnarray}

\begin{eqnarray}
V^{(6)}\ =\ t_2\Big[s_2s_6F_1-s_2(s_1-s_5-t_3)F_3+
(s_4+s_5-t_1)\big\{s_6(F_2-F_3)-\ \ \ \ \ \ \ \ \ \ \ \ \ 
\nonumber\\[5pt]
(s_3-s_5+t_1-t_3)(F_3+F_4)-
(s_1-s_3+s_5-t_1)F_5+(s_1+s_3-s_5-t_3)F_6   \big\}  \Big]+
\nonumber\\[5pt]
\frac{1}{2}(s_2s_5+s_2s_5+s_2s_5+s_2s_5+s_2s_5+s_2s_5)P_1^{(6)}
\nonumber\\[5pt]
+ \frac{1}{2}(s_2s_5+s_2s_5+s_2s_5+s_2s_5+s_2s_5+s_2s_5+s_2s_5+s_2s_5+s_2s_5+s_2s_5)P_2^{(6)}
\nonumber\\[5pt]
+ \frac{1}{2}(s_2s_5+s_2s_5+s_2s_5+s_2s_5+s_2s_5+s_2s_5+s_2s_5+s_2s_5+s_2s_5+s_2s_5+s_2s_5+s_2s_5)P_3^{(6)}
\nonumber\\[5pt]
+ \frac{1}{2}(s_2s_5+s_2s_5+s_2s_5+s_2s_5+s_2s_5+s_2s_5+s_2s_5+s_2s_5+s_2s_5+s_2s_5)P_4^{(6)}
\nonumber\\[5pt]
+ \frac{1}{2}(s_2s_5+s_2s_5+s_2s_5+s_2s_5+s_2s_5+s_2s_5)P_5^{(6)}+s_2s_5P_2^{(6)}+(s_2s_5+s_2s_5)P_3^{(6)}
\nonumber\\
\end{eqnarray}

Where,
\begin{eqnarray}
F_1\ \equiv \ F
\bigg[\begin{array}{c}
\scriptstyle 3, 2, 1 \\ \scriptstyle 0, 0, 0, 0, 0, 0
\end{array} \bigg]\ \ \ \ , \ \ \ \ 
F_3\ \equiv \ F
\bigg[\begin{array}{c}
\scriptstyle 4, 3, 2 \\ \scriptstyle 0, 0, 0, -1, 0, 0
\end{array} \bigg]\ \ \ \ , \ \ \ \ 
F_5\ \equiv \ F
\bigg[\begin{array}{c}
\scriptstyle 4, 3, 2 \\ \scriptstyle 0, -1, 0, -1, 0, 0
\end{array} \bigg]\ \ \ \ , \ \ 
\nonumber\\ \nonumber\\
F_2\ \equiv \ F
\bigg[\begin{array}{c}
\scriptstyle 4, 3, 1 \\ \scriptstyle 0, -1, 0, 0, 0, 0
\end{array} \bigg]\ \ \ \ , \ \ \ \ 
F_4\ \equiv \ F
\bigg[\begin{array}{c}
\scriptstyle 4, 4, 2 \\ \scriptstyle 0, -1, 0, 0, -1, 0
\end{array} \bigg]\ \ \ \ , \ \ \ \ 
F_6\ \equiv \ F
\bigg[\begin{array}{c}
\scriptstyle 4, 3, 2 \\ \scriptstyle 0, 0, 0, -1, -1, 0
\end{array} \bigg]  \ \ \ \ 
\end{eqnarray}

\subsubsection{Squaring the amplitude}
\label{subsubsec:6gluesquare}
Having possession of the MHV amplitude we can square it and sum the colors as in the 5 gluon case. This time though, it will not be the full squared amplitude because of the non-MHV part. For $n$ gluons there are $\frac{(n-1)!}{2}$ permutations (after using reflection symmetry) giving rise to a  $\frac{(n-1)!}{2}\times \frac{(n-1)!}{2}$ matrix:
\begin{eqnarray} 
\mathcal{S}_{ \lambda \lambda'}^{(n)}\ \equiv \ \sum_{a_1,...a_n}t^{a_{1_\lambda}\ldots a_{n_\lambda}}\  \big[t^{a_{1_{\lambda'}}\dots a_{n_{\lambda'}}}\big]^*
\end{eqnarray}

For 5 gluons it was a $12\times 12$ matrix with 3 independent entries. For 6 gluons it is $60\times 60$ with 10 independent entries\footnote{In field theory, the squaring of the amplitude is very much simplified by the use of the dual Ward identity (DWI) Eq.~(\textcolor{zz}{\ref{eq:dwi7}}). In the squared amplitude, the $\mathcal{O}(1/N^2)$ correction terms vanish completely for 4 and 5 gluons (see Eqs.~(\textcolor{zz}{\ref{eq:b7}}), (\textcolor{zz}{\ref{eq:4gs}}), (\textcolor{zz}{\ref{eq:5gs}}), (\textcolor{zz}{\ref{eq:6gluons8}})). This simplification does not occur in string amplitudes since these do NOT satisfy the DWI. The DWI follows by replacing one gluon with a photon and demanding that the amplitude vanish. Since in string theory the gluon can mix with the photon, the identity does not hold.}.
We calculate now the diagonal entry for $n=5,6$. The diagonal is the term of leading order in $1/N$, and is relatively simple to calculate. Using the standard notation:
$(a_1\ldots a_n)\equiv Tr\big(T^{a_1}\cdots T^{a_n}\big)$,
The diagonal term is:
\begin{eqnarray}
\label{eq:entrc}
\mathcal{S}_{ \lambda \lambda}^{(n)}\ \equiv \ \sum_{a_1,...a_n}t^{a_1\ldots a_n}\ \big[t^{a_1\ldots a_n}\big]^*= \ \ \ \ \ \ \ \ \ \ \ \ \ \ \ \ \ \ \ \ \ \ \ \ \ \ \ \ \ \ \ \ \ \ 
\nonumber\\
\sum_{a_1,...a_n}\big[(a_1\ldots a_n)-(a_n\ldots a_1)\big]\ \big[(a_1\ldots a_n)-(a_n\ldots a_1)\big]^*=
\nonumber\\[6pt]
2\sum_{a_1,...a_n}\Big[(a_1\ldots a_n)(a_1\ldots a_n)^*\ -\ (a_1\ldots a_n)(a_n\ldots a_1)^*\Big]=
\nonumber\\[6pt]
2\sum_{a_1,...a_n}\Big[(a_1\ldots a_n)(a_n\ldots a_1)\ -\ (a_1\ldots a_n)(a_1\ldots a_n)\Big]
\end{eqnarray}

The first term gives the leading order in $1/N$, and has a closed formula for arbitrary $n$ (see Eq. 25 of \rr{\cite{kleiss}}):
\begin{eqnarray}
\sum_{a_1,...a_n} (a_1\ldots a_n)(a_n\ldots a_1)\ =\  \frac{(N^2-1)^n+(-1)^n(N^2-1)}{(2N)^n}
\end{eqnarray}
For $n=6$ this gives:
\begin{eqnarray}
\label{eq:sixyou}
\sum_{a_1,...a_6} (a_1\ldots a_6)(a_6\ldots a_1)\ =\  (N^2-1)\cdot \frac{N^8-5N^6+10N^4-10N^2+5}{64N^4}
\end{eqnarray}

As for the second term in Eq.~(\textcolor{zz}{\ref{eq:entrc}}), we calculated it for $n=6$ using Eqs.~(\textcolor{zz}{\ref{eq:redu5}}), (\textcolor{zz}{\ref{eq:redu6}}). We got:
\begin{eqnarray}
\label{eq:sixme}
\sum_{a_1,...a_6}(a_1\ldots a_6)(a_1\ldots a_6)\ =\  (N^2-1)\cdot\frac{N^4+10N^2+5}{64N^4}
\end{eqnarray}

Plugging Eqs.~(\textcolor{zz}{\ref{eq:sixyou}}), (\textcolor{zz}{\ref{eq:sixme}}) in (\textcolor{zz}{\ref{eq:entrc}}) gives:
\begin{eqnarray}\boxed{
\mathcal{S}_{ \lambda \lambda}^{(6)}\ =\  2(N^2-1)\cdot \frac{N^8-5N^6+9N^4-20N^2}{64N^4}}
\end{eqnarray}

\textbf{\underline{Notes}}:
\begin{itemize}
\item We checked this procedure for $n=5$ and got the correct value for the diagonal $\mathcal{S}_{ \lambda \lambda}^{(5)}$ as in Table~\textcolor{yy}{\ref{groupfact}}:
\begin{eqnarray}
D\ \equiv \  \mathcal{S}_{ \lambda \lambda}^{(5)}\ =\  (N^2-1)\cdot \frac{N^4-4N^2+10}{16N} 
\end{eqnarray}

\item This procedure can be continued for larger $n$. In general:
\begin{eqnarray}
\mathcal{S}_{ \lambda \lambda'}^{(n)}\ =\ \frac{1}{2^{n-1}}N^{n-2}(N^2-1)\ \Big[ \alpha_0+\sum_{i=1}^{2[n/2-1]}\frac{\alpha_i}{N^{2i}}\Big]
\end{eqnarray}
Where $\alpha_i$ is an integer and $[a]$ is the entire of $a$. See \rr{\cite{berends4}}.
\end{itemize}

\subsection{$n$ gluons: low energy expansion}

In \rr{\cite{stieberger2}} a method for obtaining the leading stringy correction to the $n$-gluon MHV amplitude was introduced.

The result is:
\begin{eqnarray} 
\boxed{m_{string}(1^-, 2^-, 3^+, 4^+\ldots, n^+)\ =\ \Big(1-\frac{\pi^2}{12}Q^{(n)}\Big)\ m_{QCD}^{(n)}\ +\  \mathcal{O}(\alpha'^3)}
\end{eqnarray}
So that the stringy form factor is:  
\begin{eqnarray}
\mathfrak{B}_n^{(MHV)}\ \approx \ \Big(1-\frac{\pi^2}{12}Q^{(n)}\Big)\ +\  \mathcal{O}(\alpha'^3)
\end{eqnarray}

For $n=4,5,6$ we have:
\begin{eqnarray} 
Q^{(4)}\ =\  s_1s_2
\end{eqnarray}
\begin{eqnarray} 
Q^{(5)}\ =\  s_1s_2+s_1s_2+s_1s_2+s_1s_2+s_1s_2+4i\epsilon(1,2,3,4)
\end{eqnarray}
\begin{eqnarray} 
Q^{(6)}\ =\ s_1s_2+s_1s_2+s_1s_2+s_1s_2+s_1s_2+s_1s_2+t_1t_2+t_1t_2+t_1t_2+s_1s_2+s_1s_2+s_1s_2
\nonumber\\[6pt]
+4i[\epsilon(1,2,3,4)+\epsilon(1,2,3,4)+\epsilon(1,2,3,4)+\epsilon(1,2,3,4)+\epsilon(1,2,3,4)]\nonumber\\
\end{eqnarray}
See Eqs.~(\textcolor{zz}{\ref{eq:ghgp2}}), (\textcolor{zz}{\ref{eq:c1low}}).

For any $n$: 
\begin{eqnarray} 
Q^{(n)}\ =\  \sum_{k=1}^{E(\frac{n}{2}-1)}\{\llbracket1\rrbracket_k\llbracket2\rrbracket_k\}-\sum_{k=3}^{E(\frac{n}{2}-1)} \{\llbracket1\rrbracket_k\llbracket2\rrbracket_{k-2}\}+C^{(n)}+4i\sum_{k<l<i<j<N} \epsilon(k,l,i,j) 
\end{eqnarray}

Where
\begin{eqnarray} 
C^{(n)}\ =\  \begin{Bmatrix} -\{ \llbracket 1\rrbracket_{\frac{n}{2}-2} \llbracket \frac{n}{2}+1\rrbracket_{\frac{n}{2}-2}\}, & \textrm{$n>4$, even $n$}\\ 
 -\{\llbracket 1\rrbracket_{\frac{n-5}{2}}\llbracket \frac{n+1}{2}\rrbracket_{\frac{n-3}{2}}\}, & \textrm{$n>5$, odd $n$} \end{Bmatrix}
\end{eqnarray}

\clearpage


\section{\textcolor{ww}{Direct production of Regge states}}
\label{sec:a2}

\emph{References:} We follow ~\textcolor{rr}{\cite{feng}}.\\

\begin{figure}
\centering
\includegraphics[width= 60mm]{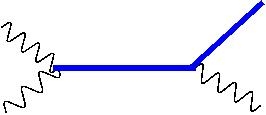}
\caption{Direct production of one Regge state.\label{rdir}}
\end{figure} 

Until now we considered amplitudes where the outgoing particles were standard model particles and the string states were internal states. When the center of mass energy of the collision exceeds the mass of the first Regge state, these particles may be directly produced as external particles (Fig.~\textcolor{yy}{\ref{rdir}}). In this section we write down the 4-point squared amplitudes for the direct production of one Regge state from the first excited level ($n=1$).

The external Regge states from the $n=1$ level which appear in the amplitudes below are\footnote{There is also the particle $\Omega(J=0)$ which does not couple to purely gluonic processes or to two quarks, thus it does not appear in the amplitudes below. $\Omega$ does however couple to two gluinos.}: $B(J=2)$,\ \ $\Phi(J=0)$,\ \ $W(J=1)$,\ \  $Q(J=1/2)$ and $Q^*(J=3/2)$. $B_0$,\ $\Phi_0$,  and $W_0$ are the corresponding $U(1)$ fields.
The $B$ particle should not be confused with the $B$ particles of sections~\ref{subsubsec:5squared} and \ref{subsubsec:a7}. Note that $W$ and $W_0$ do not couple to purely gluonic processes at tree level.

\subsection{The squared amplitudes}

\begin{itemize}
\item {\large $gg$ initial state}
\end{itemize}

\begin{eqnarray} 
|\mathcal{M}|^2(\textcolor{blue}{gg \to gB})=\ \frac{5g^4}{8}\Big(V_s^2+V_t^2+V_u^2\Big)\ \frac{(\hat{s}-1)^4+(\hat{t}-1)^4+(\hat{u}-1)^4}{\hat{s}\hat{t}\hat{u}}
\end{eqnarray}

\begin{eqnarray} 
|\mathcal{M}|^2(\textcolor{blue}{gg \to gB_0})=\ \frac{3}{4}g^4Q_A^2\Big(V_s+V_t+V_u\Big)^2\ \frac{(\hat{s}-1)^4+(\hat{t}-1)^4+(\hat{u}-1)^4}{\hat{s}\hat{t}\hat{u}}
\end{eqnarray}

\begin{eqnarray} 
|\mathcal{M}|^2(\textcolor{blue}{gg \to g\Phi})=\ \frac{5g^4}{8}\Big(V_s^2+V_t^2+V_u^2\Big)\ \frac{\hat{s}^4+\hat{t}^4+\hat{u}^4+1}{\hat{s}\hat{t}\hat{u}}
\end{eqnarray}

\begin{eqnarray} 
|\mathcal{M}|^2(\textcolor{blue}{gg \to g\Phi_0})=\ \frac{3}{4}g^4Q_A^2\Big(V_s+V_t+V_u\Big)^2\ \frac{\hat{s}^4+\hat{t}^4+\hat{u}^4+1}{\hat{s}\hat{t}\hat{u}}
\end{eqnarray}

\begin{eqnarray} 
|\mathcal{M}|^2(\textcolor{blue}{gg \to \bar{q}Q})= \frac{g^4}{4}\bigg[(\frac{3}{32}\big(V_t+V_u\big)^2+\Big(\frac{5}{96}+\frac{Q_A^2}{8}\Big)\big(V_t-V_u\big)^2\bigg]\frac{(\hat{s}-1)+(\hat{t}-1)\hat{u}^3+(\hat{u}-1)\hat{t}^3}{\hat{s}\hat{t}\hat{u}}\nonumber\\
\end{eqnarray}

\begin{eqnarray} 
|\mathcal{M}|^2(\textcolor{blue}{gg \to \bar{q}Q^*})= g^4\bigg[(\frac{3}{32}\big(V_t+V_u\big)^2+\Big(\frac{5}{96}+\frac{Q_A^2}{8}\Big)\big(V_t-V_u\big)^2\bigg]\frac{(\hat{s}-1)^3+(\hat{t}-1)^3\hat{u}+(\hat{u}-1)^3\hat{t}}{\hat{s}\hat{t}\hat{u}}\nonumber\\
\end{eqnarray}
\\

\begin{itemize}
\item {\large $gq$ initial state}
\end{itemize}

\begin{eqnarray} 
|\mathcal{M}|^2(\textcolor{blue}{qg \to qB})= -\frac{g^4}{16}\bigg[\big(V_s-V_u\big)^2+\Big(\frac{5}{9}+\frac{4Q_A^2}{3}\Big)\big(V_s+V_u\big)^2\bigg]\frac{\big[(\hat{s}-1)^2+(\hat{u}-1)^2\big](\hat{t}+4\hat{s}\hat{u})}{\hat{s}\hat{t}\hat{u}}\nonumber\\
\end{eqnarray}

\begin{eqnarray}
|\mathcal{M}|^2(\textcolor{blue}{qg \to qB_0})=\  -\frac{g^4Q_A^2}{12}\big(V_s+V_u\big)^2\frac{\big[(\hat{s}-1)^2+(\hat{u}-1)^2\big](\hat{t}+4\hat{s}\hat{u})}{\hat{s}\hat{t}\hat{u}}
\end{eqnarray}

\begin{eqnarray} 
|\mathcal{M}|^2(\textcolor{blue}{qg \to q\Phi})=\  -\frac{g^4}{4}\bigg[\big(V_s-V_u\big)^2+\Big(\frac{5}{9}+\frac{4Q_A^2}{3}\Big)\big(V_s+V_u\big)^2\bigg]\frac{\hat{s}^2+\hat{u}^2}{\hat{t}}
\end{eqnarray}

\begin{eqnarray} 
|\mathcal{M}|^2(\textcolor{blue}{qg \to q\Phi_0})=\  -\frac{g^4Q_A^2}{3}\big(V_s+V_u\big)^2\frac{\hat{s}^2+\hat{u}^2}{\hat{t}}
\end{eqnarray}

\begin{eqnarray} 
|\mathcal{M}|^2(\textcolor{blue}{qg \to qW})= -\frac{3g^4}{16}\bigg[\big(V_s-V_u\big)^2+\Big(\frac{5}{9}+\frac{4Q_A^2}{3}\Big)\big(V_s+V_u\big)^2\bigg]\frac{(\hat{s}-1)^2+(\hat{u}-1)^2}{\hat{s}\hat{u}}
\end{eqnarray}

\begin{eqnarray} 
|\mathcal{M}|^2(\textcolor{blue}{qg \to qW_0})=\  -\frac{g^4Q_A^2}{4}\big(V_s+V_u\big)^2\frac{(\hat{s}-1)^2+(\hat{u}-1)^2}{\hat{s}\hat{u}}
\end{eqnarray}

\begin{eqnarray} 
|\mathcal{M}|^2(\textcolor{blue}{qg \to gQ})= -\frac{g^4}{16}\bigg[\big(V_s+V_u\big)^2+\Big(\frac{5}{9}+\frac{4Q_A^2}{3}\Big)\big(V_s-V_u\big)^2\bigg]\frac{(\hat{t}-1)^3+(\hat{u}-1)^3\hat{s}+(\hat{s}-1)^3\hat{u}}{\hat{s}\hat{t}\hat{u}}\nonumber\\
\end{eqnarray}

\begin{eqnarray} 
|\mathcal{M}|^2(\textcolor{blue}{qg \to gQ^*})= -\frac{g^4}{4}\bigg[\big(V_s+V_u\big)^2+\Big(\frac{5}{9}+\frac{4Q_A^2}{3}\Big)\big(V_s-V_u\big)^2\bigg]\frac{(\hat{t}-1)+(\hat{u}-1)\hat{s}^3+(\hat{s}-1)\hat{u}^3}{\hat{s}\hat{t}\hat{u}}\nonumber\\
\end{eqnarray}
\\

\begin{itemize}
\item {\large $q\bar{q}$ initial state}
\end{itemize}

\begin{eqnarray} 
|\mathcal{M}|^2(\textcolor{blue}{\bar{q}q\to gX})= |\mathcal{M}|^2(\textcolor{blue}{\bar{q}q\to gX})= -\frac{8}{3}|\mathcal{M}|^2(\textcolor{blue}{qg\to qX})(s\to u,u\to t,t\to s)
\end{eqnarray}
Where $X= B, \Phi , W$.

\begin{eqnarray} 
|\mathcal{M}|^2(\textcolor{blue}{\bar{q}q \to \bar{q}Q})= 
g^4\bigg\{-\frac{t^2}{4}\Big[su|Q_{su}+\tilde{Q}_{su}|^2+ ut|Q_{su}|^2+st|\tilde{Q}_{su}|^2\Big]+
\nonumber\\[6pt]
\frac{t^2}{12}\Big[su(Q_{su}+\tilde{Q}_{su})(Q_{su}+\tilde{Q}_{su})^*+ utQ_{su}\tilde{Q^*}_{su}+ st\tilde{Q}_{su}Q^*_{su}\Big]
\nonumber\\[6pt]
-\frac{u^2}{4}\Big[st|R_{st}+\tilde{R}_{st}|^2+ ut|R_{st}|^2+su|\tilde{R}_{st}|^2\Big] \bigg\}+\ \big\{s\leftrightarrow u \big\}
\end{eqnarray}

\begin{eqnarray} 
|\mathcal{M}|^2(\textcolor{blue}{\bar{q}q \to \bar{q'}Q'})=\ \ \ \ \ \ \ \ \ \ \ \ \ \ \ \ \ \ \ \ \ \ \ \ \ \ \ \ \ \ \ \ \ \ \ \ \ \ \ \ \ \ \ \ \ \ \ \ \ \ \ \ \ \ \ \ \ \ \ \ \ \ 
\nonumber\\[6pt]
 -\frac{g^4t^2}{4}\Big(su|Q_{su}+\tilde{Q}_{su}|^2+ ut|Q_{su}|^2+st|\tilde{Q}_{su}|^2\Big)+\big(Q\to R ;u\leftrightarrow t\big)
\end{eqnarray}

\begin{eqnarray} 
|\mathcal{M}|^2(\textcolor{blue}{\bar{q}q \to \bar{q}Q^*})=\ \ \ \ \ \ \ \ \ \ \ \ \ \ \ \ \ \ \ \ \ \ \ \ \ \ \ \ \ \ \ \ \ \ \ \ \ \ \ \ \ \ \ \ \ \ \ \ \ \ \ \ \ \ \ \ \ \ \ \ \ \ \ \ \ \ \ \ \ \ \ \ \ \ \ \  \ \ \ \ \ \ \ \ 
\nonumber\\[6pt]
g^4\bigg\{-\frac{(M_s^2-t)^2}{4}\Big(su|Q_{su}+\tilde{Q}_{su}|^2+ ut|Q_{su}|^2+st|\tilde{Q}_{su}|^2\Big)+ \frac{t(M_s^2-t)}{6}\big|uQ_{su}-s\tilde{Q}_{su}\big|^2 
\nonumber\\[6pt]
+\frac{(M_s^2-t)^2}{12}\Big[su(Q_{su}+\tilde{Q}_{su})(Q_{su}+\tilde{Q}_{su})^*+ utQ_{su}\tilde{Q^*}_{su}+ st\tilde{Q}_{su}Q^*_{su}\Big]
\nonumber\\[6pt]
+\frac{t(M_s^2-t)}{18}(uQ_{su}-s\tilde{Q}_{su})(u\tilde{Q}_{su}-sQ_{su})^*
\nonumber\\[6pt]
-\frac{(M_s^2-u)^2}{4}\big(st|R_{st}+\tilde{R}_{st}|^2+ ut|R_{st}|^2+su|\tilde{R}_{st}|^2\big)+\frac{u(M_s^2-u)}{6} \big|tR_{st}-s\tilde{R}_{st}\big|^2 \bigg\}
+\big\{s\leftrightarrow u\big\} \nonumber\\
\end{eqnarray}

\begin{eqnarray} 
|\mathcal{M}|^2(\textcolor{blue}{\bar{q}q \to \bar{q'}Q'^*})=
 g^4\bigg\{-\frac{(M_s^2-t)^2}{4}\Big(su|Q_{su}+\tilde{Q}_{su}|^2+ ut|Q_{su}|^2+st|\tilde{Q}_{su}|^2\Big)+
\nonumber\\[6pt]
\frac{t(M_s^2-t)}{6} \big|uQ_{su}-s\tilde{Q}_{su}\big|^2 +\big(Q\to R ;u\leftrightarrow t\big) \bigg\}
\end{eqnarray}

Where we defined: 
\begin{eqnarray}
Q_{s,u}\ \equiv \ \alpha'e^{\phi_{10}}\int_0^1 dxx^{-s}(1-x)^{-u-1}Z^{ba}_{inst}(x)I(x,\theta^j)
\end{eqnarray}
\begin{eqnarray}
\tilde{Q}_{s,u}\ \equiv \ \alpha'e^{\phi_{10}}\int_0^1 dxx^{-s-1}(1-x)^{-u}Z^{ba}_{inst}(x)I(x,\theta^j)
\end{eqnarray}
\begin{eqnarray}
R_{s,u}\ \equiv \ \alpha'e^{\phi_{10}}\int_0^1 dxx^{-s}(1-x)^{-u-1}Z^{ba}_{inst}(x)I(x,\theta^j,\nu^j)
\end{eqnarray}
\begin{eqnarray}
\tilde{R}_{s,u}\ \equiv \ \alpha'e^{\phi_{10}}\int_0^1 dxx^{-s-1}(1-x)^{-u}Z^{ba}_{inst}(x)I(x,\theta^j,\nu^j)
\end{eqnarray}

sults from the LHC and other experiments, and hoping that they correspond to \underline{new} physics.

\newpage

\textbf{Acknowledgments}

I am very grateful to my advisor professor Cobi Sonnenschein.


\appendix

\section{\textcolor{ww}{4-point helicity amplitudes}}
\label{sec:apphelamp}
\emph{References:} ~\textcolor{rr}{\cite{lust}}.\\

We write the four point helicity amplitudes near an s-channel pole. We do the algebra of expanding the Veneziano amplitudes and simplifying the group theoretical factors.
We recall that the $(++++)$ and $(-+++)$ helicity amplitudes vanish, which leaves only the MHV amplitudes: $(-++-)$.
The simple kinematics of the process $ij \to kl$ give:
\begin{eqnarray} 
\label{eq:bra200}
\langle ij\rangle \ =\  \langle kl\rangle \ =\  \sqrt{s}
\nonumber\\[5pt]
\langle ik\rangle \ =\  \langle jl\rangle \ =\  \sqrt{t}
\nonumber\\[5pt]
\langle il\rangle \ =\  \langle jk\rangle \ =\  \sqrt{u} ,
\end{eqnarray}

In the following, we use the notation $\mathcal{M}_{m, m'}$ for the helicity amplitudes.
\subsection{$\mathcal{M}(g,g,g,g)$}

We have three helicity configurations: $(m,m')=(0,0)$, $(2,-2)$, $(2,2)$.

\begin{itemize}
\item $\mathcal{M}_{0,0} \ = \ \mathcal{M} (g_1^{-}\ g_2^{-} \to g_3^{+}\ g_4^{+})$
\end{itemize}

\begin{eqnarray} 
\mathcal{M}_{0,0}\ =\ 
4g^2 \langle 12\rangle^4 
\bigg\{ \frac{V_t}{\langle 12\rangle \langle 23\rangle \langle 34\rangle \langle 41 \rangle} \ Tr\big[T^{a_1}T^{a_2}T^{a_3}T^{a_4}+T^{a_4}T^{a_3}T^{a_2}T^{a_1}\big]
\nonumber\\[7pt]
+\frac{V_s}{\langle 14 \rangle \langle 42 \rangle \langle 23 \rangle \langle 31 \rangle} \ Tr\big[T^{a_2}T^{a_1}T^{a_3}T^{a_4}+T^{a_4}T^{a_3}T^{a_1}T^{a_2}\big]
\nonumber\\[7pt]
+\frac{V_u}{\langle 13 \rangle \langle 34 \rangle \langle 42 \rangle \langle 21 \rangle} \ Tr\big[T^{a_1}T^{a_3}T^{a_2}T^{a_4}+T^{a_4}T^{a_2}T^{a_3}T^{a_1}\big]\bigg\}
\end{eqnarray}
Note the $(4-1)!= 6$ permutations, and the reflection symmetry: $m(1,2,3,4)= m(4,3,2,1)$.

Eq.~(\zz{\ref{eq:bra200}}) gives:
\begin{eqnarray} 
\label{eq:fd456}
\mathcal{M}_{0,0}\ \ =\ \  4g^2\bigg\{ \frac{s}{u}V_t        \ Tr\big[T^{a_1}T^{a_2}T^{a_3}T^{a_4}+T^{a_4}T^{a_3}T^{a_2}T^{a_1}\big]
\nonumber\\[7pt]
+\frac{s}{t}V_u\ Tr\big[T^{a_2}T^{a_1}T^{a_3}T^{a_4}+T^{a_4}T^{a_3}T^{a_1}T^{a_2}\big]
\nonumber\\[7pt]
+\frac{s^2}{ut}V_s\ Tr\big[T^{a_1}T^{a_3}T^{a_2}T^{a_4}+T^{a_4}T^{a_2}T^{a_3}T^{a_1}\big]  \bigg\}
\end{eqnarray}

Using (Eq.~(\textcolor{zz}{\ref{eq:b4}})):
\begin{eqnarray}
V_t\ \ \ \   \xrightarrow{\hat{s} \to n}\ \ \ \  \frac{1}{(n-1)!}\ \frac{1}{\hat{s}-n}\ \Bigg[\hat{u} \prod_{K=1}^{n-1}(\hat{u}+ K)\Bigg]
\end{eqnarray}
\begin{eqnarray}
V_u\ \ \ \  \xrightarrow{\hat{s} \to n}\ \ \ \  \frac{(-1)^{n-1}}{(n-1)!}\ \frac{1}{\hat{s}-n}\ \Bigg[\hat{t} \prod_{K=1}^{n-1}(\hat{u}+ K)\Bigg]
\end{eqnarray}

gives:
\begin{eqnarray}
\mathcal{M}_{0,0}\ \simeq \ \ \frac{\hat{s}}{(n-1)!}\frac{1}{\hat{s}-n} 4g^2\ \Bigg[\prod^{n-1}_{K=1}(\hat{u}+K)\Bigg]\times \ \ \ \ \ \ \ \ \ \ \ \ \ \ \ \ \ \ \ \ \ \ \ \ \ \ \ \ \ \ \ \ \ \ \ \ \ \ \ \ \ \ 
\nonumber\\[10pt] Tr\Big[T^{a_1}T^{a_2}T^{a_3}T^{a_4}+T^{a_4}T^{a_3}T^{a_2}T^{a_1}+(-1)^{n-1}\big(T^{a_2}T^{a_1}T^{a_3}T^{a_4}+T^{a_4}T^{a_3}T^{a_1}T^{a_2}\big)\Big]
\nonumber\\ \nonumber\\[7pt]
= 4g^2\frac{n}{(n-1)!}\frac{1}{\hat{s}-n}\Bigg[\prod^{n-1}_{K=1}(\hat{u}+K)\Bigg]\cdot  \begin{Bmatrix} Tr\big(\{T^{a_1},T^{a_2}\}\{T^{a_3},T^{a_4}\}\big), & \textrm{odd $n$}\\  Tr\big([T^{a_1},T^{a_2}][T^{a_3},T^{a_4}]\big), & \textrm{even $n$} \end{Bmatrix} \nonumber\\
\end{eqnarray}
We see that near a pole, the amplitude has the nice property of being factorized as a color part times a kinematical part.

Using (see Appendix~\ref{sec:appc})
\begin{eqnarray} 
\{T^{a_3},T^{a_4} \}\ =\  4\sum_a \texttt{d}^{a_3a_4a}\ T^{a}
\end{eqnarray}
\begin{eqnarray} 
[T^{a_3},T^{a_4}]\ =\  \sum_a \texttt{f}^{a_3a_4a}\ T^{a}\ \ \ \ 
\end{eqnarray}

finally gives:
\begin{eqnarray}
\label{eq:apphela5}
\boxed{\mathcal{M}_{0,0}\ =\  
4g^2 \frac{n}{(n-1)!}\ \frac{1}{\hat{s}-n}\ \Bigg[\prod^{n-1}_{K=1}(\hat{u}+K)\Bigg]\cdot \begin{Bmatrix} 8\sum_a \texttt{d}^{a_1a_2a}\texttt{d}^{a_3a_4a}, & \textrm{odd $n$}\\ \sum_a \texttt{f}^{a_1a_2a} \texttt{f}^{a_3a_4a}, & \textrm{even $n$} \end{Bmatrix}   }
\end{eqnarray}

The minus sign in the following equation (which is valid near a pole $\hat{s}\to n$), see  Eq.~(\zz{\ref{eq:int76}}):
\bea
V_t\ =\ (-1)^{n-1}\ \frac{u}{t}\ V_u\ \ \ \ ,\ \ \ \ 
\eea
and the fact that the amplitude contained only the combination $\big( uV_u \times \text{color}+tV_t \times \text{color} \big)$,
caused the even (odd) $n$ resonances to contain only $\texttt{f}^{a_1a_2a_3}$ ($\texttt{d}^{a_1a_2a_3}$).
This will happen also for the amplitudes: $\mathcal{M}_{2,\pm 2}$ and $\mathcal{M}_{2,\pm1}$. $\texttt{f}^{a_1a_2a_3}$ vanishes when at least one of the indices are $U(1)$ fields, Eq.~(\zz{\ref{eq:one137}}). Therefore $\mathcal{M}_{0,0}$, $\mathcal{M}_{2,\pm 2}$, and $\mathcal{M}_{2,\pm1}$ will not have exchanges of $n= even$ resonances for these amplitudes.

\begin{itemize}
\item $\mathcal{M}_{2,-2} \ = \ \mathcal{M} (g_1^{-}\ g_2^{+}\to g_3^{-}\ g_4^{+})$
\end{itemize}

$\mathcal{M}_{2,-2}$ can easily be obtained by crossing $(s\leftrightarrow t)$ of Eq.~(\zz{\ref{eq:fd456}}):
\begin{eqnarray} 
\mathcal{M}_{2,-2}\ =\ 
4g^2 \bigg\{ \frac{t}{u}V_s\ Tr\big[T^{a_1}T^{a_2}T^{a_3}T^{a_4}+T^{a_2}T^{a_1}T^{a_4}T^{a_3}\big]
\nonumber\\[7pt]  
+\frac{t}{s}V_u\ Tr\big[T^{a_2}T^{a_1}T^{a_3}T^{a_4}+T^{a_1}T^{a_2}T^{a_4}T^{a_3}\big]
\nonumber\\[7pt] 
+\frac{t^2}{us}V_t\ Tr\big[T^{a_1}T^{a_3}T^{a_2}T^{a_4}+T^{a_3}T^{a_1}T^{a_4}T^{a_2}\big]  \bigg\}
\end{eqnarray}

So that: 
\begin{eqnarray} 
\label{eq:appheld1}
\boxed{
\mathcal{M}_{2,-2}\ =\ 4g^2 \frac{n}{(n-1)!}\ \frac{1}{\hat{s}-n}\ \Bigg[\frac{t^2}{s^2} \prod^{n-1}_{K=1}(\hat{u}+K)\Bigg]\cdot 
\begin{Bmatrix} 8\sum_a \texttt{d}^{a_1a_2a}\texttt{d}^{a_3a_4a}, & \textrm{odd $n$}\\  \sum_a \texttt{f}^{a_1a_2a}\texttt{f}^{a_3a_4a}, & \textrm{even $n$} \end{Bmatrix}       }
\end{eqnarray}
Note that this is equal to Eq.~(\zz{\ref{eq:apphela5}}) multiplied by $\frac{\hat{t}^2}{\hat{s}^2}$.

\begin{itemize}
\item $\mathcal{M}_{2,2} \ = \ \mathcal{M} (g_1^{-}\ g_2^{+}\to g_3^{+}\ g_4^{-})$
\end{itemize}

$\mathcal{M}_{2,2}$ can easily be obtained from the previous equation by the substitution $\frac{t^2}{s^2}\to \frac{u^2}{s^2}$ in the square brackets:
\begin{eqnarray} 
\boxed{
\mathcal{M}_{2,2}\ =\ 4g^2 \frac{n}{(n-1)!}\ \frac{1}{\hat{s}-n}\ \Bigg[\frac{u^2}{s^2} \prod^{n-1}_{K=1}(\hat{u}+K)\Bigg]\cdot 
\begin{Bmatrix} 8\sum_a \texttt{d}^{a_1a_2a}\texttt{d}^{a_3a_4a}, & \textrm{odd $n$}\\  \sum_a \texttt{f}^{a_1a_2a}\texttt{f}^{a_3a_4a}, & \textrm{even $n$} \end{Bmatrix}      }
\end{eqnarray}

\subsection{$\mathcal{M}(g,g,q,q)$}
We have four helicity configurations: $(m,m')=(2,-1),\ (2,1),\ (\frac{1}{2},\frac{1}{2}),\ (\frac{3}{2},-\frac{3}{2})$.

\begin{itemize}
\item $\mathcal{M}_{2,-1} \ = \ \mathcal{M} (q_3^{-}\ \bar{q}_4^{+}\to g_1^{-}\ g_2^{+})$
\end{itemize}

\begin{eqnarray} 
\mathcal{M}_{2,-1}\ =\ 
2g^2\delta^{\beta_4}_{\beta_3}\frac{\langle 13\rangle^2}{\langle 23\rangle \langle 24 \rangle} \Big[(T^{a_1}T^{a_2})^{\alpha_3}_{\alpha_4}\frac{t}{s}V_t+(T^{a_2}T^{a_1})^{\alpha_3}_{\alpha_4}\frac{u}{s}V_u\Big]
\nonumber\\ \nonumber\\[7pt] =2g^2\delta^{\beta_4}_{\beta_3}\sqrt{\frac{t}{u}}\Big[(T^{a_1}T^{a_2})^{\alpha_3}_{\alpha_4}\frac{t}{s}V_t+(T^{a_2}T^{a_1})^{\alpha_3}_{\alpha_4}\frac{u}{s}V_u\Big]
\nonumber\\ \nonumber\\[7pt]
\simeq 
\frac{n}{(n-1)!}\frac{1}{\hat{s}-n}2g^2\delta^{\beta_4}_{\beta_3}\Big[(T^{a_1}T^{a_2})^{\alpha_3}_{\alpha_4}+(-1)^{n-1}(T^{a_2}T^{a_1})^{\alpha_3}_{\alpha_4}\Big]\Bigg[\frac{u^{\frac{1}{2}}t^{\frac{3}{2}}}{s^2} \prod^{n-1}_{K=1}(\hat{u}+K)\Bigg]
\nonumber\\ \nonumber\\[7pt]
= 2g^2\delta^{\beta_4}_{\beta_3} \frac{n}{(n-1)!}\frac{1}{\hat{s}-n}\Bigg[\frac{u^{\frac{1}{2}}t^{\frac{3}{2}}}{s^2}\prod^{n-1}_{K=1}(\hat{u}+K)\Bigg]
\nonumber\\[8pt]
\times \begin{Bmatrix} \{ T^{a_1},T^{a_2}\}^{\alpha_3}_{\alpha_4}, & \textrm{odd}\ n\\  [T^{a_1},T^{a_2}]^{\alpha_3}_{\alpha_4}, & \textrm{even}\ n \end{Bmatrix}. 
\end{eqnarray}

So we get:
\begin{eqnarray} 
\label{eq:appheld2}\boxed{
\mathcal{M}_{2,-1}\ =\  2g^2\delta^{\beta_4}_{\beta_3}\frac{n}{(n-1)!}\ \frac{1}{\hat{s}-n} \ \Bigg[\frac{u^{\frac{1}{2}}t^{\frac{3}{2}}}{s^2}\prod^{n-1}_{K=1}(\hat{u}+K)\Bigg]
\begin{Bmatrix} 4\sum_a \texttt{d}^{a_1a_2a}T^a_{\alpha_3\alpha_4}, & \textrm{odd $n$}\\  \sum_a \texttt{f}^{a_1a_2a}T^a_{\alpha_3\alpha_4}, & \textrm{even $n$} \end{Bmatrix}      } 
\end{eqnarray}

\begin{itemize}
\item $\mathcal{M}_{2,1} \ = \ \mathcal{M} (q_3^{-}\ \bar{q}_4^{+}\to g_1^{+}\ g_2^{-})$
\end{itemize}

$\mathcal{M}_{2,1}$ can easily be obtained from the previous equation by the substitution $\frac{u^{\frac{1}{2}}t^{\frac{3}{2}}}{s^2} \to \frac{t^{\frac{1}{2}}u^{\frac{3}{2}}}{s^2}$ in the square brackets:
\begin{eqnarray} 
\boxed{
\mathcal{M}_{2,1}\ =\  2g^2\delta^{\beta_4}_{\beta_3}\frac{n}{(n-1)!}\ \frac{1}{\hat{s}-n} \ \Bigg[ \frac{t^{\frac{1}{2}}u^{\frac{3}{2}}}{s^2}\prod^{n-1}_{K=1}(\hat{u}+K) \Bigg]
 \begin{Bmatrix} 4\sum_a \texttt{d}^{a_1a_2a}T^a_{\alpha_3\alpha_4}, & \textrm{odd $n$}\\  \sum_a \texttt{f}^{a_1a_2a}T^a_{\alpha_3\alpha_4}, & \textrm{even $n$}  \end{Bmatrix} } 
\end{eqnarray}

\begin{itemize}
\item $\mathcal{M}_{\frac{1}{2},\frac{1}{2}} \ = \ \mathcal{M} (q_3^{-}\ g_1^{-}\to q_4^{+}\ g_2^{+})$
\end{itemize}

Can be obtained by crossing $\mathcal{M}_{2,-1} \xrightarrow{s \leftrightarrow t} \mathcal{M}_{\frac{1}{2},\frac{1}{2}}$\ :
\begin{eqnarray} 
\mathcal{M}_{\frac{1}{2},\frac{1}{2}}\ =\  
2g^2\delta^{\beta_4}_{\beta_3}\sqrt{\frac{s}{u}} \Big[(T^{a_1}T^{a_2})^{\alpha_3}_{\alpha_4}\ \frac{s}{t}V_s\ +\ (T^{a_2}T^{a_1})^{\alpha_3}_{\alpha_4}\ \frac{u}{t}V_u\Big]
\end{eqnarray} 

This gives:
\begin{eqnarray}
\label{eq:appheld3} \boxed{
\mathcal{M}_{\frac{1}{2},\frac{1}{2}}\ =\ (-1)^{n-1}2g^2\delta^{\beta_4}_{\beta_3} \frac{n}{(n-1)!}\ \frac{1}{\hat{s}-n}\ \Bigg[\frac{u^{\frac{1}{2}}}{s^{\frac{1}{2}}}\prod^{n-1}_{K=1}(\hat{u}+K)\Bigg]
(T^{a_2}T^{a_1})^{\alpha_3}_{\alpha_4}   }
\end{eqnarray} 

\begin{itemize}
\item $\mathcal{M}_{\frac{3}{2}, -\frac{3}{2}} \ = \ \mathcal{M} (q_4^{-}\ g_1^{+}\to q_3^{+}\ g_2^{-})$
\end{itemize}

Can be obtained by crossing $\mathcal{M}_{2,-1} \xrightarrow{s \leftrightarrow u} \mathcal{M}_{\frac{3}{2},\frac{3}{2}}$\ :
\begin{eqnarray}
\mathcal{M}_{\frac{3}{2},-\frac{3}{2}}\ =\ 
2g^2\delta^{\beta_4}_{\beta_3}\sqrt{\frac{t}{s}} \Big[(T^{a_1}T^{a_2})^{\alpha_3}_{\alpha_4}\frac{t}{u}V_t+(T^{a_2}T^{a_1})^{\alpha_3}_{\alpha_4}\frac{s}{u}V_s\Big]
\end{eqnarray}

\begin{eqnarray}
\label{eq:appheld4} \boxed{
\mathcal{M}_{\frac{3}{2},-\frac{3}{2}}\ =\ 2g^2\delta^{\beta_4}_{\beta_3} \frac{n}{(n-1)!}\ \frac{1}{\hat{s}-n}\ \Bigg[\frac{t^{\frac{3}{2}}}{s^{\frac{3}{2}}}\prod^{n-1}_{K=1}(\hat{u}+K)\Bigg]
(T^{a_1}T^{a_2})^{\alpha_3}_{\alpha_4}}
\end{eqnarray}
Note that this is equal to Eq.~(\zz{\ref{eq:appheld3}}) multiplied by $(-1)^{n-1}\frac{t^{\frac{3}{2}}}{s u^{\frac{1}{2}}}$.

\subsection{$\mathcal{M}(g,g,q,B)$}
We have four helicity configurations: $(m,m')=(2,-1),\ (2,1),\ (\frac{1}{2},\frac{1}{2}),\ (\frac{3}{2},-\frac{3}{2})$.

\begin{itemize}
\item $\mathcal{M}^{(B)}_{2,-1} \ = \ \mathcal{M} (q_3^{-}\ \bar{q}_4^{+}\to g_1^{-}\ B_2^{+})$
\end{itemize}

\begin{eqnarray} 
\label{eq:appheld76}
\mathcal{M}^{(B)}_{2,-1}\ =\ 2g_{D_{p_b}}g\frac{\langle 13 \rangle^2}{\langle 23 \rangle \langle 24 \rangle}\ (T^{a_1})^{\alpha_3}_{\alpha_4}(T^{a_2})^{\beta_4}_{\beta_3}\ V_s
\nonumber\\[7pt]
=2g_{D_{p_b}}g \sqrt{\frac{t}{u}}\ (T^{a_1})^{\alpha_3}_{\alpha_4}(T^{a_2})^{\beta_4}_{\beta_3}\ V_s
\end{eqnarray}
Which exhibits no poles since $V_s$ does not. 

\begin{itemize}
\item $\mathcal{M}^{(B)}_{2,1} \ = \ \mathcal{M} (q_3^{-}\ \bar{q}_4^{+}\to g_1^{+}\ B_2^{-})$
\end{itemize}

\begin{eqnarray} 
\mathcal{M}^{(B)}_{2,1}\ =\ 
2g_{D_{p_b}}g \sqrt{\frac{u}{t}}\ (T^{a_1})^{\alpha_3}_{\alpha_4}(T^{a_2})^{\beta_4}_{\beta_3}\ V_s
\end{eqnarray}
Which exhibits no poles.

\begin{itemize}
\item $\mathcal{M}^{(B)}_{\frac{1}{2},\frac{1}{2}} \ = \ \mathcal{M} (q_3^{-}\ g_1^{-}\to q_4^{+}\ B_2^{+})$
\end{itemize}

Can be obtained by crossing $\mathcal{M}^{(B)}_{2,-1} \xrightarrow{s \leftrightarrow t} \mathcal{M}^{(B)}_{\frac{1}{2},\frac{1}{2}}$:
\begin{eqnarray} 
\mathcal{M}^{(B)}_{\frac{1}{2},\frac{1}{2}}\ =\ 
2g_{D_{p_b}}g \sqrt{\frac{s}{u}}\ (T^{a_1})^{\alpha_3}_{\alpha_4}(T^{a_2})^{\beta_4}_{\beta_3}\ V_t
\end{eqnarray} 

\begin{eqnarray} 
\label{eq:appheld6}
\boxed{
\mathcal{M}^{(B)}_{\frac{1}{2},\frac{1}{2}}\ =\ 2g_{D_{p_b}}g\ \frac{n}{(n-1)!}\ \frac{1}{\hat{s}-n}\ \Bigg[\frac{u^{\frac{1}{2}}}{s^{\frac{1}{2}}}\prod^{n-1}_{K=1}(\hat{u}+K)\Bigg]\ (T^{a_1})^{\alpha_3}_{\alpha_4}(T^{a_2})^{\beta_4}_{\beta_3}   }
\end{eqnarray}

\begin{itemize}
\item $\mathcal{M}^{(B)}_{\frac{3}{2},-\frac{3}{2}} \ = \ \mathcal{M} (q_4^{-}\ g_1^{+}\to q_3^{+}\ B_2^{-})$
\end{itemize}

Can be obtained by crossing $\mathcal{M}^{(B)}_{2,-1} \xrightarrow{s \leftrightarrow u} \mathcal{M}^{(B)}_{\frac{3}{2},\frac{3}{2}}$\ :
\begin{eqnarray} 
\mathcal{M}^{(B)}_{\frac{3}{2},-\frac{3}{2}}\ =\ 
2g_{D_{p_b}}g \sqrt{\frac{t}{s}}\ (T^{a_1})^{\alpha_3}_{\alpha_4}(T^{a_2})^{\beta_4}_{\beta_3}\ V_u
\end{eqnarray}

\begin{eqnarray}
\label{eq:appheld68}
\boxed{
\mathcal{M}^{(B)}_{\frac{3}{2},-\frac{3}{2}}\ =\ 2g_{D_{p_b}}g\ \frac{n}{(n-1)!}\ \frac{1}{\hat{s}-n}\ \Bigg[\frac{t^{\frac{3}{2}}}{s^{\frac{3}{2}}}\prod^{n-1}_{K=1}(\hat{u}+K)\Bigg]\ (T^{a_1})^{\alpha_3}_{\alpha_4}(T^{a_2})^{\beta_4}_{\beta_3}     }
\end{eqnarray}
$\mathcal{M}^{(B)}_{\frac{1}{2},\frac{1}{2}}$ and $\mathcal{M}^{(B)}_{\frac{3}{2},-\frac{3}{2}}$ have the same structure as $\mathcal{M}_{\frac{1}{2},\frac{1}{2}}$ and $\mathcal{M}_{\frac{3}{2},-\frac{3}{2}}$ apart from the color factors and the replacement of $g\to g_{D_{p_b}}$.

\clearpage

\section{\textcolor{ww}{Calculation of $\mathcal{M}(ggqq)$: a detailed example}}
\label{sec:b8}

\emph{References:} This calculation was given in~\textcolor{rr}{\cite{lust}}. We fill in some of the details.\\

The goal of this section is to illustrate the details of a string amplitude calculation using the techniques of sections \ref{subsubsec:b1} - \ref{subsubsec:b2}. In section~\ref{subsubsec:tach7} we showed a derivation of the 4-tachyon amplitude. Some of the new features which will apppear in this section are: correlation functions of world sheet fermions, color factors, and polarization vectors.

Recalling Eq. (\textcolor{zz}{\ref{eq:b3}}), we begin by calculating the correlation function of two quarks and two gluons using the equations of Appendix \ref{sec:appa}. In the start we do not write the color matrices and coupling constants, but they will be inserted afterwords.

From Eqs. (\textcolor{zz}{\ref{eq:appb1}})-(\textcolor{zz}{\ref{eq:appb4}}) we have:
\begin{eqnarray}
\langle \ldots \rangle \equiv \langle\  V^{(0)}_{A^x}(z_1,\xi_1, k_1)\ \ V^{(-1)}_{A^y}(z_2,\xi_2, k_2)\ \ V^{(-1/2)}_{\psi^{\alpha_3}_{\beta_3}}(z_3,u_3,k_3)\ \ 
V^{(-1/2)}_{\bar{\psi}^{\beta_4}_{\alpha_4}}(z_4,\bar{u}_4,k_4)\ \rangle=
\nonumber\\ \nonumber\\ 
\frac{1}{\sqrt{2\alpha'}}\ \xi_{1\mu}\xi_2^{\nu}u^{\lambda_3}\bar{u}_{\lambda_4}\ \ 
\langle \ \Xi^{x \cap y}(z_3)\ \bar{\Xi}^{x \cap y}(z_4)\ \rangle \ \langle \  e^{-\phi(z_2)}\ e^{-\phi(z_3/2)}\ e^{-\phi(z_4/2)}\ \rangle
\nonumber\\[6pt]
\times \bigg[ i \langle \ \psi _{\nu}(z_2)\ S_{\lambda_3}(z_3)\ S^{\dot{\lambda_4}}(z_4)\ \rangle \ \ \langle \  \partial X^{\mu}(z_1)\ \prod_{i=1}^4 e^{ik_i \cdot X(z_i)}\ \rangle+
\nonumber\\[6pt]
2\alpha'\langle \ \big(k_1\cdot \psi(z_1)\big)\  \psi^{\mu}(z_1)\psi_{\nu}(z_2)\ S_{\lambda_3}(z_3)S^{\dot{\lambda_4}}(z_4)\ \rangle \ \ 
\langle \prod_{i=1}^4 e^{ik_i \cdot X(z_i)}\rangle \bigg] 
\end{eqnarray}

From Eqs. (\textcolor{zz}{\ref{eq:la9}}), (\textcolor{zz}{\ref{eq:appb10}}), (\textcolor{zz}{\ref{eq:appb5}})- (\textcolor{zz}{\ref{eq:appb6}}),  this equals:
\begin{eqnarray}
\langle \ldots \rangle\ \ =\ \  \frac{1}{\sqrt{2\alpha'}}
z_{34}^{-3/4}\big(z_{23}^{-1/2}z_{24}^{-1/2}z_{34}^{-1/4}\big)\ \prod_{i<j}^4|z_{ij}|^{2\alpha'k_ik_j}\ \ \ \ \ \ \ \ \ \ \ \ \ \ \ \ \ \ \ \ \ \ \ \ \ \ \ \ \ \ \ \ \ \ \ \ \ \ \ \ \ \ \ \ \ 
\nonumber\\[7pt]
\times \ \xi_{1\mu}\xi_2^{\nu}\ u_3 \cdot \Bigg( i\big(2z_{23}z_{24}\big)^{-1/2}\sigma_\nu(-2i\alpha')\sum_{r=1}^4\frac{k_r^\mu}{z_1-z_r}+ \ \ \ \ \ \ \ \ \ \ \ \ \ \ \ \ \ \ \ \ \ \ \ \ \ \ \ \ \ \ \ \ \ \ \ \ \ \ \ \ \ \ \ 
\nonumber\\[7pt]
2\alpha'k_{1\rho} \big(2z_{13}^2z_{14}^2z_{23}z_{24}\big)^{-1/2}\Big\{ \frac{z_{34}}{2}(\sigma^\rho \sigma^\mu \sigma^\nu)+\eta^{\rho \mu} \sigma^\nu \frac{z_{13}z_{14}}{z_{11}}- \eta^{\rho \nu}\sigma^\mu \frac{z_{13}z_{24}}{z_{12}}+ \eta^{\mu \nu}\sigma^\rho \frac{z_{13}z_{24}}{z_{12}} \Big\} \Bigg)\cdot \bar{u}_4
\nonumber\\
\end{eqnarray}

\begin{eqnarray}
\langle \ldots \rangle \ \ =\ \ \ \frac{1}{\sqrt{2\alpha'}}
2^{1/2}\alpha' \big(z_{23}z_{24}z_{34}\big)^{-1}\ \prod_{i<j}^4|z_{ij}|^{2\alpha'k_ik_j}\ \times \ \ \ \ \ \ \ \ \ \ \ \ \ \ \ \ \ \ \ \ \ \ \ \ \ \ \ \ \ \ \ \ \ \ \ \ \ \ \ \ \ \ \ \ \ \ \ \ \ \ \ 
\nonumber\\[7pt]
\xi_{1\mu}\xi_2^{\nu} u_3 \cdot \Bigg( \sigma_\nu \sum_{r=1}^4\frac{k_r^\mu}{z_1-z_r}+
k_{1\rho} \Big\{ \frac{z_{34}}{2z_{13}z_{14}}(\sigma^\rho \sigma^\mu \sigma^\nu)+\frac{\eta^{\rho \mu} \sigma^\nu}{z_{11}} - \eta^{\rho \nu}\sigma^\mu \frac{z_{24}}{z_{12}z_{14}}+ \eta^{\mu \nu}\sigma^\rho \frac{z_{24}}{z_{12}z_{14}} \Big\} \Bigg)\cdot \bar{u_4}
\nonumber\\
\end{eqnarray}

The two terms with $\frac{1}{z_{11}}$ cancel each other
\begin{eqnarray}
\langle \ldots \rangle \ \ =\ \  \sqrt{\alpha'} \big(z_{23}z_{24}z_{34}\big)^{-1}\ \prod_{i<j}^4|z_{ij}|^{2\alpha'k_ik_j}\ \times \ \ \ \ \ \ \ \ \ \ \ \ \ \ \ \ \ \ \ \ \ \ \ \ \ \ \ \ \ \ \ \ \ \ \ \ \ \ \ \ \ \ \ \ \ \ \ \ \ \ \ \ \ \ \ \ 
\nonumber\\[7pt]
\bigg\{ \Big[ \underbrace{\frac{1}{z_{12}}\xi_{2\rho}(\xi_1k_2)+ \frac{1}{z_{13}}\xi_{2\rho}(\xi_1k_3)+ \frac{1}{z_{14}}\xi_{2\rho}(\xi_1k_4)}_{\text{terms  from the sum}} -\frac{z_{24}}{z_{12}z_{14}}\xi_{1\rho}(\xi_2k_1)+\frac{z_{24}}{z_{12}z_{14}}k_{1\rho}(\xi_1\xi_2)\Big] \big(u_3\sigma^\rho \bar{u}_4\big)
\nonumber\\[7pt]
+\frac{1}{2}\frac{z_{34}}{z_{13}z_{14}}k_{1\lambda}\xi_{1\mu}\xi_{2\rho}\big(u_3\sigma^{\lambda}\bar{\sigma}^{\mu}\sigma^{\rho}\bar{u_4}\big) \bigg\}  \nonumber\\
\end{eqnarray}

Now we make the choice $z_4=\text{constant}\to \infty$. This has the effect of making the third term vanish, and $\prod_{i<j}^4|z_{ij}|^{2\alpha'k_ik_j}= \text{const}\cdot |z_{12}|^s |z_{13}|^t |z_{23}|^u$. We get:
\begin{eqnarray}
\langle \ldots \rangle \ \  = \ \ \ \sqrt{\alpha'} \ |z_{12}|^s |z_{13}|^t |z_{23}|^u |z_{23}|^{-1}\ \ \ \ \ \ \ \ \ \ \ \ \ \ \ \ \ \ \ \ \ \ \ \ \ \ \ \ \ \ \ \ \ \ \ \ \ \ \ \ \ 
\nonumber\\[7pt]
\times\ \bigg\{ \Big[ \frac{1}{z_{12}}\xi_{2\rho}(\xi_1k_2)+ \frac{1}{z_{13}}\xi_{2\rho}(\xi_1k_3) -\frac{1}{z_{12}}\xi_{1\rho}(\xi_2k_1)+\frac{1}{z_{12}}k_{1\rho}(\xi_1\xi_2)\Big]\big(u_3\sigma^\rho \bar{u}_4\big)
\nonumber\\[7pt]
+\frac{1}{2}\frac{1}{z_{13}}k_{1\lambda}\xi_{1\mu}\xi_{2\rho}\big(u_3\sigma^{\lambda}\bar{\sigma}^{\mu}\sigma^{\rho}\bar{u_4}\big) \bigg\}  \nonumber\\
\end{eqnarray}

Now we can set $z_1=0$ and $z_3=1$, and get: 
\begin{eqnarray}
\langle \ldots \rangle \ \  = \ \ 2\alpha'g_{D_{p_x}} g_{D_{p_y}}\ z_{12}^{s-1} z_{23}^{u-1} \ \ \ \ \ \ \ \ \ \ \ \ \ \ \ \ \ \ \ \ \ \ \ \ \ \ \ \ \ \ \ \ \ \ \ 
\nonumber\\[6pt]
\times \ \bigg\{ \Big[ \xi_{2\rho}(\xi_1k_2) -\xi_{1\rho}(\xi_2k_1)+k_{1\rho}(\xi_1\xi_2)+ z_{12}\ \xi_{2\rho}(\xi_1k_3)\Big]\big(u_3\sigma^{\rho} \bar{u}_4\big)
\nonumber\\
+\frac{z_{12}}{2}k_{1\lambda}\xi_{1\mu}\xi_{2\rho}\big(u_3\sigma^{\lambda}\bar{\sigma}^{\mu}\sigma^{\rho}\bar{u_4}\big) \bigg\}  \nonumber\\
\end{eqnarray}
Where we have inserted the coupling constants.

Integrating over $z_2$ we get:
\begin{eqnarray}
\langle \ldots \rangle \ \ =\ \  -2\alpha' g_{D_{p_x}} g_{D_{p_y}}\ \mathcal{K}\ B(s,u)
\end{eqnarray}

Where $\mathcal{K}$ is a kinematic factor:
\begin{eqnarray}
\mathcal{K}\equiv \Big\{ \big[ k_{1\rho}(\xi_1\xi_2)-\xi_{1\rho}(\xi_2k_1)+\xi_{2\rho}(\xi_1k_2)- \frac{s}{t}\xi_{2\rho}(\xi_1k_3) \big](u_3\sigma^\rho \bar{u}_4)-
\nonumber\\
\frac{1}{2}\frac{s}{t}k_{1\lambda}\xi_{1\mu}\xi_{2\rho}(u_3\sigma^{\lambda}\bar{\sigma}^{\mu}\sigma^{\rho}\bar{u_4}) \Big\}
\end{eqnarray}

If both gauge bosons are from stack $a$, then there are two permutations that should be considered. After inserting back the color matrices we have:
\begin{eqnarray}
\mathcal{M}(A^{a_1}, A^{a_2},\psi^{\alpha_3}_{\beta_3},\bar{\psi}^{\beta_4}_{\alpha_4})\ =\  -2\alpha'g^2\mathcal{K}\ \ \ \ \ \ \ \ \ \ \ \ \ \ \ \ \ \ \ \ \ \ \ \ \ \ \ \ \ \ \ \ \ \ 
\nonumber\\[7pt]
\times \Big[Tr(T^{a_1}T^{a_2}T^{\alpha_3}_{\beta_3}T^{\beta_4}_{\alpha_4})\ B(s,u)+Tr(T^{a_2}T^{a_1}T^{\alpha_3}_{\beta_3}T^{\beta_4}_{\alpha_4})\ \frac{t}{u}\ B(s,t)\Big]
\end{eqnarray}
Where $g \equiv g_{D_{p_a}}$.

On the other hand, if the gauge bosons are from two different stacks $a$ and $b$, then there is only one permutation which gives:
\begin{eqnarray}
\mathcal{M}(A^{a},A^{b},\psi^{\alpha_3}_{\beta_3},\bar{\psi}^{\beta_4}_{\alpha_4})\ =\  
-2\alpha'g\ g_{D_{p_b}}\ \mathcal{K}\  Tr(T^aT^{\alpha_3}_{\beta_3}T^b T_{\alpha_4}^{\beta_4})\ \frac{t}{s}B(t,u)
\end{eqnarray}

Choosing $k_2$ as the reference momentum for $\xi_1$ and $k_1$ for $\xi_1$, it is seen that $\mathcal{K}$ vanishes if $\xi_1$ and $\xi_2$ have the same helicity. If they have opposite helicity, we use the following identities: 
\begin{eqnarray}
\xi_1^{\pm}\xi_2^{\mp}= 0
\end{eqnarray}
\begin{eqnarray}
\xi_{2\rho}^+(\xi_1^-k_3) (u_3\sigma^\rho \bar{u}_4)= \frac{1}{\alpha's}\langle13\rangle^2[23][24]
\end{eqnarray}
\begin{eqnarray}
\xi_{2\rho}^-(\xi_1^+k_3)(u_3\sigma^\rho \bar{u}_4)= \frac{1}{\alpha's}\langle23\rangle^2[13][14] 
\end{eqnarray}
(see Eqs~(\textcolor{zz}{\ref{eq:firste8}}), (\textcolor{zz}{\ref{eq:laste8}}) ),
and obtain:

\begin{eqnarray}\boxed{
\mathcal{M}(g_1^-,g_2^+,q_3^-,q_4^+)= 2g^2\delta^{\beta_4}_{\beta_3}\ \frac{\langle13\rangle^2}{\langle23\rangle \langle24\rangle} \bigg[(T^{a_1}T^{a_2})^{\alpha_3}_{\alpha_4}\ \frac{t}{s}\ V_t+ (T^{a_2}T^{a_1})^{\alpha_3}_{\alpha_4}\ \frac{u}{s}\ V_u\bigg]}
\end{eqnarray}

\begin{eqnarray}\boxed{
\mathcal{M}(g_1^-,B_2^+,q_3^-,q_4^+)= 2g_{D_{p_b}}g\ \frac{\langle13\rangle ^2}{\langle23\rangle \langle24\rangle}(T^{a})^{\alpha_3}_{\alpha_4}(T^{b})^{\beta_4}_{\beta_3}\ V_s}
\end{eqnarray}
Where $B$ is the vector boson from stack $b$.
We clearly see a factorized form of a standard model sub-amplitude times a Veneziano ampltude.

We square the amplitude
\begin{eqnarray}
|\mathcal{M}|^2(g_1^-,g_2^+,q_3^-,q_4^+)= \ \ \ \ \ \ \ \ \ \ \ \ \ \ \ \ \ \ \ \ \ \ \ \ \ \ \ \ \ \ \ \ \ \ \ \ \  \ \ \ \ \ \ \ \ \ \ \ \ \ \ \ \ \ \ \ \ \ \ \ \ \ \ \ \ \ \ \ \ \ 
\nonumber\\[7pt]
4g_a^4N_b \bigg[\sum_{a_1,a_2} Tr(T^{a_1}T^{a_1}T^{a_2}T^{a_2})\ \frac{t}{us^2}\ (tV_t+uV_u)^2- \sum_{a_1,a_2,i} \frac{1}{2}\texttt{f}^{a_1a_2i}\texttt{f}^{a_1a_2i}\ \frac{t^2}{s^2}\ V_tV_u\bigg]
\end{eqnarray}

\begin{eqnarray}
|\mathcal{M}|^2(g_1^-,B_2^+,q_3^-,q_4^+)= 
4g_a^2g_b^2 \sum_{a,b} Tr(T^{a}T^{a})\ Tr(T^{b}T^{b})\ \frac{t}{u}\ V_s^2
\end{eqnarray}

Using
\begin{eqnarray}
\sum_a Tr(T^{a}T^{a})= \frac{N^2-1}{2N}\mathbf{1}_N
\end{eqnarray}

We get:
\begin{eqnarray}\boxed{
|\mathcal{M}|^2(gg\to q\bar{q})= 
g^4\frac{N_f}{2N}\ \frac{t^2+u^2}{s^2}\ \bigg[ \frac{1}{ut}(tV_t+uV_u)^2- \frac{2N^2}{N^2-1}V_tV_u\bigg]}
\end{eqnarray}
Which for $N=3$ colors coincides with Eq.~(\textcolor{zz}{\ref{eq:treeate2}}), and

\begin{eqnarray}\boxed{
|\mathcal{M}|^2( q\bar{q}'\to gB)= 
\frac{4}{9}g^4\ |T^B_{q\bar{q}'}|^2 Q_A^2\ \frac{t^2+u^2}{ut}\ V_s^2}
\end{eqnarray}
Eq.~(\zz{\ref{eq:bnlmp}}).

\clearpage

\section{\textcolor{ww}{Vertex operators}}
\label{sec:appa}

\emph{References:} ~\textcolor{rr}{\cite{lust, lust2, feng, koh}}.

\subsection{Massles particles}

We list below some vertex operators for massless fields. We supress the $z$ dependence of all the fields (for example $\psi^\mu \equiv \psi^\mu(z)$). The coupling constants are:
\begin{eqnarray}
g\equiv g_{D_{p_a}}\ \ \ ,\ \ \ g_{A}= g_{\phi}^A= (2\alpha')^{1/2}g_{D_{p_a}}\ \ \ ,\ \ \ g_{\lambda}= (2\alpha')^{1/2}\alpha'^{1/4}g_{D_{p_a}}
\nonumber\\[6pt]
g_{\psi}= (2\alpha')^{1/2}\alpha'^{1/4}e^{\phi_{10}/2}\ \ \ ,\ \ \ g_{\phi}= (2\alpha')^{1/2}e^{\phi_{10}/2}
\end{eqnarray}

\textbf{Gauge bosons} in the $(-1)$ and $(0)$ ghost picture:
\begin{eqnarray}
\label{eq:appb1}
V^{(-1)}_{A^a}(z,\xi, k)=\ g_A[T^a]^{\alpha_1}_{\alpha_2}\ e^{-\phi}\xi^{\mu}\psi_{\mu}e^{ik\cdot X}
\end{eqnarray}

\begin{eqnarray}
\label{eq:appb2}
V^{(0)}_{A^a}(z,\xi, k)=\ \frac{g_A}{(2\alpha')^{1/2}}[T^a]^{\alpha_1}_{\alpha_2}\xi_{\mu}\ \big[i\partial X^{\mu}+2\alpha'(k\psi)\psi^{\mu}\big]e^{ik\cdot X}
\end{eqnarray}
These are independent of the internal part of the SCFT.\\

\textbf{Chiral fermions} (These depend on the internal field $\Xi^{a \cap b}$):
\begin{eqnarray}
\label{eq:appb3}
V^{(-1/2)}_{\psi^{\alpha}_{\beta}}(z,u,k)=\ g_{\psi}[T^{\alpha}_{\beta}]^{\beta_1}_{\alpha_1}\    e^{-\phi/2}u^{\lambda}S_{\lambda}\Xi^{a \cap b}e^{ik\cdot X}
\end{eqnarray}

\begin{eqnarray}
\label{eq:appb4}
V^{(-1/2)}_{\bar{\psi}^{\beta}_{\alpha}}(z,\bar{u},k)=\ g_{\psi}[T^{\beta}_{\alpha}]^{\alpha_1}_{\beta_1}\ e^{-\phi/2}\bar{u}_{\dot{\lambda}}S^{\dot{\lambda}}\bar{\Xi}^{a \cap b}e^{ik\cdot X}
\end{eqnarray}

\textbf{Gauginos}:
\begin{eqnarray}
V^{(-1/2)}_{\lambda^{a,I}}(z,u,k)=\ g_{\lambda}[T^a]^{\alpha_1}_{\alpha_2}\ e^{-\phi/2}u^{\lambda}S_{\lambda}\Sigma^Ie^{ik\cdot X}
\end{eqnarray}

\begin{eqnarray}
V^{(-1/2)}_{\bar{\lambda}^{a,I}}(z,\bar{u}, k)=\ g_{\lambda}[T^a]^{\alpha_1}_{\alpha_2}\ e^{-\phi/2}\bar{u}_{\dot{\lambda}}S^{\dot{\lambda}}\bar{\Sigma}^Ie^{ik\cdot X}
\end{eqnarray}

\textbf{Adjoint scalars}:
\begin{eqnarray}
V^{(-1)}_{\phi^{a,i}}(z, k)=\ g_{\phi}^A[T^a]^{\alpha_1}_{\alpha_2}\ e^{-\phi}\Psi^i e^{ik\cdot X}
\end{eqnarray}

\begin{eqnarray}
V^{(-1)}_{\bar{\phi}^{a,i}}(z, k)=\ g_{\phi}^A[T^a]^{\alpha_1}_{\alpha_2}\ e^{-\phi}\bar{\Psi}^i e^{ik\cdot X}
\end{eqnarray}

\textbf{Chiral scalars}:
\begin{eqnarray}
V^{(-1)}_{\phi^{\alpha}_{\beta}}(z,k)=\ g_{\phi}[T^{\alpha}_{\beta}]^{\beta_1}_{\alpha_1}\ e^{-\phi}\Pi^{a \cap b}e^{ik\cdot X}
\end{eqnarray}

\begin{eqnarray}
V^{(-1)}_{\bar{\phi}^{\beta}_{\alpha}}(z,k)=\ g_{\phi}[T^{\beta}_{\alpha}]^{\alpha_1}_{\beta_1}\ e^{-\phi}\bar{\Pi}^{a \cap b}e^{ik\cdot X}
\end{eqnarray}

\subsection{First excited state}
We list below the vertex operators of the first excited state of the open string. These are used in section~\ref{sec:a2}.

\subsubsection{Bosons of the NS sector}

In $D=10$:
\begin{eqnarray}
V^{(-1)}_{NS,a}(z,k)= \frac{g_A}{\sqrt{2\alpha'}}T^ae^{-\phi}\Big[E_{mnp}\psi^m\psi^n\psi^p+B_{mn}i\partial X^m\psi^n+H_m\partial\psi^m\Big]e^{ik\cdot X}
\end{eqnarray}

In $D=4$ we have: \\

\textbf{Spin} $\mathbf{J=0}$:
\begin{eqnarray}
V^{(-1)}_{\Phi^{\pm}}(z,k)= \frac{g_A}{2\sqrt{2\alpha'}}T^ae^{-\phi} \Big[(g_{\mu \nu}+2\alpha'k_\mu k_\nu)i\partial X^\mu\psi^\nu + 2\alpha'k_\mu \partial\psi^\mu \pm\frac{i}{6}2\alpha'\epsilon_{\mu \nu \rho \lambda}k^\lambda\psi^\mu\psi^\nu\psi^\rho \Big]e^{ik\cdot X}
\nonumber\\
\end{eqnarray}

\textbf{Spin}  $\mathbf{J=2}$:
\begin{eqnarray}
V^{(-1)}_{B^a}(z,\alpha,k)=\ \frac{g_A}{\sqrt{2\alpha'}}T^a\ e^{-\phi}\alpha_{\mu \nu}i\partial X^\mu\psi^\nu e^{ik\cdot X}
\end{eqnarray}

\textbf{Spin} $\mathbf{J=0}$:
\begin{eqnarray}
V^{(-1)}_{\Omega^a}(z,k)=\ g_AT^a\ e^{-\phi}\mathcal{O} e^{ik\cdot X}
\end{eqnarray}

\textbf{Spin} $\mathbf{J=1}$:
\begin{eqnarray}
\label{eq:ww11}
V^{(-1)}_{W^a}(z,\xi,k)=\ g\sqrt{\frac{\alpha'}{6}}T^a\ e^{-\phi}\xi_\mu \psi^\mu \mathcal{J} e^{ik\cdot X}
\end{eqnarray}

\subsubsection{Fermions of the R sector}

In $D=10$:
\begin{eqnarray}
V^{(-1/2)}_{R,a}(z,v,\bar{\rho},k)=\ C_\Lambda T^a \Big[v_m^Ai\partial X^m+2\alpha'\bar{\rho}^m_{\dot{B}}\psi_m\psi^n\Gamma_n^{\dot{B}A}\Big] \Theta_A e^{-\phi/2}e^{ik\cdot X}
\end{eqnarray}

In $D=4$:\\

\textbf{Spin} $\mathbf{J=1/2}$ and $\mathbf{J=3/2}$:
\begin{eqnarray}
V^{(-1/2)}_{Q^\alpha_\beta}(z,v,\bar{\rho},k)= \alpha'^{1/4}e^{\phi_{10}/2} (T^\alpha_\beta)^{\beta_1}_{\alpha_1} \Big[iv_\mu^\beta \partial X^\mu - \sqrt{\alpha'}\bar{\rho}^\mu_{\dot{\alpha}}\psi_\mu\psi^\nu\sigma_\nu^{\dot{\alpha}\beta}\Big] S_\beta e^{-\phi/2}\Xi^{a \cap b}e^{ik\cdot X}
\end{eqnarray}

\begin{eqnarray}
V^{(-1/2)}_{\bar{Q}^\beta_\alpha}(z,\bar{v},\rho,k)= \alpha'^{1/4}e^{\phi_{10}/2} (T^\beta_\alpha)^{\alpha_1}_{\beta_1} \Big[i\bar{v}^\mu_{\dot{\beta}} \partial X_\mu - \sqrt{\alpha'}\rho_\mu^{\alpha}\psi^\mu\psi_\nu\sigma^\nu_{\alpha \dot{\beta}}\Big] S^{\dot{\beta}} e^{-\phi/2}\bar{\Xi}^{a \cap b}e^{ik\cdot X}
\nonumber\\
\end{eqnarray}

For $J=1/2$:  
\begin{eqnarray}
\bar{v}^\mu_{\dot{\alpha}}(J=1/2)\ =\  -\frac{\sqrt{\alpha'}}{2\sqrt{2}}\bar{\chi}_{\dot{\beta}} (\bar{\sigma}^\mu k)^{\dot{\beta}}_{\ \dot{\alpha}}\ \ \ ,\ \ \ \rho_\mu^{\alpha}(J=1/2)\ =\  -\frac{1}{6\sqrt{2}}\bar{\chi}_{\dot{\beta}}\bar{\sigma}_\mu ^{\dot{\beta}\alpha}
\end{eqnarray}

While for $J=3/2$:
\begin{eqnarray}
\bar{v}^\mu_{\dot{\alpha}}(J=3/2)\ =\  \bar{\chi}_{\dot{\alpha}}^\mu\ \ \ \ ,\ \ \ \ \text{such that}\ \ \ \ \ \  \bar{\chi}_{\dot{\alpha}}^\mu \bar{\sigma}_\mu ^{\dot{\alpha}\beta}\ =\  k_\mu\bar{\chi}_{\dot{\alpha}}^\mu= 0
\nonumber\\[6pt]
 \rho^{\mu\alpha}(J=3/2)\ =\  \eta^{\mu\alpha}\ =\ \sqrt{\alpha'} \bar{\chi}_{\dot{\beta}}^\mu \bar{\sigma}_\nu ^{\dot{\beta}\alpha}k^\nu
\end{eqnarray}
  
\clearpage

\section{\textcolor{ww}{Correlation functions}}
\label{sec:appcorr}
\emph{References:} ~\textcolor{rr}{\cite{lust2, lust, hartle, tong}}.

\subsection{An example calculation}

We rewrite the definition of the correlation function Eq. (\textcolor{zz}{\ref{eq:b3}}):
\begin{eqnarray}
\langle V(z_1,k_1)\cdot \cdot \cdot V(z_m,k_m)\rangle \equiv \int \mathcal{D}X \exp \bigg(-\frac{1}{2\pi \alpha '}\int d^2z\partial X\cdot \bar{\partial} X\bigg) \prod_{i=1}^m V(z_1,k_1)\cdot \cdot \cdot V(z_m,k_m) \nonumber\\
\end{eqnarray}

We now calculate the correlator for a chain of tachyon vertex operators $V(z_j, k_j)= e^{ik_j\cdot X(z_j)}$:
\begin{eqnarray}
\langle e^{ik_1\cdot X}\cdots e^{ik_m\cdot X}\rangle \ =\ \ \ \ \ \ \ \ \ \ \ \ \ \ \ \ \ \ \ \ \ \ \ \ \ \ \ \ \ \ \ \ \ \ \ \ \ \ \ \ \ \ \ \ \ \ \ \ \ \ 
\nonumber\\[6pt]
\int \mathcal{D}X \exp \bigg(-\frac{1}{2\pi \alpha '}\int d^2z\partial X\cdot \bar{\partial} X\bigg) 
\exp \bigg(i\sum_{j=1}^m k_j\cdot X(z_j)\bigg)
\end{eqnarray}

This is a gaussian integral which is computed as:
\begin{eqnarray}
\label{eq:gau78}
\int \mathcal{D}X \exp \bigg(\frac{1}{2\pi \alpha '}\int d^2z X\cdot \partial \bar{\partial} X+i J\cdot X \bigg) \approx \exp \bigg(\frac{\pi \alpha '}{2}\int d^2zd^2z' J(z,\bar{z})\frac{1}{\partial \bar{\partial}}J(z',\bar{z}') \bigg)\nonumber\\ 
\end{eqnarray}

The operator $\frac{1}{\partial \bar{\partial}}$ is the propagator $G(z,\bar{z},z';\bar{z}')$ which obeys the equation:
\begin{eqnarray}
\partial \bar{\partial}G(z,\bar{z},z';\bar{z}')= \delta (z-z', \bar{z}-\bar{z}')
\end{eqnarray}

Since we are in two dimensions, the solution for open and closed strings is:
\begin{eqnarray}
G^{(closed)}(z,\bar{z};z',\bar{z}')= \frac{1}{\pi}\ln |z-z'|= \frac{1}{2}G^{(open)}(z,\bar{z};z',\bar{z}')
\end{eqnarray}

Since,
\begin{eqnarray}
J(z,\bar{z})= \sum_{j=1}^m k_j\delta (z-z_j, \bar{z}-\bar{z_j})
\end{eqnarray}

Eq.~(\textcolor{zz}{\ref{eq:gau78}}) gives the result:
\begin{eqnarray}
\label{eq:appbc1}
\langle e^{ik_1\cdot X}\cdots e^{ik_m\cdot X}\rangle ^{(closed)}= \exp \bigg( \frac{\alpha'}{2}\sum_{j,l}k_j\cdot k_l \ln |z_j-z_l| \bigg)= \prod_{j<l}|z_j-z_l|^{\alpha' k_j\cdot k_l}
\end{eqnarray}

\begin{eqnarray}
\label{eq:appbc2}
\langle e^{ik_1\cdot X}\cdots e^{ik_m\cdot X}\rangle ^{(open)}= \prod_{j<l}|z_j-z_l|^{2\alpha' k_j\cdot k_l}
\end{eqnarray}

\subsection{List of some correlation functions}

We list some of the correlation functions of the SCFT fields which appear in the calculation of 4 and 5 point amplitudes. We define $z_{ij}\equiv z_i-z_j$.

\begin{itemize}
\item $X^\mu(z)$
\end{itemize}

\begin{eqnarray}
\langle X^\mu(z_1)\ X^\nu(z_2)\rangle=\ -2\alpha'\delta^{\mu \nu}\ln (z_{12}) 
\\[15pt]
\langle \partial X^\mu(z_1)\ X^\nu(z_2)\rangle=\ -\frac{2\alpha'\delta^{\mu\nu}}{z_{12}}
\\[15pt]
\langle \partial X^\mu(z_1)\ \partial X^\nu(z_2)\rangle=\ -\frac{2\alpha'\delta^{\mu\nu}}{z_{12}^2} 
\\[15pt]
\langle e^{ik_\mu X^\mu(z_1)}\ e^{ik_\nu X^\nu(z_2)}\rangle=\ |z_{ij}|^{2\alpha'k_1k_2} 
\end{eqnarray}

\begin{eqnarray}
\label{eq:la9}
\langle e^{ik_1\cdot X(z_1)}\cdots e^{ik_n\cdot X(z_n)}\rangle=\  \prod_{i<j}^n|z_{ij}|^{2\alpha'k_ik_j}  
\end{eqnarray} 

\begin{eqnarray}
\label{eq:appb10}
\langle \partial X^\mu(z_A)\ \prod_{i=1}^n e^{ik_i \cdot X(z_i)}\rangle=\  \Big(-2i\alpha'\sum_{r=1}^n\frac{k_r^\mu}{z_{A,r}}\Big)\ \langle \prod_{i=1}^ne^{ik_i \cdot X(z_i)}\rangle
\end{eqnarray} 

\begin{eqnarray}
\langle \partial X^\mu(z_A)\partial X^\nu(z_B)\prod_{i=1}^n e^{ik_i \cdot X(z_i)}\rangle= \Big(-4\alpha'^2 \alpha'\sum_{r,s=1}^n\frac{k_r^\mu}{z_{A,r}z_{B,s}}-2\alpha'\frac{\eta^{\mu\nu}}{z^2_{AB}}\Big)\langle \prod_{i=1}^ne^{ik_i \cdot X(z_i)}\rangle
\nonumber\\
\end{eqnarray} 

\begin{itemize}
\item $\phi(z)$
\end{itemize}

\begin{eqnarray} 
\langle e^{-\phi(z_1)}\ e^{-\phi(z_2)}\rangle=\ \frac{1}{z_{12}} 
\end{eqnarray}
\begin{eqnarray}
\label{eq:appb5}
\langle e^{-\phi(z_1)}\ e^{-\phi(z_2)/2}\ e^{-\phi(z_3)/2}\rangle =\  z_{12}^{-1/2} z_{13}^{-1/2} z_{23}^{-1/4}
\end{eqnarray} 

\begin{eqnarray}
\langle e^{-\phi(z_1)/2}\ e^{-\phi(z_2)/2}\ e^{-\phi(z_3)/2}\ e^{-\phi(z_4)/2}\rangle =\ 
\frac{1}{(z_{12}z_{13}z_{14}z_{23}z_{24}z_{34})^{1/4}}
\end{eqnarray} 

\begin{itemize}
\item $\psi^\mu(z)$ and $S_\alpha(z)$
\end{itemize}

\begin{eqnarray} 
\langle \psi^\mu(z_1)\ \psi^\nu(z_2)\rangle=\ \frac{\delta^{\mu\nu}}{z_{12}}
\end{eqnarray}
\begin{eqnarray}
\label{eq:appb8}
\langle \psi^\mu(z_1)\ S_\alpha(z_2)\ S_{\dot{\beta}}(z_3)\rangle =\  (2z_{12}z_{13})^{-1/2}\sigma^\mu_{\alpha\dot{\beta}}
\end{eqnarray} 

\begin{eqnarray}
\label{eq:appb9}
\langle \psi^\mu(z_1)\ \psi^\nu(z_2)\ \psi^\lambda(z_3)\ S_\alpha(z_4)\ S_{\dot{\beta}}(z_5)\rangle =\  (2z_{14}z_{15}z_{24}z_{25}z_{34}z_{35})^{-1/2}
\nonumber\\ \nonumber\\
\times \Big\{ \frac{z_{45}}{2}(\sigma^\mu \bar{\sigma}^\nu \sigma^\lambda)+\eta^{\mu \nu} \sigma^\lambda \frac{z_{14}z_{25}}{z_{12}}- \eta^{\mu \lambda}\sigma^\nu \frac{z_{14}z_{35}}{z_{13}}+ \eta^{\nu \lambda}\sigma^\mu \frac{z_{24}z_{35}}{z_{23}} \Big\}
\end{eqnarray} 

\begin{eqnarray} 
\langle S_{\alpha}(z_1)\  S_{\beta}(z_2)\rangle=\ -\frac{\epsilon_{\alpha\beta}}{z_{12}^{1/2}}
\end{eqnarray} 
\begin{eqnarray} 
\langle S_{\dot{\alpha}}(z_1)\  S_{\dot{\beta}}(z_2)\rangle=\ \frac{\epsilon_{\dot{\alpha}\dot{\beta}}}{z_{12}^{1/2}}
\end{eqnarray} 

\begin{itemize}
\item $\Xi^{a \cap b}(z)$
\end{itemize}

\begin{eqnarray}
\label{eq:appb6}
\langle\  \Xi^{a \cap b}(z_1)\ \bar{\Xi}^{a \cap b}(z_2)\ \rangle= z_{12}^{-3/4}
\end{eqnarray} 

\begin{eqnarray}
\langle\  \Xi^{a \cap b}(z_1)\ \bar{\Xi}^{b \cap d}(z_2)\ \Xi^{d \cap c}(z_3)\ \bar{\Xi}^{c \cap a}(z_4)\ \rangle=\  \Big(\frac{z_{13}z_{24}}{z_{12}z_{14}z_{23}z_{34}}\Big)^{3/4}\ I_\rho\ (\{ z_i\};\theta^j)
\end{eqnarray} 

\begin{eqnarray} 
\langle\  \mathcal{J}(z_1)\ \Xi^{a \cap b}(z_2)\ \bar{\Xi}^{a \cap b}(z_3)\ \rangle =\  \langle\  \mathcal{J}(z_1)\ \Sigma^{a \cap b}(z_2)\ \bar{\Sigma}^{a \cap b}(z_3)\ \rangle =\  \frac{3z_{23}^{1/4}}{2z_{12}z_{13}}
\end{eqnarray} 
Where $\mathcal{J}$ is the field in Eq.~(\textcolor{zz}{\ref{eq:ww11}}).

\clearpage
\section{\textcolor{ww}{Helicity notation}}
\label{sec:appb}

\emph{References:} We follow:~\textcolor{rr}{\cite{mangano, dixon}}.\\

We review the spinor helicity formalism.

Massless fermions with a definite helicity, which solve Dirac's equation are
\begin{eqnarray}
u_{\pm}(k)= \frac{1}{2}(1\pm \gamma_5)u(k),\ \ \  v_{\pm}(k)= \frac{1}{2}(1\mp \gamma_5)u(k)
\end{eqnarray}

These are chosen as follows
\begin{eqnarray}
u_{+}(k)= v_{-}(k)=\frac{1}{\sqrt{2}}(\sqrt{k^+},\sqrt{k^-}e^{\varphi_k},\sqrt{k^+},\sqrt{k^-}e^{\varphi_k})
\end{eqnarray}
\begin{eqnarray}
u_{-}(k)= v_{+}(k)=\frac{1}{\sqrt{2}}(\sqrt{k^+},-\sqrt{k^-}e^{-\varphi_k},-\sqrt{k^+},\sqrt{k^-}e^{-\varphi_k})
\end{eqnarray}

where
\begin{eqnarray}
k^{\pm}= k^{0} \pm k^{3},\ \ \  e^{\pm i\varphi_k}(k)= \frac{k_1\pm ik_2}{\sqrt{k_+k_-}}
\end{eqnarray}

Introducing the notation
\begin{eqnarray}
|i^{\pm}\rangle \equiv u_{\pm}(k_i)= v_{\mp}(k_i), \ \ \ \langle i^{\pm}|\equiv \bar{u}_{\pm}(k_i)= \bar{v}_{\mp}(k_i)
\end{eqnarray}

we get the products
\begin{eqnarray}
\langle ij\rangle \equiv \langle i^-|j^+\rangle = \bar{u}_-(k_i)u_+(k_j) = \sqrt{k_i^-k_j^+}e^{i\varphi_{k_i}}-\sqrt{k_i^+k_j^-}e^{i\varphi_{k_j}}= \sqrt{|s_{ij}|}e^{i\phi_{ij}}
\end{eqnarray}

\begin{eqnarray}
[ij]\equiv \langle i^+|j^-\rangle= \bar{u}_+(k_i)u_-(k_j)= -\sqrt{k_i^-k_j^+}e^{i\varphi_{k_i}}+\sqrt{k_i^+k_j^-}e^{i\varphi_{k_j}}=  -\sqrt{|s_{ij}|}e^{-i\phi_{ij}}
\end{eqnarray}

where
\begin{eqnarray}
s_{ij}\equiv(k_i+k_j)^2= 2k_i\cdot k_j
\end{eqnarray}
\begin{eqnarray}
\cos\phi_{ij}= \frac{k_i^1 k_j^+ - k_j^1 k_i^+}{\sqrt{|s_{ij}|k_i^+k_j^+}}, \ \ \ \sin\phi_{ij}= \frac{k_i^2 k_j^+ - k_j^2 k_i^+}{\sqrt{|s_{ij}|k_i^+k_j^+}}
\end{eqnarray}

Lets write down some usefull identities
\begin{eqnarray}
\langle ij\rangle[ji]= s_{ij}
\end{eqnarray}

Antisymmetry
\begin{eqnarray}
\langle ij\rangle^*= -[ij]= [ji]
\end{eqnarray}

\emph{Schouten identity}
\begin{eqnarray}
\langle ij\rangle \langle kl\rangle= \langle ik\rangle \langle jl\rangle- \langle il\rangle \langle jk\rangle
\end{eqnarray}

\emph{Fierz rearangement }
\begin{eqnarray}
\label{eq:fierz6}
\langle i^+|\gamma_\mu|j^+\rangle \langle k^+|\gamma^\mu|l^+\rangle=2[ik]\langle lj\rangle
\end{eqnarray}

Identities involving the anti-symmetric tensor $\epsilon(i,j,l,m)\equiv \epsilon_{\mu\nu\sigma\rho}k_i^\mu k_j^\nu  k_l^\rho k_m^\sigma$:

\begin{eqnarray}
\langle ij\rangle[jl]\langle lm\rangle[mi]= \frac{1}{2}\big[s_{ij}s_{lm}-s_{il}s_{jm}+s_{im}s_{jl}-4i\epsilon(i,j,l,m)\big]
\end{eqnarray}
\begin{eqnarray}
[ij]\langle jl\rangle[lm]\langle mi\rangle= \frac{1}{2}\big[s_{ij}s_{lm}-s_{il}s_{jm}+s_{im}s_{jl}+4i\epsilon(i,j,l,m)\big]
\end{eqnarray}

Subtracting the last two:
\begin{eqnarray}
4i\epsilon(i,j,l,m)= [ij]\langle jl\rangle[lm]\langle mi\rangle- \langle ij\rangle[jl]\langle lm\rangle[mi]
\end{eqnarray}

Momentum conservation
\begin{eqnarray}
\sum_{n\neq i,j}\langle in\rangle[nj]=0
\end{eqnarray}

A spinor representation for the polarization vector of a massless gauge boson:
\begin{eqnarray}
\label{eq:firste8}
\xi_{\mu}^+(k,q)= \frac{\langle q^-|\gamma_\mu|k^-\rangle}{\sqrt{2}\langle qk\rangle},\ \ \ \ \ \xi_{\mu}^-(k,q)= - \frac{\langle q^+|\gamma_\mu|k^+\rangle}{\sqrt{2}[qk]}
\end{eqnarray}
Where $k$ is the momentum of the gauge boson and $q$ is called the \emph{reference momentum}.

We write down some identities involving polarisation vectors. In each one, the expression after the arrow is obtained by choosing the reference vectors $q_1=k_2$, $q_2=k_1$ as in Appendix~\ref{sec:b8}. This choice causes much simplification.
\begin{eqnarray}
\xi_{\mu}^-(k_1,q_1)\xi^{\mu -}(k_2,q_2)= \frac{2\langle q_1^+|\gamma_\mu|k_1^+\rangle \  \langle q_2^+|\gamma^\mu|k_2^+\rangle}{2[q_1k_1]\ [q_2k_2]}= \frac{[q_1q_2]\langle k_2k_1\rangle}{[q_1k_1][q_2k_2]}\to \frac{\langle k_1k_2 \rangle}{[k_1k_2]}
\end{eqnarray}
Where we used Eq.~(\ref{eq:fierz6}).

\begin{eqnarray}
\xi_{\mu}^+(k_1,q_1)\xi^{\mu +}(k_2,q_2)= \frac{[k_1k_2]\langle q_2q_1\rangle}{\langle q_1k_1\rangle \langle q_2k_2\rangle}\to \frac{[k_1k_2]}{\langle k_1k_2 \rangle}
\end{eqnarray}

\begin{eqnarray}
\xi_{\mu} ^{-} (k_1,q_1)\xi^{\mu +} (k_2,q_2)= \frac{[q_1k_2]\langle q_2k_1\rangle}{[q_1k_1]\langle q_2k_2\rangle}\to 0
\end{eqnarray}

\begin{eqnarray}
p^{\mu} \xi_{\mu}^+(k_1,q_1)= \frac{\langle q_1p\rangle[pk_1]}{\sqrt{2}\langle q_1k_1\rangle}
\end{eqnarray}

\begin{eqnarray}
p^\mu \xi_{\mu}^-(k_1,q_1)= -\frac{\langle k_1p\rangle[pq_1]}{\sqrt{2}[q_1k_1]}
\end{eqnarray}

\begin{eqnarray}
\xi_{\mu}^+(k_2,q_2)\big[\bar{u}(p_4)\gamma^\mu u(p_3)\big]= \sqrt{2}\frac{[k_2p_4]\langle p_3q_2\rangle}{\langle q_2k_2\rangle}\to \sqrt{2}\frac{[k_2p_4]\langle p_3k_1\rangle}{\langle k_1k_2\rangle}
\end{eqnarray}

\begin{eqnarray}
\label{eq:laste8}
\xi_{\mu}^-(k_2,q_2)\big[ \bar{u}(p_4)\gamma^\mu u(p_3)\big]= -\sqrt{2}\frac{[q_2p_4]\langle p_3k_2\rangle}{[q_2k_2]} \to -\sqrt{2}\frac{[k_1p_4]\langle p_3k_2\rangle}{[k_1k_2]}
\end{eqnarray}  

\clearpage

\section{\textcolor{ww}{Color factors }}
\label{sec:appc}

\emph{References:} ~\textcolor{rr}{\cite{lust, lust2, anc5, vanritbergen}}.\\

We collect in this appendix group theory results and color sums. We write $\texttt{f}^{abc}$ and $\texttt{d}^{abc}$ in typewriter text because $d$ and $f$ are already in use as the Wigner $d$-functions and hypergeometric functions. 

The symmetrized trace is defined as:
\begin{eqnarray} 
STr\big(T^{a_1}\cdots T^{a_n}\big)\  \equiv \ \frac{1}{n!}\sum_{\pi} Tr\big(T^{a_{\pi_1}}\cdots T^{a_{\pi_n}}\big)
\end{eqnarray} 

Proceeding:
\begin{eqnarray} 
\label{eq:dire19}
\texttt{d}^{a_1a_2a_3}\ =\ STr(T^{a_1}T^{a_2}T^{a_3})
\end{eqnarray}
\begin{eqnarray} 
\label{eq:du12}
\texttt{d}^{000}\ =\  \frac{1}{\sqrt{8N}}\ \ ,\ \ \texttt{d}^{00A}\ =\  0\ \ ,\ \ \texttt{d}^{0AB}\ =\  \frac{1}{\sqrt{8N}}\delta^{AB}
\end{eqnarray}
\begin{eqnarray} 
\texttt{d}^{a_1a_2a_3a_4}\ =\ STr(T^{a_1}T^{a_2}T^{a_3}T^{a_4})
\end{eqnarray}
\begin{eqnarray} 
\label{eq:appd202}
\{T^{a_3},T^{a_4} \}\ =\  4\sum_a \texttt{d}^{a_3a_4a}T^{a}
\end{eqnarray}
\begin{eqnarray} 
\sum^{N^2-1}_{b,c=1}\texttt{d}^{abc}\texttt{d}^{abc}\ =\  \frac{N^2-4}{16N}
\end{eqnarray}
\begin{eqnarray} 
\sum^{N^2-1}_{a,b,c=1}\texttt{d}^{abc}\texttt{d}^{abc}\ =\  (N^2-1)\frac{N^2-4}{16N}
\end{eqnarray}
\begin{eqnarray} 
\sum^{N^2-1}_{a_1,a_2,a_3,a_4=1}\texttt{d}^{a_1a_2a_3a_4}\texttt{d}^{a_1a_2a_3a_4}\ =\  \frac{(N^2-1)(N^4-6N^2+18)}{96N^2}
\end{eqnarray}
\begin{eqnarray} 
2\sum^{N^2-1}_{b=1}\texttt{d}^{ab0}\texttt{d}^{ab0}\ =\  \frac{1}{4N}
\end{eqnarray}
\begin{eqnarray} 
\sum^{N^2-1}_{b,c=1}\texttt{d}^{bc0}\texttt{d}^{bc0}\ =\  \frac{N^2-1}{8N}
\end{eqnarray}
\begin{eqnarray} 
\texttt{d}^{000}\texttt{d}^{000}\ =\  \frac{1}{8N}
\end{eqnarray}

\begin{eqnarray} 
\label{eq:appd201}
[T^{a_3},T^{a_4}]\ =\  \sum_a \texttt{f}^{a_3a_4a}T^{a}
\end{eqnarray}
\begin{eqnarray} 
\sum^{N^2-1}_{a_1,a_2,a_3=1}\texttt{f}^{i_1a_1a_2}\texttt{f}^{i_2a_2a_3}\texttt{f}^{i_3a_3a_1}\ =\  \frac{N}{2}\texttt{f}^{i_1i_2i_3}
\end{eqnarray}
\begin{eqnarray} 
\label{eq:one137}
\texttt{f}^{ab0}\ =\ \texttt{f}^{a00}\ =\ \texttt{f}^{000}\ =\  0
\end{eqnarray}

\begin{eqnarray}
\sum_a (T_R^aT_R^a)_{ij}\ =\  C_R\delta_{ij}
\end{eqnarray}

For the adjoint representation ($C_R=C_A$) of $SU(N)$ this equation gives:
\begin{eqnarray} 
\sum^{N^2-1}_{b=1}\sum^{N^2-1}_{c=1}\texttt{f}^{a_1bc}\texttt{f}^{a_2bc}\ = \  N\delta^{a_1a_2}
\end{eqnarray}
For the fundamental representation ($C_R=C_F$) of $SU(N)$ it gives:
\begin{eqnarray}
\label{eq:dire7} 
\sum_a (T_F^aT_F^a)_{ij}\ =\ \sum^{N^2-1}_{a=1}\sum^{N}_{\beta=1}T^{a}_{i \beta}T^{a}_{ \beta j}\ =\  \frac{N^2-1}{2N}\delta_{ij}
\end{eqnarray}

\begin{eqnarray}
\label{eq:dire77} 
(T_F^0T_F^0)_{ij} \ =\ \sum^{N}_{\beta =1}T^{0}_{i \beta}T^{0}_{ \beta j}\ =\  \frac{1}{2N} \delta_{ij}
\end{eqnarray}

\begin{eqnarray}
Tr(T^{a}T^{b}T^{c})\ =\  d^{abc}+\frac{i}{4}f^{abc}
\end{eqnarray}

\begin{eqnarray}
\label{eq:appd1}
Tr(T^{a_1}T^{a_2}T^{a_3}T^{a_4})\ \ \ \ \ \ \ \ \ \ \ \ \ \ \ \ \ \ \ \ \ \ \ \ \ \ \ \ \ \ \ \ \ \ \ \ \ \ \ \ \ \ \ \ \ \ \ \ \ \ \ \ \ \ \ \ \ \ \ \ \ \ \ \ \ \ \ \ \ \ \ \ \ \  
\nonumber\\[6pt]  
=\texttt{d}^{a_1a_2a_3a_4}+\frac{i}{2}\big(\texttt{d}^{a_1a_4n}\texttt{f}^{a_2a_3n}-\texttt{d}^{a_2a_3n}\texttt{f}^{a_1a_4n}\big)+\frac{1}{12}\big(\texttt{f}^{a_1a_4n}\texttt{f}^{a_2a_3n}-\texttt{f}^{a_1a_2n}\texttt{f}^{a_3a_4n}\big)
\end{eqnarray}

\begin{eqnarray}
t^{a_1a_2a_3a_4}\ \equiv \  Tr(T^{a_1}T^{a_2}T^{a_3}T^{a_4})- Tr(T^{a_4}T^{a_3}T^{a_2}T^{a_1})=
\nonumber\\[6pt]
i(\texttt{d}^{a_1a_4j}\texttt{f}^{a_2a_3j}- \texttt{d}^{a_2a_3j}\texttt{f}^{a_1a_4j})
\end{eqnarray}

\begin{eqnarray}
t^{a_1a_2a_3a_4a_5}\ \equiv \ Tr(T^{a_1}T^{a_2}T^{a_3}T^{a_4}T^{a_5})-Tr(T^{a_5}T^{a_4}T^{a_3}T^{a_2}T^{a_1})=\ \ \ \ \ \ \ \ 
\nonumber\\[7pt]
i\texttt{f}^{a_1a_2n}\big(\texttt{d}^{a_3a_4a_5n}-\frac{1}{12}\texttt{f}^{a_3a_4m}\texttt{f}^{a_5nm}\big)+
i\texttt{f}^{a_1a_3n}\big(\texttt{d}^{a_2a_4a_5n}-\frac{1}{12}\texttt{f}^{a_2a_4m}\texttt{f}^{a_5nm}\big)+
\nonumber\\[7pt]
i\texttt{f}^{a_2a_3n}\big(\texttt{d}^{a_1a_4a_5n}-\frac{1}{12}\texttt{f}^{a_1a_5m}\texttt{f}^{a_4nm}\big)+
i\texttt{f}^{a_4a_5n}\big(\texttt{d}^{a_1a_2a_3n}-\frac{1}{12}\texttt{f}^{a_2a_3m}\texttt{f}^{a_1nm}\big)\ \ \ \ \ 
\end{eqnarray}

\begin{eqnarray}
t^{\alpha_4\beta_5a_1a_2a_3}_{\alpha_5\beta_4}\ \equiv \  (T^{a_1}T^{a_2}T^{a_3})^{\alpha_4}_{\alpha_5}\delta^{\beta_5}_{\beta_4}
\end{eqnarray}
\begin{eqnarray}
t^{\alpha_4\beta_5a_1a_2b}_{\alpha_5\beta_4}\ \equiv \ (T^{a_1}T^{a_2})^{\alpha_4}_{\alpha_5}(T^{b})^{\beta_5}_{\beta_4}
\end{eqnarray}

\begin{eqnarray}
\sum_{a=1}^{N^2-1} (T^{a})_{i_1j_1}(T^{a})_{i_2j_2}\ =\  \delta_{i_1j_2}\delta_{i_2j_1}-\frac{1}{N}\delta_{i_1j_1}\delta_{i_2j_2}
\end{eqnarray}

Orthogonality of the Chan-Paton basis to leading order in $1/N$:
\begin{eqnarray}
\label{eq:sq12}
\sum_{a_1,\dots a_n}^{N^2-1} Tr(T^{a_1}\ldots T^{a_n})\big[Tr(T^{b_1}\ldots T^{b_n})\big]^*= N^{n-2}(N^2-1)
\Big[ \delta_{\{a\}\{b\}}+\mathcal{O}(1/N^2)\Big]
\end{eqnarray}

Reduction of traces:
\begin{eqnarray}
\label{eq:redu5}
\sum_{a=1}^{N^2-1} Tr(T^aX)Tr(T^aY)=\frac{1}{2}Tr(XY)-\frac{1}{2N}Tr(X)Tr(Y)
\end{eqnarray}
\begin{eqnarray}
\label{eq:redu6}
\sum_{a=1}^{N^2-1} Tr(T^aXT^aY)=\frac{1}{2}Tr(X)Tr(Y)-\frac{1}{2N}Tr(XY)
\end{eqnarray}
Where $X$ and $Y$ are general chains of the matrices $T.$

\clearpage

\section{\textcolor{ww}{Collider phenomenology}}
\label{sec:phenapp}
\emph{References:} ~\textcolor{rr}{\cite{perelstein, eichten, anc2, harris22}}.\\

\begin{figure}
\centering
\includegraphics[width= 100mm]{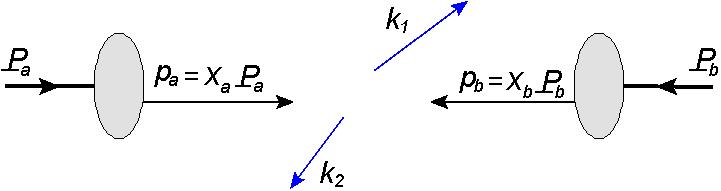}
\caption{A collision of two protons, and the partonic process.\label{proton}}	
\end{figure}

\begin{figure}
\centering
\includegraphics[width= 60mm]{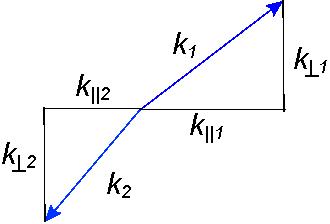}
\caption{Two final state parton kinematics.\label{dijetkin}}	
\end{figure}

Figs.~\textcolor{yy}{\ref{proton}}, \textcolor{yy}{\ref{dijetkin}} show the collision of two protons, and two partons in the final state.

The \emph{rapidities} $y_1$, $y_2$ obey:
\begin{eqnarray}
\tanh y_1 = \frac{k_{|| 1}}{E_1}\ \ \ ,\ \ \   \sinh y_1 =  \frac{k_{|| 1}}{k_{\bot 1}}\ \ \ ,\ \ \ 
\cosh y_1 =  \frac{E_1}{k_{\bot 1}} \ \ \ ,\ \ \  E_1 =  \sqrt{k_{|| 1}^2 +k_{\bot 1}^2}
\nonumber\\[9pt]
\tanh y_2 = \frac{k_{|| 2}}{E_2}\ \ \ ,\ \ \   \sinh y_2 =  \frac{k_{|| 2}}{k_{\bot 2}}\ \ \ ,\ \ \ 
\cosh y_2 =  \frac{E_2}{k_{\bot 2}} \ \ \ ,\ \ \  E_2 =  \sqrt{k_{|| 2}^2 +k_{\bot 2}^2}
\end{eqnarray}

So that
\begin{eqnarray}
y_1\ =\  \frac{1}{2}\ln \Big(\frac{E_1+k_{|| 1}}{E_1-k_{|| 1}}\Big)\ \ \ \ ,\ \ \ \ y_2\ =\  \frac{1}{2}\ln \Big(\frac{E_2+k_{|| 2}}{E_2-k_{|| 2}}\Big)
\end{eqnarray}

In the c.o.m frame the rapidities will be:
\begin{equation}
y_1\ =\  y_2\ =\  -\ln\big(\tan\frac{\theta}{2}\big)\ =\  \frac{1}{2}\ln\Big(\frac{1+|\cos\theta|}{1-|\cos\theta|}\Big)
\end{equation}
Where $\theta$ is the scattering angle as seen in the c.o.m frame.

We define
\begin{equation} 
y\equiv \frac{1}{2}(y_1-y_2)\ \ \ \ ,\ \ \ \ \ \ Y\equiv \frac{1}{2}(y_1+y_2)\ \ \ \ ,\ \ \ \ \ \ k_{\bot}\equiv k_{\bot 1}+k_{\bot 2}
\end{equation}

The variables $y_i$ and $p_{\bot i}$ are useful because of their simple boost transformations along the beamline: 
\begin{eqnarray}
y_1 \ \longrightarrow \  y_1- \tanh^{-1}(\beta)\ \ \ \ \  ,\ \ \ \ \  k_{\bot 1} \ \longrightarrow \ k_{\bot 1} 
\nonumber\\[7pt]
y_2 \ \longrightarrow \  y_2- \tanh^{-1}(\beta)\ \ \ \ \  ,\ \ \ \ \  k_{\bot 2} \ \longrightarrow \  k_{\bot 2} 
\end{eqnarray}

So that,
\begin{eqnarray}
Y \to Y- \tanh^{-1}(\beta)\ \ \ \ \ ,\ \ \ \ \  y \to y
\end{eqnarray}

We define the Mandelstam variables for the protons: $\widetilde{s}, \widetilde{t}, \widetilde{u}$.
Likewise for the partons: $s, t, u$.

We have:
\begin{eqnarray} 
\tau \equiv \frac{s}{\widetilde{s}}= x_a x_b\ \ \ ,\ \ \ 
s\ =\  4k_\bot^2\cosh^2y
\end{eqnarray}

\begin{eqnarray} 
x_a= \sqrt{\tau}e^Y= \frac{2k_\bot}{\sqrt{\widetilde{s}}}\cosh y e^Y\ \ \ \ , \ \ \ \ x_b= \sqrt{\tau}e^{-Y}= \frac{2k_\bot}{\sqrt{\widetilde{s}}}\cosh y e^{-Y}
\end{eqnarray}

The Mandelstam variables:
\begin{eqnarray}
t\ =\  -\frac{s}{2}\frac{e^{-y}}{\cosh y}\ =\  -2k^2_{\bot}\cosh ye^{-y}\ =\ -\frac{s}{2}(1-\cos\theta)
\nonumber\\[6pt]
u\ =\  -\frac{s}{2}\frac{e^{y}}{\cosh y}\ = \ -2k^2_{\bot}\cosh ye^{y}\ =\ -\frac{s}{2}(1+\cos\theta)
\end{eqnarray}

The following variable is useful for angular distributions analysis
\begin{equation} 
\chi \ \equiv \  e^{2y}\ =\   \frac{1+\cos\theta}{1-\cos\theta}
\end{equation}

Then
\begin{eqnarray} 
t= -s\frac{1}{1+\chi}\ \ \ ,\ \ \ u= -s\frac{\chi}{1+\chi}  
\nonumber\\
\cos \theta= -\frac{1-\chi}{1+\chi}\ \ \ ,\ \ \ \sin \theta= \frac{2\chi^{1/2}}{1+\chi}
\end{eqnarray}

The calculation of a cross section $d\widetilde{\sigma}$ is done by convoluting the partonic cross section $d\sigma (ij\to kl)$ with the \emph{parton distribution functions} of the two colliding protons:
\bea
\boxed{
d\widetilde{\sigma} = \int_0^1 \int_0^1 dx_a\ dx_b \sum_{ijkl}\ f_i(x_a, M)\ f_j(x_b, M)\ d\sigma (ij\to kl)   }
\eea

The partonic cross section and the squared amplitude are related:
\begin{eqnarray} 
\boxed{
|\mathcal{M}(ij \to kl)|^2\ =\  64\pi^2 s \ \frac{d\sigma}{d \Omega} \ =\ 16\pi s^2\ \frac{d\sigma}{d t}   }
\end{eqnarray}

Now we discuss the dijet final state (The formalism for e.g. $\gamma+ jet$ final state is similar.).

The \emph{mass of a dijet} is:
\begin{equation} 
M\ \equiv \  \sqrt{(E^{j_1}+E^{j_2})^2+(\vec{k}^{j_1}+\vec{k}^{j_2})^2}
\end{equation}

The dijet cross section is:
\begin{eqnarray} 
\frac{d\widetilde{\sigma}}{dM^2}\ =\  \frac{1}{(2\pi)^2}  \int  \int \frac{d^3k_1}{2E_1} \frac{d^3k_2}{2E_2}\ \sum_{ijkl}\int_0^1 \int_0^1 dx_a dx_b\ \ \ \ \times \ \ \ 
\nonumber\\
f_i(x_a, M)\ f_j(x_b, M)\ \delta^4(p_a+p_b-k_1-k_2)\ \delta((p_a+p_b)^2-M^2)\  8\pi s \ \frac{d\sigma}{dt}
\end{eqnarray}

It will be usefull, though, to write the cross section in terms of the variables $y,Y$.
We can transform this equation by using:
\bea
\frac{d^3k_1}{2E_1}\ =\  \frac{\pi}{2}\ dk^2_{1\bot}\ dy_1 \ \ \ \ \ ,\ \ \ \ \ \frac{d^3k_2}{2E_2}\ =\ \frac{\pi}{2}\ dk^2_{2\bot}\ dy_2
\eea 
\begin{eqnarray} 
s\ =\ (p_a+p_b)^2\ =\ 4k_\bot^2\cosh^2y
\end{eqnarray}
\bea
\delta^4(p_a+p_b-k_1-k_2)= \delta(E_a+E_b-E_1-E_2)\ \delta(p_{a ||}+p_{b ||}-k_{|| 1}-k_{|| 2})\delta(\vec{k}_{\bot 1}+\vec{k}_{\bot 2})
\eea

and we get:
\begin{eqnarray} 
\frac{d\widetilde{\sigma}}{dM^2}\ =\  M^2 \int \int dy_1 dy_2 \ \ \sum_{ijkl}\int  \int dx_a dx_b\ \ \ \ \  \times \ \ \ \ \ \ \ \ 
\nonumber\\
f_i(x_a, M)\ f_j(x_b, M)\ \frac{1}{4\cosh^2 y}\ \delta(E_a+E_b-E_1-E_2)\ \delta(p_{a ||}+p_{b ||}-k_{|| 1}-k_{|| 2})\ \frac{d\sigma}{dt}
\end{eqnarray}

Using:
\bea
\delta(E_a+E_b-E_1-E_2)\ \delta(p_{a ||}+p_{b ||}-k_{|| 1}-k_{|| 2})\ =\ 2\ \delta(\sqrt{s}x_a-Me^Y)\ \delta(\sqrt{s}x_b-Me^{-Y}) \nonumber\\
\eea

we get:
\bea
\frac{d\widetilde{\sigma}}{dM}\ =\  \frac{1}{2}M\tau \int dy_1 dy_2\sum_{ijkl} \  \frac{1}{\cosh^2 y}\ f_i(\sqrt{\tau} e^Y, M)\ f_j(\sqrt{\tau} e^{-Y}, M)\ \frac{d\sigma}{dt}
\eea

Using 
\begin{eqnarray}
 \int dy_1  \int  dy_2 \ =\  \int_{-Y_{max}}^0 dY \int_{-(y_{max}+Y)}^{y_{max}+Y} dy \ +\  \int_0^{Y_{max}} dY  \int_{-(y_{max}-Y)}^{y_{max}-Y} dy 
\end{eqnarray}

we finally get:
\bea
\label{eq:dijetcross}
\frac{d\widetilde{\sigma}}{dM}\ =\ 
M\tau \sum_{ijkl}\int dY\ f_i(\sqrt{\tau} e^Y, M)\ f_j(\sqrt{\tau} e^{-Y}, M)\int dy\ \frac{1}{\cosh y}\ \frac{d\sigma}{dt}+
\nonumber\\[6pt]
M\tau \sum_{ijkl}\int dY\ f_i(\sqrt{\tau} e^Y, M)\ f_j(\sqrt{\tau} e^{-Y}, M)\int dy\ \frac{1}{\cosh y}\ \frac{d\sigma}{dt}
\eea

\clearpage

\section{\textcolor{ww}{Mathematical functions}}
\label{sec:appe}

\emph{References:} ~\textcolor{rr}{\cite{gsw, weisstein2, weisstein, abramowitz, biedenharn, erdelyi}}.

We review the properties of the mathematical functions which appear in this work.

\subsection{Gamma function and Pochammer symbol}

The \emph{gamma function} is
\begin{eqnarray}
\Gamma(a)= \int_0^\infty dx\ x^{a-1}e^{-x}
\end{eqnarray}

It satisfies the following properties
\begin{eqnarray}
\Gamma(a+1)=a!\ \ ,\ \  if\ \   a=\text{integer}
\\[13pt]
\Gamma(a+1)=a\ \Gamma(a)
\\[13pt]
\Gamma(a+n)=a(a+1)\ .\ .\ .\ (a+n-1)\Gamma(a)
\\[13pt]
\Gamma(a)\Gamma(1-a)=\frac{\pi}{\sin\pi a}
\\[13pt]
\Gamma(a)\Gamma(-a)=\frac{-\pi}{a\sin\pi a}
\end{eqnarray}

There is a simple pole at each non-positive integer:
\begin{eqnarray}
\Gamma(a)\approx \frac{1}{a+n}\frac{(-1)^n}{n!}
\end{eqnarray}

\emph{Stirling's formula} is:
\begin{eqnarray}
\Gamma(a+1)\approx \sqrt{2\pi a} \left(\frac{a}{e}\right)^a\Big[1+\frac{1}{12a}+\ldots \Big]\ ,\ \text{for}\ \  a>>1.
\end{eqnarray}
Hence besides having poles, $\Gamma(a)$ diverges extremely fast for large values of $a$.

Gauss multiplication formula:
\bea
\label{eq:gaussmult}
\Gamma(na)=(2\pi)^{\frac{1}{2}(1-n)}n^{na-\frac{1}{2}}\prod_{k=0}^{n-1}\Gamma \big(a+\frac{k}{n}\big)
\eea

The \emph{Pochammer symbol} (or rising factorial) is:
\begin{eqnarray}
(a)_n\equiv a(a+1)\ .\ .\ .\ (a+n-1)=\frac{\Gamma(a+n)}{\Gamma(a)}
\end{eqnarray}

Eq.~(\zz{\ref{eq:gaussmult}}) applied to the Pochammer symbol gives:
\bea
\big(m_1a\big)_{m_1m_2}=m_1^{m_1m_2} \prod_{k=0}^{m_1-1}\Big(a+\frac{k}{m_1}\Big)_{m_2}
\eea
Where $m_1$ and $m_2$ are integers.

The Pochammer symbol satisfies:
\bea
(a+b)_n= \sum_{p=0}^n \bigg(\begin{array}{c} n \\ p \end{array} \bigg)(a)_{n-p}\ (b)_{p}
\eea
\bea
(a)_m(a)_n= \sum_{p=0}^m
\bigg(\begin{array}{c} m \\ p \end{array} \bigg)
\bigg(\begin{array}{c} n \\ p \end{array} \bigg)
\ p!\ (a)_{m+n-k}
\eea

\subsection{Beta function}

The \emph{Beta function} is:
\begin{eqnarray}
B(a,b)\equiv \int_0^1dx\ x^{a-1}(1-x)^{b-1}=\frac{\Gamma{(a)}\Gamma{(b)}}{\Gamma{(a+b)}}
\end{eqnarray}

It satisfies the following properties
\begin{eqnarray}
B(a,b)=B(b,a)
\\[13pt]
B(a+1,b)=\frac{a}{a+b}B(a,b)
\\[13pt]
B(a+1,b)+B(a,b+1)=B(a,b)
\end{eqnarray}

The singularities of $B(a,b)$ are:
\begin{eqnarray}
B(a,b)\approx \frac{1}{b+n}\frac{(-1)^n}{n!}(a-1)(a-2)\cdots(a-n)\ ,\ \text{for}\ \  b\approx -n.
\end{eqnarray}

$B(a,b)$ has the remarkable property of being the sum of its singularities:
\begin{eqnarray}
B(a,b)= \sum_{n=0}^\infty \frac{1}{b+n}\frac{(-1)^n}{n!}(a-1)(a-2)\cdots(a-n)\ ,\ \text{for}\ \  Re(a)>0.
\end{eqnarray}

\underline{Proof}:
The sum on the right hand side reproduces all of the singularities of $B(a,b)$, so that it can differ from it only by an entire function of $b$. It is seen that the sum on the RHS vanishes at $|b|\to \infty$, and the LHS does too. Thus the entire function must also vanish at this limit. The only asymptotically vanishing entire function is 0, therefore the two sides are equal.

At small values:
\begin{eqnarray}
B(a,b)\approx \frac{a+b}{ab}\ ,\ \text{for}\ \  a,b << 1.
\end{eqnarray}

Two limits for large values:
\begin{eqnarray}
\label{eq:betaf1}
B(a,b)\approx a^{-b}\ ,\ \text{for}\ \  a>>1,\ b=\text{const}.
\end{eqnarray}

and
\begin{eqnarray}
\label{eq:betaf2}
B(a,\epsilon a)\approx e^{-f(\epsilon)a}\ ,\ \text{for}\ \  a>>1.
\end{eqnarray}

Where
\begin{eqnarray}
\label{eq:betaf3}
f(\theta)=\epsilon \ln \epsilon + (1-\epsilon) \ln (1-\epsilon)
\end{eqnarray}

The first is a polynomial divergence (for $a>0,\ b<0$). The second is an exponential vanishing.

\subsection{Stirling numbers}
Abramowitz and Stegun \rr{\cite{abramowitz}} define the Stirling numbers of the first kind through the expansion of the \underline{falling} factorial. 
\bea
x(x-1)\cdots(x-1+n)= \sum_{m=0}^n S^{(m)}_n x^m
\eea
Since the rising factorial (Pochammer symbol) is related to the falling factorial through:
\bea
x(x-1)\cdots(x-1+n)\ =\ (-1)^n(-x)_n\ \ \ \ ,
\eea
the expansion for the Pochammer symbol will be:
\bea
\label{eq:pochstir1}
(x)_n\ =\  \sum_{m=0}^n (-1)^{n-m}\ S^{(m)}_n\ x^m
\eea

Another generating function for the Stirling numbers is:
\bea
\big(\ln(1+x)\big)^m\ =\  m! \sum_{m=m}^{\infty} S^{(m)}_n\ \frac{x^n}{n!}\ \ ,\ \ \text{for}\ |x|<1 \ .
\eea

Special values:
\bea
S^{(0)}_n=\delta_{0,n}\ \ \ \ \ \ \ \ \ \ \ \ \ \ \ ,\ \ \ \ \ \ \ \ \ \ \ \ \ \ \ S^{(1)}_n= (-1)^{n-1}\ (n-1)!
\eea
\bea
S^{(n-1)}_n= - \bigg(\begin{array}{c} n \\ 2 \end{array} \bigg)\ \ \ \ \ \ \ \ \ \ , \ \ \ \ \ \ \ \ \ S^{(n)}_n= 1
\eea
 
An explicit form for the Stirling numbers is: 
\bea
S^{(m)}_n\ =\ \sum_{k=0}^{n-m}\sum_{p=0}^{k}\ \frac{(-1)^p}{k!} \ 
\bigg(\begin{array}{c} n-1+k \\ n-m+k \end{array} \bigg)
\bigg(\begin{array}{c} 2n-m \\ n-m-k \end{array} \bigg)
\bigg(\begin{array}{c} k \\ p \end{array} \bigg)
\ p^{n-m+k}
\eea

The Stirling numbers satisfy the following relations:
\bea
S_{n+1}^{(m)}\ =\ S^{(m-1)}_n-nS^{(m)}_n
\eea
\bea
\bigg(\begin{array}{c} m\\ r \end{array} \bigg) 
S^{(m)}_n\ =\ \sum_{k=m-r}^{n-r}
\bigg(\begin{array}{c} n\\ k \end{array} \bigg) 
\ S_{n-k}^{(r)}\ S_{k}^{(m-r)}
\eea
\bea
\sum_{m=1}^{n}S^{(m)}_n\ =\ 0
\eea
\bea
\sum_{m=0}^{n}(-1)^{n-m}\ S^{(m)}_n\ =\  n!
\eea
\bea
\sum_{k=m}^{n}S^{(k+1)}_{n+1}n^{k-m}\ =\ S^{(m)}_n
\eea

\subsection{Hypergeometric functions}

The generalized hypergeometric function is:
\begin{eqnarray}
{_pF_q}
\bigg[\begin{array}{c}
\scriptstyle a_1, a_2 \ldots a_p \\ \scriptstyle b_1, b_2 \ldots b_q
\end{array} \bigg]= \sum_{k=0}^\infty \frac{(a_1)_k(a_2)_k\ldots (a_p)_k}{(b_1)_k(b_2)_k\ldots (b_q)_k}\ \ \frac{x^k}{k!}
\end{eqnarray}
Where $(a)_n$ is the Pochammer symbol.

The function
\begin{eqnarray}
{_{p+1}F_p}
\bigg[\begin{array}{c}
\scriptstyle a_1, a_2\ldots a_{p+1} \\
\scriptstyle b_1, b_2\ldots b_p
\end{array} ; x \bigg] 
\end{eqnarray}

obeys the differrential equation
\begin{eqnarray}
\Big[ \vartheta(\vartheta+b_1-1)\cdots (\vartheta+b_p-1)\ -\ x\ \vartheta(\vartheta+a_1-1)\cdots (\vartheta+a_{p+1}) \Big]\ y= 0
\end{eqnarray}

Where
\begin{eqnarray}
\vartheta \equiv x\frac{d}{dx}
\end{eqnarray}

In particular, the first hypergeometric function
\begin{eqnarray}
{_2F_1}
\bigg[\begin{array}{c}
\scriptstyle a, b \\
\scriptstyle c
\end{array} ; x \bigg] 
\end{eqnarray}

obeys:
\begin{eqnarray}
z(1-z)y''+\big[c-(a+b+1)z\big]y'-aby=0
\end{eqnarray}

Gauss's hypergeometric theorem is:
\begin{eqnarray}
{_2F_1}
\bigg[\begin{array}{c}
\scriptstyle a, b \\
\scriptstyle c
\end{array} ; 1 \bigg] = \frac{\Gamma(c)\Gamma(c-a-b)}{\Gamma(c-a)\Gamma(c-b)}
\end{eqnarray}

A recursion relation:
\begin{eqnarray}
{_{p+1}F_{q+1}}=
\bigg[\begin{array}{c}
\scriptstyle a_1, \ldots a_p, c \\
\scriptstyle b_1, \ldots b_q, d
\end{array} ; x \bigg] = \frac{\Gamma(d)}{\Gamma(c)\Gamma(d-c)}\int_0^1 dtt^{c-1}(1-t)^{d-c-1}{_{p}F_{q}}\bigg[\begin{array}{c}
\scriptstyle a_1, \ldots a_p \\
\scriptstyle b_1, \ldots b_q
\end{array} ; tx \bigg]
\end{eqnarray}

\subsection{Jacobi polynomials}
\label{subsec:Jacobipol8}

The Jacobi polynomials are:
\begin{eqnarray}
\label{eq:jaco1}
P_{k}^{(\alpha, \beta)}(x)\ =\  
\frac{1}{2^k}\sum_p 
\bigg(\begin{array}{c}
k+\alpha \\
p
\end{array}  \bigg)\ \ 
\bigg(\begin{array}{c}
k+\beta \\
k-p
\end{array}  \bigg)\ \ 
\big(x-1\big)^{k-p}\ \big(x+1\big)^{p} =
\nonumber\\[7pt]
\frac{(k+\alpha)!\ (k+\beta)!}{2^k}\ \sum_p \frac{1}{p!\ (k+\alpha-p)!\ (\beta+p)!\ (k-p)!}\ \ \ \big(x-1\big)^{k-p}\ \big(x+1\big)^{p} \nonumber\\
\end{eqnarray}

They are solutions of the following differential equation:
\bea
(1-x^2)y''+\big[\beta-\alpha-(\alpha+\beta+2)x  \big]y'+ n(n+\alpha+\beta+1)y\ =\ 0
\eea 

These polynomials are orthogonal in the following sense:
\begin{eqnarray}
\label{eq:orthjaco}
\int_{-1}^{1}dx\ P_{p}^{(\alpha, \beta)}(x)P_{k}^{(\alpha, \beta)}(x)(1-x)^\alpha(1+x)^\beta=\  \frac{2^{\alpha+\beta+1}}{2k+\alpha+\beta+1}\frac{(k+\alpha)!\ (k+\beta)!}{k!\ (k+\alpha+\beta)!}\ \delta_{kp} \nonumber\\
\end{eqnarray}

They satisfy the recursion relations:
\begin{eqnarray}
(2k+1)(k+\alpha+\beta+1)(2k+\alpha+\beta)\ P_{k+1}^{(\alpha, \beta)}(x)= 
\nonumber\\[7pt]
\Big[(2k+\alpha+\beta+1)(\alpha^2-\beta^2)+(2k+\alpha+\beta)(2k+\alpha+\beta+1)(2k+\alpha+\beta+2) x\Big] P_{k}^{(\alpha, \beta)}(x)
\nonumber\\[7pt]
-2(k+\alpha)(k+\beta)(2k+\alpha+\beta+2)\ P_{k-1}^{(\alpha, \beta)}(x)
\nonumber\\ 
\end{eqnarray}

Special values:
\begin{eqnarray}
P_{k}^{(\alpha, \beta)}(x=1)= \frac{(k+\alpha)!}{k!\ \alpha!}
\end{eqnarray}
\begin{eqnarray}
\label{eq:reflec}
P_{k}^{(\alpha, \beta)}(-x)= (-1)^k\ P_{k}^{(\beta, \alpha)}(x)
\end{eqnarray}
\begin{eqnarray}
P_{k}^{(0, 0)}(x)=  P_{k}(x)
\end{eqnarray}
Where  $P_{k}(x)$ are the Legendre polynomials.

The Jacobi polynomials will be used to express the Wigner $d$ functions, and we will have $x= \cos\theta$.
The expansion Eq.~(\textcolor{zz}{\ref{eq:jaco1}}) contains powers of $\frac{1+x}{2}= -\frac{\hat{u}}{\hat{s}}$ and of $\frac{1-x}{2}= -\frac{\hat{t}}{\hat{s}}$. We would prefer having only powers $\hat{u}$ though. Reference \rr{\cite{abramowitz}} gives the following expansion in terms of $\hat{t}$:

\begin{eqnarray}
P_{k}^{(\alpha, \beta)}(x)\ =\  
\frac{\Gamma(\alpha+k+1)}{k!\ \Gamma(\alpha+\beta +k+1)}\sum_{p=0}^k 
\bigg(\begin{array}{c}
k\\
p
\end{array}  \bigg)\ \  
\frac{\Gamma(\alpha+\beta+k+p+1)}{\Gamma(\alpha+p+1)}\ \Big(\frac{x-1}{2}\Big)^{p}
\end{eqnarray}

We transform this to an expression with $\hat{u}$ by using Eq.~(\zz{\ref{eq:reflec}}):
\begin{eqnarray} 
P_{k}^{(\alpha, \beta)} (x)= (-1)^k P_{k}^{(\beta, \alpha)} (-x)=
\nonumber\\[9pt] 
(-1)^k\frac{\Gamma(\beta+k+1)}{k!\ \Gamma(\alpha+\beta +k+1)}\sum_{p=0}^k 
\bigg(\begin{array}{c}
k\\
p
\end{array}  \bigg)\ \  
\frac{\Gamma(\alpha+\beta+k+p+1)}{\ \Gamma(\beta+p+1)}\  \bigg(\frac{-(x+1)}{2}\bigg)^{p}=
\nonumber\\[9pt] 
\frac{(-1)^k\ \Gamma(\beta+k+1)}{k!\ \Gamma(\alpha+\beta +k+1)}\sum_{p=0}^k 
\bigg(\begin{array}{c}
k\\
p
\end{array}  \bigg)\ \  
\frac{\Gamma(\alpha+\beta+k+p+1)}{\ \Gamma(\beta+p+1)}\  \bigg(\frac{\hat{u}}{\hat{s}}\bigg)^{p}
\end{eqnarray}  

Which can be written as:
\begin{eqnarray}
\label{eq:kunine}
\boxed{
P_{k}^{(\alpha, \beta)} (\hat{u})= \ \frac{(-1)^k\ (\beta+k)!}{k!\ (\alpha+\beta +k)!}\sum_{p=0}^k 
\bigg(\begin{array}{c}
k\\
p
\end{array}  \bigg)\ \  
\frac{(\alpha+\beta+k+p)!}{\ (\beta+p)!\ \hat{s}^p}\ \ \hat{u}^{p}}
\end{eqnarray}  

\clearpage

\section{\textcolor{ww}{Wigner $d$-functions}}
\label{sec:appd}

In order to deal with collider phenomenology of particles with spin, it is necessary to get acquainted with the \emph{Wigner $d$-functions}. These can be viewed as generalizations of the Legendre functions, see Eq.~(\zz{\ref{eq:ylm7}}).

\subsection{Review}

\emph{References:} ~\textcolor{rr}{\cite{rose, erdelyi, abramowitz, biedenharn, hara, weisstein}}.

We review here properties of the \emph{Wigner $d$-functions}.\\

The rotation matrix is:
\begin{equation} 
R\ =\  e^{-i\alpha J_z}e^{-i\beta J_y}e^{-i\gamma J_z}
\end{equation}
Where $\alpha, \beta$, $\gamma$ are the \emph{Euler angles}.

Acting with $R$ on an angular momentum state $|Jm \rangle$, gives a linear combination of states with different $m$: 
\begin{eqnarray}
\label{eq:zb3} 
R|Jm\rangle \ =\ \sum_{m'} D^J_{m',m}(\alpha \beta \gamma)|Jm' \rangle
\end{eqnarray}

For integral $J=l$ this takes the form
\begin{eqnarray}
Y_{lm}(\theta, \phi)\ =\ \sum_{m'} D^{l}_{m',m}(\alpha\beta\gamma)Y_{lm'}(\theta', \phi')
\end{eqnarray}

We isolate $D^J_{m',m}$ by operating with $\langle Jm'|$ on Eq. (\textcolor{zz}{\ref{eq:zb3}}), 
\begin{eqnarray} 
D^J_{m',m}(\alpha \beta \gamma)\ =\  \langle Jm'|e^{-i\alpha J_z}e^{-i\beta J_y}e^{-i\gamma J_z}| Jm \rangle = 
\nonumber\\[10pt]
e^{-im' \alpha}\langle Jm'|e^{-i \beta J_y}| Jm \rangle e^{-im\gamma}\ \equiv \ e^{-im'  \alpha } d^J_{m',m}(\beta)e^{-im \alpha}
\end{eqnarray}

In the last line we defined:
\begin{eqnarray}
d^J_{m',m}(\beta)\ \equiv \ \langle Jm'|e^{-i \beta J_y}| Jm \rangle
\end{eqnarray}

From now we will use $\theta$ instead of $\beta$ as the argument of $d$.
The $d$'s satisfy:
\begin{eqnarray} 
d^J_{m',m}(\theta)\ =\  (-1)^{m-m'}d^J_{m,m'}(\theta)\ =\ d^J_{-m,-m'}(\theta)\ =\ d^J_{m,m'}(-\theta)
\end{eqnarray}
\begin{eqnarray} 
\label{eq:denden}
d^J_{m',m}(\pi- \theta)\ =\  (-1)^{J+m'} d^J_{m', -m}(\theta)\ 
\end{eqnarray}

\begin{eqnarray} 
\label{eq:ylm7}
d^l_{m,0}(\theta)\ =\  \sqrt{\frac{4\pi}{2l+1}}Y^l_m(\theta, 0)
\end{eqnarray}

A special case:
\begin{eqnarray}
d^J_{JJ}(\theta)\ =\  \Big(\cos \frac{\theta}{2} \Big)^{2J}
\end{eqnarray}

Orthogonality:
\begin{eqnarray}
\label{eq:orthds}
\int d \cos\theta \ d^J_{m',m}(\theta)d^{J'}_{m',m}(\theta)\ =\ \frac{2}{2J+1}\delta^{JJ'}
\end{eqnarray}
This is equivalent to the orthogonality of the Jacobi polynomials Eq.~(\zz{\ref{eq:orthjaco}}). 

\begin{eqnarray}
\sum_\lambda  d^J_{m',\lambda}(\theta_1)\ d^{J'}_{\lambda,m}(\theta _2)\ =\ d^J_{m',m}(\theta _1+\theta _2)
\end{eqnarray}

The $d's$ satisfy the following recursion relations:
\begin{eqnarray}
2(m\cos\theta-m')\ d^J_{m',m}\ =\ \ \ \ \ \ \ \ \ \ \ \ \ \ \ \ \ \ \ \ \ \ \ \ \ \ \ \ \ \ \ \ \ \ \ \ \ \ \ \ \ \ \ \ \ \ \ \ \ \ \ \ \ \ \ \ \ \ \ \ \ \ 
\nonumber\\[7pt]
\sqrt{(J-m)(J+m+1)}\ \sin \theta \ d^J_{m',m+1}\ +\ \sqrt{(J-m')(J+m'+1)}\ \sin\theta \ d^J_{m',m-1}
\end{eqnarray}
\begin{eqnarray}
(m-m')\cot\frac{\theta}{2}\  d^J_{m',m}\ =\ \ \ \ \ \ \ \ \ \ \ \ \ \ \ \ \ \ \ \ \ \ \ \ \ \ \ \ \ \ \ \ \ \ \ \ \ \ \ \ \ \ \ \ \ \ \ \ \ \ \ \ \ \ \ \ \  
\nonumber\\
\sqrt{(J+m)(J-m+1)}\ d^J_{m',m-1}\ +\ \sqrt{(J+m')(J-m'+1)}\ d^J_{m'-1,m}
\end{eqnarray}
\begin{eqnarray}
\sqrt{J-m'+1}\  d^J_{m',m}\ =\ \ \ \ \ \ \ \ \ \ \ \ \ \ \ \ \ \ \ \ \ \ \ \ \ \ \ \ \ \ \ \ \ \ \ \ \ \ \ \ \ \ \ \ \ \ \ \ \ \ \ \ \ \ \ \ 
\nonumber\\[7pt]
\sqrt{J-m+1}\ \cos\frac{\theta}{2}\ d^{J+1/2}_{m'-1/2,m-1/2}\ +\ \sqrt{J+m+1}\ \sin\frac{\theta}{2}\ d^{J+1/2}_{m'-1/2,m+1/2}
\nonumber\\
\end{eqnarray}
\begin{eqnarray}
\sqrt{J+m'+1}\  d^J_{m',m}\ =\ \ \ \ \ \ \ \ \ \ \ \ \ \ \ \ \ \ \ \ \ \ \ \ \ \ \ \ \ \ \ \ \ \ \ \ \ \ \ \ \ \ \ \ \ \ \ \ \ \ \ \ \ \ \ \ \ 
\nonumber\\[7pt]
-\sqrt{J-m+1}\ \sin\frac{\theta}{2}\ d^{J+1/2}_{m'+1/2,m-1/2}\ +\ \sqrt{J+m+1}\ \cos\frac{\theta}{2}\ d^{J+1/2}_{m'+1/2,m+1/2}
\end{eqnarray}
\begin{eqnarray}
\sqrt{J+m'}\  d^J_{m',m}\ =\ \ \ \ \ \ \ \ \ \ \ \ \ \ \ \ \ \ \ \ \ \ \ \ \ \ \ \ \ \ \ \ \ \ \ \ \ \ \ \ \ \ \ \ \ \ \ \ \ \ \ \ \ \ 
\nonumber\\[7pt]
\sqrt{J+m}\ \cos\frac{\theta}{2}\ d^{J-1/2}_{m'-1/2,m-1/2}\ -\ \sqrt{J-m}\ \sin\frac{\theta}{2}\ d^{J-1/2}_{m'-1/2,m+1/2}
\end{eqnarray}
\begin{eqnarray}
\sqrt{J-m'}\  d^J_{m',m}\ =\ \ \ \ \ \ \ \ \ \ \ \ \ \ \ \ \ \ \ \ \ \ \ \ \ \ \ \ \ \ \ \ \ \ \ \ \ \ \ \ \ \ \ \ \ \ \ \ \ \ \ \ \ 
\nonumber\\[7pt]
\sqrt{J+m}\ \sin\frac{\theta}{2}\ d^{J-1/2}_{m'+1/2,m-1/2}\ +\ \sqrt{J-m}\ \cos\frac{\theta}{2}\ d^{J-1/2}_{m'+1/2,m+1/2}
\end{eqnarray}

The following explicit form is useful for calculating the $d$'s. We used this equation to calculate the Tables in section~\ref{subsec:tblds}.
\begin{eqnarray}
\label{eq:zz1}
\boxed{d^{J}_{m',m}(\theta)\ =\ 
A \sum_{p}B_p \big[\cos(\theta/2)\big]^{2J+m-m'-2p}\ \big[\sin(\theta/2)\big]^{m'-m+2p}}
\end{eqnarray}

Where,
\begin{eqnarray}
A\ =\  \sqrt{(J+m')!\ (J-m')!\ (J+m)!\ (J-m)!}\ \ \ \ \ \ 
\nonumber\\[8pt]
B_p\ =\  \frac{(-1)^{m'-m+p}}{(J+m-p)!\ p!\ (m'-m+p)!\ (J-m'-p)!}
\end{eqnarray}

Or alternatively in terms of $\hat{u}$ and $\hat{t}$:
\begin{eqnarray}
\label{eq:utjaco}
d^{J}_{m,m'}(u,t)\ =\ 
\frac{(-1)^{J}}{\hat{s}^J}A \sum_{p}B_p\ \hat{u}^{J+m/2-m'/2-p}\ \hat{t}^{m'/2-m/2+p}
\end{eqnarray}

\subsection{$d$-functions and Jacobi polynomials}
\label{subsec:jacobi}

In this section we express the $d$-functions in terms of \emph{Jacobi polynomials}. 
 
A given $d$-function is some factor times a Jacobi polynomial:
\begin{eqnarray}
\label{eq:zb4}\boxed{
d^{J}_{m', m}(\theta)\ =\ 
\Omega^{(J-m)}\ \big[\cos(\theta/2)\big]^{m+m'}\ \big[\sin(\theta/2)\big]^{m-m'}\ P_{J-m}^{(m-m',\ m+'m)}(\cos\theta)} \nonumber\\
\end{eqnarray}
Where
\bea
\Omega^{(J-m)}\ \equiv \ \sqrt{\frac{(J+m)!\ (J-m)!}{(J+m')!\ (J-m')!}}
\eea

Compare with Eq.~(\textcolor{zz}{\ref{eq:zz1}}). We can understand the factor in front of the Jacobi polynomial from Eqs.~(\zz{\ref{eq:orthjaco}}) and (\zz{\ref{eq:orthds}}): it is the "weight function" under which the polynomials are orthogonal. The normalization $\frac{1}{2J+1}$ is the same as for the Legendre polynomials which are the special cases $P_n^{(0,0)}$.

Writing this in terms of $\hat{u}$ and $\hat{t}$ instead of $\theta$ gives:
\begin{eqnarray}
\label{eq:dean11}
\boxed{
d^{J}_{m', m}(\hat{u}, \hat{t})\ =\ 
\Omega^{(J-m)}\ \frac{(-1)^m}{\hat{s}^m}\ \hat{u}^{\frac{1}{2}(m+m')}\ \hat{t}^{\frac{1}{2}(m-m')}\ P_{J-m}^{(m-m',\ m+m')}(\hat{u})} \nonumber\\
\end{eqnarray}

Eq.~(\textcolor{zz}{\ref{eq:kunine}}) gives:
\begin{eqnarray}
\label{eq:gaco9}
\boxed{
P_{J-m}^{(m-m',\ m+m')}(\hat{u}) \ = \ \frac{(-1)^{J-m}\ (J+m')!}{(J-m)!\ (J+m)!}\ \ \sum_{p=0}^{J-m} 
\bigg(\begin{array}{c}
J-m\\
p
\end{array}  \bigg)\ \  
\frac{(J+m+p)!}{(m+m'+p)!\ \hat{s}^p}\ \ \hat{u}^{p}}
\end{eqnarray}  
We write this equation in shorthand notation:
\begin{eqnarray}
\label{eq:dean12}
P_{J-m}^{(m-m',\ m+m')}(\hat{u})\ = \ \sum_{p=0}^{J-m} \ \Delta_{m,m'}^{(p)}\ \hat{u}^{p}
\end{eqnarray}  

We will use the coefficient of the highest power in the sum:
\bea
\label{eq:dean13}
\Delta_{m,m'}^{(J-m)}\ =\  \frac{(-1)^{J-m}\ (2J)!}{(J-m)!\ (J+m)!\ \hat{s}^{J-m}}
\eea
Notice that this coefficient is independent of $m'$.
 
Thus the 5 combinations for $m$ and $m'$ can be written as:
\begin{eqnarray}
\label{eq:d0000}
d^{J}_{0,0}(\theta)\ =\ 
P_{J}^{(0,\ 0)}
\end{eqnarray}

\begin{eqnarray}
\label{eq:overtime1}
d^{J}_{2, -2}(\theta)\ =\ 
\Big(\frac{1-\cos\theta}{2}\Big)^2\ P_{J-2}^{(4,\ 0)}\ =\  \frac{t^2}{s^2}\ P_{J-2}^{(4,\ 0)}
\end{eqnarray}

\begin{eqnarray}
d^{J}_{2, -1}(\theta)\ =\ 
\sqrt{\frac{J+2}{J-1}}\ \frac{\sin\theta(1-\cos\theta)}{4}\ P_{J-2}^{(3,\ 1)}\ =\  \sqrt{\frac{J+2}{J-1}}\ \frac{t^{3/2}u^{1/2}}{s^2}P_{J-2}^{(3,\ 1)}
\end{eqnarray}

\begin{eqnarray}
d^{J}_{1/2, 1/2}(\theta)\ =\ 
\cos\Big(\frac{\theta}{2}\Big)\ P_{J-1/2}^{(0,\ 1)}\ =\ \frac{u^{1/2}}{s^{1/2}}\ P_{J-1/2}^{(0,\ 1)}
\end{eqnarray} 
 
\begin{eqnarray}
\label{eq:d3232}
d^{J}_{3/2, -3/2}(\theta)\ =\ 
\sin\Big(\frac{\theta}{2}\Big)\Big(\frac{1-\cos\theta}{2}\Big)\ P_{J-3/2}^{(3,\ 0)}\ =\ \frac{t^{3/2}}{s^{1/2}}\ P_{J-3/2}^{(3,\ 0)}
\end{eqnarray} 

From Eq.~(\zz{\ref{eq:gaco9}}), the power expansions of the 5 Jacobi polynomials are:
\begin{eqnarray} 
\label{eq:p0000}
P^{(0,0)}_{J}(\hat{u})\ =\  (-1)^{J}\sum_{p=0}^J  \frac{(J+p)!}{(J-p)!\ p!^2\ \hat{s}^p}\ \  \hat{u}^{p}
\end{eqnarray}  

\begin{eqnarray} 
\label{eq:overtime2}
P^{(4,0)}_{J-2}(\hat{u})\ =\  
\frac{(-1)^J(J-2)!}{(J+2)!}\sum_{p=0}^{J-2}  
\frac{(J+p+2)!}{p!^2\ (J-p-2)!\ \hat{s}^p}\  \hat{u}^{p}
\end{eqnarray} 

\begin{eqnarray} 
P^{(3,1)}_{J-2}(\hat{u})\ =\  
\frac{(-1)^J}{J(J+1)(J+2)}\sum_{p=0}^{J-2}   
\frac{(J+p+2)!}{(p+1)!\ p!\ (J-p-2)! \hat{s}^p}\  \hat{u}^{p}
\end{eqnarray} 

\begin{eqnarray} 
P^{(0,1)}_{J-1/2}(\hat{u})\ =\  
(-1)^{J-1/2}\sum_{p=0}^{J-1/2} 
\frac{(J+p+1/2)!}{p!\ (p+1)!(J-p-1/2)!\ \hat{s}^p}\  \hat{u}^{p}
\end{eqnarray} 

\begin{eqnarray} 
\label{eq:p3232}
P^{(3,0)}_{J-3/2}(\hat{u})\ =\  
\frac{(-1)^{J-3/2}}{(J-1/2)(J+1/2)(J+3/2)}\sum_{p=0}^{J-3/2}  
\frac{(J+p+3/2)!}{p!^2\ (J-p-3/2)!\ \hat{s}^p}\  \hat{u}^{p}
\end{eqnarray}

\subsection{Tables}
\label{subsec:tblds}

From Eq. (\textcolor{zz}{\ref{eq:zz1}}) we calculated some $d's$. For each combination of $m$ and $m'$, we calculated up to the fifth $J$. Since there are 5 combinations of $m$ and $m'$, there are 5 tables each containing 5 $d$'s.  We used \emph{MATHEMATICA}.

\begin{table}[h]
\centering
\begin{tabular}{|c|}
\hline 
$d^0_{0,0}\ =\ 1$    \\[7pt]  
$d^1_{0,0}\ =\ \cos \theta$     \\[7pt]  
$d^2_{0,0}\ =\ \frac{3}{2}\cos^2 \theta-\frac{1}{2}$   \\[7pt]
$d^3_{0,0}\ =\ \frac{5}{2}\cos^3 \theta-\frac{3}{2}\cos \theta$     \\[7pt]
$d^4_{0,0}\ =\ \frac{1}{8}(35\cos^4 \theta-30\cos^2\theta+3)$     \\[7pt] 
\hline   
\end{tabular} 
\caption{$d^J_{0,0}(\theta)$} 
\end{table}

\begin{table}[h]
\centering
\begin{tabular}{|c|}
\hline 
$d^2_{2,-2}\ =\ (\frac{1-\cos \theta}{2})^2$   \\[9pt] 
$d^3_{2,-2}\ =\ (\frac{1-\cos \theta}{2})^2\ \big[3\cos \theta+2\big]$     \\[9pt]  
$d^4_{2,-2}\ =\ (\frac{1+\cos \theta}{2})^2\ \big[7\cos^2\theta+7\cos\theta+1\big]$   \\[9pt]
$d^5_{2,-2}\ =\ (\frac{1+\cos \theta}{2})^2\ \big[15\cos^3\theta+18\cos^2\theta+3\cos\theta-1\big]$ \\[9pt]
$d^6_{2,-2}\ =\  (\frac{1+\cos \theta}{2})^2\ \big[\frac{495}{16}\cos^4\theta+\frac{165}{4}\cos^3\theta+ \frac{45}{8}\cos^2\theta-\frac{27}{4}\cos\theta-\frac{17}{16}\big]$   \\[9pt] 
\hline     
\end{tabular}
\caption{$d^J_{2,-2}(\theta)$}  
\end{table}

\begin{table}[h]
\centering
\begin{tabular}{|c|}
\hline 
$d^2_{2,-1}\ =\ 2(\frac{1-\cos \theta}{4})\ \sin\theta$     \\[9pt] 
$d^3_{2,-1}\ =\ \sqrt{\frac{5}{2}}(\frac{1-\cos \theta}{4})\sin\theta\ \big[3\cos \theta+1\big]$      \\[9pt]  
$d^4_{2,-1}\ =\ \frac{1}{\sqrt{2}}(\frac{1-\cos \theta}{4})\sin\theta\ \big[14\cos^2\theta+7\cos\theta-1\big]$    \\[9pt]
$d^5_{2,-1}\ =\ \frac{\sqrt{7}}{2}(\frac{1-\cos \theta}{4})\sin\theta\ \big[15\cos^3\theta+9\cos^2\theta-3\cos\theta-1\big]$  \\[9pt]
$d^6_{2,-1}=\sqrt{\frac{5}{2}}(\frac{1-\cos \theta}{4})\sin\theta\ \big[\frac{99}{4}\cos^4\theta+\frac{33}{2}\cos^3\theta- 9\cos^2\theta-\frac{9}{2}\cos\theta+\frac{1}{4}\big]$    \\[9pt]
\hline        
\end{tabular}
\caption{$d^J_{2,-1}(\theta)$}  
\end{table}

\begin{table}[h]
\centering
\begin{tabular}{|c|}
\hline 
$d^{1/2}_{1/2,1/2}\ =\ \cos\theta/2$     \\[9pt] 
$d^{3/2}_{1/2,1/2}\ =\ \frac{1}{2}\cos\theta/2\ \big[3\cos\theta-1\big]$       \\[9pt]  
$d^{5/2}_{1/2,1/2}\ =\ \frac{1}{2}\cos\theta/2\ \big[5\cos^2\theta-2\cos\theta-1\big]$   \\[9pt]
$d^{7/2}_{1/2,1/2}\ =\ \frac{1}{8}\cos\theta/2\ \big[35\cos^3\theta-15\cos^2\theta-15\cos\theta+3\big]$   \\[9pt] 
$d^{9/2}_{1/2,1/2}\ =\ \cos\theta/2\ \big[\frac{63}{8}\cos^4\theta-\frac{7}{2}\cos^3\theta-\frac{21}{4}\cos^2\theta+\frac{3}{2}\cos\theta+\frac{3}{8}\big]$    \\[9pt]
\hline         
\end{tabular} 
\caption{$d^J_{1/2,1/2}(\theta)$} 
\end{table}

\begin{table}[h]
\centering
\begin{tabular}{|c|}
\hline
$d^{3/2}_{3/2,-3/2}\ =\ (\frac{1-\cos \theta}{2})\sin\theta/2$     \\[9pt]  
$d^{5/2}_{3/2,-3/2}\ =\ \frac{1}{2}(\frac{1-\cos \theta}{2})\sin\theta/2\ \big[5\cos\theta+3\big]$       \\[9pt]  
$d^{7/2}_{3/2,-3/2}\ =\ \frac{1}{4}(\frac{1-\cos \theta}{2})\sin\theta/2\ \big[21\cos^2\theta+18\cos\theta+1\big]$    \\[9pt]
$d^{9/2}_{3/2,-3/2}\ =\ \frac{1}{2}(\frac{1-\cos \theta}{2})\sin\theta/2\ \big[ 21\cos^3\theta+ 21\cos^2-2\big]$   \\[9pt] 
$d^{11/2}_{3/2,-3/2}=(\frac{1-\cos \theta}{2})\sin\theta/2\ \big[\frac{165}{8}\cos^4\theta+ \frac{45}{2}\cos^3\theta- \frac{9}{4}\cos^2\theta-\frac{11}{2}\cos\theta-\frac{3}{8}\big]$    \\[9pt]
\hline        
\end{tabular}  
\caption{$d^J_{3/2,-3/2}(\theta)$}
\end{table}
 
\clearpage

\section{\textcolor{ww}{Tables of the coefficients $C_{m,m'}^{n,J}$}}
\label{sec:appf}

The following tables result from the calculation of the coefficients $C_{m,m'}^{nJ}$, by approach 1 of section~\ref{subsubsec:approach1}.

\begin{table}[h]
\centering
\begin{tabular}{|c|c|c|}
\hline 
$n=1$ & $n=2$ & $n=3$ \\  
\whline 
   $1$ &  $\hat{u}+1$ &      $(\hat{u}+1)(\hat{u}+2)$   \\[2pt]  
\hline 
     $1$ &   $-\cos \theta$ &  $\frac{1}{4}(9\cos^2 \theta-1)$  \\[4pt]
\hline
$\textcolor{brick}{d^0_{0,0}}$ & $\textcolor{brick}{-d^1_{0,0}}$ & $\textcolor{brick}{\frac{1}{2}[3d^2_{0,0}+d^0_{0,0}]}$  \\[2pt] 
\hline         
\end{tabular} 
\caption{$C_{0,0}^{n,J}$} 
\end{table}

\begin{table}[h]
\centering
\begin{tabular}{|c|c|}
\hline
 $n=4$   & $n=5$\\  
\whline
        $(\hat{u}+1)(\hat{u}+2)(\hat{u}+3)$ &  $(\hat{u}+1)(\hat{u}+2)(\hat{u}+3)(\hat{u}+4)$  \\[2pt]  
\hline
       $-8\cos^3 \theta+ 2\cos \theta$ &  $\frac{1}{16}(625\cos^4 \theta-250\cos^2 \theta+9)$ \\[4pt]
\hline
 $\textcolor{brick}{-\frac{2}{5}[8d^3_{0,0}+7d^1_{0,0}]}$ & $\textcolor{brick}{\frac{125}{14}d^4_{0,0}+\frac{250}{21}d^2_{0,0}+\frac{19}{6}d^0_{0,0}}$  \\[2pt] 
\hline         
\end{tabular}
\caption{$C_{0,0}^{n,J}$}  
\end{table}

\begin{table}[h]
\centering
\begin{tabular}{|c||c|c|c|c|c|c|c|c|}
\hline
$ $ & $J=0$ & $1$ & $2$ & $3$ & $4$ \\[2pt]  
\whline
$n=1$ & $1 $ & $$ & $$ & $$ & $$ \\[2pt]
\hline
$ 2$ & $0 $ & $-1$ & $$ & $$ & $$ \\[2pt]
\hline
$ 3$ & $\frac{1}{2} $ & $0$ & $\frac{3}{2}$ & $$ & $$ \\[2pt]
\hline
$4 $ & $0 $ & $\frac{-14}{5}$ & $0$ & $\frac{-16}{5}$ & $$ \\[2pt]
\hline
$ 5$ & $\frac{19}{6} $ & $0$ & $\frac{250}{21}$ & $0$ & $\frac{125}{14}$ \\[2pt]
\hline
\end{tabular}  
\caption{collection of $C_{0,0}^{n,J}$'s \label{c1n} }
\end{table}

\begin{table}[h]
\centering
\begin{tabular}{|c|c|c|c|c|c|}
\hline
$n=1$  & $n=2$  & $n=3$  \\  
\whline
  $\frac{\hat{t}^2}{\hat{s}^2}$ &   $\frac{\hat{t}^2}{\hat{s}^2}\ (\hat{u}+1)$ &   $\frac{\hat{t}^2}{\hat{s}^2}\ (\hat{u}+1)(\hat{u}+2)$  \\[5pt]  
\hline
   $(\frac{1-\cos\theta}{2})^2$ &  $(\frac{1-\cos\theta}{2})^2\ \big[-\cos \theta\big]$ &  $(\frac{1-\cos\theta}{2})^2\ \big[\frac{1}{4}(9\cos^2 \theta-1)\big]$  \\[4pt] 
\hline
 $\textcolor{brick}{d^2_{2,-2}}$ & $\textcolor{brick}{\frac{1}{3}\big[-d^3_{2,-2}+2d^2_{2,-2}\big]}$ & $\textcolor{brick}{\frac{1}{4}\big[\frac{9}{7}d^4_{2,-2}-3d^3_{2,-2}+\frac{26}{7}d^2_{2,-2}\big]}$  \\[5pt] 
\hline           
\end{tabular}  
\caption{$C_{2,-2}^{n,J}$}
\end{table}

\begin{table}[h]
\centering
\begin{tabular}{|c|c|c|c|c|c|}
\hline
 $n=4$   \\  
\whline
          $\frac{\hat{t}^2}{\hat{s}^2}\ (\hat{u}+1)(\hat{u}+2)(\hat{u}+3)$   \\[5pt]  
\hline
       $(\frac{1-\cos\theta}{2})^2\ \big[-8\cos^3 \theta+2\cos \theta \big]$ \\[4pt]
\hline
 $\textcolor{brick}{2\big[-\frac{4}{15}d^5_{2,-2}+\frac{24}{35}d^4_{2,-2}-d^3_{2,-2}+\frac{22}{21}d^2_{2,-2}\big]}$   \\[5pt] 
\hline           
\end{tabular}  
\caption{$C_{2,-2}^{n,J}$}
\end{table}

\begin{table}[h]
\centering
\begin{tabular}{|c|c|c|c|c|c|}
\hline
 $n=5$ \\
 \whline 
 $\frac{\hat{t}^2}{\hat{s}^2}\ (\hat{u}+1)(\hat{u}+2)(\hat{u}+3)(\hat{u}+4)$  \\[5pt] 
 \hline
 $(\frac{1-\cos\theta}{2})^2\ \big[\frac{1}{16}(625\cos^4 \theta-250\cos^2 \theta+9)\big]$ \\[4pt] 
 \hline
$\textcolor{brick}{\frac{125}{99}d^6_{2,-2}-\frac{1375}{396}d^5_{2,-2}+\frac{125}{22}d^4_{2,-2}-\frac{125}{18}d^3_{2,-2}+\frac{239}{36}d^2_{2,-2}}$  \\[5pt] 
\hline           
\end{tabular}  
\caption{$C_{2,-2}^{n,J}$}
\end{table}

\begin{table}[h]
\centering
\begin{tabular}{|c||c|c|c|c|c|c|c|c|}
\hline
$ $ & $J=2$ & $3$ & $4$ & $5$ & $6$ \\[2pt]  
\whline
$n=1$ & $1$ & $$ & $$ & $$ & $$ \\[2pt]
\hline
$ 2$ & $\frac{2}{3} $ & $\frac{-1}{3}$ & $$ & $$ & $$ \\[2pt]
\hline
$ 3$ & $\frac{13}{14} $ & $\frac{-3}{4}$ & $\frac{9}{28}$ & $$ & $$ \\[2pt]
\hline
$4 $ & $\frac{44}{21}$ & $-2$ & $\frac{48}{35}$ & $\frac{-8}{15}$ & $$ \\[2pt]
\hline
$ 5$ & $\frac{239}{36} $ & $\frac{-125}{18}$ & $\frac{125}{22}$ & $\frac{-1375}{396}$ & $\frac{125}{99}$ \\[2pt]
\hline
\end{tabular}  
\caption{Collection of $C_{2,-2}^{n,J}$'s \label{c2n} }
\end{table}

\begin{table}[h]
\centering
\begin{tabular}{|c|c|c|c|c|c|}
\hline
$n=1$  & $n=2$ & $n=3$  \\  
\whline
  $\frac{\hat{t}^{\frac{3}{2}}\hat{u}^{\frac{1}{2}}}{s^2}$ &   $\frac{\hat{t}^{\frac{3}{2}}\hat{u}^{\frac{1}{2}}}{s^2}\ (\hat{u}+1)$ &      $\frac{\hat{t}^{\frac{3}{2}}\hat{u}^{\frac{1}{2}}}{s^2}\ (\hat{u}+1)(\hat{u}+2)$   \\[5pt]  
\hline
$(\frac{1-\cos\theta}{2})\frac{\sin\theta}{2}$  & $(\frac{1-\cos\theta}{2})\frac{\sin\theta}{2}\ [-\cos\theta]$ & $(\frac{1-\cos\theta}{2})\frac{\sin\theta}{2}\ \big[\frac{1}{4}(9\cos^2 \theta-1)\big]$   \\[4pt]
\hline
$\textcolor{brick}{\frac{1}{2}d^2_{2,-1}}$ & $\textcolor{brick}{-\frac{1}{6}\big[\sqrt{\frac{8}{5}}d^3_{2,-1}-d^2_{2,-1}\big]}$ & $\textcolor{brick}{\frac{1}{2}\big[\frac{9\sqrt{2}}{28}d^4_{2,-1}-\frac{3}{4}\sqrt{\frac{2}{5}}d^3_{2,-1}+\frac{2}{7}d^2_{2,-1}\big]}$  \\[5pt] 
\hline              
\end{tabular}  
\caption{$C_{2,-1}^{n,J}$ }
\end{table}


\begin{table}[h]
\centering
\begin{tabular}{|c|}
\hline
 $n=4$   \\  
\whline
 $\frac{\hat{t}^{\frac{3}{2}}\hat{u}^{\frac{1}{2}}}{s^2}\ (\hat{u}+1)(\hat{u}+2)(\hat{u}+3)$  \\[5pt]  
\hline
$(\frac{1-\cos\theta}{2})\frac{\sin\theta}{2}\ \big[-8\cos^3 \theta+ 2\cos \theta\big]$   \\[4pt]
\hline $\textcolor{brick}{-\frac{16}{15\sqrt{7}}d^5_{2,-1}+\frac{12\sqrt{2}}{35}d^4_{2,-1}-\frac{2\sqrt{2}}{3\sqrt{5}}d^3_{2,-1}+\frac{5}{21} d^2_{2,-1}} $  \\[5pt]
\hline              
\end{tabular}  
\caption{$C_{2,-1}^{n,J}$ }
\end{table}

\begin{table}[h]
\centering
\begin{tabular}{|c|}
\hline
$n=5$\\
\whline
$\frac{\hat{t}^{\frac{3}{2}}\hat{u}^{\frac{1}{2}}}{s^2}\ (\hat{u}+1)(\hat{u}+2)(\hat{u}+3)(\hat{u}+4)$ \\[5pt]  
\hline
$(\frac{1-\cos\theta}{2})\frac{\sin\theta}{2}\ \big[\frac{1}{16}(625\cos^4 \theta-250\cos^2 \theta+9)\big]$   \\[4pt]
\hline
$\textcolor{brick}{\sqrt{\frac{2}{5}}\frac{625}{396}d^6_{2,-1}-\frac{125}{72}\frac{2}{\sqrt{7}}d^5_{2,-1}+\frac{625}{616}\sqrt{2}d^4_{2,-1}-\sqrt{\frac{2}{5}}\frac{125}{72}d^3_{2,-1}+\frac{596}{63}d^2_{2,-1}}$ \\[5pt] 
\hline              
\end{tabular} 
\caption{$C_{2,-1}^{n,J}$ } 
\end{table}

\begin{table}[h]
\centering
\begin{tabular}{|c||c|c|c|c|c|c|c|c|}
\hline
$ $ & $J=2$ & $3$ & $4$ & $5$ & $6$ \\[2pt]  
\whline
$n=1$ & $\frac{1}{2}$ & $$ & $$ & $$ & $$ \\[2pt]
\hline
$ 2$ & $\frac{1}{6} $ & $\frac{-\sqrt{8}}{6\sqrt{5}}$ & $$ & $$ & $$ \\[2pt]
\hline
$ 3$ & $\frac{1}{7} $ & $\frac{-3\sqrt{2}}{8\sqrt{5}}$ & $\frac{9\sqrt{2}}{56}$ & $$ & $$ \\[2pt]
\hline
$4 $ & $\frac{5}{21}$ & $\frac{-2\sqrt{2}}{3\sqrt{5}}$ & $\frac{12\sqrt{2}}{35}$ & $\frac{-16}{15\sqrt{7}}$ & $$ \\[2pt]
\hline
$5 $ & $\frac{596}{63}$ & $\frac{-125\sqrt{2}}{72\sqrt{5}}$ & $\frac{625}{616}\sqrt{2}$ & $\frac{-125}{72}\frac{2}{\sqrt{7}}$ & $\frac{625\sqrt{2}}{396\sqrt{5}}$ \\[2pt]
\hline
\end{tabular}  
\caption{Collection of $C_{2,-1}^{n,J}$'s \label{c3n} }
\end{table}

\begin{table}[h]
\centering
\begin{tabular}{|c|c|c|}
\hline
$n=1$  & $n=2$  & $n=3$  \\  
\whline
 $\frac{\hat{u}^{\frac{1}{2}}}{\hat{s}^{\frac{1}{2}}}$ & $\frac{\hat{u}^{\frac{1}{2}}}{\hat{s}^{\frac{1}{2}}}\ (\hat{u}+1)$ &      $\frac{\hat{u}^{\frac{1}{2}}}{\hat{s}^{\frac{1}{2}}}\ (\hat{u}+1)(\hat{u}+2)$   \\[7pt]  
\hline
$\cos\frac{\theta}{2}$  & $\cos\frac{\theta}{2}\ [-\cos \theta]$  & $\cos\frac{\theta}{2}\ \big[\frac{1}{4}(9\cos^2 \theta-1)\big]$     \\[4pt] 
\hline
$\textcolor{brick}{d^{1/2}_{1/2,1/2}}$ & $\textcolor{brick}{-\frac{1}{3}\big[2d^{3/2}_{1/2,1/2}+d^{1/2}_{1/2,1/2}\big]}$ &    $\textcolor{brick}{\frac{9}{10}d^{5/2}_{1/2,1/2}+\frac{3}{5}d^{3/2}_{1/2,1/2}+\frac{1}{2}d^{1/2}_{1/2,1/2}}$   \\[5pt] 
\hline      
\end{tabular} 
\caption{$C_{\frac{1}{2},\frac{1}{2}}^{n,J}$} 
\end{table}

\begin{table}[h]
\centering
\begin{tabular}{|c|}
\hline
$n=4$  \\  
\whline
      $\frac{\hat{u}^{\frac{1}{2}}}{\hat{s}^{\frac{1}{2}}}\ (\hat{u}+1)(\hat{u}+2)(\hat{u}+3)$   \\[7pt]  
\hline
$\cos\frac{\theta}{2}\ [-8\cos^3 \theta+ 2\cos \theta]$      \\[4pt] 
\hline $\textcolor{brick}{-\frac{2}{35}\big[32d^{7/2}_{1/2,1/2}+24d^{5/2}_{1/2,1/2}+\frac{98}{3}d^{3/2}_{1/2,1/2}+\frac{49}{3}d^{1/2}_{1/2,1/2}\big]}$  \\[5pt] 
\hline      
\end{tabular}  
\caption{$C_{\frac{1}{2},\frac{1}{2}}^{n,J}$}
\end{table}

\begin{table}[h]
\centering
\begin{tabular}{|c|}
\hline
$n=5$\\  
\whline
$\frac{\hat{u}^{\frac{1}{2}}}{\hat{s}^{\frac{1}{2}}}\ (\hat{u}+1)(\hat{u}+2)(\hat{u}+3)(\hat{u}+4)$ \\[7pt]
\hline
$\cos\frac{\theta}{2}\ \big[\frac{1}{16}(625\cos^4 \theta-250\cos^2 \theta+9)\big]$    \\[4pt]
\hline
$\textcolor{brick}{\frac{625}{126}d^{9/2}_{1/2,1/2}+\frac{250}{63}d^{7/2}_{1/2,1/2}+\frac{450}{63}d^{5/2}_{1/2,1/2}+\frac{100}{21}d^{3/2}_{1/2,1/2}+\frac{19}{6}d^{1/2}_{1/2,1/2}}$
\\[5pt]
\hline      
\end{tabular} 
\caption{$C_{\frac{1}{2},\frac{1}{2}}^{n,J}$} 
\end{table}

\begin{table}[h]
\centering
\begin{tabular}{|c||c|c|c|c|c|c|c|c|}
\hline
$ $ & $J=\frac{1}{2}$ & $\frac{3}{2}$ & $\frac{5}{2}$ & $\frac{7}{2}$ & $\frac{9}{2}$ \\[2pt]  
\whline
$n=1$ & $1$ & $$ & $$ & $$ & $$ \\[2pt]
\hline
$ 2$ & $\frac{-1}{3} $ & $\frac{-2}{3}$ & $$ & $$ & $$ \\[2pt]
\hline
$ 3$ & $\frac{1}{2} $ & $\frac{3}{5}$ & $\frac{9}{10}$ & $$ & $$ \\[2pt]
\hline
$4 $ & $\frac{-14}{15}$ & $\frac{-28}{15}$ & $\frac{-48}{35}$ & $\frac{-64}{35}$ & $$ \\[2pt]
\hline
$5 $ & $\frac{19}{6}$ & $\frac{100}{21}$ & $\frac{450}{63}$ & $\frac{250}{63}$ & $\frac{625}{126}$ \\[2pt]
\hline
\end{tabular}  
\caption{Collection of $C_{\frac{1}{2},\frac{1}{2}}^{n,J}$'s \label{c4n} }
\end{table}

\begin{table}[h]
\centering
\begin{tabular}{|c|c|c|}
\hline
$n=1$  &$n=2$  & $n=3$  \\  
\whline
 $\frac{\hat{t}^{\frac{3}{2}}}{\hat{s}^{\frac{3}{2}}}$ & $\frac{\hat{t}^{\frac{3}{2}}}{\hat{s}^{\frac{3}{2}}}\ (\hat{u}+1)$ &      $\frac{\hat{t}^{\frac{3}{2}}}{\hat{s}^{\frac{3}{2}}}\ (\hat{u}+1)(\hat{u}+2)$   \\[7pt]  
\hline   
$(\frac{1-\cos\theta}{2})\sin\frac{\theta}{2}$  & $(\frac{1-\cos\theta}{2})\sin\frac{\theta}{2}\ [-\cos\theta]$  & $(\frac{1-\cos\theta}{2})\sin\frac{\theta}{2}\ \big[\frac{1}{4}(9\cos^2 \theta-1)\big]$   \\[4pt]
\hline
$\textcolor{brick}{d^{3/2}_{3/2,-3/2}}$ & $\textcolor{brick}{-\frac{1}{5}\big[2d^{5/2}_{3/2,-3/2}-3d^{3/2}_{3/2,-3/2}\big]}$ &    $\textcolor{brick}{\frac{3}{7}d^{7/2}_{3/2,-3/2}-\frac{27}{35}d^{5/2}_{3/2,-3/2}+\frac{4}{5}d^{3/2}_{3/2,-3/2}}$   \\[5pt] 
\hline
\end{tabular} 
\caption{$C_{\frac{3}{2},-\frac{3}{2}}^{n,J}$} 
\end{table}

\begin{table}[h]
\centering
\begin{tabular}{|c|}
\hline
  $n=4$   \\  
\whline
       $\frac{\hat{t}^{\frac{3}{2}}}{\hat{s}^{\frac{3}{2}}}\ (\hat{u}+1)(\hat{u}+2)(\hat{u}+3)$   \\[7pt] 
\hline
$(\frac{1-\cos\theta}{2})\sin\frac{\theta}{2}\ [-8\cos^3 \theta+ 2\cos \theta]$       \\[4pt]
\hline $\textcolor{brick}{-\frac{16}{21}d^{9/2}_{3/2,-3/2}+\frac{32}{21}d^{7/2}_{3/2,-3/2}-\frac{204}{105}d^{5/2}_{3/2,-3/2}+\frac{62}{35}d^{3/2}_{3/2,-3/2}}$  \\[5pt] 
\hline
\end{tabular}  
\caption{$C_{\frac{3}{2},-\frac{3}{2}}^{n,J}$}
\end{table}

\begin{table}[h]
\centering
\begin{tabular}{|c|}
\hline
$n=5$\\  
\whline
$\frac{\hat{t}^{\frac{3}{2}}}{\hat{s}^{\frac{3}{2}}}\ (\hat{u}+1)(\hat{u}+2)(\hat{u}+3)(\hat{u}+4)$  \\[7pt]  
\hline
$(\frac{1-\cos\theta}{2})\sin\frac{\theta}{2}\ \big[\frac{1}{16}(625\cos^4 \theta-250\cos^2 \theta+9)\big]$  \\[4pt]
\hline
$\textcolor{brick}{\frac{125}{66}d^{11/2}_{3/2,-3/2}-\frac{625}{154}d^{9/2}_{3/2,-3/2}+\frac{125}{21}d^{7/2}_{3/2,-3/2}-\frac{275}{42}d^{5/2}_{3/2,-3/2}+\frac{233}{42}d^{3/2}_{3/2,-3/2}}$ \\[5pt] 
\hline
\end{tabular} 
\caption{$C_{\frac{3}{2},-\frac{3}{2}}^{n,J}$} 
\end{table}

\clearpage

\begin{table}[h]
\centering
\begin{tabular}{|c||c|c|c|c|c|c|c|c|}
\hline
$ $ & $J=\frac{3}{2}$ & $\frac{5}{2}$ & $\frac{7}{2}$ & $\frac{9}{2}$ & $\frac{11}{2}$ \\[2pt]  
\whline
$n=1$ & $1$ & $$ & $$ & $$ & $$ \\[2pt]
\hline
$ 2$ & $\frac{3}{5} $ & $\frac{-2}{5}$ & $$ & $$ & $$ \\[2pt]
\hline
$ 3$ & $\frac{4}{5} $ & $\frac{-27}{35}$ & $\frac{3}{7}$ & $$ & $$ \\[2pt]
\hline
$4 $ & $\frac{62}{35}$ & $\frac{-204}{105}$ & $\frac{32}{21}$ & $\frac{-16}{21}$ & $$ \\[2pt]
\hline
$5 $ & $\frac{233}{42}$ & $\frac{-275}{42}$ & $\frac{125}{21}$ & $\frac{-625}{154}$ & $\frac{125}{66}$ \\[2pt]
\hline
\end{tabular}  
\caption{Collection of $C_{\frac{3}{2},-\frac{3}{2}}^{n,J}$'s \label{c5n} }
\end{table}

\clearpage

Most of the drawings were created using the program JaxoDraw \cite{binosi}.

\bibliographystyle{plain}

\end{document}